\documentclass[twoside,11pt]{thesis2side}

\usepackage[pdftex]{graphicx,color}
\usepackage[ragged]{sidecap} 
    
\usepackage[T1]{fontenc}  
\usepackage{verbatim} 
\usepackage{rotating} 
\usepackage{float}  
\usepackage{amsmath} 
	\numberwithin{equation}{section}
\usepackage{color} 
\usepackage{amssymb} 
\usepackage{paralist}  
\usepackage{pdfsync}  

\DeclareGraphicsExtensions{.jpg}

\usepackage{import}

\title{Microfabrication techniques for trapped ion quantum information processing}
\author{Joe}{Britton}
\otherdegrees{B.A., The University of Chicago, 1999}
\degree{Doctor of Philosophy}{Ph.D.}
\dept{Department of}{Physics}
\advisor{}{Dr.~David~Wineland}
\reader{Prof.~Frank~Barnes}
\readerThree{Prof.~Eric~Cornell}
\readerFour{Prof.~Emanuel~Knill}
\readerFive{Prof.~John~Price and Prof.~Jun~Ye}
\abstract{\OnePageChapter  Quantum-mechanical principles can be used to process information. In one
approach, linear arrays of trapped, laser cooled ion qubits (two-level quantum
systems) are confined in segmented multi-zone electrode structures. Strong
Coulomb coupling between ions is the basis for quantum gates mediated by phonon
exchange.  Applications of Quantum Information Processing (QIP) include solution
of problems believed to be intractable on classical computers.  The ion trap
approach to QIP requires trapping and control of numerous ions in electrode
structures with many trapping zones.

In support of trapped ion QIP, I investigated microfabrication of structures to
trap, transport and couple large numbers of ions. Using $^{24}$Mg$^+$ I
demonstrated loading and transport between zones in microtraps made of boron
doped silicon.  This thesis describes the fundamentals of ion trapping, the
characteristics of silicon-based traps amenable to QIP work and apparatus to trap
ions and characterize traps. Microfabrication instructions appropriate for
nonexperts are included.  A key characteristic of ion traps is the rate at which
ion motional modes heat. In my traps upper bounds on heating were determined;
however, heating due to externally injected noise could not be
completely ruled out.  Noise on the RF potential responsible for providing
confinement was identified as one source of injected noise.

Using the microfabrication technology developed for ion traps, I made a
cantilevered micromechanical oscillator and with coworkers demonstrated a method
to reduce the kinetic energy of its lowest order mechanical mode via its
capacitive coupling to a driven RF resonant circuit. Cooling results from a RF
capacitive force, which is phase shifted relative to the cantilever motion. The
technique was demonstrated by cooling a 7~kHz fundamental mode from room
temperature to $45$~K. Ground state cooling of the mechanical modes of motion of harmonically
trapped ions is routine; equivalent cooling of a macroscopic harmonic oscillator
has not yet been demonstrated. Extension of this method to devices with higher
motional frequencies in a cryogenic system, could enable ground state cooling and
may prove simpler than related optical experiments.

I also discuss an implementation of the semiclassical quantum Fourier transform
(QFT) using three beryllium ion qubits.  The QFT is a crucial step in a number of
quantum algorithms including Shor's algorithm, a quantum approach to integer
factorization which is exponentially faster than the fastest known classical
factoring algorithm. This demonstration incorporated the key elements of a
scalable ion-trap architecture for QIP.  

}

  \ToCisShort  
  \LoFisShort  
  \LoTisShort  
  
\usepackage[draft]{varioref} 
 
\newcommand{\href}[2]{#2}

\begin{document}

 	\def\textmu{$\mu$} 

 \singlespacing
    \clearpage
\chapter{Ion Trapping and Microfabricated Ion Traps}
\section{Introduction}
\label{ions:sec:intro}

Ion traps are versatile tools found in applications including mass spectroscopy
~\cite{paul1990a}, precision atomic \cite{schmidt2005a} and molecular
spectroscopy \cite{vogelius2004a,schmidt2006b}, quantum information science
\cite{bible,steane1997a,kielpinski2002a,blatt2008a} and tests of fundamental
physics \cite{van-dyck-jr1987a,russell-stutz2004a}.

In this chapter\footnote{In Summer 2008 I coauthored a book chapter titled
Microfabricated Chip Traps for Ions with J.~Amini, D.~Leibfried and
D.~J.~Wineland.  It appears in the book \underline{Atom Chips} edited by D. Jakob
Reichel (MIT) and Prof. Vladan Vuletic (ENS)~\cite{amini2008a}.  This section of
my thesis is a modified version of this chapter and serves as an introduction to
microfabricated ion traps and the ion trapping portion of my thesis work.}, I
discuss the trapping of atomic ions for quantum information
 purposes. This is of current interest because individual ions can be the
 physical representations of qubits for quantum information processing
 \cite{cirac1995a,bible,monroe2008a,blatt2008a}. To support this application I
 investigated microfabricated structures to trap, transport and couple many ions.
 Microfabrication holds the promise of forming large arrays of traps that would
 allow the scaling of current quantum information processing efforts to the level
 needed  to implement useful algorithms \cite{bible,blatt2008a}.
 
 There are two primary types of ion traps: Penning traps and Paul traps. In a
 Penning trap charged particles are trapped by a combination of electrostatic and
 magnetic fields~\cite{dehmelt1990a}.  In a Paul trap, a spatially-varying
 time-dependent electric field, typically in the radio-frequency (RF) domain,
 confines charged particles in space~\cite{paul1990a}. In this review only
  Paul traps will be considered.
  
I will begin this chapter with an introduction to the dynamics of ions confined
in Paul traps based on the pseudopotential approximation.  Subsequent topics
include numeric and analytic models for various Paul trap geometries, a list of
considerations for practical trap design and finally an overview of
microfabricated trapping structures.  The details of microtraps built and tested
for this thesis are discussed in Chapter~\vref{sec:thetrapschapter}.
  
 Some readers may be familiar with neutral atom traps. Neutral atoms are trapped
 by a coupling between external trapping fields and atoms' electric or magnetic
 moments.  As this coupling depends on the atom's internal atomic states, 
 these states can become entangled with the atom's motion (unless special
 precautions are taken). Trap depths of 1~meV or less (several Kelvin) are
 achieved.  In ion traps an ion is confined by a coupling between external
 electric trapping fields and the atom's net charge.  This coupling does not
 depend on the ion's internal electronic state, leaving it unperturbed. Typical
 ion trap depths are 1~eV (many times room temperature).

\section{Radio-frequency ion traps}
\label{ions:sec:ModOfR}

In this section I discuss the equations of motion of a charged
particle in a spatially inhomogeneous RF field based on the pseudopotential model.
Several suitable electrode geometries are discussed.

\subsection{Motion of ions in a spatially inhomogeneous RF field}
\label{ions:subsec:EquMot}

\begin{figure}[htb]
  \centering
  \includegraphics[width=0.8\textwidth]{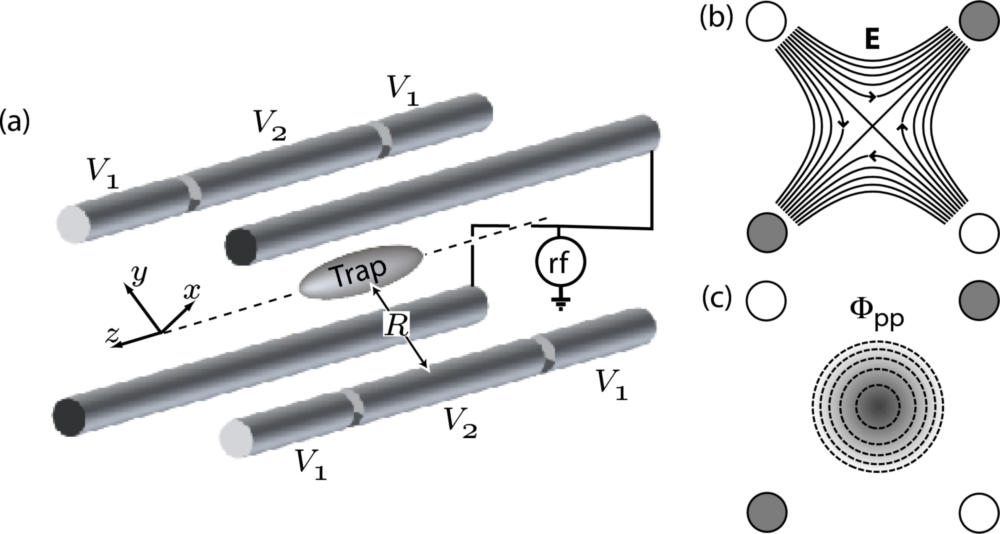}
  \caption
  [Schematic drawing of the electrodes for a linear Paul trap.]
  { Schematic drawing of the electrodes for a linear Paul trap (a). A common
  RF~potential $V_0\cos(\omega_{\rm RF}t)$ is applied to the two
  continuous electrodes as indicated. The other electrodes are held at RF ground
  through capacitors (not shown) connected to ground. In (b) is
  the radial ($x$-$y$) instantaneous electric fields from the applied RF potential. The
  pseudopotential due to this RF-field is shown in (c). 
  A static trapping potential is created  along
  the z-axis by applying a positive potential $V_1>V_2$ (for positive ions) 
  to the outer segments relative to the center segments. (Figure by J. Amini)}
  \label{ions:fig:LinEle}
  \label{ions:fig:2layerTrap}
\end{figure}

Most schemes for quantum information processing with trapped ions are based on
the linear RF trap shown schematically in Figure~\vref{ions:fig:LinEle}. This
trap is essentially a linear quadrupole mass filter with its ends plugged by
static potentials. Radial confinement (the $x$-$y$ plane in
Figure~\vref{ions:fig:LinEle}) is provided by a radio frequency (RF) potential
applied to two of the electrodes (with the other electrodes held at RF ground).
The RF potential alone can not generate full 3D confinement; static potentials
$V_1$ and $V_2$ applied to ``end cap'' control electrodes provide weaker axial
($z$-axis) confinement.

If potential $V_0\cos(\Omega_{\rm RF} t)$ is applied to the RF electrodes with
the others grounded ($V_1=V_2=0$), the potential near the geometric center of the
four rods takes the form
\begin{equation}\label{ions:eq:fullPotential}
  \Phi\approx\frac{1}{2}V_0\cos(\Omega_{\rm RF}
  t)(1+\frac{x^2-y^2}{R^2}),
\end{equation}
where $R$ is a distance scale that is approximately the distance from the trap axis
to the nearest surface of the electrodes
\cite{bible,paul1990a}. The resulting electric field is
shown in Figure~\vref{ions:fig:LinEle}. There is a field null at the trap center and
the field magnitude increases linearly with distance from the center 
in the radial direction~\cite{paul1990a}. 

We can think of the RF electric field as analogous to the electric field from the
trapping laser in an optical dipole trap \cite{grimm2000a}.  For a neutral atom,
the laser's electric field induces a dipole moment, classically a pair of
opposite charges bound elastically together with a resonant eigenfrequency
$\omega_0$ in the optical range. If the electric field is inhomogeneous, the
force on the dipole, averaged over one cycle of the radiation, can give a
trapping force.  For detunings red of the atom's resonant frequency $\omega_0$,
this potential is a minimum at high fields, while for detunings blue of
$\omega_0$ it is a minimum at low fields. An ion, however, is a free particle in
the absence of a trapping field and its eigenfrequency is zero. The RF trapping
potential is therefore analogous to a blue detuned light field and the ion seeks
the position of lowest intensity. In the case of
Equation~\vref{ions:eq:fullPotential}, that corresponds to $x=y=0$.

The equation of motion for an ion placed in this field can be treated in two
ways: as exact solutions of the Mathieu differential equation or as solutions of
a static effective potential called the `pseudopotential'~\cite{ghosh1995a}. The
Mathieu solutions provide insights on trap stability and high frequency motion,
while the pseudopotential approximation is more straightforward and is convenient
for the analysis of trap designs. The Mathieu solutions will be relegated to
Section~\vref{sec:mathieuEquation} and in what follows I will use the
pseudopotential~\cite{ghosh1995a,paul1990a}\footnote{An alternate derivation of the force
on an ion in an inhomogeneous field (aka the pseudopotential) 
is in Section~\vref{sec:rfam:inhomoEfieldForce}.}.  

We define the pseudopotential that governs the secular motion as follows~\cite{dehmelt1967a}. 
The motion of the ion in the RF field is combination of fast `micromotion' at
the RF frequency on top of a slower `secular' motion.
For a particle of charge $q$ and mass $m$ in a
uniform electric field $E=E_0\cos(\Omega_{\rm RF} t)$, the ion motion takes the form
\begin{equation} \label{ions:eq:ionmotion}
  x(t)=-\tilde{x} \cos(\Omega_{\rm RF} t) + v_s t,
\end{equation}
where $\tilde{x}=q E_0/(m \Omega_{\rm RF}^2)$ is the amplitude of what we will call
micromotion and $v_s$ is the velocity of the secular motion.
Along the x direction, if the RF field has a spatial dependence $E_0(x)$, there is a non-zero
net force on the ion when we average over an RF cycle
\begin{equation}
  F_{\rm net} = \left<qE(x)\right> \approx -\frac{1}{2}q\left.\frac{dE_0(x)}{dx}\right|_{x\rightarrow x_s}\tilde{x}
  = -\frac{q^2}{4m\Omega_{\rm RF}^2}\left.\frac{dE_0^2(x)}{dx}\right|_{x\rightarrow x_s}
  = -\frac{d}{dx} (q\Phi_{\rm pp})
\end{equation}
where $x$ was evaluated at position $x_s$ and the pseudopotential 
$\Phi_{\rm pp}$ is defined by
\begin{equation}
  \Phi_{\rm pp}(x_s)\equiv\frac{1}{4}\frac{q E_0^2(x_s)}{m \Omega_{\rm RF}^2}.
\end{equation}
We have made the approximation
that the solution in Equation~\vref{ions:eq:ionmotion} holds over an RF cycle and have
dropped higher order terms in the Taylor expansion of $E_0(x)$ around $x_s$.  For regions
near the center of the trapping potential, these approximations hold.
In three dimensions, we make the substitution $E_0^2 \rightarrow |E|^2=E_{0,x}^2+E_{0,y}^2+E_{0,z}^2$. 
Note that the pseudopotential is dependent on the magnitude of the
electric field, not its direction. 

The pseudopotential has another interpretation in terms of conservation of energy.
From Equation~\vref{ions:eq:ionmotion}, the kinetic energy averaged over a single RF cycle is 
\begin{equation}
  \mathcal{E}_{\rm total} = q\Phi_{\rm pp} + \frac{1}{2}m v_s^2
\end{equation}
where the first term is the average kinetic energy of the micromotion and the second term
is the kinetic energy of the secular motion. The pseudopotential therefore
represents the average kinetic energy of the micromotion.  In the limit where the 
approximations we made are valid, this total energy remains constant: any increase (decrease) in the secular
kinetic energy is accompanied by a corresponding decrease (increase) in the average
micromotion kinetic energy. 

For the quadrupole field given in Equation~\vref{ions:eq:fullPotential}, the
pseudopotential is that of a 2D harmonic potential (see
Figure~\vref{ions:fig:LinEle}c)
\begin{equation}\label{ions:eq:harmpot}
  q\Phi_{\rm pp} = \frac{1}{2}m\omega_r^2(x^2+y^2)
\end{equation}
where $\omega_r \simeq qV_0/(\sqrt{2}m\Omega R^2)$ is the secular frequency. 
As an example, for $^{24}Mg^+$ in a Paul trap with $V_0=50$~V, $\Omega_{\rm
RF}/2\pi=100$~MHz and
$R=50$~{\textmu m}, typical for a microfabricated trap, 
the secular frequency is $\omega_r/2\pi=14$~MHz. 

The effect of the RF pseudopotential is to confine
the ion in the $x$-$y$ plane.   Axial trapping is obtained by the addition of the static
``end cap'' control potentials $V_1$ and $V_2$ in Figure~\vref{ions:fig:LinEle}.  

When the axial potential is weak compared to the overall radial potential,
multiple ions can form a linear crystal along the trap axis due to mutual Coulomb
repulsion. The inter-ion spacing is determined by the axial potential curvature
($\omega_z$). The characteristic length scale of ion-ion spacing is
\begin{equation}\label{ions:eq:ionspacing}
  s=\left(\frac{q^2}{4 \pi \epsilon_0 m \omega_z^2}\right)^{1/3}.
\end{equation}
For a three ion crystal the adjacent separation of the ions is $s_{3}=(5/4)^{1/3}s$
\cite{bible}. For example, $s_3=5.3$~{\textmu m} for $\,^{24}\text{Mg}^{+}$ and
$\left.\omega_{z}\right/2\pi=1.0$~MHz. For multiple ions in a linear Paul trap,
$\omega_z$ is the frequency of the lowest vibrational center 
of mass mode along the trap axis.

A single ion's radial  motion  in the potential given by
Equation~\vref{ions:eq:harmpot} can be decomposed into uncoupled harmonic motion
in the $x$ and $y$ directions, both with the same trap frequency $\omega_r$.
Because the potential is cylindrically symmetric about $z$, we could
choose the decomposition about any two orthogonal directions, called the
principle axes.  We will see in section~\vref{ions:sec:doppler} when discussing
Doppler cooling that we need to break this cylindrical symmetry by the
application of static electric fields. In that case,  choice of the principle
axes becomes fixed and there are now two radial trapping frequencies $\omega_1$
and $\omega_2$, one for each principle axis.

\subsection{Electrode geometries for linear quadrupole traps}
\label{ions:sec:EleGeo}

\begin{figure}[htb]
  \centering
  \includegraphics[width=1\textwidth]{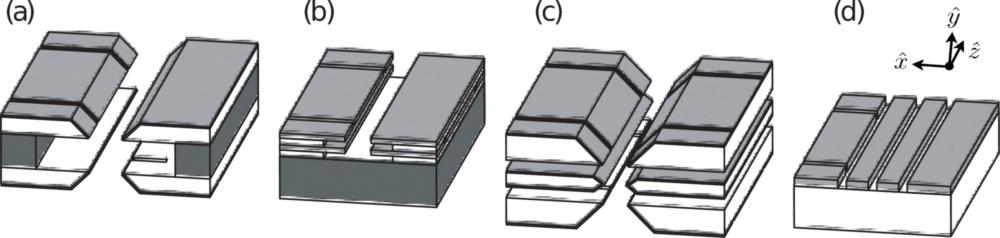}
  \caption [Examples of microfabricated trap structures.] {Examples of
  microfabricated trap structures: (a) two wafers mechanically clamped over a
  spacer, (b) two layers of electrodes fabricated onto a single support wafer,
  (c) three wafers clamped with spacers (spacers not shown) and  (d) surface
  electrode construction. (Figure by J. Amini)}
  \label{ions:fig:examples}
\end{figure}

Designs for miniaturized ion traps conserve the basic features of the Paul trap
shown in Figure~\vref{ions:fig:LinEle}.  Figure~\vref{ions:fig:examples} shows 
several microtrap geometries that have been experimentally realized. All these
geometries generate a radial quadratic potential near the trap axis, though the
extent of deviations from the ideal quadrupole potential away from the axis will
depend on the exact design.

\begin{figure}[htb]
  \centering
  \includegraphics[width=1.0\textwidth]
  {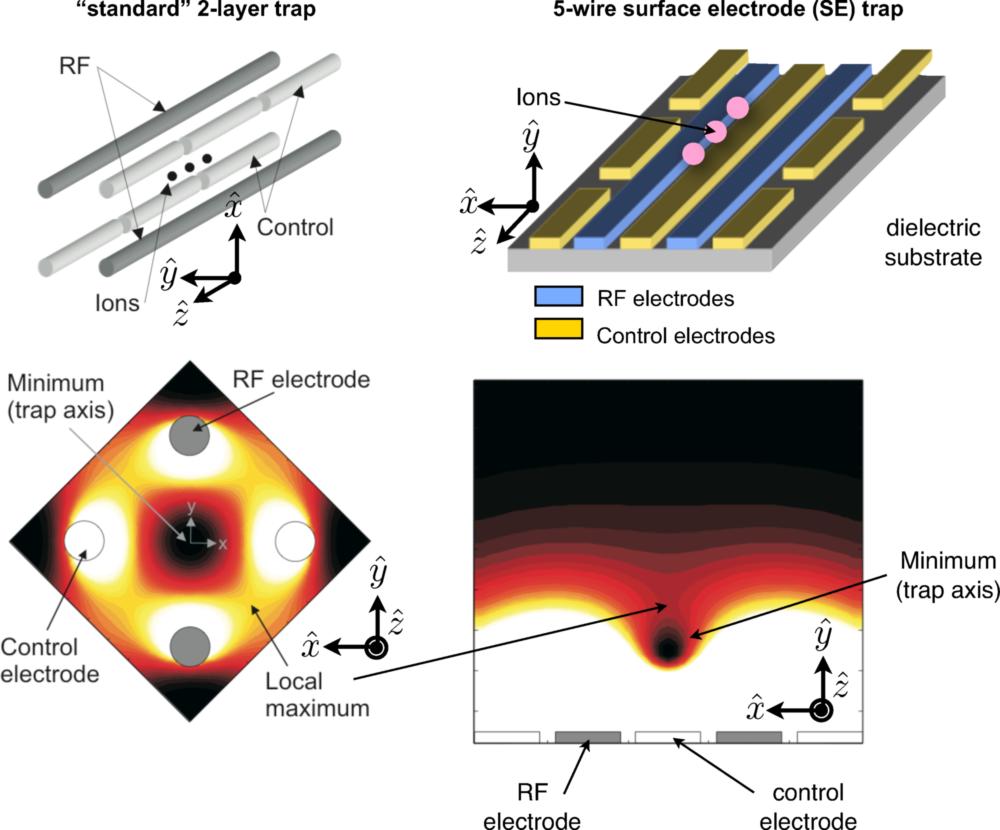}
  \caption[Figure illustrating the trapping potential for a standard 2-layer
  trap and a 5-wire surface electrode trap.]
  {Figure illustrating the trapping potential for a standard 2-layer
  trap and a 5-wire surface electrode trap (SET). In the top row are schematics
  illustrating the trap geometries. Below each is a false color contour
  plot of the radial pseudopotential (small: black, large: white).}
  \label{fig:traptheory:2layerGeometry}
\end{figure}

\label{ions:sec:SETs}
In one particular geometry called a surface electrode trap (SET), the electrodes
are positioned to lie on a single plane with the ion suspended above the plane as
shown in Fig.~\vref{ions:fig:examples}d. Such SETs were investigated fairly
recently~\cite{chiaverini05b} and trapping of atomic ions in a SET was first
demonstrated in 2006~~\cite{seidelin2006a}. Trapping in SETs is possible over a
wide range of geometries albeit with $1/6-1/3$ the motional frequencies and
$1/30-1/200$ the trap depth
of more conventional quadrupolar traps (e.g. Figure~\vref{ions:fig:LinEle}) at comparable RF potentials and
ion-electrode distances~\cite{chiaverini05b}.  A comparison between the trapping
potentials and geometries for 4-wire traps and SETs is in
Figure~\vref{fig:traptheory:2layerGeometry}.  

Advantages of surface electrode traps over two-layer 
traps~\cite{madsen2004a,rowe2002a,stick2006a} 
include easier MEMS fabrication and the possibility
of integrating control electronics on the same trap wafer \cite{kim2005a}.
A SET at cryogenic temperature was implemented
at MIT in 2008~\cite{labaziewicz2008a}.  

Research on SET designs is ongoing and holds promise to yield complex geometries
that would be difficult to realize in non-surface electrode designs.

\section{Design Considerations for Paul traps}
\label{ions:sec:considerations}
In this section, we will discuss the requirements that need to be addressed 
when designing a practical ion trap. 

\subsection{Doppler cooling}
\label{ions:sec:doppler}

Ions are loaded into a trap by ionizing neutral atoms passing near the trap
center and cooling them with a Doppler laser beam.  For Doppler laser cooling,
only a single laser beam is needed; trap strengths far exceed the laser beam
radiation pressure~\cite{metcalf1999a}. However, for motion perpendicular to the
laser beam, cooling is offset by heating due to recoil. Therefore, to cool in all directions,
the laser beam k-vector must have a component along all three principle axes. This also
implies that the trap frequencies are not degenerate, otherwise one principle axis could be 
chosen normal to the laser beam's k-vector. 

\begin{figure}[htb]
  \centering
  \includegraphics[width=0.8\textwidth]{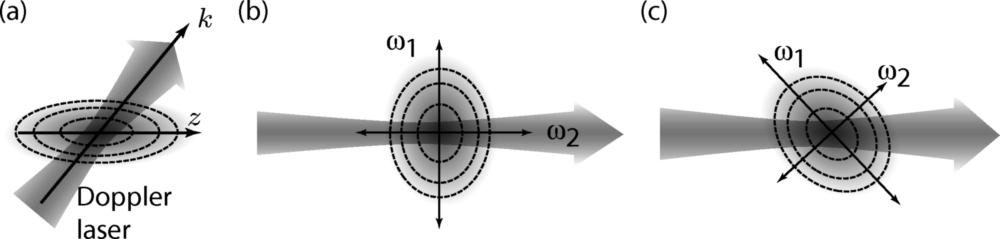}
  \caption[Schematic showing laser beam geometry for 
  Doppler cooling with a single laser beam.]
  {Doppler cooling with a single laser beam. The dashed lines are
  equipotential curves for the
  pseudopotential. The overlap with the axial direction is fairly straightforward
  as in (a), but care has to be taken that the orientation of the two radial modes
  does not place one of the mode axes perpendicular to the laser (b). (c) The radial
  axes must be at an angle with respect to the laser beam k-vector for efficient cooling. (Figure by J. Amini)}
  \label{ions:fig:cooling}
\end{figure}

Meeting the first condition is usually straightforward for non-SET type traps,
where laser beam access is fairly open (see Figure~\vref{ions:fig:cooling}). For
SETs, where laser beams are typically confined to running parallel to the chip
surface,  care has to be taken in designing the trap so that neither radial
principle axis is perpendicular to the trap surface. Alternately, for SETs, we
could bring the Doppler laser beam at an angle to the surface but the beam would
then strike the surface.  This can cause problems with scattered light affecting
fluorescence detection of the ion and with charging of exposed dielectrics (see
Section~\vref{ions:sec:exposedDielectric}).

\begin{SCfigure}[5]
  \centering \includegraphics[width=0.5\textwidth]
  {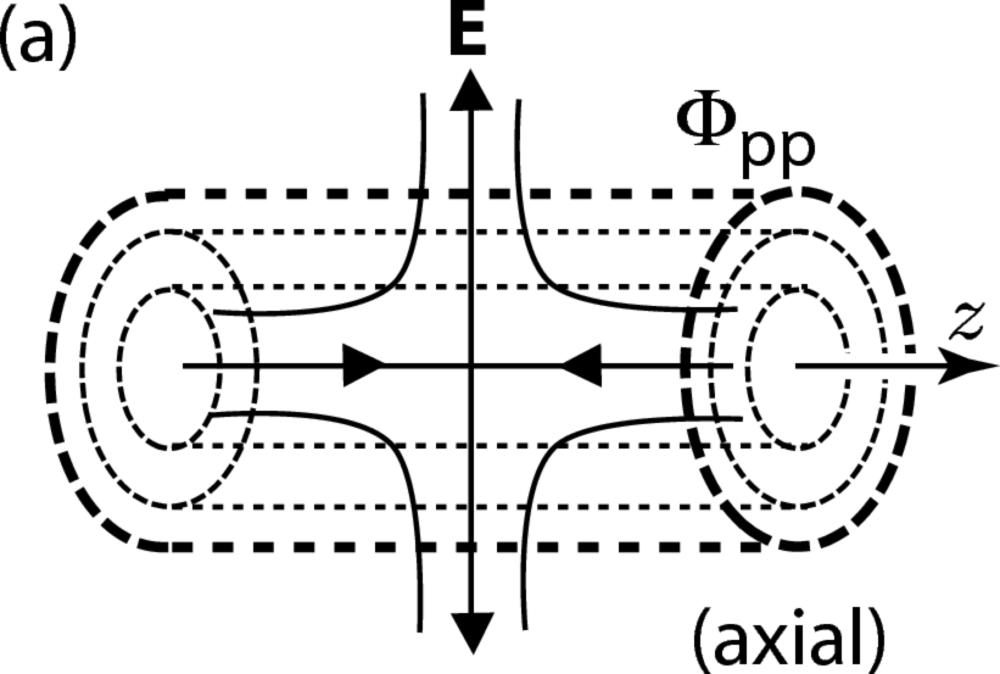} \caption [The degeneracy in the radial
  trap modes is lifted by the radial component of the static axial confinement.]
  {The degeneracy in the radial trap modes is lifted by the radial component of
  the static axial confinement. The RF pseudopotential which provides radial
  confinement is cylindrically symmetric.  A static quadrupole is used to provide
  axial confinement.  This static field is overlaid on radial component; it
  deforms the net potential seen by the ion, breaking the cylindrical symmetry.
  (Figure by J. Amini)} 
  \label{ions:fig:trap_rotation}
\end{SCfigure}

If any two trap frequencies are degenerate, then the trap axes in the plane
containing these modes are not well defined and the motion in a direction
perpendicular to the Doppler laser beam k-vector will not be cooled. The axial
trap frequency can be set independently of the radial frequencies and can be
selected to prevent a degeneracy with either of the radial modes. There are several ways to
break this degeneracy, but usually the axial trapping potential is sufficient. 
When we apply an axial trapping potential, Laplace's equation forces us to have a
radial component to the electric field. In general, this radial field is not
cylindrically symmetric about $z$ and will distort the net trapping potential as
shown in Figure~\vref{ions:fig:trap_rotation}, lifting the degeneracy of the
radial frequencies. If this is not sufficient,  offsetting {\it all} the control
electrodes by a common potential will result in a static field that has the same
spatial dependence (i.e. the same function of $x$ and $y$) as the field generated
by the RF electrodes. This field, shown in Figure~\vref{ions:fig:LinEle}b, will
further split the radial frequencies. Finally, extra control electrodes can be
designed into the trap to generate the necessary static fields to lift the
degeneracy.  Experimental techniques to measure trap secular frequencies are
discussed in Section~\ref{sec:secularFreqMeasurement}.

\subsection{Micromotion}
\label{ions:sec:micromotion}

If the pseudopotential at the equilibrium position of a trapped ion is non-zero,
then the ion motion will include persistent micromotion at frequency $\Omega_{\rm
RF}$.  This section discusses two mechanisms that can generate a non-zero
pseudopotential minimum.

Complex trap structures break the symmetry of the Paul trap in
Figure~\vref{ions:fig:LinEle}. For such traps there can be a component of the RF field
in the axial direction. In this case, the minimum of the pseudopotential well is
non-zero. Since this effect is caused by the design of the trap, we refer to the the resulting
micromotion as \emph{intrinsic} micromotion~\cite{berkeland1998a}.

Secondly, a static electric field can shift the equilibrium position of an ion
away from the pseudopotential minimum.  This may be due to 'stray' electric
fields as from charging of a dielectric surface near the trap center. Shim
potentials applied to the control electrodes can often null these fields. We call
this sort of nullable micromotion `excess' micromotion.

\begin{SCfigure}[50][htb]
  \centering
  \includegraphics[width=0.5\textwidth]{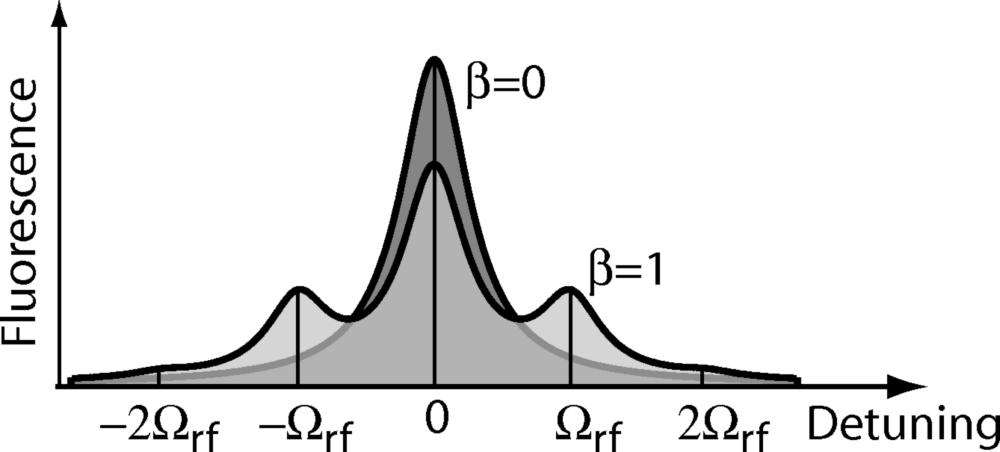}
  \caption
  [In the rest frame of the ion, micromotion induces sidebands of a probing
  laser.]
  {In the rest frame of the ion, micromotion induces sidebands on a probing
  laser.  The monochromatic laser spectrum has been convolved with the atomic
  linewidth. (Figure by J. Amini)}
  \label{ions:fig:sidebands}
\end{SCfigure}

Both intrinsic and excess micromotion can cause problems with the laser-ion
interactions such as Doppler cooling, ion fluorescence, and Raman transitions
\cite{bible,berkeland1998a}. An ion with micromotion experiences a frequency
modulated laser field due to the Doppler shift. In the rest frame of the ion,
this modulation introduces sidebands to the laser frequency (as seen by the ion)
at multiples of $\Omega_{\rm RF}$ and reduces its intensity at the carrier
frequency as shown in Figure~\vref{ions:fig:sidebands}.  The strength of these
sidebands is parametrized by the modulation index $\beta$ given by
\begin{equation}
  \beta = \frac{2\pi}{\lambda} x_{\mu\text{m}} \cos \theta,
\end{equation}
where $\lambda$ is the laser wavelength and $\theta$ is the angle the laser 
makes with micromotion.  For laser beams tuned near resonance, ion fluorescence becomes 
weaker and can disappear entirely. For example, when $\beta=1.43$, 
the carrier and first micromotion sideband have
equal strength. For $\beta<1$, the fractional loss of on-resonance fluorescence
is approximately $\beta^2/2$. As a rule of thumb, we aim for $\beta<0.25$ which
corresponds to less than five percent drop in on-resonant fluorescence.

For a given static electric field $E_{\rm dc}$ along the direction of the RF
electric field, an ion's displacement $x_d$ from trap center and the resulting
excess micromotion amplitude $x_{\mu\text{m}}$ are
\begin{equation}
  x_d=\frac{q E_{\rm dc}}{m \omega_r^2},\\
  x_{\mu\text{m}} = \sqrt{2}\frac{w_r}{\Omega_{\rm RF}}x_d,
\end{equation}
where $\omega_r$ is the radial secular frequency. 

Assuming $^{24}Mg^+$, $\Omega_{\rm RF}/2\pi=100$~MHz and $\omega_r/2\pi=10$~MHz,
a typical SET with $R\sim 50$~{\textmu m} and an excess potential of 1~V on a
control electrode will produce a radial electric field at the ion of
$\sim500$~V/m.  The resulting displacement is $x_d=500$~nm and the corresponding
micromotion amplitude is $x_{\mu\text{m}}=70$~nm.  This results in a laser
modulation index of $\beta=1.14$.

Stray electric fields can be nulled if the control electrode geometry permits
application of independent compensation fields along each radial principle axis.
For the Paul trap in Figure~\vref{ions:fig:LinEle}, a common potential applied to
the control electrodes can only generate a field at the trap center that is along
the diagonal connecting the electrodes. Compensation for other directions is possible for example by
applying a static offset to one of the time-varying potentials on an RF electrode or by
adding extra compensation electrodes.

There are several experimental approaches to detecting and minimizing excess
micromotion~\cite{berkeland1998a}. One technique uses the dependence
of the fluorescence from a cooling laser beam on the micromotion modulation
index. The micromotion can be minimized by maximizing the fluorescence when the
laser is near resonance and minimizing the fluorescence when tuned to the RF the sidebands.
For more on micromotion detection see Section~\vref{sec:micromotion:detection}.

Intrinsic micromotion can also be caused by an RF phase difference $\phi_{\rm
RF}$ between the two RF electrodes. A phase difference can arise due to a path
length difference or differential capacitive coupling to ground for the wires
supplying the electrodes with RF potential. We aim for $\beta<0.25$ (see
Section~\vref{ions:sec:micromotion}) for typical parameters, which requires
$\phi_{\rm RF}<0.5^\circ$.  This sort of micromotion may have been problematic in
a microtrap I built (see Section~\vref{sec:dv16k}).

For more on micromotion see Section~\vref{sec:micromotion}.


\subsection{Exposed dielectric}
\label{ions:sec:exposedDielectric}

Exposed dielectric surfaces near the trapping region can accumulate stray charge
and the resulting stray electric fields can cause problems
(see~Section\vref{ions:sec:micromotion}). Stray charge can be produced by
photo-emission from the cooling laser or from electron sources sometimes used to
ionize atoms by electron impact (see Section~\vref{sec:eGuns}). Depending on the
resistivity of the dielectric these charges can remain on the surfaces for
seconds or longer, requiring time-dependent micromotion nulling or waiting a
sufficient time for the charge to dissipate.

SETs are particularly prone to this problem. The metallic trapping electrodes are
often supported by an insulating substrate and the spaces between the electrodes
expose the substrate.  The effect of charging of these regions can be mitigated
by increasing the ratio of electrode conductor thickness to the inter-electrode
spacing.

\begin{SCfigure}[10][htb]
  \centering
  \includegraphics[width=0.6\textwidth]
  {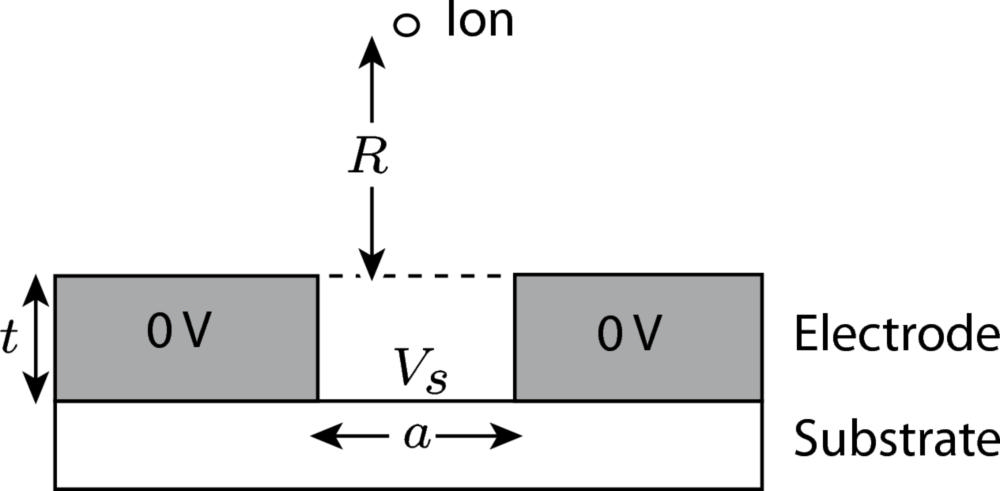}
  \caption
  [Model used for calculating the effect of stray charging.]
  {Model used for calculating the effect of stray charging. We assume
  that $R\ll a$ and $t\geq a/\pi$. (Figure by J. Amini)}
  \label{ions:fig:straycharge}
\end{SCfigure}

Figure~\vref{ions:fig:straycharge} illustrates a model for estimating how thick
electrodes can suppress the field from a strip of exposed substrate charged to a
potential $V_s$. The sidewalls are assumed conducting and grounded.  Along the
midpoint of the trench the potential drops exponentially with height
\cite{jackson1999a}. Using this solution to relate $V_s$ to the potential at
the top of the trench, and then the techniques described in
section~\vref{ions:sec:analyticsolutions} to relate the surface potential to a
field at the ions, we obtain an approximate expression for the field seen by the ion
\begin{equation}
E_{vert} = (4V_s/\pi^2) (a/R^2) e^{-\pi t/a},
\end{equation}
where $a$ is the width of the exposed strip of substrate, $t$ is the
electrode thickness, and $R$ is the distance from the trap surface to the ion. 
We have assumed $R\gg a$ and $\pi t\geq a$.  Thus, the effect
of the stray charges drops off rapidly with the ratio of
electrode thickness to gap spacing. 
  
\subsection{Loading ions}
\label{ions:sec:loading}

Ions are loaded into traps by ionizing neutral atoms as they pass through the
trapping region. The neutral atoms are usually
supplied by a heated oven, but can also come from background vapor in the vacuum
or laser ablation of a sample.  See Section~\vref{sec:ionizationTechniques} for
more on ionization techniques and Section~\vref{sec:MgOvens} for more
on ovens.

It is necessary that the neutral atom flux reach the trapping region but not deposit
on insulating spacers which might short adjacent trap electrodes.  In practice,
we do this by careful shielding and, in some SETs, undercutting of electrodes to
form a shadow mask (see Figure~\vref{ions:fig:signefab}). Alternately, for SETs, a
hole machined through the substrate can direct neutral flux from an oven on the
back side of the wafer to a small region of the trap, preventing coating of the
surface. This is called backside loading and has been demonstrated in several
traps (e.g. dv16m in Section~\vref{sec:dv16m}).

\subsection{Electrical connections}
\label{ions:sec:elecInterconnect}

Radio frequency trapping potentials and DC control potentials are delivered to
the trap electrodes by wiring that includes traces on the trap substrate.  Care
is needed to avoid several pitfalls.

The high voltage RF potential is typically by a quarter-wave RF
resonator~\cite{jefferts1995a} or LC lumped-element resonant circuit.  RF losses
in  a microtrap's insulating substrate can degrade the resonator quality factor
($Q_L$) and can cause Ohmic heating of the microtrap itself.  This can be be
mitigated by use of low loss insulators (e.g. quartz or alumina) and decreasing
the capacitive coupling of the RF electrodes to ground through the insulators. 
Typical RF parameters for our traps are $\Omega_{\rm RF}/{2\pi}=100$~MHz, $V_{\rm
RF}=100~V$ and $Q_L=200$.  See also Section~\vref{sec:apparatus:trapQL}.

\begin{figure}[htb]
  \centering
  \includegraphics[width=0.8\textwidth]{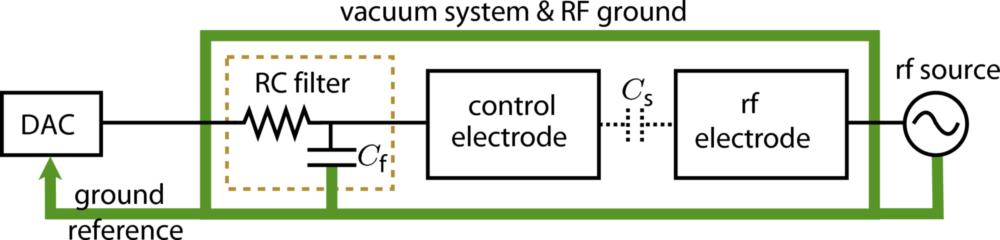}
  \caption
  [Figure showing proper filtering and grounding of a trap control
  electrode.]
  {Figure showing typical filtering and grounding of a trap control
  electrode. Inside the vacuum system are low pass RC filters which reduce noise
  from the control potential source and provide low impedance
  shorts to ground for the RF coupled to the control electrodes by stray 
  capacitances $C_{\rm s}<<C_f$. The RC filters typically lie inside the vacuum
  system, within 2~cm of the trap electrode. The control potential 
  (assumed in the figure to be derived from a DAC)
  is referenced to the trap RF ground and is supplied over a properly shielded wire.}
\label{ions:fig:properRC}
\end{figure}

The RF electrodes have a small capacitive coupling to each control electrode
(typically $<0.1~pF$) which result in unwanted RF fields at the ion. This RF
potential is shunted to ground by a low pass RC filter (typically $R=1~k\Omega$
and $C=1~nF$) on each control electrode (Figure~\vref{ions:fig:properRC}). The
impedance of the lines leading to the RC filters should be low or the filtering
will be compromised. Proper grounding, shielding and filtering of the electronics
supplying the control potentials is also important to suppress pickup and ground
loops (which can cause motional heating, Section~\vref{ions:sec:motional}).

\section{Multiple trapping zones}
\label{ions:sec:multiplezones}

The emphasis in the recent generation of traps aim to store ions in multiple
trapping zones and can transport ions between them.

\begin{figure}[htb]
  \centering
  \includegraphics[width=0.9\textwidth]{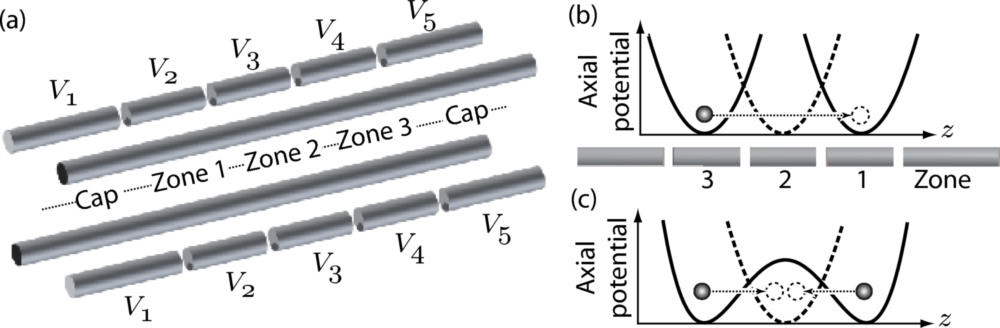}
  \caption
  [Ion transport in a multizone trap.]
  {Example of a multizone trap (a). By applying appropriate waveforms
  to the segmented control electrodes,
  ions can be shuttled from zone to zone (b) or pairs of ions
  can be merged into a single zone or split into separate zones (c). 
  (Figure by J. Amini)} 
  \label{ions:fig:multizone}
\end{figure}

We can modify the basic Paul trap in Figure~\vref{ions:fig:LinEle} to support multiple
zones and ion transport by dividing the control electrodes into a series of
segments as shown in Figure~\vref{ions:fig:multizone}a. By applying appropriate
potentials to these segments, an axial harmonic well can be moved along the length
of the trap carrying ions along with it (Figure~\vref{ions:fig:multizone}b). In
the adiabatic limit (with respect to $\omega_z^{-1}$), ions have been transported
a distance of 1.2~mm in $50~\mu$s with undetectable heating and undetectable
internal-state decoherence~\cite{rowe2002a}.

For quantum information processing, for example, we need to be able to take pairs
of ions in a single zone (e.g. zone~2)  and separate them into independent zones
(e.g. one ion in zone~1 and a second in zone~3) with minimal heating, as shown in
Figure~\vref{ions:fig:multizone}c. We must likewise be able to combine separated
ions with low heating.  Separating and combining are more difficult tasks than
ion transport; the theory is discussed by Home~\cite{home2006a} and there are
several demonstrations~\cite{rowe2002a,barrett2004a}. The basis for these
potentials is the quadratic and quartic terms of the axial potential. Proper
design of the trap electrodes can increase the strength of the quartic term and
facilitate faster ion separation and merging with less heating. Groups of two and
three ions have been separated while heating the center of mass mode less than
10~quanta and the higher order modes to less than 2~quanta~\cite{barrett2004a}.

The segmented Paul trap in Figure~\vref{ions:fig:multizone} forms a linear series
of trapping zones, but other geometries are possible.  Of particular interest are
junctions with linear trapping regions extending from each leg.  Specific
junction geometries are discussed in section~\vref{ions:sec:traps}. The broad
goal is to create large interconnected trapping structures that can store,
transport and reorder ions so that any two ions can be brought together in a
common zone.

\section{Motional heating}
\label{ions:sec:motional}
\label{ions:sec:heating}

Doppler and Raman cooling can place a trapped ion's harmonic motion into the
ground state with >99\% probability~\cite{monroe1995a,king1998a,bible}. If we are
to use the internal electronic states of an ion to store information, we have to
turn off the cooling laser beams during that period.  Unfortunately, the ions do
not remain in the ground state and this heating can reduce the fidelity of
operations performed with the ions. One source of heating comes from laser
interactions used to manipulate the electronic states~\cite{ozeri2007a}.  Another
source is ambient electric fields that have a frequency component at a secular
frequency. We expect such fields from the Johnson noise on the electrodes, but
the heating rates observed experimentally are typically several orders of
magnitude larger than the Johnson noise can account
for~\cite{turchette2000a,deslauriers2006a,bible}. Currently, this source of
anomalous heating is not explained, but recent
experiments~\cite{deslauriers2006a,labaziewicz2008a} indicate that it is
thermally activated and consistent with patches of fluctuating potentials with
size smaller than the ion-electrode spacing~\cite{turchette2000a}.

\begin{figure}[htb]
  	\centering
   \includegraphics[width=0.9\textwidth]{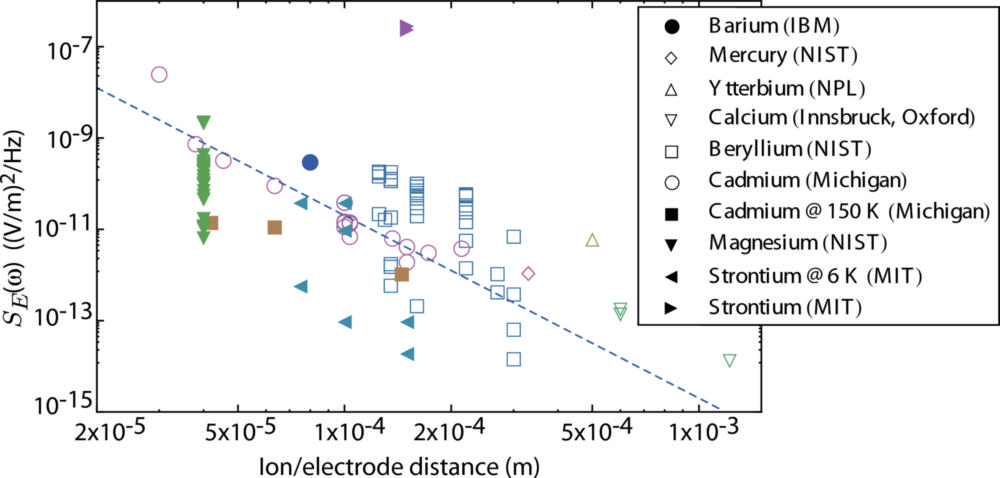}
   \caption
   [Spectral density of electric-field fluctuations inferred from
  observed ion motional heating rates.]
   {Spectral density of electric-field fluctuations inferred from
  observed ion motional heating rates. Data points show heating measurements in
  ion traps observed in different ion species by several research groups
  \cite{diedrich1989a, monroe1995a,roos1999a,tamm2000a,turchette2000a,devoe2002a,
  rowe2002a,home2006b,stick2006a,
  deslauriers2006a,epstein2007b,labaziewicz2008a,britton2008a,blakestad2008a}.  
  The several
  $Be^+$ measurements between 100~$\mu$m and 400~$\mu$m are for identical
  traps measured at different times \cite{turchette2000a}.  Unless
  specified, the data was taken with the trap at room temperature.
  The dashed line shows a $R^{-4}$ trend for ion heating vs ion-electrode
  separation. (Figure by J. Amini and J. Britton)}
   \label{ions:fig:ionHeatingScatterPlot}
   \label{fig:ionHeatingScatterPlot}
 \end{figure}

The spectral density of electric field fluctuations $S_E$ inferred from ion
heating measurement in a number of ions traps are plotted in
Figure~\vref{ions:fig:ionHeatingScatterPlot}. The dependence of $S_E$ on the
minimum ion-electrode distance $R$ and on the trap frequency $\omega$ follows a
roughly $R^{-\alpha}\omega^{-\beta}$ scaling, where $\alpha\approx3.5$
\cite{deslauriers2006a,turchette2000a} and $\beta\approx0.8-1.4$
\cite{turchette2000a,seidelin2006a,deslauriers2006a,epstein2007b,labaziewicz2008a}.
In addition to being too small to account for these measured heating rates,
Johnson noise scales as $R^{-2}$ \cite{bible,turchette2000a}.

In the context of ion quantum information processing, microtraps are advantageous
because quantum logic gate speeds and ion packing densities increase as $R$ decreases
\cite{leibfried2003b,bible, kielpinski2002a}. However, these gains are at odds
with the the highly unfavorable dependence of motional heating on ion-electrode
distance. Extrapolating from current best heating results observed at room
temperature, in a trap with $R=10$~{\textmu m} the heating rate would exceed
$1\times10^6$~quanta/sec \cite{seidelin2006a}. Heating between gate operations is
also problematic because hot ions require more time to recool to the motional
ground state \cite{king1998a}.

I discuss in greater detail motional heating in the context of a microtrap I
built in Section~\vref{sec:dv16m:heating}.

\clearpage
\section{Trap modeling}
\label{ions:sec:modeling}

Calculation of trap depth, secular frequencies, and transport and separation
waveforms requires detailed knowledge of the potential and electric fields near
the trap axis. In the pseudopotential approximation, the general time-dependent
problem is simplified to a slowly varying electrostatic one.  For simple 4-rod type traps (as
in Figure~\vref{ions:fig:LinEle}), good trap design is not difficult using numerical
simulation owing to their symmetry. However, SET design is more difficult since
the potential may have large anharmonic terms and highly asymmetric designs are
common. Fortunately, for certain SET geometries analytic solutions exist. These
closed form expressions permit efficient parametric optimization of electrode
geometries not practical by numerical methods.  In this section we will first
discuss the full 3D calculations and then introduce the analytic solutions.

\subsection{Modeling 3D geometries}
\label{ions:sec:modelinggeom}
\label{ions:sec:BEM}

There are several numerical methods for solving the general electrostatic
problem. I used the boundary element method implemented in a commercial software
package called Charged Particle Optics
 (\href{http://www.electronoptics.com/}{CPO, Ltd}).
In contrast to the finite element method,  the solutions from the boundary
element method are in principle differentiable to all orders.  A simulation
consists of calculating the potential due to each control electrode when that
electrode is set to one volt and all others are grounded. The solution for an
arbitrary set of potentials on the control electrodes is then a linear
combination of these one volt solutions. Similarly, the pseudopotential is
obtained from the field calculated for one volt on the RF electrodes and ground
on the control electrodes.

\subsection{Analytic solutions for surface-electrode traps}
\label{ions:sec:analyticsolutions}

\begin{figure}[htb]
  \centering
  \includegraphics[width=0.75\textwidth]{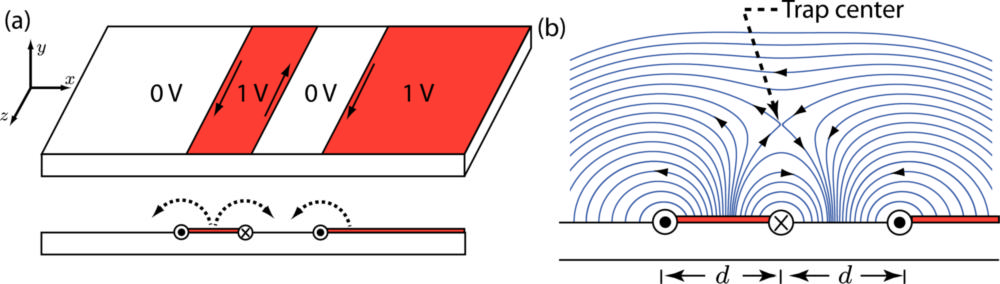}
  \caption [Surface-electrode trap composed of two RF electrodes embedded in a
  ground plane (four-wire trap).] {Surface-electrode trap composed of two RF
  electrodes embedded in a ground plane (four-wire trap) (a). All electrodes
  extend to infinity along the trap ($z$) axis. The field lines derived from an
  analytic potential  are shown in (b). (Figure by J. Amini)}
  \label{ions:fig:lincalc}
\end{figure}

Numerical calculations work for any electrode geometry, but they are are slow and
not well suited to automatic optimization of SET electrode shapes. For the
special case of SETs, an analytic solution exists subject to a few realistic
geometric constraints. Electrodes are modeled as a collection of gapless 
plates embedded in an
infinite ground plane (see Figure~\vref{ions:fig:lincalc}).  The
electric field that would be observed from a plate is proportional to the
magnetic field produced by a current flowing along its
perimeter~\cite{oliveira2001a}. The problem is then reduced from solving
Laplace's equation to integrating a Biot-Savart type integral around the patch
boundary. Furthermore, for patches that have boundaries composed of straight line
segments, the integrals have analytic solutions. 


Of pedagogical value is the application of this technique to a very simple SET
geometry in Reference~\cite{amini2008a}.  The application of this technique to
SETs is in Reference~\cite{wesenberg2008b}.

The main shortcoming of this method is the requirement that there be no gaps
between the electrodes.  Typical SET fabrication techniques produce
$1-5$~{\textmu m} gaps which can only be accounted for at the level important to
ion dynamics by full numerical simulations.

\section{Trap examples}
\label{ions:sec:traps}

Having covered the general principles for Paul trap designs,  I will now give
specific examples of microfabricated ion traps.  A number of fabrication
techniques have been used for micro-traps, starting with stacking multiple wafers
to form a traditional Paul trap type 
design~\cite{rowe2002a,barrett2004a,wineland2005a, britton2008a,blakestad2008a,huber2008a,
hensinger2006a, schulz2008a}.  
Recently, trap fabrication has been
extended to single wafer designs using substrate materials such as Si, GaAs, and
quartz~\cite{seidelin2006a,britton2006b,stick2006a, pearson2006a,
brown2007a,labaziewicz2008a,britton2008a}.  
The fabrication process flows include such microfabrication standards as
photolithography, metallization, and chemical vapor deposition as well as other
less used techniques such as laser machining.  A detailed discussion
of the microfrabrication techniques I used to build microtraps is in 
Section~\vref{sec:fab}.

\begin{SCfigure}[10][htb]
  \centering
  \includegraphics[width=0.5\textwidth]{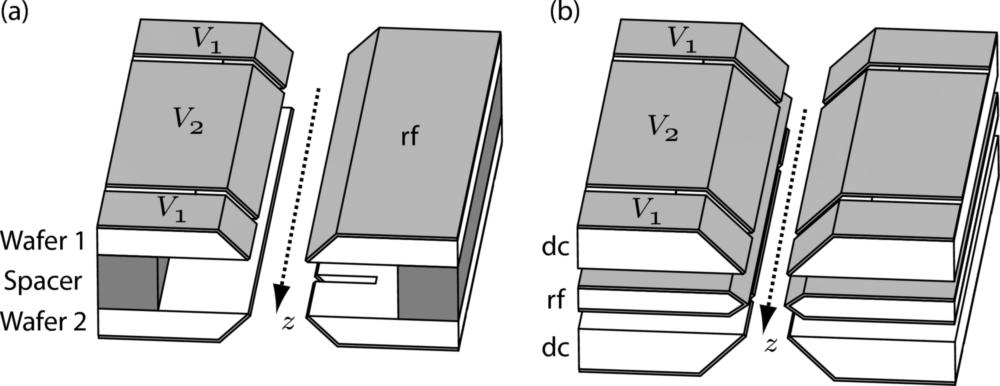}
  \caption
  [Schematics of two and three-layer Paul traps.]
  {Multi-wafer traps mechanically clamp or bond multiple substrates to
  form a classic Paul trap type structure (a) or a modified three-layer structure (b)
  ~\cite{hensinger2006a}  Typically, both the top and bottom layers include
  segmented control electrodes. (Figure by J. Amini)} 
  \label{ions:fig:multiwafer}
\end{SCfigure}

\begin{SCfigure}[10][htb]
  \centering
  \includegraphics[width=0.5\textwidth]{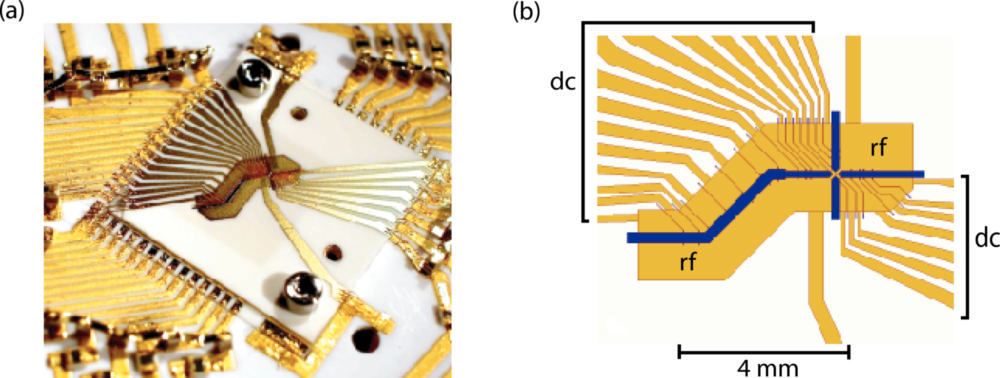}
  \caption [Example of a two wafer trap with an 'X' junction.] {Example of a two
  wafer trap with an 'X' junction. The trap conductors are evaporated and
  electroplated gold films deposited on alumina substrates.  The substrates have
  been laser machined to form the large slots where the ions reside and to
  provide the gaps ($20\mu$m) that separate the control
  electrodes~\cite{blakestad2008a}. (Figure by J. Amini)}
  \label{ions:fig:brad}
\end{SCfigure}

The microfabricated equivalent to the prototypical four-rod Paul trap
(Figure~\vref{ions:fig:LinEle}) takes two insulating substrates patterned with
electrodes and then clamps or bonds the substrates together across an insulating
spacer. This approach has been implemented in a number of traps (see
\cite{rowe2002a,barrett2004a} and Section~\vref{sec:dv10}) using two substrates
as shown in Figure~\vref{ions:fig:multiwafer}a. Most recently at NIST this
approach culminated in a trap with $18$ zones and an 'X'
junction~\cite{blakestad2008a} shown in Figure~\vref{ions:fig:brad}. The
University of Michigan demonstrated a three wafer trap
design like that shown Figure~\vref{ions:fig:multiwafer}b incorporating a 'T'
shaped junction~\cite{hensinger2006a}.

Another approach demonstrated recently used two patterned wafers mounted with the
conducting layers facing each other and without any slots in the
wafers~\cite{debatin2008a}.  The trapping region was then in the gap between the
wafers and laser beam access was in the plane perpendicular to the plane of the
wafers.

The direct analog to the four-rod Paul trap can also be realized in single wafer
microfabrication by depositing two conducting layers separated by an insulting
layer and then etching the electrode profiles
\cite{stick2006a}.

\begin{figure}[htb]
  \centering
  \includegraphics[width=0.8\textwidth]{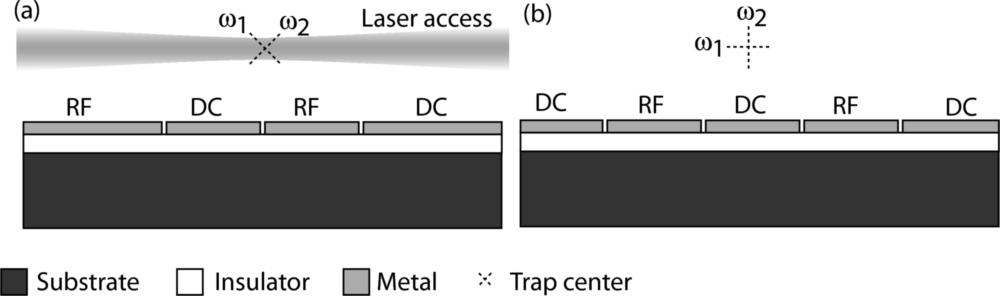}
  \caption [Comparison of trap principle axes for two surface electrode
  geometries.] {Comparison of trap principle axes for two surface electrode
  geometries. (a) four-wire geometry and (b)
  five-wire geometry.  In practice, the five-wire geometry is not used
  because of the difficulty of cooling the vertical motion of the trapped ions.
  (Figure by J. Amini)}
  \label{ions:fig:planar}
\end{figure}

Surface-electrode traps (SETs) have the benefit of using standard
microfabrication methods where layers of metal and insulator are deposited on the
surface of the wafer without the need for milling of the substrate itself. There
are two general versions of the surface trap electrode geometry as shown in
Figure~\vref{ions:fig:planar}.  We looked at the fields from these geometries in
Section~\vref{ions:sec:analyticsolutions}. The four-wire geometry has the trap
axis rotated at 45$^\circ$ to the substrate plane, which allows for efficient
cooling of the ion (Figure~\vref{ions:fig:planar}a).  The five-wire geometry has
one trap axis perpendicular to the surface, which makes that axis difficult to
Doppler cool because the cooling laser beams propagate parallel to the surface 
(Figure~\vref{ions:fig:planar} and Section~\vref{ions:sec:doppler}).

Surface-electrode traps (SETs) are fairly new, and only a few designs have been
demonstrated~\cite{seidelin2006a, britton2006b, labaziewicz2008a}. The first
surface-electrode trap for atomic ions was constructed on a fused quartz
substrate with electroplated gold electrodes~\cite{seidelin2006a}. In addition,
meander line resistors were fabricated on the chip as part of the DC filtering.
Surface mount capacitors gap welded to the chip provide a low impedance shunt to
ground for the RF (see Section~\vref{ions:sec:elecInterconnect}).  The process
sequence is shown in Figure~\vref{ions:fig:signefab} (see Section~\vref{sec:fab}
for more on microfabrication). The bonding pads and the thin meander line
resistors were formed by liftoff of evaporated gold. Charging of the exposed
substrate between the electrodes was a concern, so the trapping structure was
thick 6~\textmu m electroplated gold with 8~\textmu m gaps.

A SET for microscopic aminopolystyrene spheres was also
demonstrated~\cite{pearson2006a}.

\begin{SCfigure}[10][htb]
  \centering
  \includegraphics[width=0.6\textwidth]{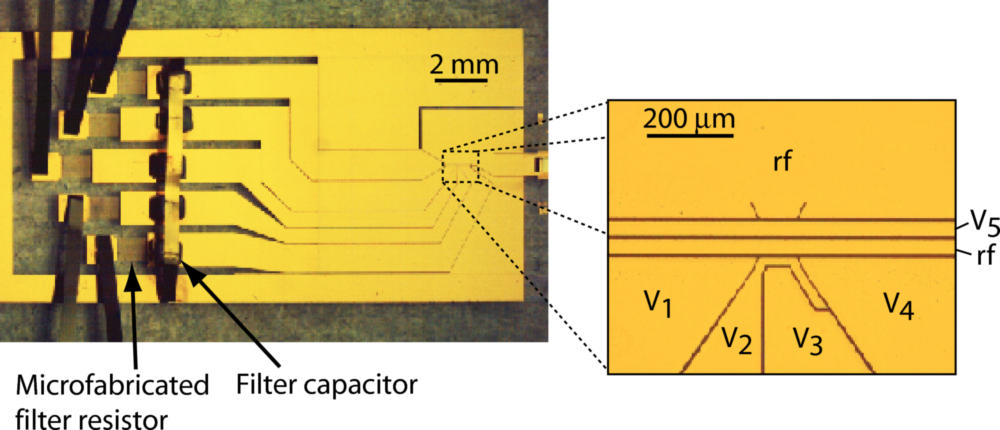}
  \caption [Micrographs of a four wire surface electrode trap constructed of
  electroplated gold on a quartz substrate.] {An example of a four wire surface
  electrode trap constructed of electroplated gold on a quartz substrate. (Figure
  by J. Amini)}
  \label{ions:fig:signe}
\end{SCfigure}

\begin{figure}[htb]
  \centering
  \includegraphics[width=0.9\textwidth]{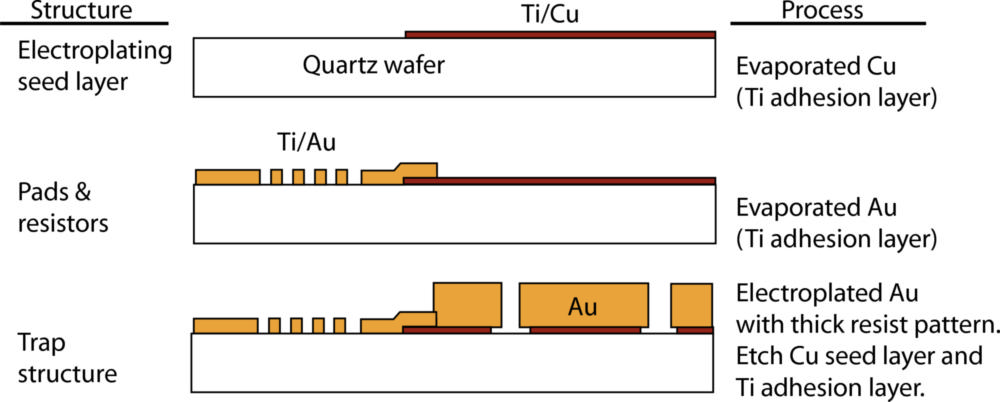}
  \caption [Fabrication steps for the example trap in
  Figure~\vref{ions:fig:signe}.] {Fabrication steps for the example trap in
  Figure~\vref{ions:fig:signe}. The copper seed layer could not extend under the
  meander line resistors because the final step of etching of the seed layer
  would fully undercut the narrow meander pattern. (Figure by J. Amini)}
  \label{ions:fig:signefab}
\end{figure}

A similar design  was built for low temperature testing using 1~\textmu m
evaporated silver on quartz. They reported
a strong dependence of the anomalous heating on temperature (see
Section~\vref{ions:sec:motional})~\cite{labaziewicz2008a}.

\begin{SCfigure}[5][htb]
  \centering \includegraphics[width=.5\textwidth]
  {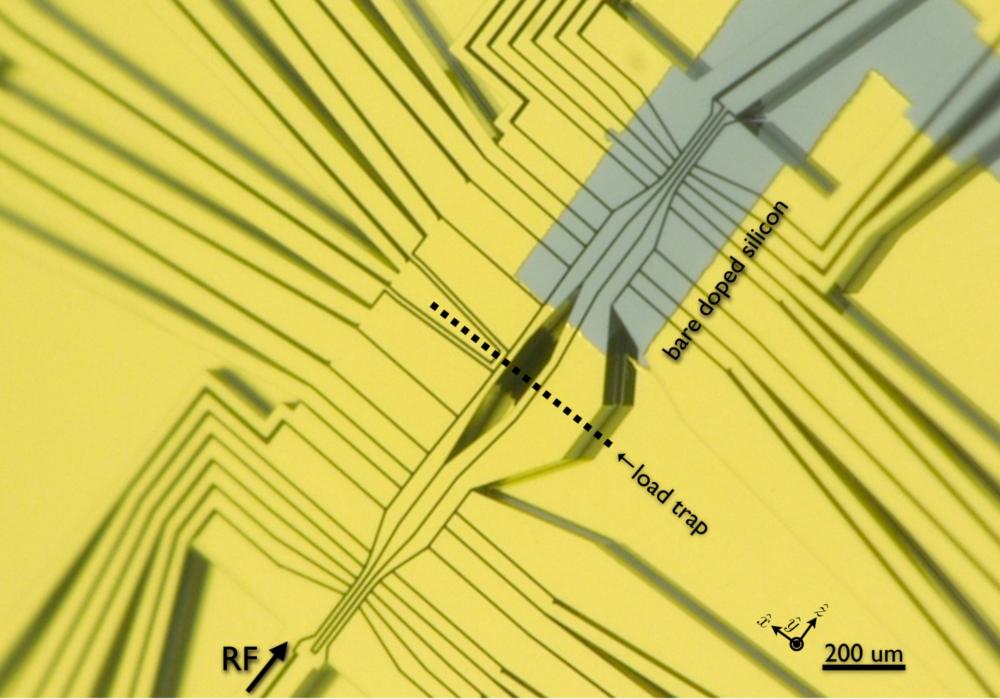} \caption[Optical micrograph of
  dv16m.]{Optical micrograph of a SOI SET ion trap.  The bright gold regions are
  gold coated; the dark regions are bare boron doped silicon. The camera is
  tilted, showing a perspective view of the trap which emphasizes the height of
  the SOI device layer ($100~\mu$m thick). This trap demonstrated multiple
  trapping zones in a SET linear array and backside loading ions.}
  \label{ions:fig:dv16mOpticalPerspectivePhoto}
\end{SCfigure}

The construction of these traps was based on adding conducting layers to an
insulating substrate.  In my thesis work I employed an alternate fabrication
method based on removal of material from a conducting substrate.  In these traps
the conducting electrode surfaces were boron doped silicon; it has a low
electrical resistance.  This silicon was deep etched to form electrically
isolated electrodes.  The electrodes were fixed in place by anodic bonding to a
glass substrate (see Section~\vref{sec:dv10} and \cite{britton2006b}) or because
it was part of a silicon-on-insulator structure (see Section~\vref{sec:dv16m}).
The SOI design depicted in Figure~\vref{ions:fig:dv16mOpticalPerspectivePhoto}
demonstrated multiple trapping zones in a SET linear array and backside loading
of ions.  These traps are the subject of my thesis work and 
are discussed in greater detail in Chapter~\vref{sec:theTraps}.

\begin{SCfigure}[10][htb]
  \centering
  \includegraphics[width=0.65\textwidth]
  {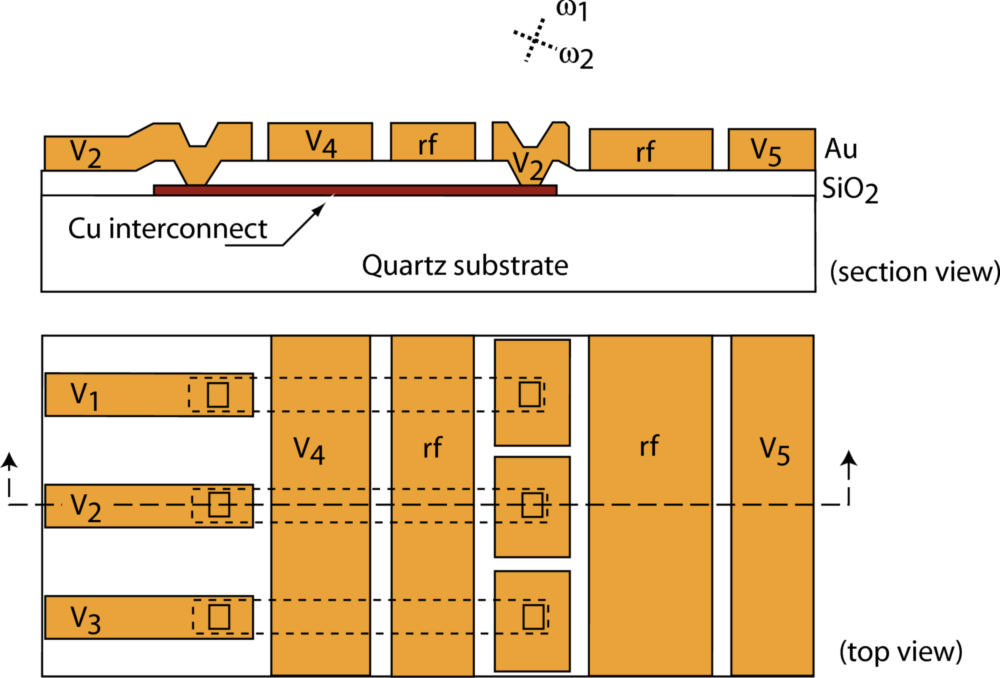}
  \caption
  [Fabrication of a multilayer gold on quartz surface-electrode trap.]
  {Fabrication of a multilayer gold on quartz surface-electrode trap.  A CVD
  oxide insulates the surface electrodes from the second layer of
  interconnects.  Plasma etched holes in the insulated layer connect the two
  conducting layers. (Figure by J. Amini)}
  \label{ions:fig:multilayerfab}
\end{SCfigure}

\begin{SCfigure}[10][htb]
  \centering
  \includegraphics[width=0.6\textwidth]{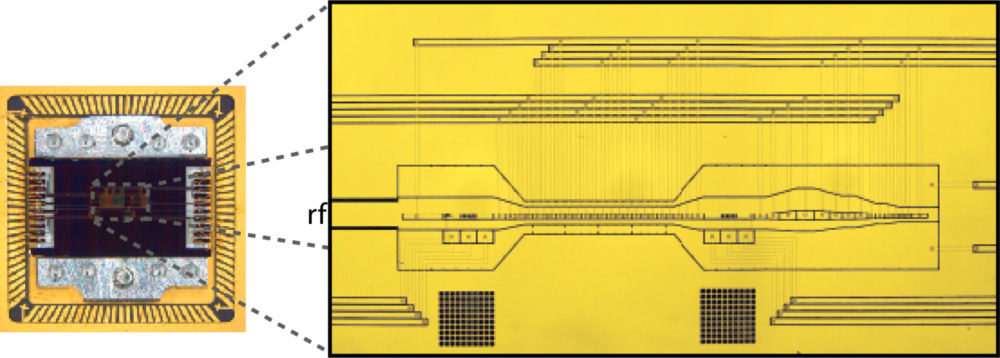}
  \caption
  [Multilayer trap mounted in its carrier and an enlargement of the
  active region.]
  {Multilayer trap mounted in its carrier and an enlargement of the
  active region. (Figure by J. Amini)} 
  \label{ions:fig:multilayer}
\end{SCfigure}

Surface-electrode traps allow for complex arrangements of trapping zones, but making
electrical connections to these electrodes quickly becomes intractable as the
complexity grows.  This problem can be addressed by incorporating multiple
conducting layers into the design with only the field from the top layer
affecting the ion \cite{kim2005a,amini2007b}. An example of such a multilayer trap
was fabricated at NIST on an amorphous quartz substrate is shown in
Figure~\vref{ions:fig:multilayer}. The metal layers were separated by chemical
vapor deposited (CVD) oxide and connections between metal layers were made by vias
that are plasma etched through the oxide as shown in
Figure~\vref{ions:fig:multilayerfab}.  The fabrication process for the surface
gold layer was similar to the electroplating shown in
Figure~\vref{ions:fig:signefab}.

Table~\vref{tab:theTrapComparisionTable} is a detailed comparison of the
performance of several ion trap fabrication approaches which may meet the needs
of future ion trap quantum computing applications. Ion traps developed for other
applications like mass spectroscopy are not included.
\begin{sidewaystable}
  \centering
  \footnotesize
  \begin{tabular}{l|lll|llll|llll|llll|l}
      &&&&($\mu$m)&&(MHz)&&(V)&&(MHz)&(MHz)&$\left((V/m)^{2}/Hz\right)$&(min)&
      (sec)&(eV)&\tabularnewline
      
       Year  & Fabricator& Geom.& Materials& R & Ion& $\frac{\Omega_{\rm
       RF}}{2\pi}$& N &$V_{\rm RF}$& $Q_L$ & $\omega_{z}/2\pi$ &
     $\omega_{x,y}/2\pi$&
$S_{E}(\omega_{z})$ & 
$\tau_{Dop}$ &
$\tau_{dark}$ &
$\phi_{eV}$ & BL\tabularnewline
\hline
2002-8  & NIST  & 2-layer & alumina/Au & 140   & $^{9}Be^{+}$   & 80   & 16   &
500   &40-200  & 4   & 12   & $2.22\times10^{-13}$  & $>600$   & $>100$  &
$\sim1$ & \footnote{See Section~\ref{sec:aluminaTraps} and
\cite{rowe2002a,blakestad2008a}}
\tabularnewline 

2004  & NIST dv10  & 2-layer  & glass/B{*}Si & 122   &
$^{24}Mg^{+}$   & 87   & 1 & 125   & 372   &  &  &  & $>60$   & $>20$   &  &
\footnote{See Section~\ref{sec:dv10} and \cite{wineland2005a}} \tabularnewline

2006 & NIST dv14 & SET  & glass/B{*}Si & 78 & $^{24}Mg^{+}$   & 85   & 1   & 100 
& 373   & 0.77   & 3.1, 4.3   &  & $>60$   &  &  &n
\footnote{This trap suffered from mechanical oscillation of its trapping
electrodes.  See Section~\ref{sec:dv14} and~\cite{britton2006b}}
\tabularnewline

2006-8  & NIST  & SET &
quartz/Au & 42   & $^{24}Mg^{+}$   & 87 & 1   & 103   &  & 2.83   & 16   &
$7\times10^{-12}$  & $>120$   & 10   & 0.117 &y 
\footnote{See \cite{seidelin2006a,amini2008a}} \tabularnewline 

2006  & U. Mich.  & 2-layer &
GaAs/AlGaAs & 30   & $^{111}Cd^{+}$   & 16   & 2   & 55   &  & 0.9   & 4   &
$2\times10^{-8}$   & 10   & 0.1   & 0.08 & 
\footnote{This trap had problems with breakdown that prevented higher trap
frequencies.  See \cite{stick2006a}}
\tabularnewline

2007 & NIST dv16m& SET & B{*}SOI/Au & 40   & $^{24}Mg^{+}$   & 67   & 9   &
$\sim60$   & 72 & 1.1   & 11   & $5\times10^{-11}$  & $>60$   & 10   & 0.1 &y
\footnote{See Section~\ref{sec:dv16}}\tabularnewline

2007  & Lucent  & SET &
Si/SiO2/Al &60  &$^{24/25}Mg^{+}$  &50  &5  &  &80  &1.9  &14 
&$5.7\times10^{-11}$ &$>60$ &30 & & y \footnote{Exposed SiO2 in this trap made
ion transport difficult. }\tabularnewline

$2007$  & Sandia  & SET
& Si/SiN/W &100 &$^{24}Mg^{+}$ &45 & 5 &66 &67 &0.71  &5.3  &  &$>60$  &60 
&$>0.15$ & y \footnote{Fabrication
process was Sandia's
\href{http://mems.sandia.gov/tech-info/summit-v.html}{SUMMiT V}. Ion lifetime 
was limited by background pressure.}

\tabularnewline
  \end{tabular}
  
  \caption[Table comparing microtrap technologies designed for ion trap quantm
  information processing.] {\small Table comparing room temperature microtrap
  technologies designed for ion trap quantum information 
  processing.  All traps
  were tested at room temperature. NIST Boulder tested all the traps except for
  the U. Michigan's trap. SET denotes a surface electrode (ion) trap. B{*}Si
  denotes boron doped silicon.
  \\
  {\bf Year} is the year the trap was tested.\\
  {\bf Fabricator} is the research group that developed a fabrication technology and
  built the traps.\\
  {\bf Geom.} is the trap geometry.\\ 
  {\bf Materials} are the insulating and conducting materials used.\\
  {\bf R} is the characteristic ion-electrode separation. \protect \\
  {\bf Ion} is the ion species. \protect \\
  {\bf $\Omega_{\rm RF}/2\pi$} is the RF drive frequency. \protect \\
  {\bf N} is the number of trapping zones demonstrated to work.\\
  {\bf $V_{\rm RF}$} is the peak RF drive potential amplitude. \protect \\
  {\bf $Q_L$} is the loaded trap resonator quality factor at $\Omega_{\rm RF}$.\\
  {\bf $\omega_z/2\pi$} is the axial trap frequency.\\
  {\bf $\omega_{\rm x,y}/2\pi$} are the radial trap frequencies.  These frequencies are
  approximately equal; sometimes the splitting was not recorded and so
  only one frequency is listed for some traps.\\   
  {\bf $S_{\rm E}(\omega_z)$} is the electric field noise spectral density at
  {\bf $\tau_{\rm Dop}$} is the typical ion lifetime with continuous Doppler laser
  cooling.\\
  {\bf $\tau_{\rm dark}$} is the maximum observed ion lifetime without any laser
  cooling.\\
  {\bf $\phi_{\rm{eV}}$} is the ion trap depth in electron volts according to
  simulation. \\
  {\bf BL} indicates if the trap demonstrated backside loading (only if it is a SET). }
  \label{tab:theTrapComparisionTable}
\end{sidewaystable}


\section{Future traps}
\label{ions:sec:future}

As trapping structures become smaller and trap complexity increases, features
such as junctions will expand the capabilities traps. While two of the multiwafer
traps we looked at in Section~\vref{ions:sec:traps} incorporated a junction,
the slots in multi-layer traps, the required alignment and assembly steps make it
difficult to scale such structures.

\begin{SCfigure}[10][htb]
  \centering
  \includegraphics[width=0.4\textwidth]
  {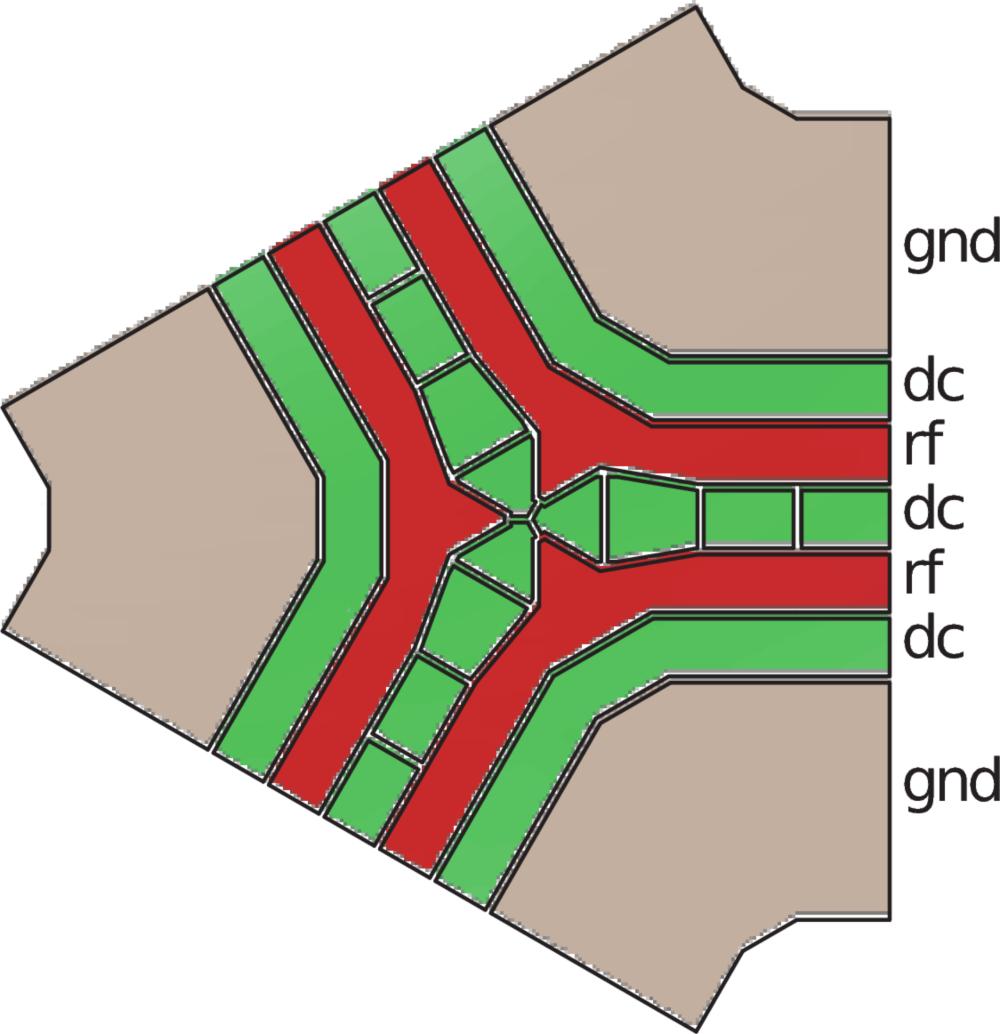}
  \caption
  [Example of a SET 'Y' junction.]
  {Example of a SET 'Y' junction. (Figure by J. Amini)}
  \label{ions:fig:setjunction}
\end{SCfigure}

Figure~\vref{ions:fig:setjunction} shows an example design of a 'Y' version of an SET
junction. The shape of the RF junction is an example of optimization
using analytic solutions to the fields (see
Section~\vref{ions:sec:analyticsolutions}). One feature of this geometry is that
the pseudopotential does not have large axial micromotion `bumps' (see
Section~\vref{ions:sec:micromotion}).  Components such as this 'Y' can be
assembled into larger structures.
SETs fabricated using standard recipes in a foundry and using standard patterns
may eventually make ion traps more accessible to research groups that do not have the
considerable time or resources needed to develop their own.

With increased trap complexity, there are several issues that arise. One of them
is the question of how to package traps and provide all the electrical
connections needed to operate them. Another issue is that of corresponding
complexity of the lasers used in manipulating the ions. Beyond cooling, lasers
are needed to manipulate the internal states of the ions and couple pairs or
groups of ions.  Multiplexing sets of lasers to a number of trapping zones has
its difficulties and complicates simultaneous manipulation of several ions.
Alternately, instead of laser optical fields magnetic structures, active wire
loops and passive magnetic layers could be 
used~\cite{mintert2001a,leibfried2007a,ospelkaus2008a}. If viable,
this would switch much of the experimental complexity from large laser systems to
electronic packages, which can be reliably engineered and scaled~\cite{kim2005a}.

The field of microfabricated ion traps is developing rapidly. In this chapter, I
looked at the current state-of-the-art ion traps and considerations related to
their design.  I look forward to continued development of microfabricated ion
traps which anticipates expanded single and multi-ion control and, some day,
useful ion quantum information processing.  In the chapter that follows I
discuss several microfabricated ion traps that I built and the apparatus 
used to test them.

    \clearpage
\chapter{Boron Doped Silicon Microtraps}
	\label{sec:thetrapschapter}
  	This chapter discusses several microtraps I made as part of my thesis work.
  	The trap geometries were of the two layer and surface electrode trap (SET) types.  
  	
\paragraph{introduction}
\label{sec:theTraps}In $2001$ it was recognized that a fundamental change in ion
trap technology was needed to satisfy the needs anticipated by some ion trap
quantum information processing (QIP) schemes \cite{kielpinski2002a,steane2004a}.
At the time the only ion traps used for QIP were two-layer linear gold-coated
laser-machined alumina ion traps \cite{rowe2001a}. These devices were serviceable
for demonstration experiments but have no clear path for scaling to the smaller
electrode geometries and larger number of traps per chip needed for QIP.

The use of heavily doped silicon in lieu of gold-coated alumina was proposed by
Kielpinski in $2001$ \cite{kielpinski2001a}. In $2005$ Chiaverini, \textit{et
al.} proposed a more fundamental departure: a trap geometry with all electrodes
lying in a single plane \cite{chiaverini05b}. As part of my thesis work I
developed these ideas, and fabricated and trapped ions in several doped silicon
structures. The two-layer boron-doped silicon trap I demonstrated in $2004$
\cite{britton2006b} was the first in a series of experiments exploring new
fabrication technologies at NIST which culminated in a microfabricated surface
electrode ion traps with many zones in $2007$. Also at NIST, Chiaverini, Seidelin
and Amini pursued gold surface electrode ion traps on fused quartz
\cite{seidelin2006a}. Other groups also exploring scalable trapping technology
include the Monroe group at the
\href{http://www.jqi.umd.edu}{Joint Quantum Institute} (JQI) at the
\href{http://www.umd.edu}{University of Maryland} and the
\href{http://www.nist.gov}{National Institute of Standards and Technology}
(NIST) in Gaithersburg, MD., the Slusher group at the
\href{http://www.gtri.gatech.edu}{Georgia Tech Research Institute}, the Kim
Group at \href{http://www.duke.edu}{Duke University}. and
\href{http://www.sandia.gov}{Sandia National Labs}. It is hoped that this
these efforts will some day permit very large trap arrays.

This section starts with a description of the tried and true gold-on-alumina trap
used at NIST since 2000.  Then I describe the boron doped ion traps I made
including trap geometry, fabrication steps and trap performance.  The section
concludes with a comparison of competing microtrap trap technologies.

A variety of microfabrication techniques were used to make my devices. Since
many techniques are common to several devices they are separately detailed in
Section~\vref{sec:microfabTechniques}. Refer to that section for the following topics.
\begin{compactitem}
  \item silicon deep etch by deep reactive ion etching (DRIE) including photolithography
  \item silicon oxide etch including plasma etching and buffered oxide etch
  (BOE)
  \item wafer backside alignment
  \item metallization and Ohmic contacts in silicon
  \item anodic and wafer bonding including silicon on insulator (SOI) wafers
  \item wafer cleaning including Piranha etch
  \item electrical interconnect using wire bonding, gap welding and resistive
  welding
  \item suppliers and specifications for doped silicon wafers, glass wafers,
  and ultrasonic milling of glass
  \item shadow masks and wafer dicing saw blades
\end{compactitem}
Section \vref{sec:trapTestingApparatus} discusses the apparatus used to
test the boron doped silicon traps.

\paragraph{trap technologies}

Table~\vref{tab:tableTrapsByDate} is a chronology of the several ion trap
fabrication approaches which may meet the needs of future ion trap quantum computing applications. Ion traps
developed for other applications like mass spectroscopy are not included.

\begin{table}
  \begin{tabular}{ll|lll}
     Year  & Fabrication & Geometry & Materials & Citation
     \tabularnewline
    \hline
    2002-8  & NIST & 2-layer& alumina/Au&
    Sec.~\ref{sec:aluminaTraps} \cite{rowe2002a,blakestad2008a} \\ 
    $2004$  & NIST & 2-layer & B{*}Si &
    Sec.~\ref{sec:dv10} \cite{britton2006b} \\ 
    $2006$ & NIST &SET & B{*}SOI & Sec.~\ref{sec:dv14} \\ 
    $2006$  & NIST &SET& quartz-Au&\cite{seidelin2006a} \\ 
    $2006$  & U. Mich. &2-layer&GaAs/AlGaAs& \cite{stick2006a} \\ 
    $2007$ & NIST&SET&B{*}SOI/Au& Sec.~\ref{sec:dv16}\\
    $2007$  & Lucent &SET&Si/$SiO_2$/Al&\\
    $2007$  & Sandia &SET&Si/SiN/W&\href{http:http://mems.sandia.gov/tech-info/summit-v.html}{SUMMiT V} \\
  \end{tabular}

\caption[Table comparing trap technologies exploring scalable ion trap fabrication
for quantm information processing.]{Table comparing trap technologies exploring scalable ion trap fabrication
for quantm information processing. SET is a surface electrode (ion) trap. 
B{*}Si is boron doped silicon.}
\label{tab:tableTrapsByDate}
\end{table}

\section{2-layer gold coated laser machined alumina ion traps}
\label{sec:aluminaTraps}
\begin{figure}
  \centering
  \includegraphics[width=0.9\textwidth]{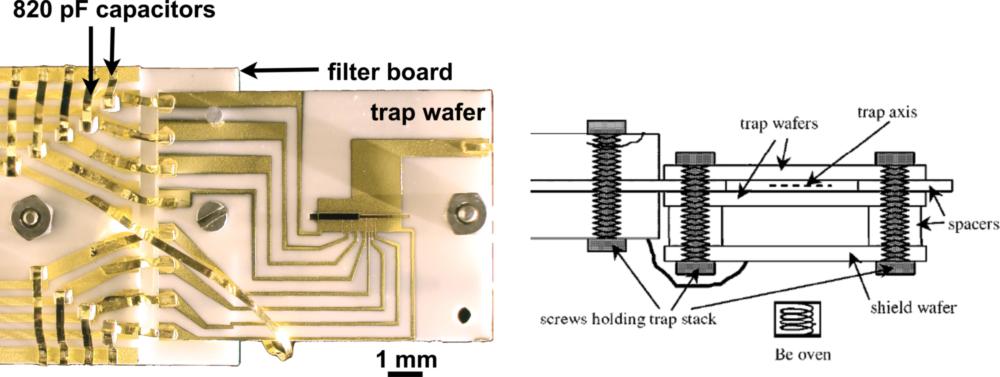}
  \caption
  [Photograph showing assembly of gold coated laser machined alumina
  wafers.]
  {Photograph showing assembly of gold coated laser machined alumina
  wafers. The left hand photograph shows a top view of a trap. Gold
  traces are visible on the trap wafer, the filter board, 820 pF capacitors
  and gold ribbon used to make electrical connections
  (see Section~\vref{sec:electricalInterconnect}).
  The right hand figure shows a cross
  section view \cite{rowe2002a}.A pair of
  trap wafers and ancillary spacer wafers were held together with stainless
  screws. A NIST trap tested in 2008 using this technology had alignment errors
  ($\sim$$1^{\circ}$ rotation, $\sim10~\mu$m horizontal, $\sim40~\mu$m
  vertical) which caused ion transport difficulties~\cite{blakestad2008a}.  }
  \label{fig:AluminaScrews}
\end{figure}

\begin{figure}
  \centering
  \includegraphics[width=0.8\textwidth]{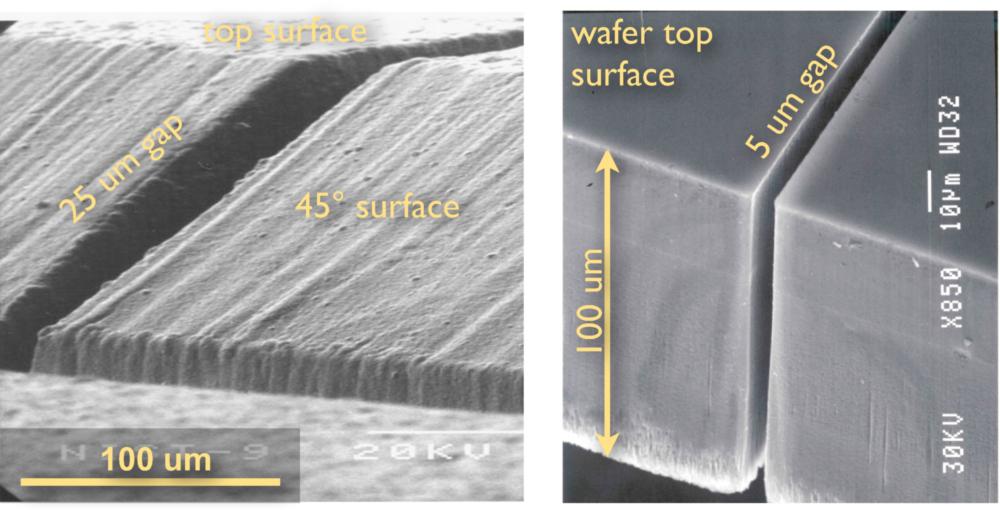}
  \caption[Scanning electron micrographs of alumina trap surface roughness.]
  {Scanning electron micrographs comparing the surface roughness of
  gold coated laser machined alumina (left) with deep etched silicon
  (right). \\(left) The laser machined wafer is patterned with a vertical trench
  cut fully through its bulk. The laser machining was done by
  \href{https://www.resonetics.com}{Resonetics, Inc}. The roughness is dominated
  by the step size of the Excimer laser used to ablate the alumina. The cuts
  had a $5^{\circ}$ sidewall slope. The best available tolerance for
  15~$\mu$m wide slits is $\pm3~\mu$m.\\(right)  The deep etched silicon wafer
  is patterned with a 5~$\mu$m trench (gap) that cuts fully thru the 
  wafer's bulk.  The etching was done using
  the STS DRIE discussed in Section~\vref{sec:DRIE}. Tolerances are
  $1-2^{\circ}$ sidewall slope and a pattern resolution of 1~$\mu$m
  for cuts $<20~\mu$m deep and $3~\mu$m for cuts $<200~\mu$m
  deep (see Section~\vref{sec:DRIE} for more on silicon etch).}
  \label{fig:laserMachiningVsDIRE}
\end{figure}

Gold coated laser machined linear ion traps were used by
NIST since the late $1990$'s \cite{turchette2000a}. As of $2008$ only traps made
using this technology have been used for quantum information processing
experiments at NIST \cite{turchette2000a,rowe2002a,blakestad2008a}. The traps were
made by laser machining alumina wafers to define the electrode structure. They
were then coated with a 1-3~$\mu$m layer of gold by e-beam deposition, RF
sputter deposition or electroplating. Wafers comprising the two trapping
electrodes were then aligned by hand and held together with metal screws (see
Figure~\vref{fig:AluminaScrews}). This technology has several limitations
including exposed dielectric surfaces (bare alumina) which can accumulate surface
charges that perturb trapped ions, large wafer-wafer alignment errors, and rough
surfaces due to the laser machining (see Figure~\vref{fig:laserMachiningVsDIRE}
). With currently available laser machining trap electrodes smaller than $25~\mu$m
suffer from large fabrication errors. Alumina traps are also difficult to
fabricate and the turn-around time for a new trap is at least a year. This
technology is difficult to scale beyond several~$10$'s of electrodes.

\clearpage
\section{dv10: 2-layer doped silicon trap}
\label{sec:dv10} 
\begin{figure}
  \centering
  \includegraphics[width=0.6\textwidth]{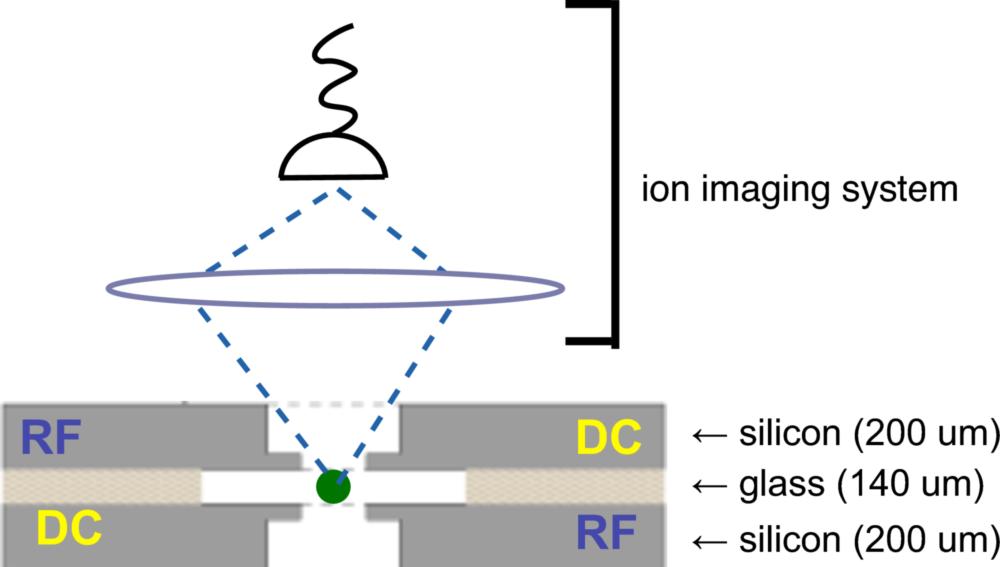}
  \caption[Schematic of a 2-layer boron doped silicon trap dv10 in cross
  section.] {Schematic of a 2-layer boron doped silicon trap in cross section.
  The ion location is marked by a green dot. Due to the L shaped electrode geometry, the
  trap's numerical aperture is approximately F1 as viewed from above. 
  This matches the numerical aperture of the UV imaging
  system, maximizing ion fluorescence collection. It also minimizes
  scattered light from the cooling laser beams. The drawing is not to
  scale: the glass insulator is at least laterally 5~mm away from the trapping
  region.}
  \label{fig:dv10crossSectionView}
\end{figure}

The first doped silicon trap was a variant of previous 2-layer
gold coated alumina traps (see Section~\vref{sec:aluminaTraps}). The geometry was
the same (see Figure~\vref{fig:dv10crossSectionView}) but the structural material
and conductor differed: glass and boron doped silicon were used instead of
alumina and gold. Also, instead of screws the wafers were held together by anodic
bonding (see Section~\vref{sec:waferBonding}). I call this first incremental
design dv10.  It was first presented in~\cite{wineland2005a}.

Advantages of this approach include bonding without screws, smoother
trapping surfaces and high resolution photolithographic patterning
of trap features. Moreover, because the trap structural material
is a conductor there are no insulating surfaces anywhere near the ion. See
Figure~\vref{fig:laserMachiningVsDIRE} for a comparison with laser machined
alumina.

\subsection{Fabrication}
\label{sec:dv10fabSteps}
\begin{figure}
  \centering
  \includegraphics[width=6in]{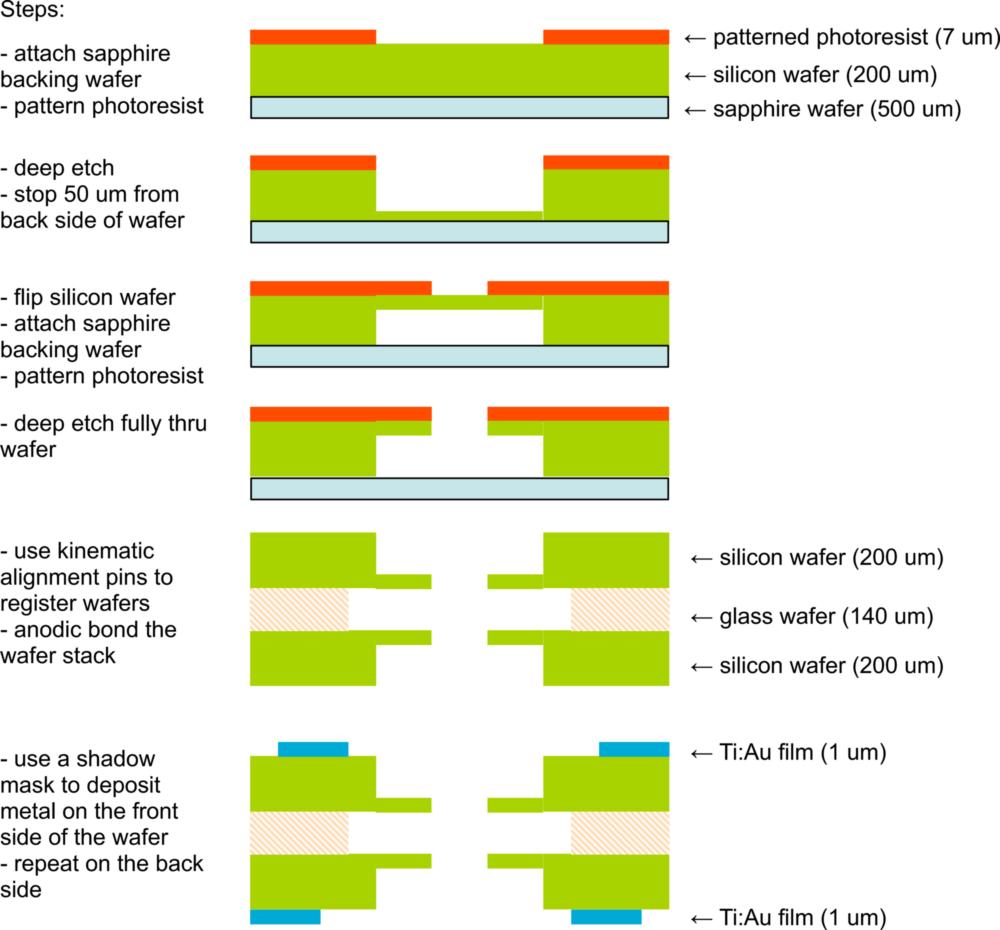}
  \caption[Figure outlining the process steps to fabricate trap dv10.]
  { Figure outlining the process steps to fabricate trap dv10. The individual
  steps are discussed in the text. Note that the dimensions are not
  to scale.}
  \label{fig:dv10FabSteps}
\end{figure}

\begin{SCfigure}[10]
  \centering
  \includegraphics[width=0.4\textwidth]{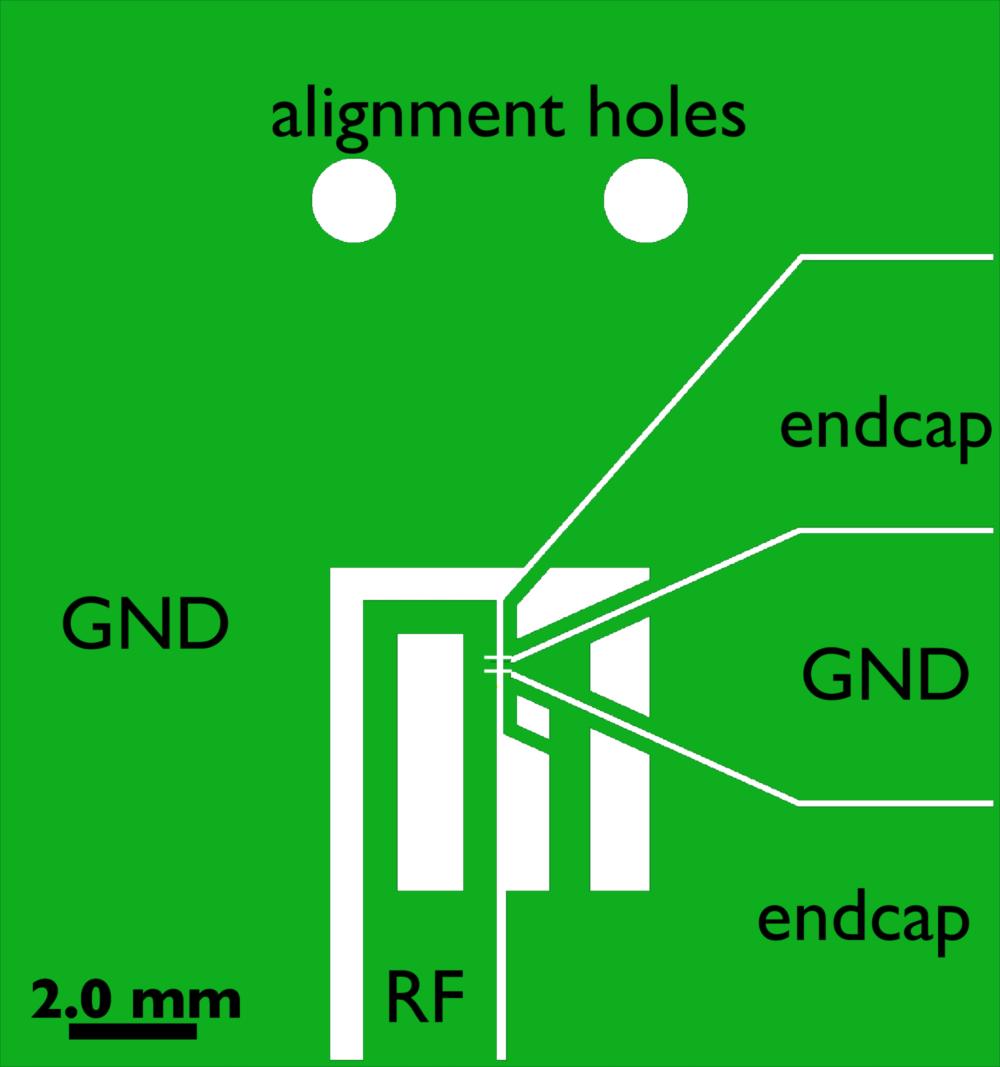}
  \caption[Figure showing the deep etch pattern for a dv10 silicon wafer.]
  {Figure showing the deep etch pattern for a dv10 silicon wafer. The
  white regions are apertures etched fully through the silicon wafer.}
  \label{fig:dv10etchPattern}
\end{SCfigure}

The fabrication steps were as follows. 
\begin{compactenum}
  \item double sided deep etch - defines trap geometry
  \item dice wafers into chips
  \item clean the chips
  \item align and bond
  \item metallization
  \item dice the chips to electrically isolate the electrodes
  \item vacuum processing
\end{compactenum}
The trap structure illustrated in Figure~\vref{fig:dv10crossSectionView}
was deep etched with the STS DRIE (Section~\vref{sec:DRIE}) into
a $200~\mu$m double side polished born doped silicon wafer (see
Section~\vref{sec:dopedSilicon} for more on the wafers). The fabrication steps
are shown in Figure~\vref{fig:dv10FabSteps}. Nine trap chips were simultaneously
etched into a single 3~inch wafer, each patterned to appear as in 
Figure~\vref{fig:dv10etchPattern}.
The wafer was etched from both the front and back sides. Care was taken
to avoid surface contamination which can cause needles to form during
deep etching (see Figure~\vref{fig:DRIE:needles}) and may prevent bonding.

\paragraph{double sided deep etch }
\begin{compactenum}
  \item Measure the process wafer thickness with a micrometer.
  \item Attach a sapphire backing wafer with wax (see
  Section~\vref{sec:DRIE:thruWaferAdvice}).
  \item Spin photoresist on the top side of the process wafer using the $7~\mu$m
  recipe. Transfer to the photoresist the trap pattern from a mask using the recipe in Section~\vref{sec:DRIE:deepEtchPR}.
  \item Etch using DIRE recipe SPECB to etch the pattern in Figure~\vref{fig:dv10etchPattern}.
  Stop etching when there is 40-50~$\mu$m material remaining of the 200~$\mu$m
  wafer. As a rough estimate assume an etch rate of 3 $\mu$m/min.
  Use an optical microscope with a shallow ($<2~\mu$m) depth of
  field and z-axis indicator to measure the etch depth.
  \item Release the backing wafer as described in Section~\vref{sec:DRIE:thruWaferAdvice}.
  \item Flip the process wafer and reattach the backing wafer.
  \item Spin photoresist on the top side of the process wafer using the
  7~$\mu$m recipe and expose as last time.
  \item Etch using DRIE recipe SPECB until large apertures fully penetrate
  the silicon wafer.
  \item Etch using the DIRE recipe for SOI (see Section~\vref{sec:DRIE:dielectric}).
  Inspect narrow trenches on an optical microscope with backlighting.
  They should be free of silicon (opaque in the visible). If not, continue
  to etch using the SOI recipe.
  \item Release the backing wafer as described in Section~\vref{sec:DRIE:thruWaferAdvice}.
\end{compactenum}

\paragraph{wafer dicing }
An automated saw was used to dice the etched 3~inch silicon wafer into
9~chips. A regular silicon-cutting blade was used. See
Section~\vref{sec:dicingSaw} for details.

\paragraph{glass spacers}
Glass spacers were made of Corning 7070 glass (see
Table~\vref{tab:lossTangentTable}). The glass was patterned by ultrasonic milling
(see Section~\vref{sec:usonicMilling}). It was then diced on the automated dicing
saw but with a special Resinoid blade (see Section~\vref{sec:dicingSaw}).

\paragraph{clean the wafers}
High strength anodic bonding requires clean surfaces.  Cleaning steps I,
II and IV in Section~\vref{sec:RCA} were followed for both the silicon and glass
chips. I used the wafer holders described in Section~\vref{sec:teflonBaskets}.
Both the silicon and glass were processed at the same time.

\paragraph{align and bond}
Anodic bonding was used to adhere the wafers into a stack as in
Figure~\vref{fig:dv10crossSectionView} and Figure~\vref{fig:dv10FabSteps}. See
Section~\vref{sec:waferBonding} on how anodic bonding works. Follow the advice in
Section~\vref{sec:bondingInPractice} on how to do wafer bonding.

\paragraph{metallization and electrical interconnect }
For electrical contact to the bare silicon electrodes, $1~\mu$m gold contacts
were deposited at the periphery of the bonded chips. As dv10 is a two layer
device, pads needed to be deposited on both sides in two deposition steps.
Contact to these pads was made using gold ribbon and a resistive welder. I
followed the recipe in Section~\vref{sec:shadowMaskRecipe} with FASTAU. The same
recipe was repeated for the back side gold pads. The chip's gold pads were
connected to pads on the filter board using resistive welding as in
Section~\vref{sec:electricalInterconnect}.

\begin{figure}
  \centering
  \includegraphics[width=0.5\textwidth]{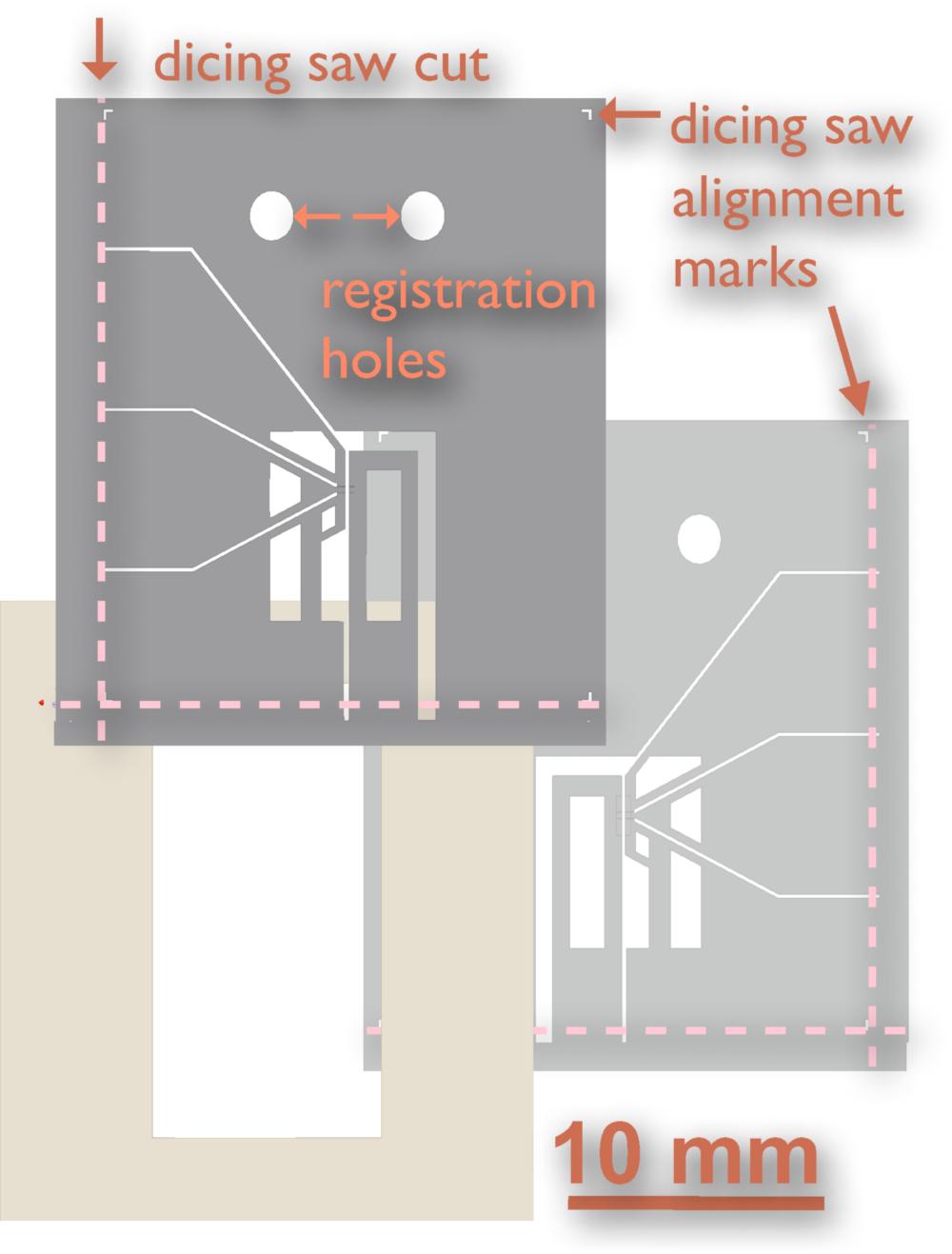}
  \caption[Schematic showing the two silicon and one glass layers comprising
  trap dv10.]{Schematic showing the two silicon and one glass layers
  that comprised trap dv10. The holes in the silicon were formed by
  deep etching; the glass was ultrasonically milled. The three layers were
  aligned and anodically bonded on a special stage (see
  Figure~\vref{fig:bondingPlatformVacuumStage}). Alignment was kinematic. The silicon
  chips were etched with high tolerance ($<5~\mu$m error) alignment holes.
  The interior edge of the glass was sized to press against the circumference of
  these holes. Alumina alignment pins on the bonding stage provided
  layer-to-layer registration. Under a microscope, the lateral alignment error
  appeard to be less than $20~\mu$m. The etched silicon and glass chips were
  oversized; all the silicon electrodes were initially shorted together. This
  excess material holds the silicon electrodes in place during bonding. 
  After bonding the electrodes were coated in wax for protection, then
  electrically isolated by cutting fully thru the wafer stack with a dicing saw. 
  Alignment marks were etched into the silicon indicating where to cut (pink
  dashed lines in this figure).}
  \label{fig:dv10alignmentAndCrossSection}
\end{figure}

\paragraph{dice to electrically isolate electrodes}
At this stage in the fabrication process all the doped silicon electrodes
are still shorted together. This was necessary for structural support
prior to bonding. At this point however they can be safely separated with a
second dicing saw cut. See Figure~\vref{fig:dv10alignmentAndCrossSection} for a
schematic and discussion of this process. Prior to dicing, the
delicate structures were protected with wax as in the Wafer Dicing steps above.
Again, a Resinoid blade was used since one of the layers was glass.

\paragraph{vacuum processing}
Prior to insertion into the final ion trap vacuum system, the thermal
silicon oxide layer was stripped off the trap. In 2003 this was done
with a 15~sec BOE dip (step II in Section~\vref{sec:RCA}) within
10~min of insertion into the vacuum. However, this is risky as
HF degrades Ti and can damage the bond pads. Success requires thorough
rinsing with deionized water. A more conservative approach (as was used in
subsequent traps) may have been a plasma etch (see
Section~\vref{sec:oxideEtch}).
\begin{SCfigure}
  \centering
  \includegraphics[width=3in]{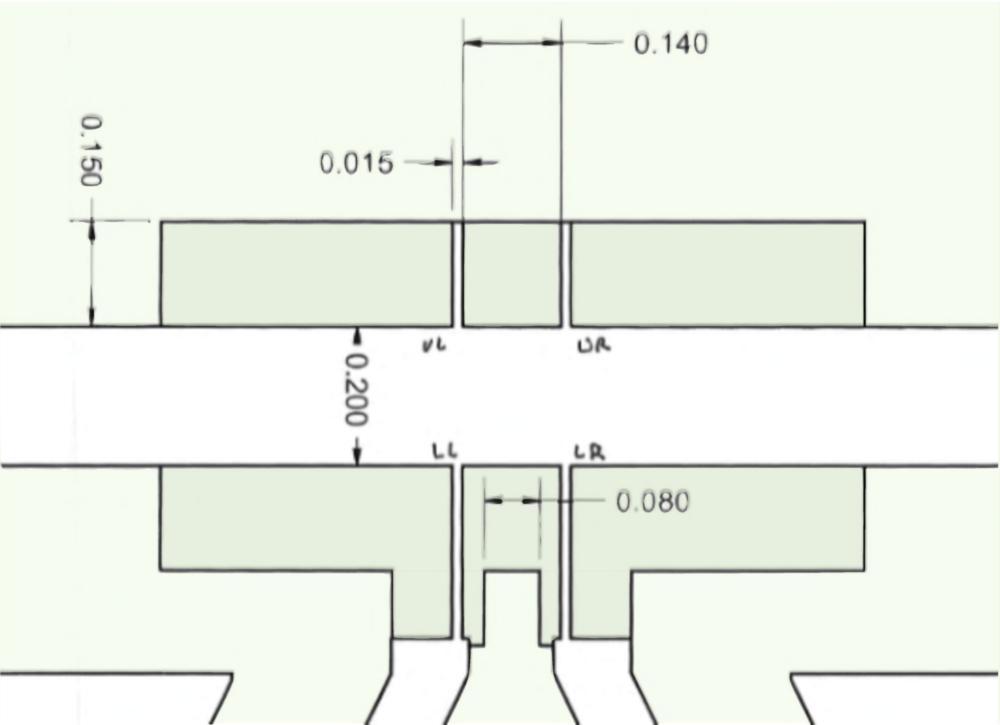}
  \caption[Schematic showing trapping region of dv10.]
  {Schematic showing trapping region of dv10. The light green surfaces
  mark the top of the wafer. The dark green surfaces were recessed by
  about 150~$\mu$m. The white regions were etched fully thru the wafer.
  Units are in mm. }
  \label{fig:dv10trappingRegionSchematic}
\end{SCfigure}

\subsection{Performance}
\label{sec:dv10performance}The trap was installed in a coaxial $\lambda$/4 style
vacuum housing (see Section~\vref{sec:vacuumApparatus}). The endcap voltages were
$+3.0$~V, the middle electrodes were grounded and the RF amplitude was $125$~V at
$87$~MHz. It loaded easily in May 2004. Linear chains of a half dozen ions could
be loaded in the trap. Ion lifetime with Doppler cooling was regularly in excess
of 1 hour. Ions were loaded from a $\,^{24}\text{Mg}$ oven and ionized by
electron impact from a hot tungsten filament biased at +50 to +100~V. The trap
was successful but time constraints on the experimental apparatus precluded
measurement of the ion oscillation frequencies and motional heating rate.
Table~\vref{tab:dv10characteristics} reviews the characteristics of this trap.
\begin{table}
  \begin{tabular}{c|ccccc|cccc|ccc}
    {\footnotesize $trap$ } & \begin{sideways}
{\footnotesize $d$ ($\mu$m)}%
\end{sideways} & \begin{sideways}
{\footnotesize $\frac{\Omega_{RF}}{2\pi}\,(MHz)$}%
\end{sideways} & \begin{sideways}
{\footnotesize $V_{RF}\,(V)$ }%
\end{sideways} & {\footnotesize $Q$ } & {\footnotesize $Ion$ } & \begin{sideways}
{\footnotesize $\frac{\omega_{z}}{2\pi}\,(MHz)$ }%
\end{sideways} & \begin{sideways}
{\footnotesize $\frac{\omega_{x,y}}{2\pi}\,(MHz)$ }%
\end{sideways} & {\footnotesize $N$ } & \begin{sideways}
{\footnotesize $S_{E}(\omega_{z})$ $\left(\frac{(V/m)^{2}}{Hz}\right)$}%
\end{sideways} & \begin{sideways}
{\footnotesize $\tau_{dop}\,(min)$ }%
\end{sideways} & \begin{sideways}
{\footnotesize $\tau_{dark}\,(sec)$ }%
\end{sideways} & \begin{sideways}
{\footnotesize $\phi_{eV}\,(eV)$ }%
\end{sideways}\tabularnewline
\hline
{\footnotesize B{*}Si 2-layer dv10} & {\footnotesize 122} & {\footnotesize 87}
& {\footnotesize 125} & {\footnotesize 372} & {\footnotesize $^{24}Mg^{+}$ } & 
&  & {\footnotesize 1} &  & {\footnotesize $>60$} & {\footnotesize $>20$} &
\tabularnewline
  \end{tabular}
  \caption[Trap characteristics of the 2-layer boron doped silicon trap,
  dv10.]
  {Trap characteristics of dv10. See Table \vref{tab:theTrapComparisionTable} 
  for an explanation of the nomenclature.  Trap depth $\phi_{\text{ev}}$ and
  $V_{\rm RF}$ were estimated by simulation.  Neither the heating rate nor the 
  secular frequencies were measured in this trap.}
  \label{tab:dv10characteristics}  
\end{table}
\begin{SCfigure}
  \centering
  \includegraphics[width=0.6\textwidth]{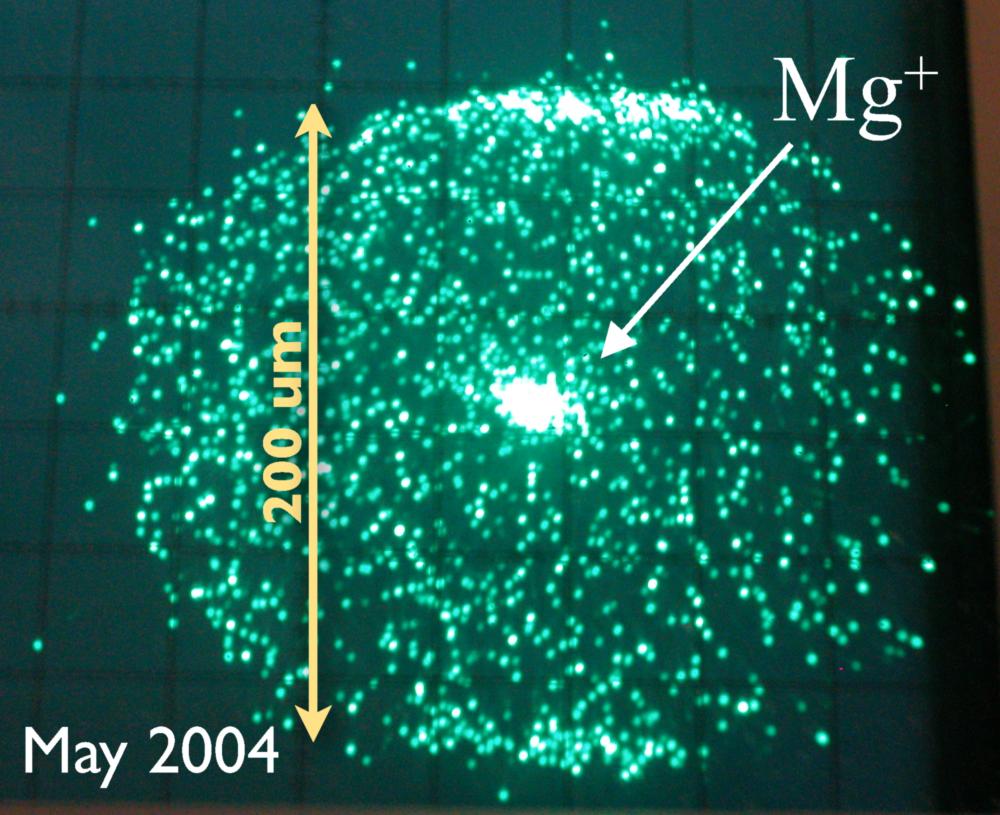}
  \caption[Oscilloscope image showing a trapped $\,^{24}\text{Mg}^{+}$ in trap
  dv10.]{Oscilloscope image showing a trapped $\,^{24}\text{Mg}^{+}$ in trap
  dv10. The image was created from collected ion fluorescence at 280~nm
  due to the Doppler cooling laser beam. The field of view was 200~$\mu$m
  wide.}
  \label{fig:dv10ionGlamorShot}
\end{SCfigure}

\clearpage

\section{dv14: surface electrode doped silicon ion trap}
\label{sec:dv14}The second doped silicon trap I built drew upon the success of
dv10 and included a departure from ion trapping convention in its electrode
geometry. In 2005 Chiaverini, \textit{ et al.} proposed a trap geometry with all
electrodes lying in a single plane \cite{chiaverini05b}. My second trap aimed to
demonstrate trapping for the first time in such a structure. I succeeded in
trapping and Doppler cooling a single $\,^{24}\text{Mg}^{+}$ ion in March 2006,
but was several months too slow. In parallel at NIST, Seidelin led an effort to
build a surface electrode ion trap on quartz. In early 2006 we published our
results in Physical Review Letters \cite{seidelin2006a}. Later in 2006, 
Pearson, \textit{ et al.} demonstrated buffer gas cooling of half-micron size
aminoploystyrene spheres in a surface electrode trap (PRA \cite{pearson2006a})
and Brown, \textit{ et al.} loaded a $\,^{88}\text{Sr}^{+}$ ion cloud in a
surface electrode trap (PRA \cite{brown2006a,brown2007c}). This section discusses
my first surface electrode trap which I call dv14.

Generic advantages of surface electrode traps (SET) over two-layer traps include
much easier MEMS fabrication and the possibility of integrating control
electronics on the same trap wafer \cite{kim2005a}. My SET fabrication approach
has several advantages over other demonstrated SETs. It provides superior
shielding of structural dielectric surfaces and addition of large through wafer
holes for backside loading is easy. These two features in particular are
difficult to obtain in, for instance, gold on fused quartz traps \cite{seidelin2006a}. 
See Section \vref{ions:sec:SETs} for more on SET geometry.

\begin{figure}
  \centering
  \includegraphics[width=0.75\textwidth]{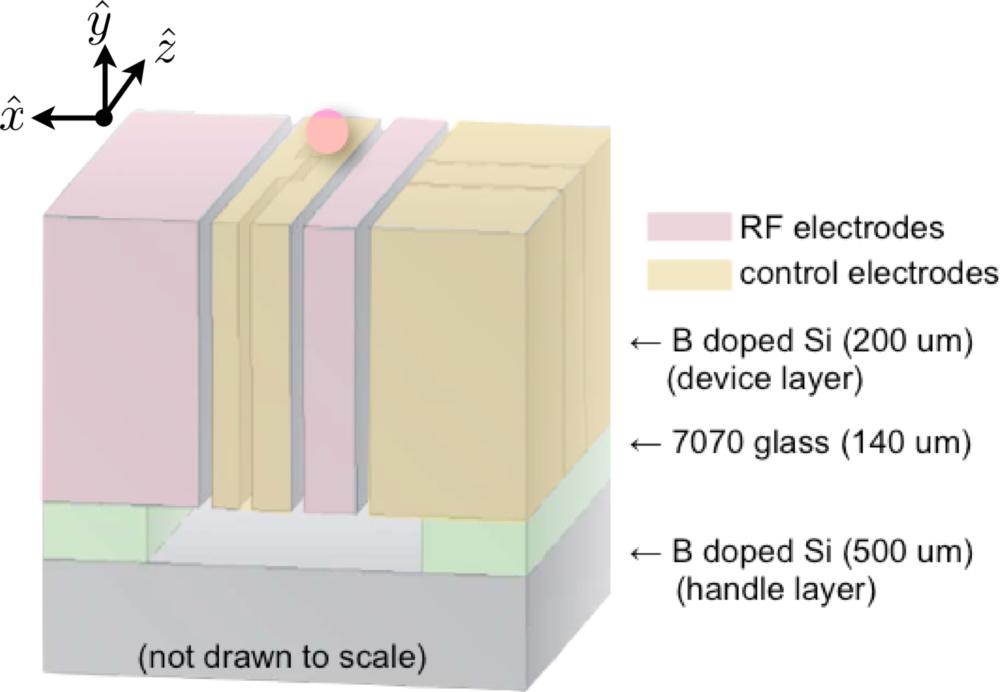}
  \caption[Perspective view of the surface electrode ion trap dv14.]
  {Perspective view of the surface electrode ion trap dv14. The conducting
  surfaces near the ion are bare doped silicon. The trap electrodes
  are cantilevered over the insulating~7070 glass layer. The ion location
  is marked by a pink dot. To the left of trap center is an adjacent
  pair of control electrodes which help null radial micromotion. The
  drawing is not to scale: the glass insulator is at least 5~mm away
  from the trapping region.}
  \label{fig:dv14perspectiveView}
\end{figure}

\subsection{Fabrication}

\label{sec:dv14fabDetails}
\begin{figure}[H]
  \centering
  \label{fig:dv14schematicSecondLabel} 
  \includegraphics[width=0.6\textwidth]{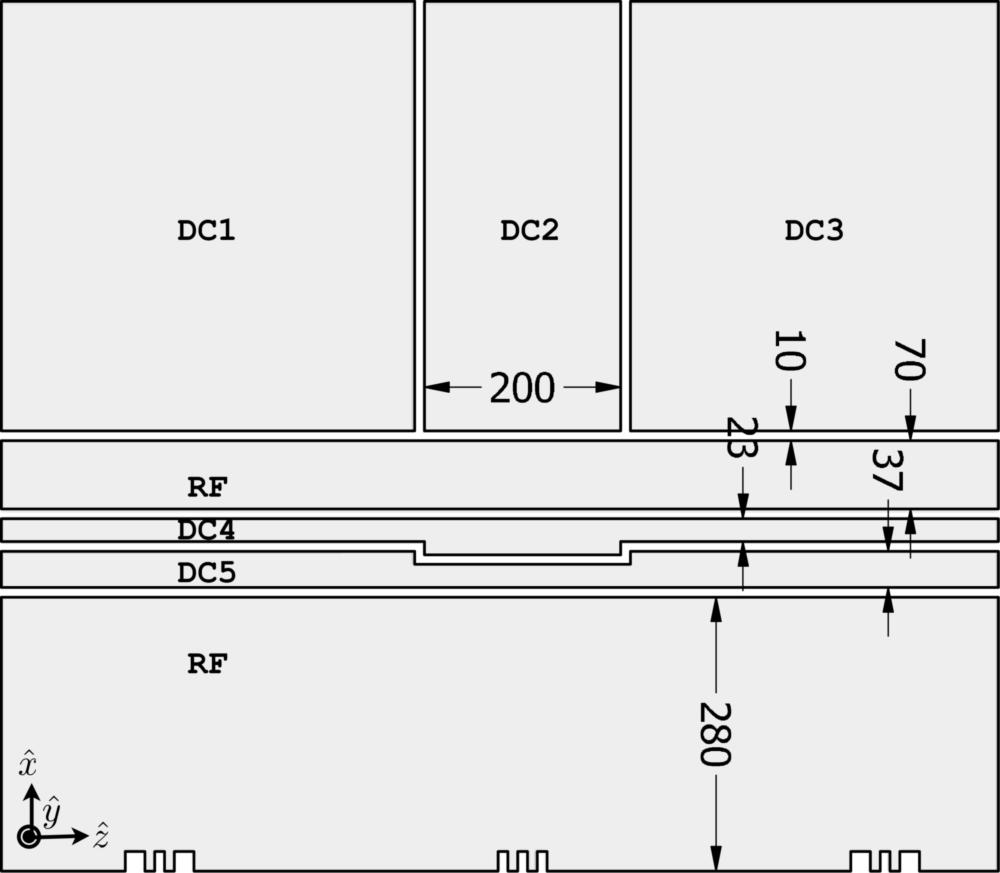}
  \caption[Schematic showing the dimensions of the trapping region of dv14.]
  {Schematic showing the dimensions of the trapping region of dv14.
  The gray surfaces are the top of the wafer. The white regions are
  etched fully thru the wafer. Units are in~$\mu$m.}
  \label{fig:dv14schematic} 
\end{figure}

dv14 was a  4-wire surface electrode ion trap made of
boron doped silicon.  The trap structure is illustrated in
Figures~\vref{fig:dv14schematicSecondLabel} and~\vref{fig:dv14perspectiveView}.  The trap
features were deep etched with the STS DRIE (see Section~\ref{sec:DRIE}) into a
200~$\mu$m double side polished boron doped silicon wafer. The trap electrodes
were cantilevered over an insulating 7070~glass layer. The conducting surfaces
near the ion were bare doped silicon. The fabrication steps were nearly
identical to those for dv10, Section~\vref{sec:dv10fabSteps}. Nine trap chips were
simultaneously etched into a single 3~inch wafer.

The fabrication steps are as follows.
\begin{compactenum}
  \item double sided deep etch - defines trap geometry
  \item dice wafers into chips
  \item clean the chips
  \item align and bond
  \item metallization
  \item dice the chips to electrically isolate the electrodes
  \item vacuum processing
\end{compactenum}

\begin{figure}[H]
  \centering
  \includegraphics[width=6in]{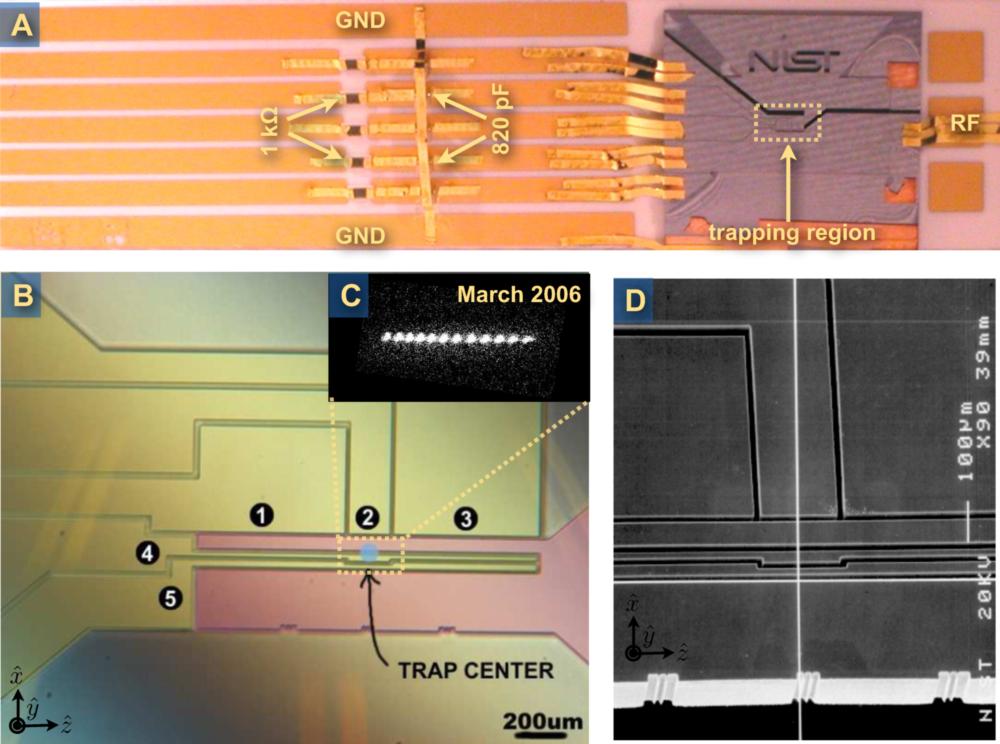}
  \caption[Micrographs of a completed dv14 ion trap.]
  {Micrographs of a completed dv14 ion trap. (A)~Trap chip mounted on
  a ceramic filter board. Also visible are the RC filter components
  and traces which provide control and RF potentials to the chip. (B)~A false
  color optical photograph of the trapping region with a blue dot at the 
  trap center. This is the same view that the ion imaging
  optics has of the trap. The the shape of control electrodes 1-5 was
  selected to produce a static harmonic electric potential well along
  the trap z-axis. Electrodes 2, 4 and 5 permit radial
  micromotion nulling by creating a static radial electric fields at
  the ion which can compensate for stray electric fields. RF electrodes
  are in red. Inset (C)~shows a linear chain of ions from the same
  perspective as photo B. (D) A scanning electron micrograph showing
  details of the trapping region. Alignment marks cut into the edge
  of the wafer (visible in B and C) assist with laser beam alignment.}
  \label{fig:dv14photoMontage}
\end{figure}

\subsection{Performance}
\label{sec:dv14trapPerformance}The results for this trap were first presented in
\cite{britton2006b}. These results and additional details follow.
Table~\vref{tab:dv14characteristics} summarizes the characteristics of this trap
using the nomenclature of Figure~\vref{tab:theTrapComparisionTable}.

Ions were loaded from a $\,^{24}\text{Mg}$ oven and photoionized by a~285~nm
laser beam. A $\lambda/2$~style vacuum system was used.  Initial trap parameters
(RF and control electrode potential amplitudes) were determined numerically by
simulation. The primary solution constraint was that the ion lie a the
pseudopotential zero.
An ion was trapped with the following potentials.
\begin{eqnarray*}
  V_{1}  =  +0.72V;
  V_{2}  =  +0.32V;
  V_{3}  =  +0.74V;
  V_{4}  =  -0.90V;
  V_{5}  =  +1.00V \\
  V_{RF}  =  125V\text{ at }87~MHz
\end{eqnarray*}

Table \vref{tab:dv14expVsSim} compares the trap characteristics predicted
by simulation with observation.

\begin{table}
  \centering
  \begin{tabular}{l|lllll}
    & $\frac{\omega_{z}}{2\pi}$ (1-ion) & $\frac{\omega_{x}}{2\pi}$ & $\frac{\omega_{y}}{2\pi}$ & $\theta_{axis}$\tabularnewline
    \hline
    endcap tickle & 0.7682 &  &  &   \tabularnewline
    ion spacing & 0.55 &  &  &   \tabularnewline
    simulation (1-ion) & 0.786 & 3.143 & 5.257 &  11.1\tabularnewline
  \end{tabular}
  \caption[Comparison of observed and simulated characteristics of dv14.]
  {Comparison of observed and simulated characteristics of dv14.  The secular
  frequencies were measured experimentally by the endcap tickle technique 
  (see Section~\vref{sec:secularFreqMeasurement:endcapTickle}) and 
  by measurement of ion spacing in a linear crystal (see
  Section~\vref{sec:secularFreqMeasurement:ionSpacing} and
  Figure~\vref{fig:dv14ionGlamorShotSecondLabel}). The potentials are those reported in the
  text. See Table~\vref{tab:theTrapComparisionTable} for an explanation of the nomenclature.}
  \label{tab:dv14expVsSim}
\end{table}
%
%

\begin{table}
  \begin{tabular}{c|ccccc|cccc|ccc}
    {\footnotesize $trap$ } & \begin{sideways}
{\footnotesize $d$ ($\mu$m)}%
\end{sideways} & \begin{sideways}
{\footnotesize $\frac{\Omega_{RF}}{2\pi}\,(MHz)$}%
\end{sideways} & \begin{sideways}
{\footnotesize $V_{RF}\,(V)$ }%
\end{sideways} & {\footnotesize $Q$ } & {\footnotesize $Ion$ } & \begin{sideways}
{\footnotesize $\frac{\omega_{z}}{2\pi}\,(MHz)$ }%
\end{sideways} & \begin{sideways}
{\footnotesize $\frac{\omega_{x,y}}{2\pi}\,(MHz)$ }%
\end{sideways} & {\footnotesize $N$ } & \begin{sideways}
{\footnotesize $S_{E}(\omega_{z})$ $\left(\frac{(V/m)^{2}}{Hz}\right)$}%
\end{sideways} & \begin{sideways}
{\footnotesize $\tau_{dop}\,(min)$ }%
\end{sideways} & \begin{sideways}
{\footnotesize $\tau_{dark}\,(sec)$ }%
\end{sideways} & \begin{sideways}
{\footnotesize $\phi_{eV}\,(eV)$ }%
\end{sideways}\tabularnewline
\hline
{\footnotesize B{*}Si 1-layer dv14} & {\footnotesize 78} & {\footnotesize 85} &
{\footnotesize 100} & {\footnotesize 373} & {\footnotesize $^{24}Mg^{+}$ } &
{\footnotesize 0.77} & {\footnotesize 3.1, 4.3} & {\footnotesize 1} &  & {\footnotesize $>60$} &  & \tabularnewline
  \end{tabular}
  \caption[Trap characteristics of dv14. ]
  {Trap characteristics of dv14 in the format of Table
  \vref{tab:theTrapComparisionTable}. Trap depth $\phi_{\text{ev}}$ and
  $V_{\rm RF}$ were estimated by simulation. }
  \label{tab:dv14characteristics}
\end{table}

\begin{figure}[H]
  \centering
  \includegraphics[width=0.45\textwidth]{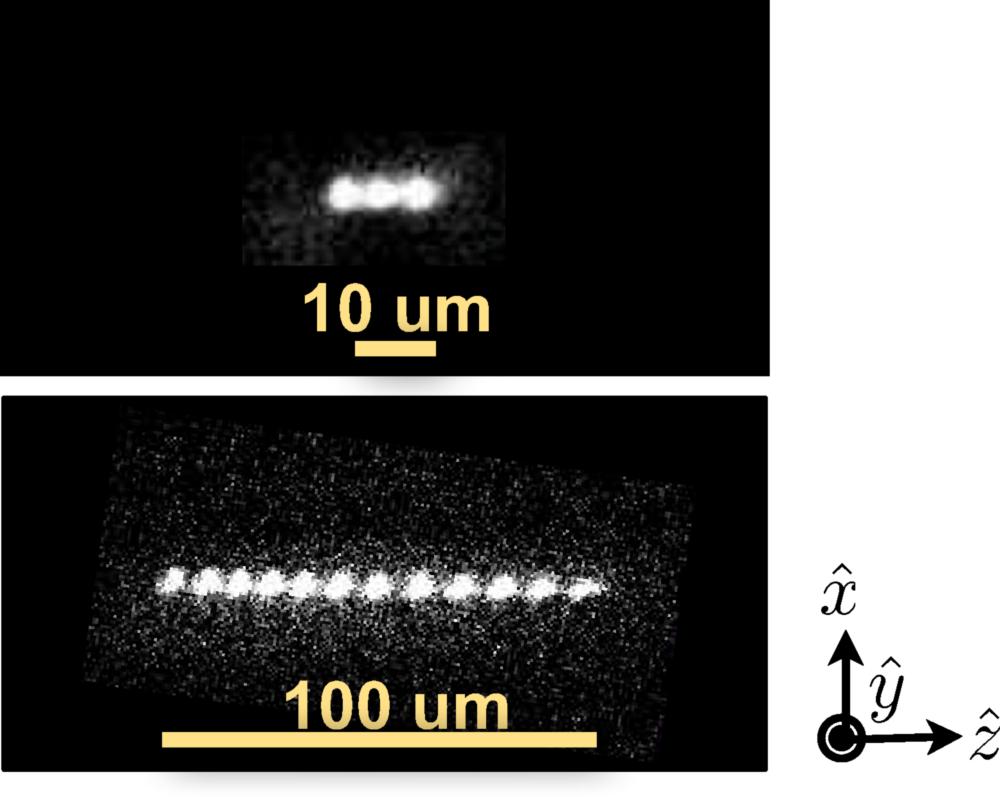}
  \caption[Linear ion crystals trapped in dv14.]
  {Linear ion crystals trapped in dv14. (Top) A linear crystal of three
  ions along the trap axis confined with the potentials listed in the
  text. The length of the crystal is about 17~$\mu$m suggesting an
  axial frequency of $\sim550$~kHz. (Bottom) A linear crystal of 12~ions in a
  very weak trap.}
  \label{fig:dv14ionGlamorShot}
  \label{fig:dv14ionGlamorShotSecondLabel}
\end{figure}
Ions are detected by observing ion fluorescence on a charge-coupled
device (CCD) camera or PMT. For QIP applications low background scatter
is required \cite{bible}. The background includes stray light scattered
from the cooling laser beam by the trap electrodes and other apparatus.
The Doppler cooling beam for dv14 had a waist of 40~$\mu$m trained
on the trap center 78~$\mu$m above the surface. The signal-to-background
for a single ion was observed to be better than~100:1, even with a
beam intensity 40~times the resonance saturation intensity (limited
by noise due the CCD chip). The camera used was an
\href{http://www.andor.com}{Andor, Inc.} iXon electron-multiplied charge-coupled
device (EMCCD). 

In the presence of Doppler laser cooling, single-ion lifetimes greater than one
hour were observed. Loading was reliable.  

Ion motional heating measurements were not conducted in this trap because it
failed before it was fully characterized. One of the electrodes snapped, shorting
an RF electrode to ground. The failure was due to excitation by the RF drive of a
mechanical mode in one of the long cantilevered electrodes. This excitation
phenomenon is of interest in its own right and is discussed in
Chapter~\vref{sec:ccool}. 

In retrospect, the trap would have been robust had I
used a continuous sheet of glass and fixed the trap electrodes in place by anodic bonding. This could be done
without exposing the ion to the dielectric glass surface, but at the cost of an
increase in the capacitance of the RF electrode (to ground thru the glass) which
causes RF losses.


\clearpage

\section{dv16: SOI silicon traps}
\label{sec:dv16}
\label{sec:dv16SOItop}

\subsection{SOI as a building material}
\label{sec:SOIasBuildingMaterial}
\begin{figure}[H]
  \centering
  \includegraphics[width=4in]{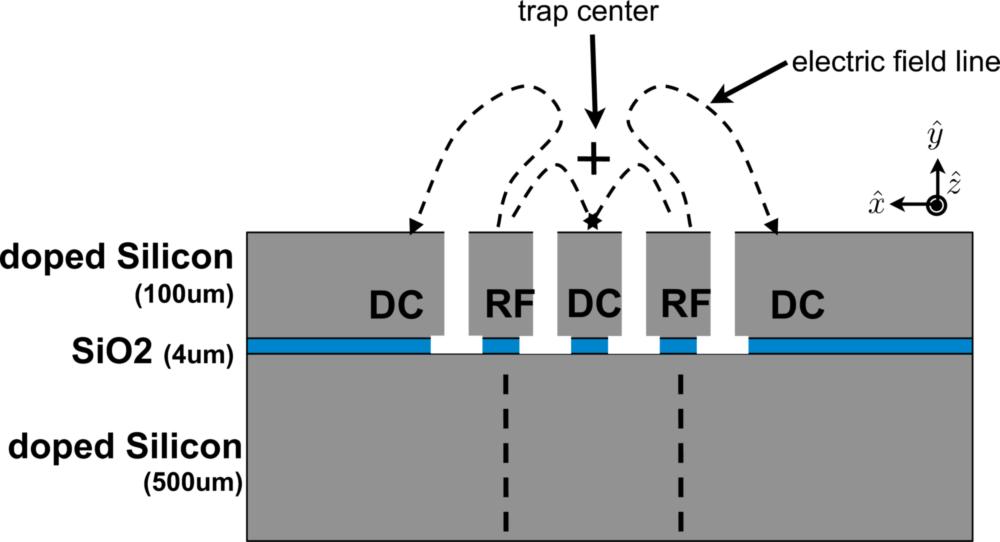}
  \caption[Figure showing an SOI ion trap chip in schematic.]
  {Figure showing an SOI ion trap chip in schematic. The dashed lines
  above the top silicon layer show electric field lines which form an
  RF quadrupole. The dashed lines on the 500~$\mu$m wafer indicate
  the location of a backside loading slot.}
  \label{fig:SOIsideview}
\end{figure}

The third and fourth doped silicon traps I built
improved upon dv14 by eliminating anodic bonding. Instead of 3~separate silicon
and glass wafers, I used a single silicon on oxide wafer (SOI). SOI is monolithic
heterostructure (Si:oxide:Si) requiring no assembly (eg anodic bonding) and it is
commercially available. The oxide is $\text{SiO}_{2}$ which has a loss tangent
which approaches that of fused silica (0.0002), making it two orders of magnitude
lower loss than the 7070~glass used in previous traps (see
Table~\vref{tab:lossTangentTable}). Use of SOI also reduced the device
fabrication time by 2/3 and greatly improved yield. This generation of traps was
also the first to use the octagon style vacuum system and UHV compatible chip
carrier (see Section~\vref{sec:vacuumApparatus}). See
Figure~\vref{fig:SOIsideview} for a schematic and Section~\vref{sec:SOI} for more
on SOI.

SOI traps also have advantages over other demonstrated surface electrode traps.
If a thick device layer is used (relative to inter-electrode gap size) the
shielding of structural dielectric surfaces is excellent. Also, fabrication and
alignment of holes for backside loading is easy. These features are demonstrated
in my traps. In the future, integration with on-chip control electronics is
possible using well established microfabrication techniques. One route for this
is to put CMOS on an ordinary silicon wafer and wafer or bump bonding it to an
SOI trap wafer \cite{kim2005a}.

The maximum oxide thickness in SOI wafers is traditionally 4~$\mu$m (for
thermally grown oxide). This small gap and its high dielectric constant
($\epsilon_{r}/\epsilon_{0}=3.9$ for thermal oxide), results in a larger
capacitance between RF electrodes and ground than in the anodically bonded traps.
The result is that the RF current which must flow to establish the trapping
fields may be significant. This can cause Ohmic heating and can degrade the RF
resonator quality factor. However, these effects did not appear to be serious
problems in my traps which had loaded quality factor of 80~and loaded easily.
Thicker oxide can be grown using plasma techniques (see Section~\vref{sec:SOI}
for more on SOI).  The thin oxide layer is also responsible for a larger
capacitance between ground and the control electrodes. Under some circumstances
this capacitance can act as a shorting capacitor while in others it can cause
intrinsic micromotion (see Section~\vref{sec:trapTestingRF} note~6).


I made two SET ion traps in SOI. Outside of demonstrating basic trapping, the
devices were designed with features to make possible tests of anomalous ion
heating \cite{turchette2000a, deslauriers2006a}. Both included tapered regions
where ion-electrode separation could be continuously varied. They also included
regions where the conducting surfaces were either bare doped silicon or
evaporated gold, permitting a heating rate comparison. The trap I call dv16k also
included a y-junction. It was fabricated and electrically tested but did not make
it to the optics table due to concerns about intrinsic micromotion (see Section
\vref{ions:sec:micromotion}). Another simpler trap, dv16m, successfully loaded
ions. A comparison between the gold and silicon traps in this device was
not conclusive due to other possible sources of ion heating.

The fabrication steps for SOI traps are discussed first. Then, the geometry and
performance of the traps is detailed.

\subsection{Fabrication}
\label{sec:SOItrapGenericFabricationSteps}
\begin{figure}[H]
  \centering
  \includegraphics[width=0.8\textwidth]{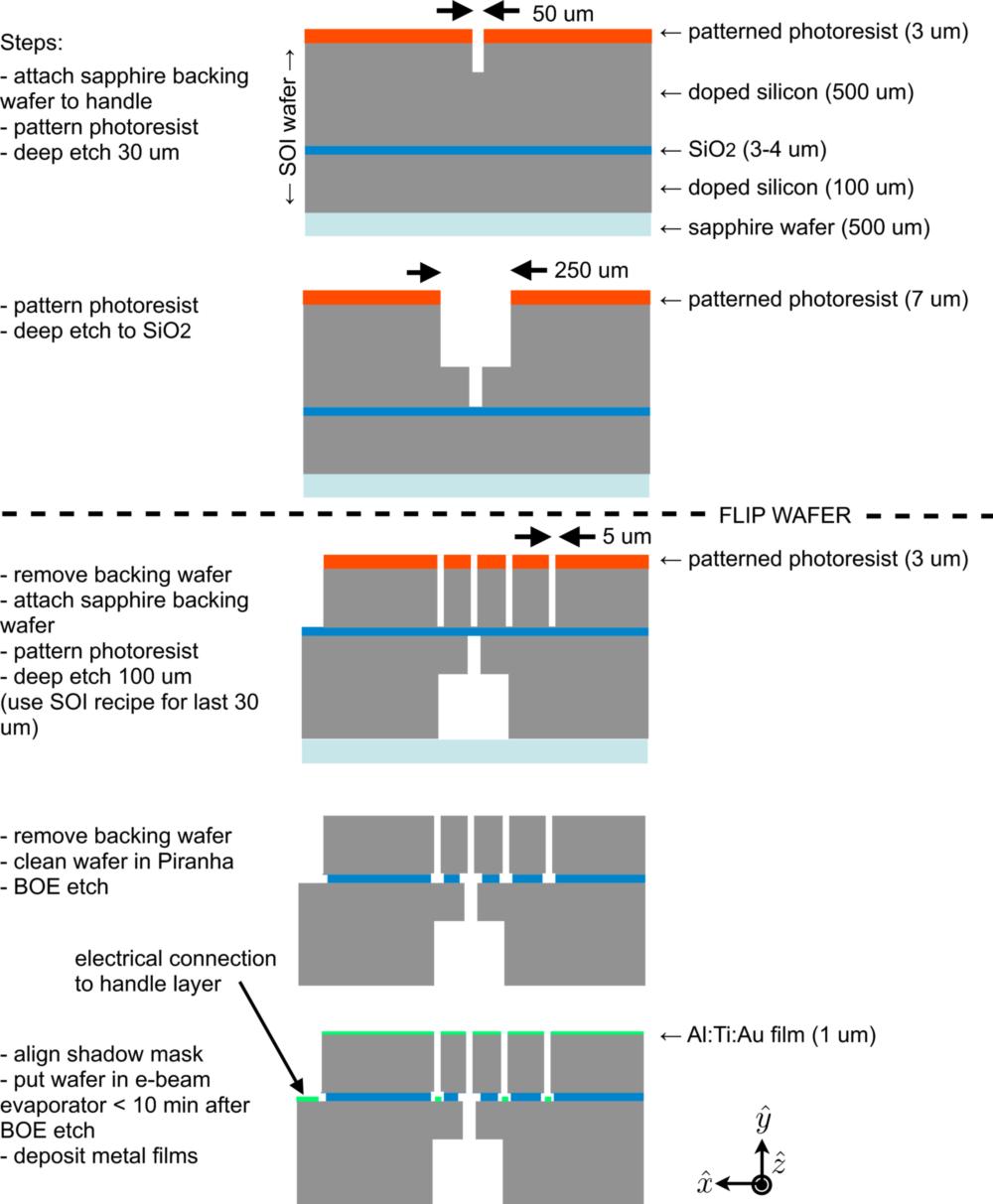}
  \caption[Overview of SOI ion trap fabrication steps.]
  {Overview of SOI ion trap fabrication steps. Note that drawings are
  not to scale. The depicted trap is an asymmetric 5-wire trap.}
  \label{fig:SOIfabOverview}
\end{figure}

The fabrication steps are as follows. All the fabrication steps were
done in the NIST clean room. 
\begin{compactenum} 
  \item etch loading slot in handle layer
  \item etch trap features in device layer
  \item clean in Piranha bath
  \item etch $SiO_2$ in BOE bath
  \item metallization
  \item dice into chips
  \item packaging: mount on chip carrier and wire bond
  \item continuity tests
  \item vacuum processing
\end{compactenum}
Steps 1-5~are illustrated in Figure~\vref{fig:SOIfabOverview}. The
trap structures were deep etched in doped silicon using an STS DRIE
(see Section~\vref{sec:DRIE}). The wafers are commercially available
silicon on insulator (SOI, see Section~\vref{sec:SOI}). Nine trap
chips were simultaneously etched and metalized on a single 3 inch
wafer. Each chip included a backside loading slot and required etching
from both the front and back sides. Care was taken to avoid surface
contamination which can cause needles to form during deep etching
(see Figure~\vref{fig:DRIE:needles}).

The SOI wafers used in my experiments consisted of the following layers. \\
\\
\begin{tabular}{l|ll}
	layer name& thickness& composition\tabularnewline
	\hline 
	device~layer& $100~\mu$m &0.005-0.020 $\Omega$-cm B doped silicon
	\tabularnewline 
	oxide~layer& $3~\mu$m &$SiO_{2}$\tabularnewline 
	handle~layer& $550~\mu$m& 0.005-0.020 $\Omega$-cm B doped
	silicon\tabularnewline
\end{tabular}\\
\\

\paragraph{deep etch of loading slot}
To provide good support for the device layer trap electrodes without causing
inconveniently sharp collimation of the neutral beam, the backside loading slot
was etched in a two step process. There was a a narrow 50~$\mu$m slot, followed
by a wide 250~$\mu$m trench. This etch sequence relied on deep etching being an
additive process. The resolution of the wide trench photolithography was reduced
due to presence the narrow slot which causes ripples in the photoresist.

\begin{compactenum}
  \item Note SOI handle layer thickness (from the manufacturer's specifications).
  \item Attach a sapphire backing wafer with wax, handle side up (see
  Section~\vref{sec:DRIE:thruWaferAdvice}).
  \item Spin on 3~$\mu$m photoresist. Follow the recipe in
  Section~\vref{sec:DRIE:deepEtchPR}.
  \item Expose the photoresist with the mask for the narrow slit pattern.
  \item Etch using DIRE recipe SPECB. Stop after 30~$\mu$m of material is
  removed. As a rough estimate, assume an etch rate of 3~$\mu$m/min.
  \item Spin on 7~$\mu$m photoresist. Follow the recipe in
  Section~\vref{sec:DRIE:deepEtchPR}. Don't worry about the slight ripple in photoresist thickness near
  the already etched features.
  \item Expose the photoresist with the mask for the wide trench pattern.
  \item Etch using DIRE recipe SPECB. Stop when 30~$\mu$m of material remains
  at the base of the wide trench. As a rough estimate, assume an etch
  rate of 3~$\mu$m/min.
  \item Etch using the DIRE recipe for SOI (see Section~\vref{sec:DRIE:dielectric}).
  Inspect narrow trenches with an optical microscope. They should be
  free of silicon. If not continue to etch.
  \item Release the backing wafer as described in Section~\vref{sec:DRIE:thruWaferAdvice}.
\end{compactenum}

\paragraph{deep etch of trap features}

\begin{compactenum}
  \item Note SOI device layer thickness (from the manufacturer's specifications).
  \item Flip the SOI wafer and reattach the sapphire backing wafer to the
  handle side (device side up).
  \item Spin on 7~$\mu$m photoresist. Follow the recipe in
  Section~\vref{sec:DRIE:deepEtchPR}.
  \item Expose the photoresist with the mask for the trap pattern.
  \item Etch using DIRE recipe SPECB. Stop when 30~$\mu$m of material remains
  at the base of the widest features. As a rough estimate, assume an
  etch rate of 3~$\mu$m/min.
  \item Etch using the DIRE recipe for SOI (see Section~\vref{sec:DRIE:dielectric}).
  Inspect narrow trenches with an optical microscope, they should be
  free of silicon. If not continue to etch. The thin $\text{SiO}_{2}$
  layer is translucent in the visible.
  \item The $\text{SiO}_{2}$ membrane may rupture due to trapped air in the
  loading slot. This does not seem to damage the trap structure.
  \item Release the backing wafer as described in Section~\vref{sec:DRIE:thruWaferAdvice}.
\end{compactenum}

\paragraph{clean the wafers}
The long deep etches in the preceding steps may result in baked-on photoresist
which can't be removed by solvents. This baked-on resist may appear as in
Figure~\vref{fig:DRIE:snakeSkin}. Remove it and other contaminants as follows.
\begin{compactenum}
  \item Clean the wafer in Piranha for 30-45~minutes as discussed in Step
  I of the recipe in Section~\vref{sec:RCA}. Use the 3~inch wafer holders
  described in Section~\vref{sec:teflonBaskets}.
  \item Dry the wafers as in Step IV.
  \item Etch the $\text{SiO}_{2}$ exposed at the base of the deep silcon 
  etch\footnote{This is done for three reasons. First, exposed dielectric is problematic
  for ion trapping. Second, electrical contact to the handle wafer from
  the top side requires removal of the insulator. Third, the etched
  features in the device layer will be used as a shadow mask to prevent
  shorting of adjacent trap electrodes during the metallization step.}.  
  The wet etch used in this step requires
  constant agitation so that fresh etchant is forced into the deep trenches.
  This is accomplished by agitation with an orbital shaker.
  
  \item Place the wafer in a Teflon or PTFE container on top of an orbital
  shaker.
  \item Immerse the wafer in buffered oxide etch (BOE) as in Step II of the
  recipe in Section~\vref{sec:RCA}.
  \item Etch for about 35~minutes. Fresh BOE at room temperature etches
  $\text{SiO}_{2}$ at a rate of about 87~nm/min (see
  Section~\vref{sec:oxideEtch:wet}).
  \item Inspect the narrow trenches with an optical microscope; the thin $\text{SiO}_{2}$
  layer is translucent in the visible. They should be free of $\text{SiO}_{2}$.
  If not continue to etch. Protect the wafer with a clean towel when
  placing on the optical microscope.
\end{compactenum}

\paragraph{metallization}
A 1~$\mu$m gold layer was deposited on much of the wafer
surface. At the periphery of each trap chip, this layer is used to
provide bond pads for connection to external potential sources. The
metal layer also lowers the impedance of the path from the bond pads
to trap electrodes. During the etch of the device side, a large region
of the handle wafer was exposed. Electrical contact to the handle
layer is made by contact to bond pads in this region. That is, electrical
contact to both the device and handle layers is achieved in a single
metallization step. Some regions of the wafer were protected from
metallization by a stainless steel shadow mask.
\begin{compactenum}
  \item Follow the FASTAU recipe in Section~\vref{sec:AuMetallization}. Use
  a shadow mask (see Section~\vref{sec:shadowMaskRecipe}). For advice
  on deposition of thick Au films see Section~\vref{sec:AuMetalThickFilms}.
  \item If the wafer is placed under vacuum in the e-beam evaporator within
  10~minutes of completing the BOE oxide etch, there's no need for the
  Axic etch step.
\end{compactenum}

\paragraph{wafer dicing}
An automated saw is used to dice the etched 3~inch silicon wafer into
9~chips. A regular silicon-cutting blade is used. See
Section~\vref{sec:dicingSaw} for details.

\paragraph{packaging}
The SOI trap chips were tested in the octagon style vacuum system.
They were mounted on special chip carriers. The mounting and wiring
methods are discussed in Section~\vref{sec:chipCarrierSocket}.

\paragraph{continuity tests}
Follow the advice in Section~\vref{sec:prebakeChecklist} prior to
insertion into the vacuum system for vacuum processing.

\paragraph{vacuum processing}
Prior to insertion into the final ion trap vacuum system the thermal
silicon oxide layer is stripped off the trap. A plasma etch was used
for this (see Section~\vref{sec:oxideEtch}). Be sure to ground all
the electrodes to prevent breakdown across the chip in the plasma.

\section{dv16k: SOI y-trap}
\label{sec:dv16k}
\label{sec:dv16kSOIytrap}
\begin{SCfigure}[40]
  \centering
  \includegraphics[width=0.6\textwidth]
  {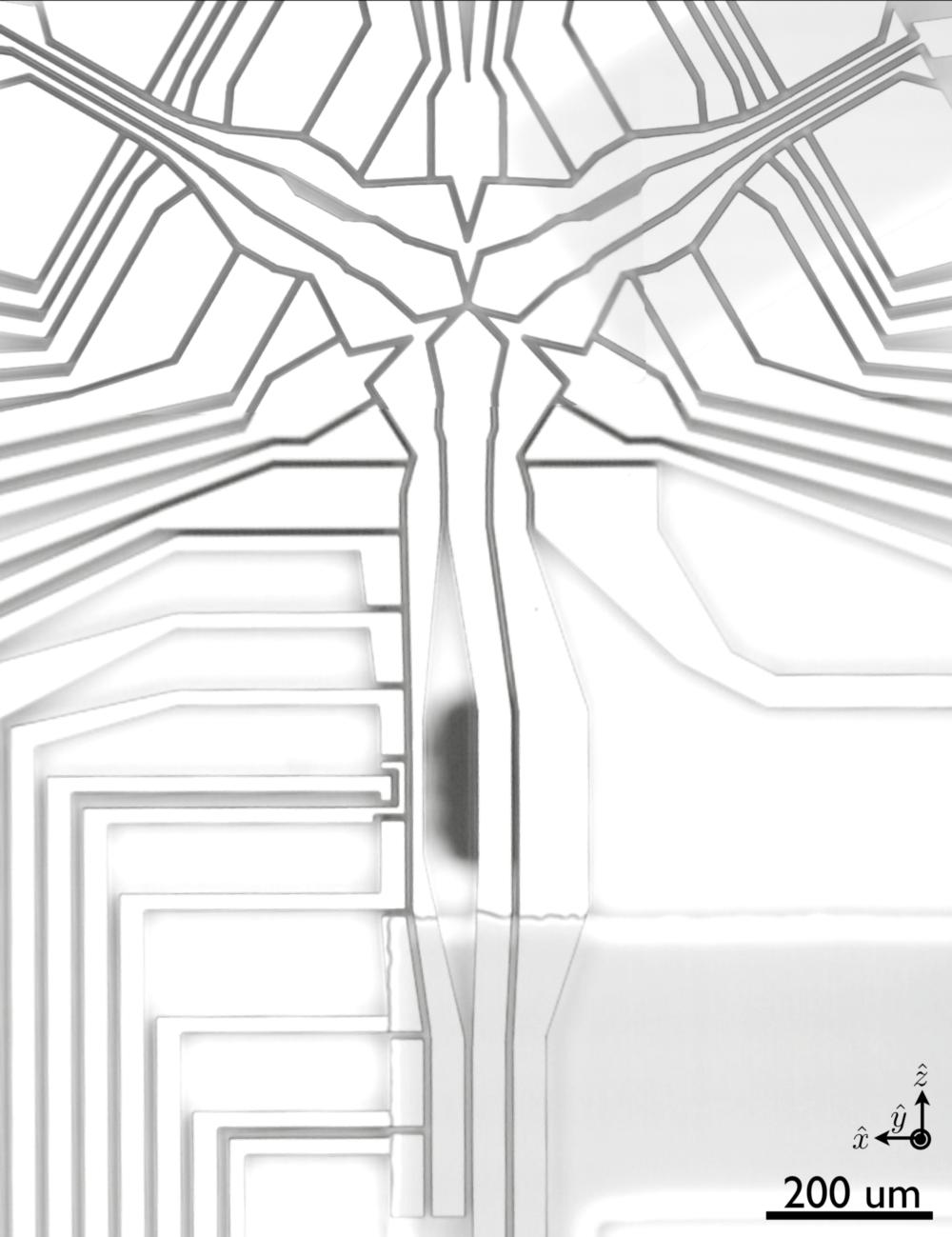}
  \caption[Optical micrograph of dv16k.]
  {Optical micrograph of dv16k.  Bright regions are gold coated.  Dark regions
  are bare boron doped silicon.}
  \label{fig:dv16kmicrograph}
\end{SCfigure}

My first SOI trap was built to demonstrate a surface
electrode y-junction. The electrode geometry was selected to minimize the height
of axial pseudopotential bumps along the length of the radial pseudopotential
minimum. The electrode shape at the y-junction was dictated by this minimization
\cite{wesenberg2008a}. This trap was named dv16k. See
Figure~\vref{fig:dv16kmicrograph} for a photo.


dv16k was fabricated and electrically tested, but not loaded with ions due to
concerns about a sizable phase difference $\phi_{\text{RF}}$ between the RF
electrodes. If $\phi_{\text{RF}}\neq0$, there is intrinsic micromotion which
can't be shimmed using control electrode potentials.  See
Section~\vref{sec:micromotion}.

\subsection{Electrical tests}
The loaded $Q_{L}$ of several trap chips was measured in an octagon style vacuum
system (see Section~\vref{sec:octagonVacSystem}). For these tests $10-20$~dBm RF
was applied to the resonator and all control electrodes were grounded. I observed
$Q_{L}=50-80$ for $\Omega/2\pi=30-35\text{MHz}$. Similar results were obtained
for the test trap chip in vacuum and in air. See Section~\vref{sec:QLchipLoss} to
put these measurements in context.

\subsection{Possible causes of $\phi_{\text{RF}}\neq0$}
\label{sec:dv16k:causesofphiRFneqzero}
\begin{figure}
  \includegraphics[width=6in]{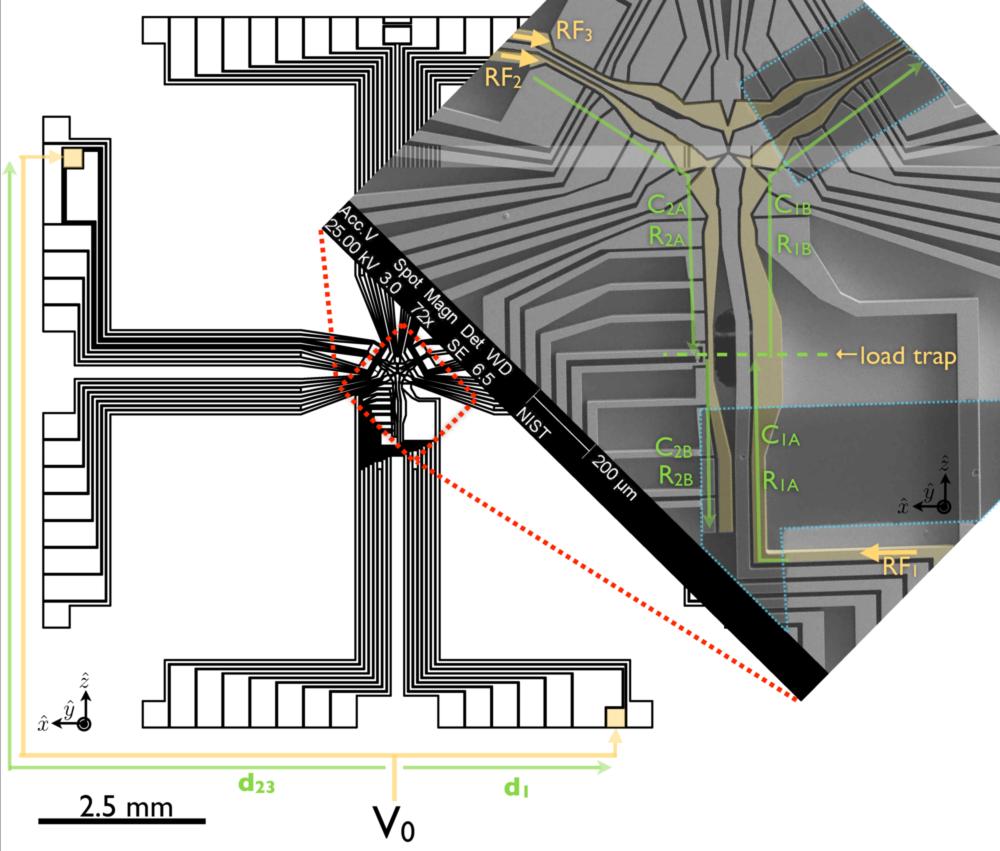}
  \caption[Schematic of SOI ion trap dv16k annotated to explain why it may have
  had excessive intrinsic micromotion.]
  {Schematic of SOI ion trap dv16k annotated to explain why it would be
  expected to have excessive intrinsic micromotion. At the center of the
  schematic (boxed in red) is the trapping region; an SEM blowup is to the right. 
  At the periphery of the schematic are wire bonding pads which supply
  the trap electrodes with RF and control potentials. The three RF electrodes
  (shaded in the SEM) are supplied RF potential by two bond pads (shaded
  in the schematic). The path length difference between them is 
  $\Delta_{d}=d_{23}-d_{1}\sim1$~cm.
  All bond pads and traces are 100~$\mu$m thick doped silicon overcoated
  by 1~$\mu$m gold, except for the two darker regions in the SEM (boxed
  in blue) which have no gold overcoating. The series resistance of the 
  electrodes is much higher without the gold. The micromotion analysis
  discussed in the text requires an estimate of the impedance of the traces (marked
  with green arrows) leading to the load trap.}
  \label{fig:dv16kRFsem}
\end{figure}

The underlying problems that contributed to $\phi_{\text{RF}}\neq0$ were that the
paths supplying potential to the RF electrodes followed different geometric
paths, with different coupling to ground. Consider the two paths ($d_{23}$ and
$d_1$) leading to the two RF electrodes at the load zone ($RF_1$ and $RF_2$). The
trap geometry is illustrated in Figure~\vref{fig:dv16kRFsem}. The path length
difference was $\Delta_{d}=d_{23}-d_{1}\sim1$~cm. For
$\Omega_{\text{RF}}/2\pi=35$~MHz, this path length difference would imply a phase
difference,
\begin{align*}
  	\phi_{\rm RF}&=360^\circ \times \frac{\Delta_{d}}{\lambda_{\rm
  RF}} 
  	=360^\circ \times 1~\text{cm}/8.6~\text{m}
	=0.4^{\circ}.
\end{align*}

\begin{SCfigure}[40]
  \centering
  \includegraphics[width=0.6\textwidth]{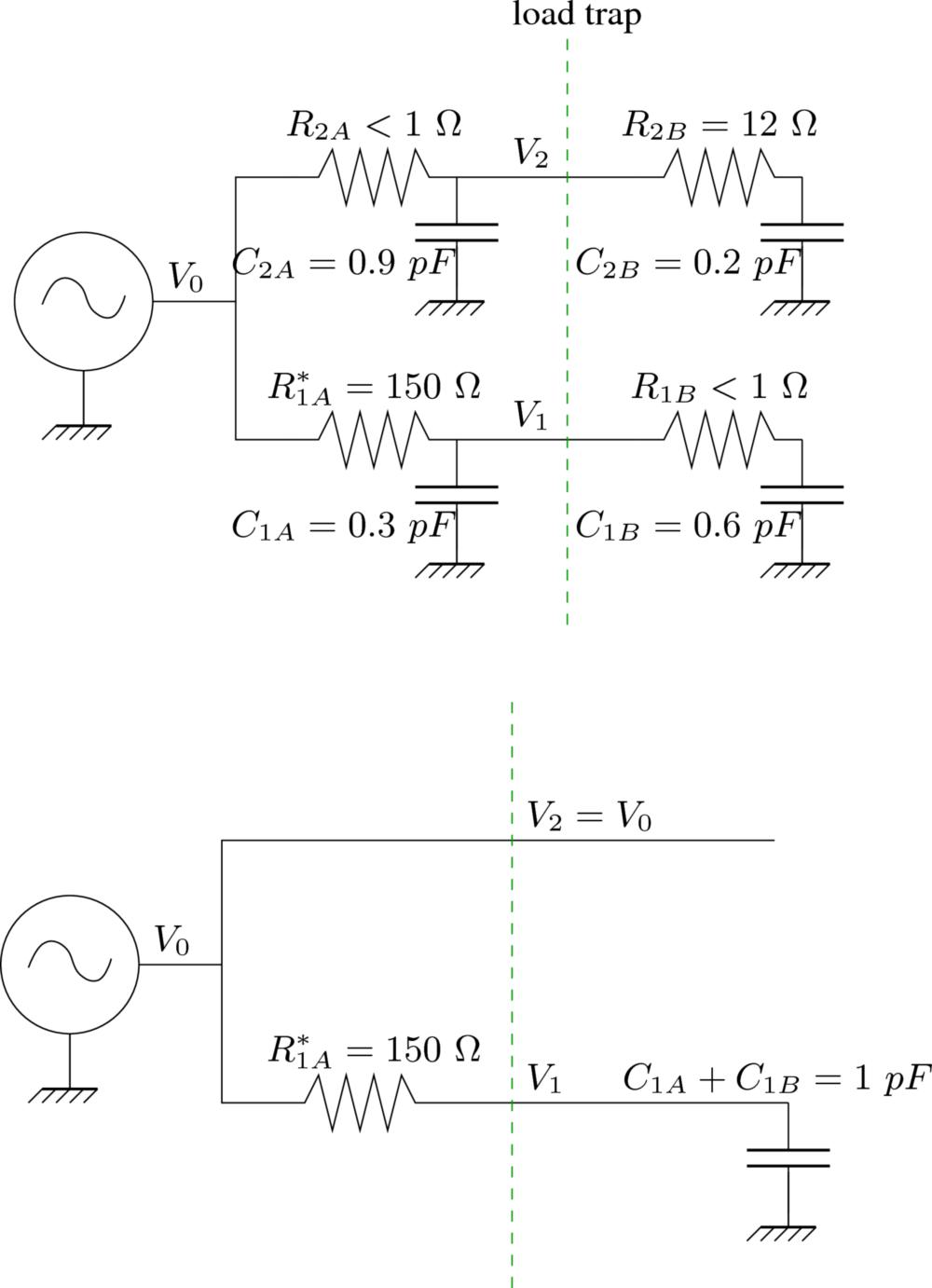}
  \caption[Schematic representation of the trap RF impedances.] {Schematic
  representation of the trap RF impedances. (top) A lumped circuit model for the
  phase difference between RF voltages at the loading region of trap dv16k.  The
  impedances were calculated assuming: $\text{SiO}_{2}$ thickness 3~$\mu$m,
  inter-electrode spacing $10~\mu$m, $\text{SiO}_{2}$ dielectric constant
  $\epsilon_{r}/\epsilon_{0}=3.9$,
  $\rho_{\text{Si}}=25,000\times10^{-6}~\Omega$-cm (experimentally determined as
  in Figure~\vref{fig:measureDeviceR}) and $\rho_{\rm
  Au}=2.27\times10^{-6}~\Omega$-cm. The resistance marked with {*} was measured
  experimentally using a pair of tungsten needle probes. (bottom)  Simplified
  model neglecting low resistance elements. This model places an upper bound on
  $\phi_{RF}$.}
  \label{fig:dv16kRFphaseCircuit}
\end{SCfigure}

The paths leading to the load zone for the two RF electrodes $\text{RF}_{1}$ and
$\text{RF}_{2}$ had different series resistances and different capacitive
couplings to their environment. A crude lumped circuit model is sketched in
Figure~\vref{fig:dv16kRFphaseCircuit} that takes into account the resistance of
the leads and their capacitive coupling thru vacuum to adjacent electrodes and
thru $\text{SiO}_{2}$ to the SOI handle. The model places distributed
capacitances at the end of a trace. A simplification to this model  is to ignore
all the elements with small impedance: keep only $R_{1A}$, $C_{1B}$ and $C_{1A}$.
This leaves a simple impedance divider.  Suppose the input voltage is
$V_{0}e^{i\omega t}$ and the impedance of the divider is $X+iY$. Then,
\begin{equation*}
  V_{1} = V_{0}e^{i\omega t}(X+iY) = V_{0}Ze^{i(\omega t+\phi)}
\end{equation*}
where $Z=\sqrt{X^{2}+Y^{2}}$ and $\phi=\tan^{-1}(Y/X)$.

\begin{SCfigure}[50] 
  \centering
  \includegraphics[width=0.4\textwidth] 
  {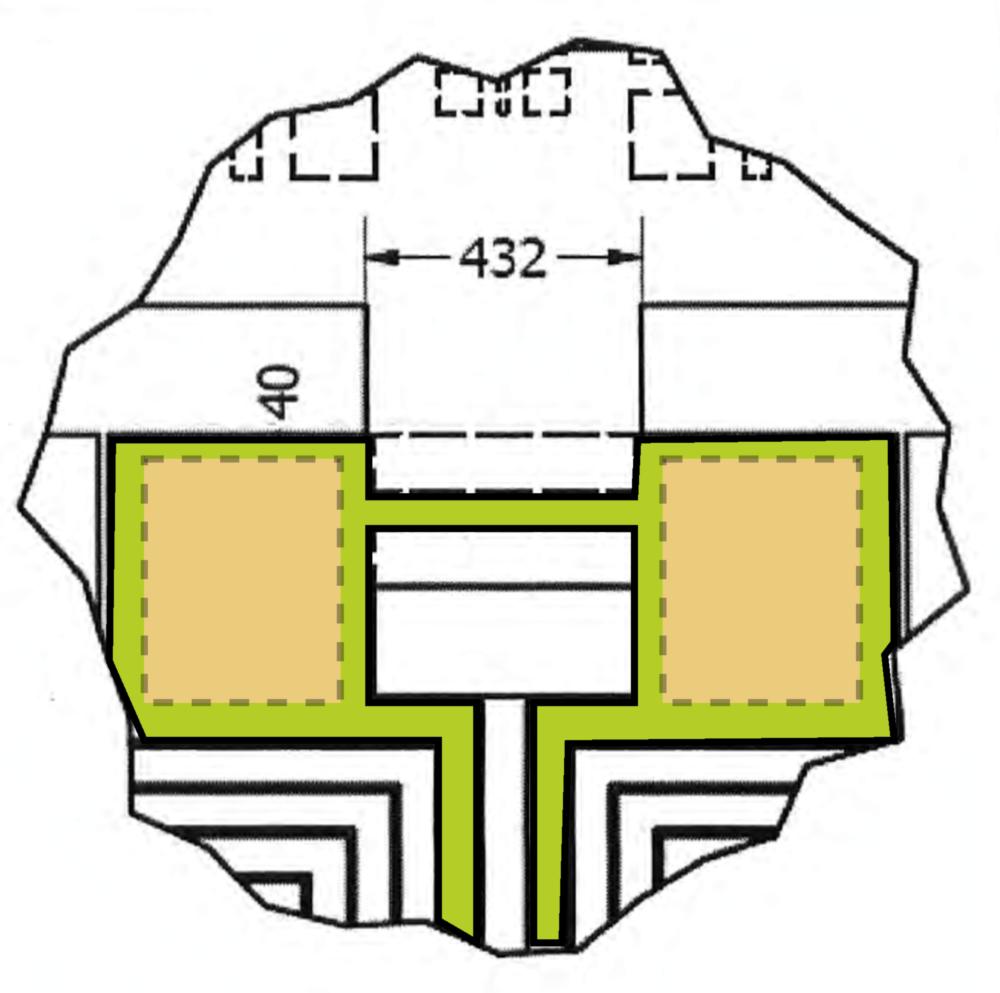}
  \caption[Doped silicon bulk resistance measurement structure.]
  {Doped silicon bulk resistance measurement test structure. Gold pads (dashed
  in the Figure) are connected by a $100~\mu$m tall, $40~\mu$m
  wide silicon bridge (shaded green in Figure). The silicon traces leading
  downward are electrically open. The resistance across this bridge
  was measured by plotting current vs voltage and confirmed to be Ohmic
  (see Section~\ref{sec:dopedSilicon}). From the resistance measurements the
  restivity was determined to be
  $\rho_{\text{Si}}=25,000\times10^{-6}~\Omega\text{cm}$. This is
  near the range ($5,000-20,000\times10^{-6}~\Omega$-cm)
  specified by the manufacturer (see Section~\vref{sec:SOI}). The dimension
  units are $\mu$m.}
  \label{fig:measureDeviceR}
\end{SCfigure}

$R_{1A}^*$ varied from chip to chip. It was not understood why it
was measured to be 6 to 22 times higher than expected from $\rho_{\text{Si}}$ (measured
independently as in Figure~\vref{fig:measureDeviceR}). For the RF phase
calculation assume the measured resistance, $R_{1A}=150~\Omega$.

For the values above and the simplified model in Figure~\vref{fig:dv16kRFphaseCircuit},
$\phi_{\text{RF}}=1.8^{\circ}+0.4^{\circ}=2.2^{\circ}$. This is an imperfect
model for $\phi_{\text{RF}}$; $C_{1A}$ and
$C_{1B}$ are distributed, not located at the tip of electrode $\text{RF}_{1}$
as the model assumes. 

The micromotion amplitude due to $\phi_{\text{RF}}$ is calculated in
\cite{berkeland1998a}.  An alternate derivation follows. Suppose there is a field
$E_{0}$ at the location of the ion due to each of the RF electrodes. Normally
these fields are in phase and cancel at the center of the quadrupole. If however,
there's a phase shift $\phi$ between them, there is a nonvanishing electric field
component at the center of the quadrupole.

\begin{eqnarray*}
E_{total} & = & E_{0}\cos(\Omega t)-E_{0}\cos(\Omega t+\phi)\\
 & = & E_{0}(\cos(\Omega t)-\cos(\Omega t)\cos(\phi)+\sin(\Omega t)\sin(\phi))\\
 &  & \mbox{assuming }\phi\ll1,\\
 & \approx & E_{0}(\cos(\Omega t)(1-1)+\phi\sin(\Omega t))\\
 & = & E_{0}\phi\sin(\Omega t)\end{eqnarray*}
 The ion's response to $E_{\text{total}}$ can be found by assuming
$x=x_{0}\sin(\Omega t)$. 

\begin{eqnarray*}
  q	E_{\text{total}}  & = & m\ddot{x}\\
  qE_{0}\phi\sin(\Omega t)  &=&  m\left(-\Omega^{2}\right)x_{0}\sin(\Omega t)\\
  \to  x_{0}  &=&  -\frac{qE_{0}\phi}{m\Omega^{2}}
\end{eqnarray*}
$x_{0}=x_{0\mu \text{m}}$ is the micromotion amplitude resulting from the
dipole term. For the dv16k load zone, 1~V on one of the RF electrodes
produces a 2000~V/m radial field at the trap center. Typical trap parameters:
$^{24}\text{Mg}^{+}$, RF potential 80~V (as determined from measurement of
secular frequencies and the electrode geometry), $\Omega/2\pi=35~\text{MHz}$.
For these parameters the expected micromotion for
$\phi_{\text{RF}}=2.2^{\circ}$ is $x_{0\text{$\mu$m}}=500$~\text{nm}.

Ion motion can modulate the laser beam phase at the position of the ion. Let
$\beta$ be the modulation depth. It is convenient to express micromotion
amplitude in terms of $\beta$,
\begin{equation*}	
	\mathbf{k\cdot x_{\mu \text{m}}}=\beta\cos(\Omega_{RF}),
\end{equation*}
where $k\cdot x_{\mu \text{m}}=k_{x}x_{0\text{$\mu$m}}$ is the overlap
of the micromotion and laser beam k-vector. Typically, the Doppler cooling
laser beam propagates at $45^\circ$ with
respect to $\hat{x}_{\mu \text{m}}$. So, with
$k_{x}=\frac{1}{\sqrt{2}}k=\frac{1}{\sqrt{2}}\frac{2\pi}{\lambda}$ and $\lambda$ = 280~nm,
$\beta=\frac{1}{\sqrt{2}}\frac{2\pi}{\lambda}x_{0\text{$\mu$m}}=8.1$. 
This micromotion amplitude would seriously affect the usefulness
of the recooling method of measuring ion heating and the usefulness of the trap
in general.  Note that when $\beta=1.43$, the carrier and first
micromotion sideband have equal strength (see
Section~\vref{sec:umDetectionByFluoresence}).

\subsection{Conclusion}
These calculations suggest that as fabricated the RF potentials were expected to
acquire an intolerable differential phase $\phi_{\text{RF}}$, leading to
considerable intrinsic micromotion. The problem was caused by a higher than
expected $R_{1A}$ (see Figure~\vref{fig:dv16kRFphaseCircuit}). This could have
been fixed by a second deposition of gold across the whole chip. However, this
would have removed the possibility of comparing the heating rate of gold with
that of bare doped silicon, an experimental priority. This problem was corrected
in this trap's successor, dv16m.

The SOI wafer used in this experiment was obtained from a distributor that
resells surplus wafers. SOI with higher conductivity is available by special
order.  See Section~\vref{sec:SOI}.

\clearpage

\section{dv16m: SOI linear trap}
\label{sec:dv16m}

\paragraph{introduction}
A second SOI trap was fabricated with a layout that minimized $\phi_{\text{RF}}$.
This was accomplished by simplifying the trap layout to include just two RF
electrodes and by making the path to the RF rails more symmetric. This trap had
two multi-zone experimental regions with and without gold overcoating the
silicon.  Part of these experimental regions were tapered so that the
ion-electrode distance $R$ varied from $45~\mu$m in the load region to $10~\mu$m.
These design features were included to test the dependence of ion motional
heating on material type and $R$.  I call this trap dv16m.

Trap dv16m successfully loaded ions. I was able to measure the ion secular
frequencies and found they agree reasonably well with simulation. Transport into
both the gold and bare doped silicon experimental regions was reproducible over a
period of several months. Heating rate measurements were made in the gold and bare silicon
experimental zones. However, a comparison of heating from these materials was not
conclusive due to other possible sources of ion heating.  This section presents
the geometry of dv16m and its trapping performance.  It then discusses some of
the tests done to identify sources of ion heating.

\begin{figure}[b]
  \centering
  \includegraphics[width=1\textwidth]
  {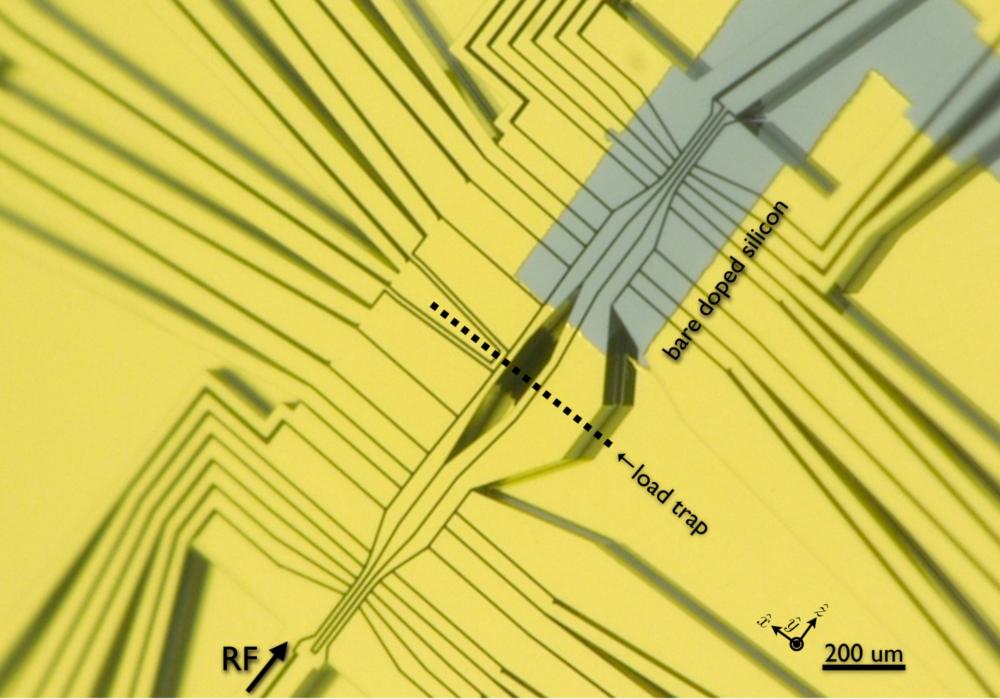}
  \caption[Optical micrograph of dv16m.]{Optical micrograph of dv16m.  The bright
  gold regions are gold coated; the darker regions are bare boron doped silicon.
  The camera is tilted, showing a perspective view of the trap which emphasizes
  the height of the SOI device layer ($100~\mu$m thick). }
  \label{fig:dv16m:OpticalPerspectivePhoto}
\end{figure}

\begin{table}
  \footnotesize
  \begin{tabular}{c|ccccc|cccc|ccc}

     trap & 
     $d$ & 
     $\frac{\Omega_{RF}}{2\pi}$ &
     $V_{RF}$  & 
     $Q$  &  
     $Ion$  & 
     $\frac{\omega_{z}}{2\pi}$ & 
 	 $\frac{\omega_{x,y}}{2\pi}$& 
     $N$  & 
	 $S_{E}(\omega_{z})$ & 
     $\tau_{dop}$ & 
     $\tau_{dark}$ & 
     $\phi_{eV}$ 
            \tabularnewline
     &$\mu$m&MHz&V& & &MHz&MHz&&$\frac{(V/m)^{2}}{Hz}$&min&sec&eV\\
\hline
 B{*}SOI  &  41 &  67 &
 50 &  90 & $^{24}Mg^{+}$  &
 1.125 & 7.80, & 9 & $1\times10^{-10}$ & 60 & 10 &$>60$ \tabularnewline
 SET&&&&&&& 9.25&&&&&\\
  dv16m&&&&&&&&&&&&
  \end{tabular}
  \caption[Trap characteristics of dv16m. ]
  {Trap characteristics of zone m370 in dv16m in the format of Table
  \vref{tab:theTrapComparisionTable}. Trap depth $\phi_{\text{ev}}$ and $V_{\rm
  RF}$ were estimated by simulation.}
  \label{tab:dv16mcharacteristics}
\end{table}

\subsection{Trap schematics}

\begin{figure}
  \centering
  \includegraphics[width=0.5\textwidth]
  {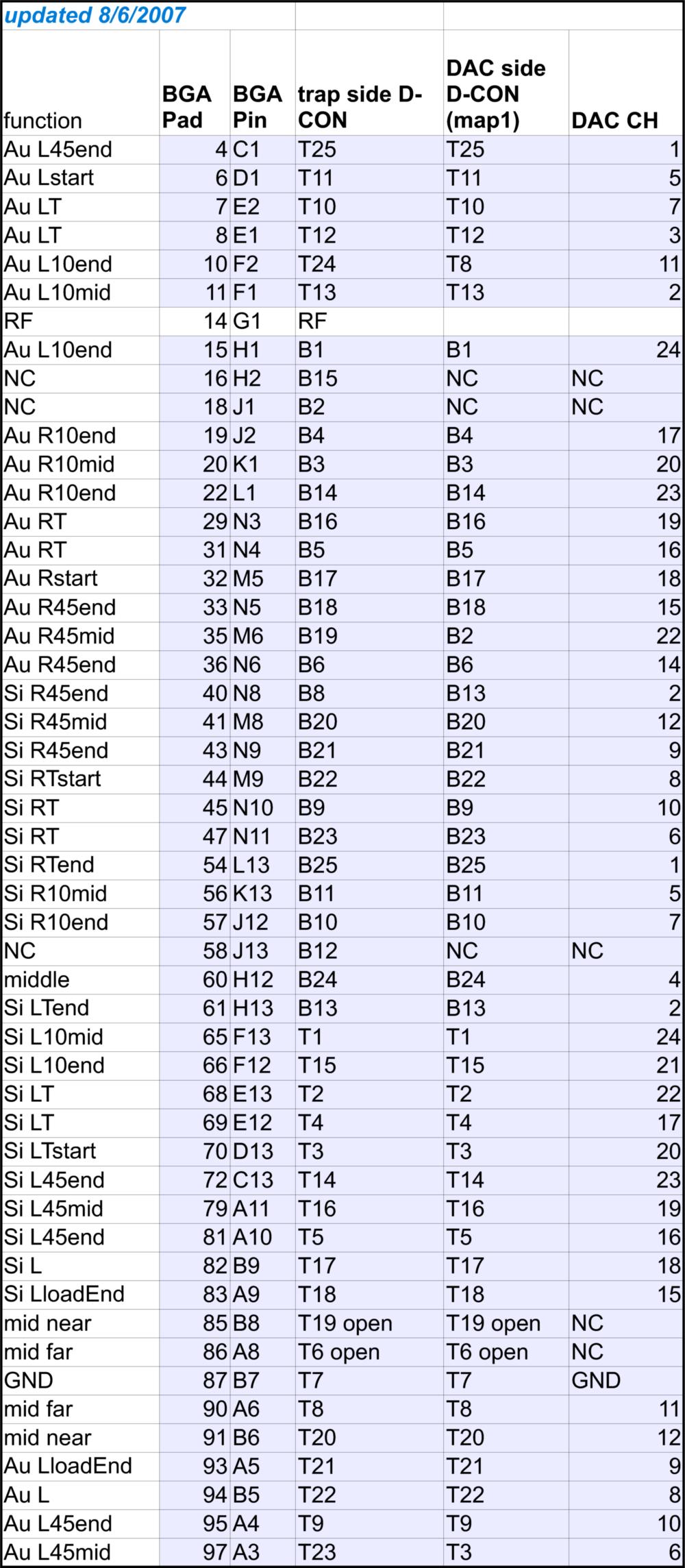}
  \caption[Wiring table for dv16m.]
  {Wiring table for dv16m.  This table shows details of the signal mapping
  between each DAC channel and the LTCC CPGA chip carrier.  See
  \vref{sec:chipCarrierSocket} for more about the chip carrier and its pin
  numbering.}
  \label{fig:dv16m:WiringTable}
\end{figure}

\begin{figure}
  \centering
  \includegraphics[width=0.9\textwidth]
  {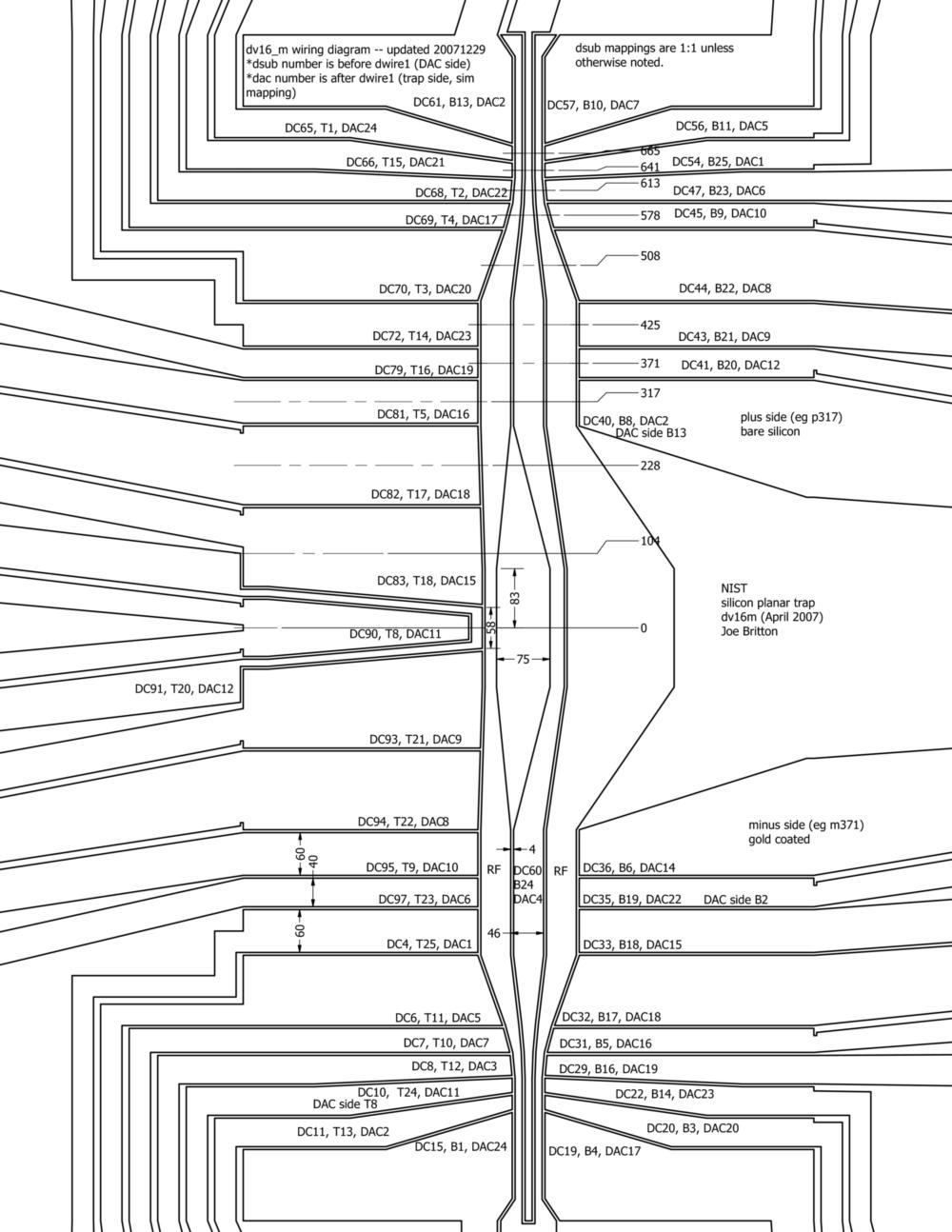}
  \caption[Electrode geometry for dv16m.]
  {Electrode geometry for dv16m.  The lengths are reported in $\mu$m.  
  Printed on each electrode is 
  the electrode name (e.g. DC83), it's corresponding pin on the UHV dsub
  vacuum feedthrough (e.g. T18) and  which DAC channel controlled its potential
  (e.g. DAC15).  As there were more electrodes than DAC channels, several of the
  electrode potentials were connected to the same DAC channel when doing so would
  not interfere with ion trapping and transport (e.g. DAC15: DC83 and DC33).  The
  nomenclature for particular trapping zones is p/m (plus/minus; the z-axis
  origin lies at the load zone) followed by the distance of
  the zone center from the center of the trapping region (e.g. p317).  The
  first bare silicon experimental zone is p317 and the first gold coated
  experimental zone is m317.}
  \label{fig:dv16m:WiringDiagram}
\end{figure}

Figure~\vref{fig:dv16m:WiringDiagram} is a dimensional schematic of the
trapping region which also shows how the 24 DAC channels are shared among the
trap electrodes. Figure~\vref{fig:dv16m:WiringTable} is a wiring table showing
the mapping between the trap chip and the dsub cables leading to the DACs.  
Table~\vref{tab:dv16m:schematics:DACtable} is a mapping of 
DAC channel to electrode number.

\begin{table}
  \centering
  \begin{tabular}{l|l|l}
      card (line)&channel&electrodes\\
      \hline 
      	1 (1)&1& DC54, DC4\\
    	1 (2)&2& DC40, DC61, DC11\\
    	1 (3)&3& DC8 \\
    	1 (4)&4& DC60\\
    	1 (5)&5& DC56, DC6\\
    	1 (6)&6& DC47, DC97\\
    	1 (7)&7& DC57, DC7\\
    	1 (8)&8& DC44, DC94\\
    	\hline
    	2 (1)&9& DC43, DC93\\
    	2 (2)&10& DC45, DC95\\
    	2 (3)&11& DC90, DC10\\
    	2 (4)&12& DC41, DC91\\
    	2 (5)&13& \\
    	2 (6)&14& DC36\\
    	2 (7)&15& DC83, DC33\\
    	2 (8)&16& DC81, DC31\\
    	\hline
    	3 (1)&17& DC69, DC25, DC19\\
    	3 (2)&18& DC82, DC32\\
    	3 (3)&19& DC79, DC29\\
    	3 (4)&20& DC70, DC20\\
    	3 (5)&21& DC66\\
    	3 (6)&22& DC68, DC35\\
    	3 (7)&23& DC72, DC22\\
    	3 (8)&24& DC65, DC15\\
  \end{tabular}
  \caption[Mapping of DAC channel to electrode number.]{Mapping of DAC channel to electrode number.
  There are three NI6733 DAC cards in this system each with 8 outputs (lines).  The 24 DAC
  channels are shared among 44 electrodes.  For example, electrodes DC54 and DC4 are 
  at the same potential supplied by DAC channel 1.}
  \label{tab:dv16m:schematics:DACtable}
\end{table}

\clearpage
\subsection{Simulation}
\label{sec:dv16m:simulation}
Calculation of the electronic potential and fields from a configuration of trap
electrodes is discussed in Section~\vref{ions:sec:BEM}.  This section presents 
typical results for these  
calculations specific to dv16m and gives numerical values for the fields in one of the 
experiment zones.  

\begin{figure}[ht]  
  \centering
  \includegraphics[width=1\textwidth]
  {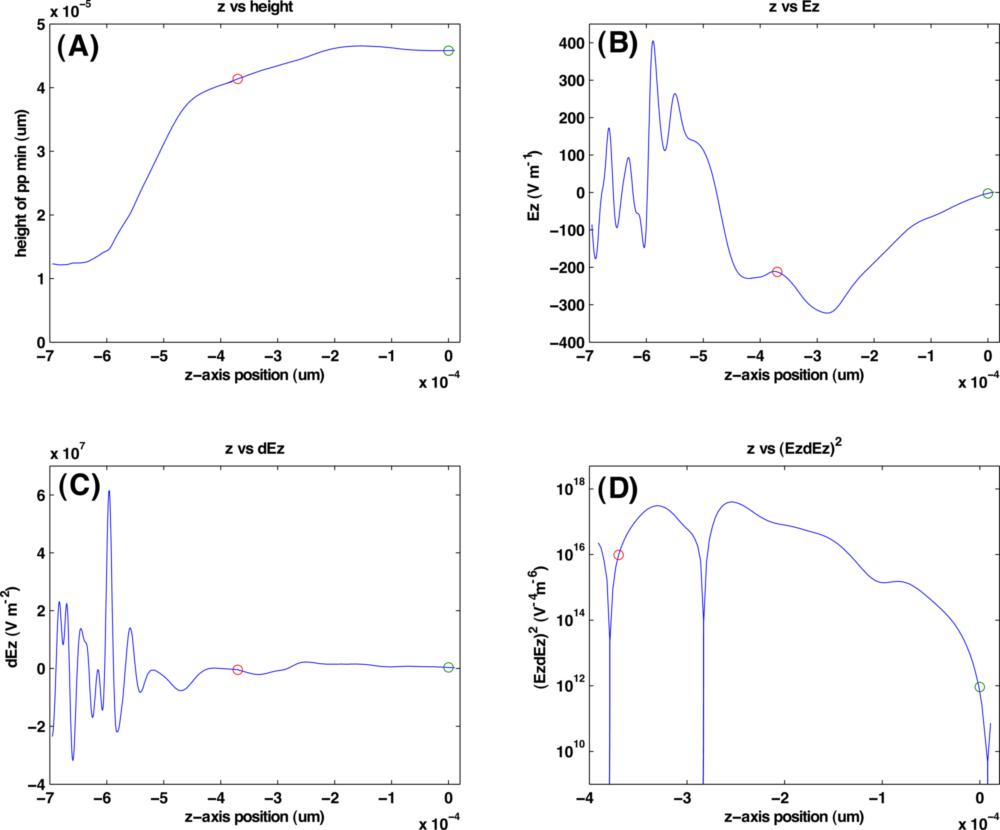}
  \caption[Simulated ion position and fields for trap dv16m.]
  {In all plots the parameters were evaluated 
  at the local pseudopotential 
  minimum in the x-y plane for each point along the z-axis.  The z-axis ranges
  from the load zone ($z=0~\mu$m) to experiment zone m370 ($z=-370~\mu$m) to
  the end of the taper ($z=-700~\mu$m).  The location of the load zone and
  m370 are marked with circles in each plot.     \\
  (A) Ion height above the wafer surface vs distance along the z-axis.\\  
  (B) z-component of the electric field $E_z(z)$ vs distance along the z-axis.\\
  (C) $\frac{\partial E_z(z)}{\partial z}$ vs distance along the z-axis.\\
  (D) $\left(E_z(z) \frac{\partial E_z(z)}{\partial z}\right)^2$ 
  vs distance along the z-axis.  
  }
  \label{fig:dv16m:simulation:4zoneFieldPlot}
\end{figure}

Figure~\vref{fig:dv16m:simulation:4zoneFieldPlot} shows some plots of the axial
component of the electric field along the trap axis.  It includes a plot of
$\left(E_z(z) \frac{\partial E_z(z)}{\partial z}\right)^2$ which can be a source
of ion heating as discussed in Section~\vref{sec:rfam:axialheating}.  Ions were
not transported successfully into zones with ion to surface distance smaller than
$40~\mu$m.  The simulation suggests that trapping would have been difficult due
to excessive motion and heating.

Figure~\vref{fig:dv16m:simulation:4zoneFieldPlot} shows data generated by
numerical simulation. It includes a plot of ion height
vs distance along the trap axis; the pseudopotential minimum draws closer to the
surface as the electrodes taper.

\begin{SCfigure}[10][ht]
  \centering
  \includegraphics[width=0.6\textwidth]
  {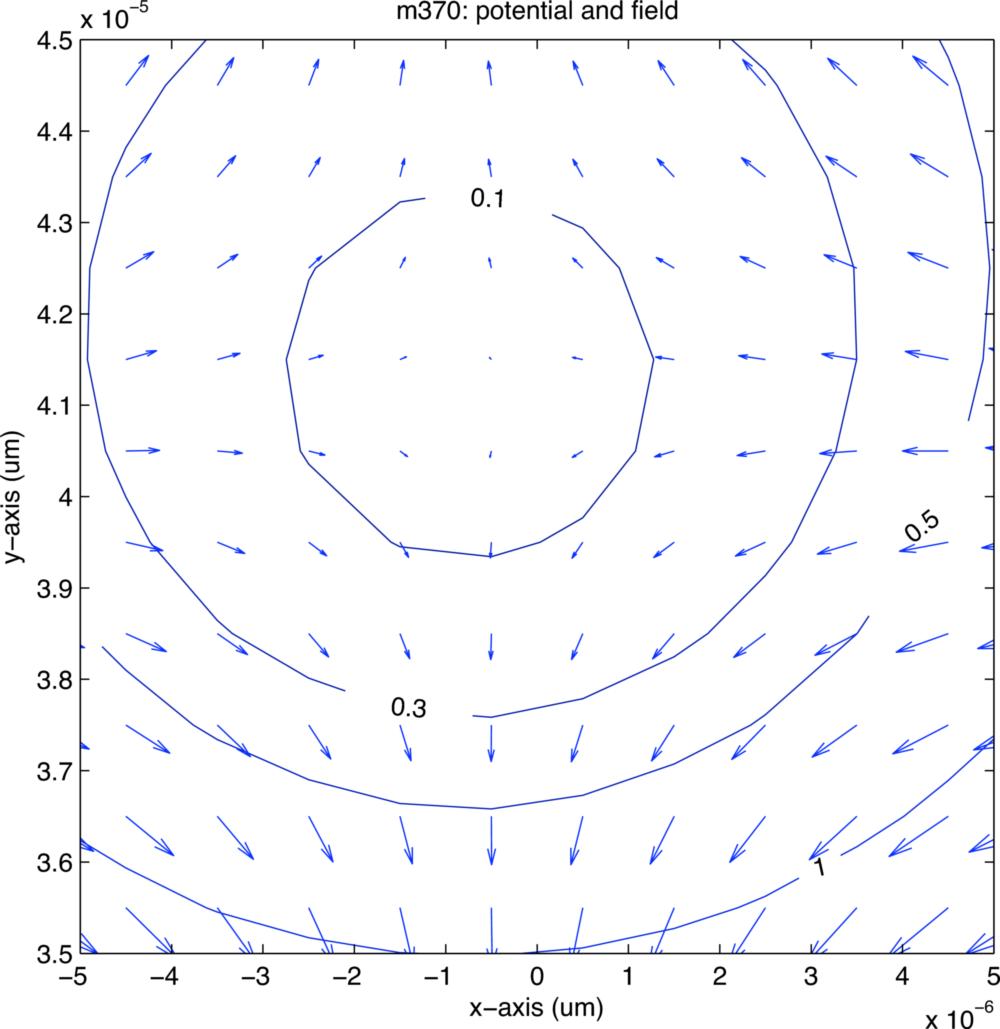}
  \caption[Plot of electric potential and fields in zone m370.]
  {Plot of electric potential and fields in zone m370 (an x-y slice at
  $z=-370~\mu$m) due to a potential of 1~V 
  on the RF electrodes.  The contour plot presents the numerical potential.
  The arrows are proportional to the electric field. 
  }
  \label{fig:dv16m:simulation:m370xyslice}
\end{SCfigure}

Figure~\vref{fig:dv16m:simulation:m370xyslice} is an equipotential plot superposed
with electric field vectors near the pseudopotential minimum in zone m370.  

\begin{figure}[b] 
  \centering
  \includegraphics[width=1\textwidth]{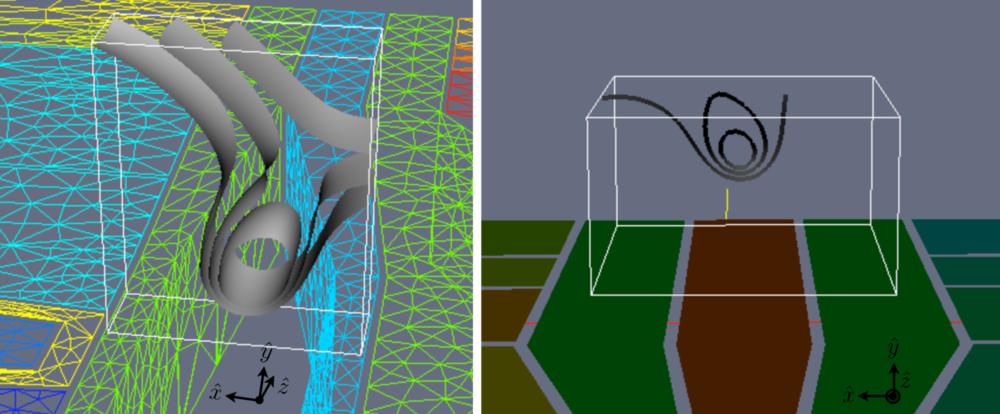}
  \caption[Plot of simulated trapping potentials for two zones in dv16m.] {Plot
  of simulated total trapping potentials (pseudopotential plus control
  potentials) for two zones in dv16m. The white cube in each plot bounds the
  volume where the potentials were extracted from the simulation. (left)
  Pseudopotential equipotential surfaces at 50, 70 and 100~meV for the load zone;
  the minimum lies 45.7~$\mu$m above the electrode surface. Visible in the
  background of this figure is the mesh grid used in the simulation.  Note that
  the mesh extends downward into the loading slot lying beneath the trapping
  region. (right) Pseudopotential equipotential surfaces at 10, 25, and 50~meV
  for the m370~zone; the minimum lies 41.6~$\mu$m above the electrode surface.
  Visible in the background of this figure are the trap electrodes; the
  simulation mesh is hidden from view.  A comparison with experiment is in
  Table~\vref{tab:dv16m:simulationVsExperiment}. The control potentials used for
  these traps are given in Tables~\vref{tab:dv16m:potentials}.}
  \label{fig:dv16m:simulationParaview}
\end{figure}

Figure~\vref{fig:dv16m:simulationParaview}
shows equipotential surfaces for the control potentials listed in
Table~\vref{tab:dv16m:potentials}.

\subparagraph{dv16m-p371 fields}
\label{sec:dv16m:p371CEi}
This section gives numerical values for the fields in the bare silicon experiment zone
p371.

Let $C_{E_x}\left(\overset{\rightharpoonup }{x}\right)$ be the the $x$ component
of the electric field at position $\overset{\rightharpoonup }{x}$ for 1~V on a
particular electrode (all other electrodes grounded). For example, for a 1~volt
potential on endcap electrode DC72 the electric field strength at the ion
position $\overset{\rightharpoonup }{x}$ (x~=~371~$\mu$m, y~=~0~$\mu$m,
z~=~41~$\mu$m) is,
\begin{align}
  C_{E_x}&=-226~V/m \label{eq:dv16mp371x}\\
  C_{E_y}&= +476~V/m \label{eq:dv16mp371y} \\
  C_{E_z}& =-457~V/m \label{eq:dv16mp371z}.
\end{align}

Let $D_{E_x}\left(\overset{\rightharpoonup }{x}\right)$ be the the $x$ component
of the electric field at position $\overset{\rightharpoonup }{x}$ for 1~V
potential on the RF electrodes.  At the same position 
$\overset{\rightharpoonup }{x}$, the electric field strength at the ion is,
\begin{align} 
  D_{E_x}&=0~V/m \label{eq:dv16mp371Ex}\\
  D_{E_y}&= 0~V/m \label{eq:dv16mp371Ey} \\
  D_{E_z}& = 231~V/m \label{eq:dv16mp371Ez}\\
  \frac{\partial E_z}{\partial z}& = 23,500~V/m^2 \label{eq:dv16mp371dEz}.
\end{align}
The axes are the same as those in Figure~\vref{fig:dv16m:simulationParaview}.
These numbers are used to calculate how the potential noise on RF and control 
electrodes can cause motional heating.  Call this case dv16m-p371. 

\begin{SCfigure}
  \centering
  \includegraphics[width=0.65\textwidth]{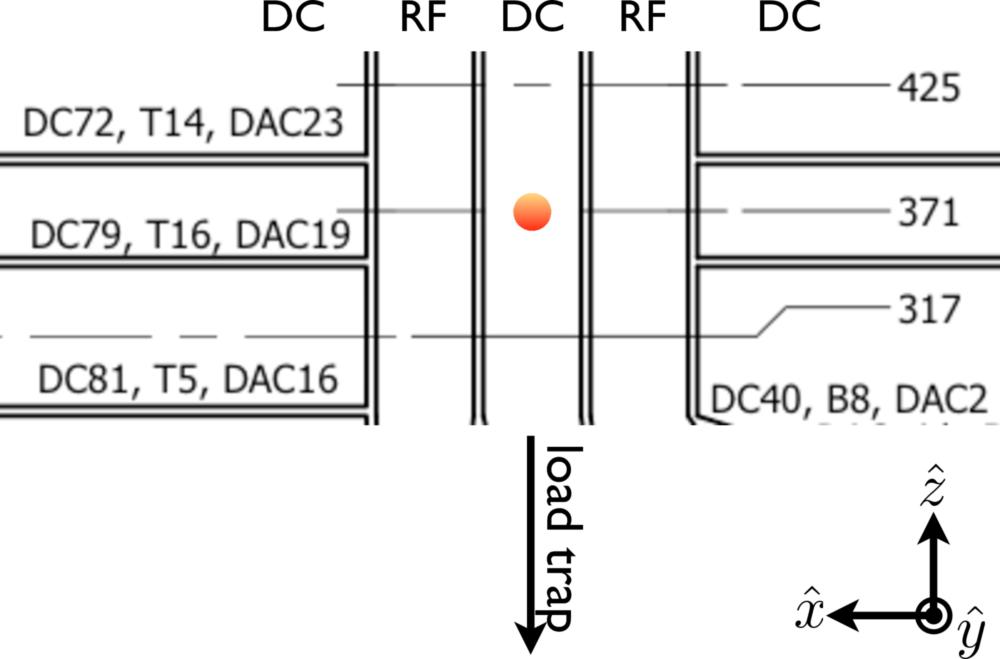}
  \caption[Ion trap schematic for first experimental zone of trap dv16m.]
  {Ion trap schematic for first experimental zone of trap dv16m.
  This zones lies $+371~\mu$m from the load zone and is $41~\mu$m above the
  trap surface.  The electric fields quoted in the text
  (Equations~\eqref{eq:dv16mp371x} to \eqref{eq:dv16mp371z}) are derived from
  simulation.}
  \label{fig:dv16mZone370}
\end{SCfigure}

\clearpage

\clearpage
\subsection{Comparison with simulation}
\label{sec:dv16m:comparisonWithSimulation}
Numerical simulation can be used to optimize trap parameters such as axis rotation
and trap depth.  This section compares trap parameters measured in the experiment
with those predicted by simulation. Figure~\vref{fig:dv16m:simulationParaview}
shows equipotential surfaces for the control potentials listed in
Table~\vref{tab:dv16m:potentials}.
Table~\vref{tab:dv16m:simulationVsExperiment} compares simulation with
experiment.

\begin{table}[b]
  \begin{tabular}{ll|ll|lll|ll}
    zone & & $\Omega_{\rm RF}$&$V_{\rm RF}$&$\omega_z/2\pi$&
    $\omega_{x}/2\pi$& $\omega_{y}/2\pi$& $\phi_{\rm ev}$& $\theta_{\rm
    tilt}$\\ & & (MHz)&(Volts)&(MHz)&(MHz)&(MHz)&(meV)&(degrees)\\
    \hline
    load& experiment&43.45&	&2.48&7.15&	8.24		&				&\\
                            &simulation&43.45&46.0&2.02&7.15&7.82&67&31$^\circ$\\
                            m370& experiment&67.20&   &1.125&7.80& 9.25  &    &\\
                            &simulation&67.20&50.0&1.082&7.80&8.43&25&2.5$^\circ$
  \end{tabular}
  \caption[Table comparing numerical simulation with experiment for
  dv16m.]{Table comparing numerical simulation with experiment for dv16m.
  In the simulation, $V_{\rm RF}$ was set by the requirement that $\omega_x$
  match the experiment.  The secular frequencies were measured by the endcap
  tickle technique (see
  Section~\vref{sec:secularFreqMeasurement:endcapTickle}).  
  For the load trap the simulation does not match experiment very well. 
  This may be because the loading slot geometry was not accurately specified
  in the simulation due to its nonuniform geometry. The control potentials for these traps are given in
  Table~\vref{tab:dv16m:potentials}.}
  \label{tab:dv16m:simulationVsExperiment}
\end{table}

\begin{table}[b]
  \centering
  \begin{tabular}{lllll}
      \multicolumn{2}{c}{endcaps}&\multicolumn{2}{c}{middle}&center\\
      DC83&DC93&DC60&DC91&DC90\\
      \hline
      +4.00&+4.00&+0.53&0.00&-9.38
  \end{tabular}
  \\
  \begin{tabular}{lllllll}
      \multicolumn{2}{c}{endcaps A}&\multicolumn{2}{c}{endcaps
      B}&\multicolumn{2}{c}{middle}&center\\ 
      DC95&DC33&DC4&DC36&DC60&DC97&DC35\\
      \hline
      +0.20&+0.20&+0.71&+0.71&+0.11&-1.81&-0.58
  \end{tabular}
  \caption[dv16m control electrode potentials]{This table lists the control 
  potentials used for the experimental data in 
  Table~\vref{tab:dv16m:simulationVsExperiment}.  (top) dv16m load zone potentials.
  (bottom) dv16m m370 zone potentials.  The endcaps
  are asymmetrically biased in order to induce a slight twist to the trap
  axes.}
  \label{tab:dv16m:potentials}
\end{table}

\subsection{Transport waveforms}

\begin{figure}
  \centering
  \includegraphics[width=0.9\textwidth] 
  {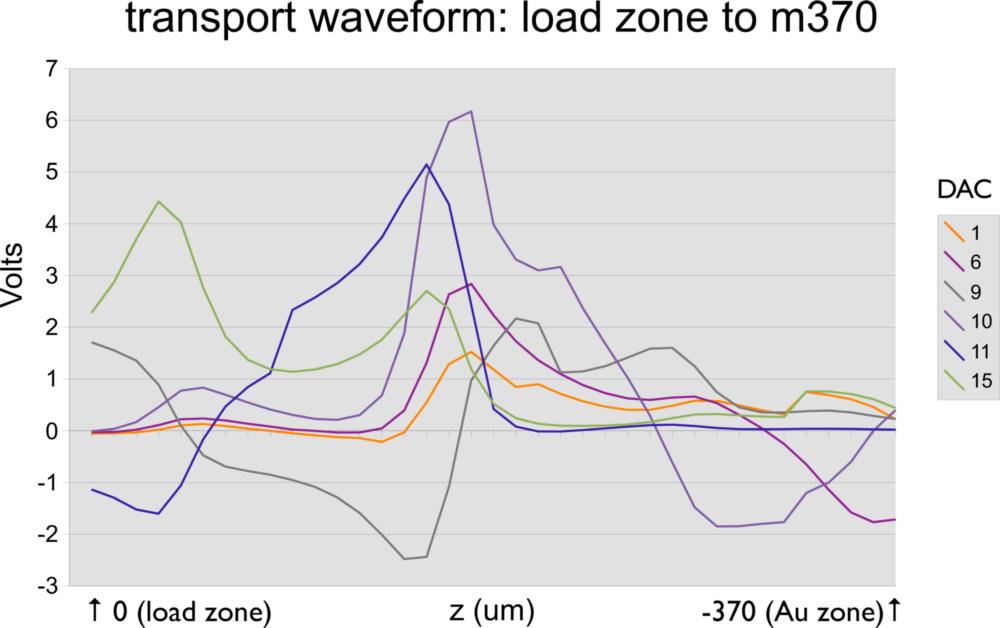}
  \caption[Transport waveform for moving ions from the load zone to zone m370.]
  {Transport waveform for moving ions from the load zone to zone m370.  To make
  the plot easier to read, only the potentials for electrodes in the load zone
  (DAC15, DAC11 and DAC9) zone and one side of the the destination zone (DAC10,
  DAC6 and DAC1) are plotted. The full waveform specifies potentials for all 24 DAC channels.
  The waveform step size is 10~$\mu$m and the full transport typically took
  about 1~ms. At $z=0~\mu$m the load zone ends caps are at $\sim2.0~$V. At
  $z>0$ the ion is initially pushed along the trap axis in the negative direction by the
  potential on DAC15. The waveform ends in a weaker trap at m370 with the
  endcaps at $\sim 200$~mV.}
  \label{fig:dv16m:LoadTom370WvfPlot}
\end{figure}

Ion transport from one zone to another was accomplished by applying appropriate
potentials to the control electrodes.  These \emph{waveform} potentials
translated the axial trapping minimum.  This waveform is generated from numerical
simulation.  For example, an excerpt of the waveform used to transport an ion
from the load zone of dv16m to the m370 experiment zone (a distance of
370~$\mu$m) is plotted in Figure~\vref{fig:dv16m:LoadTom370WvfPlot}. The initial waveform
for this transport was produced by solving for the control potentials required to
place an ion at the pseudopotential minimum with an axial frequency of
$\omega_z=1.5$~MHz at each of 38 positions along the z-axis.  This solution
didn't provide useful transport. The final waveform in the Figure was generated
by incrementally moving from one point of the simulated waveform to the next and
manually tweaking the control potentials to keep the ion's fluorescence a maximum.
The same technique was used to generate a waveform for transport from the load
zone to p370.  A possible reason the simulation alone did not produce successful
transport is imperfect representation of the trap geometry in the simulation.

Once a satisfactory waveform was determined, ions could be reliably transported
from the load zone to the experimental regions.  For example, transport from the
load zone to an experimental zone and back could be repeated hundreds of times at
a $1~Hz$ repetition rate without ion loss.  For this looping test a pair of
Doppler cooling laser beams were positioned in the starting and ending traps. 


\clearpage

\subsection{Electrical tests}
\label{sec:dv16m:elecTests}
The loaded $Q_{L}$ of several trap chips was measured in an octagon style vacuum
system (see Section~\vref{sec:octagonVacSystem}). For these tests $20$~dBm RF was
applied to the resonator with all control electrodes grounded. I observed
$Q_{L}=80-100$ for $\Omega/2\pi=43~\text{MHz}$.  The DC resistance between
ground and the control electrodes was >10~M$\Omega$, with typically one low
resistance short (10-100~k$\Omega$) between a control electrode and ground per
chip.  Motional heating tests were done with a chip where $Q_{L}=90$ for
$\Omega/2\pi=67-74~\text{MHz}$ and with R(DC22-GND)=40~k$\Omega$ (all others
>10~M$\Omega$). See Section~\vref{sec:QLchipLoss} to put these measurements in
context.


\subsection{Dark lifetime}
In well behaved traps, ions will remain trapped even in the absence of cooling
laser beams. For example, without laser cooling $^9Be^+$ ions survived in a room
temperature trap for several minutes~\cite{rowe2002a}.  The lifetime of several
ions in the dark for the load zone of dv16m is in
Table~\vref{tab:dv16m:ionDarkTime}. The longest observed lifetime was
10~seconds.\footnote{
  Note that during my first attempt at loading ions in trap dv16m I observed
  longer dark lifetimes than reported in Table~\vref{tab:dv16m:ionDarkTime}.  For
  example, one ion survived for 60~sec in the dark twice in succession. For this
  trap $\omega_z/2\pi\cong500$~kHz and $\omega_{x,y}/2\pi\cong8$~MHz. 
  Unfortunately, this trap stopped loading after about two days and I installed a
  new copy of dv16m in the vacuum system. This thesis discusses this second trap
  which worked reliably for $>6$~months.}

\begin{table}
  \centering
  \begin{tabular}{llllllll}
      ion no.& 1/2&1&2&3&4&5&10\\
      \hline
      	1&1&3&1&2&1&1&0\\
	    2&1&3&1&2&1&1&0\\
	    3&1&3&1&2&1&0&\\
	    4&1&3&1&2&1&2&0\\
	    5&1&3&1&2&1&2&1\\
	    6&1&3&1&2&1&1&0\\
  \end{tabular}
  \caption[Table reporting ion lifetime without Doppler cooling in dv16m.]
  {Table reporting ion lifetime without Doppler cooling in dv16m.  The load
  zone was used for these measurements and the control potentials were supplied
  by batteries.  For this experiment I did not measure the trap frequencies;
  from previous experiments I estimate that they were $\omega_z/2\pi\cong2.5$~MHz and
  $\omega_{x,y}/2\pi\cong8$~MHz.  Each line in the table above corresponds to an ion that was
  loaded into the trap then Doppler cooled for 3 minutes to permit the vacuum
  to recover.  Then, the cooling laser beam was blocked for durations of 1/2 to
  10~seconds.  The number in the table is the number of times the beam
  was blocked for each ion.  Ion loss is indicated by an entry of 0.  
  For example, the first ion survived for periods in the dark of 0.5~sec, 1.0~sec,
  1.0~sec, 1.0~sec, 2.0~sec, 3.0~sec, 3.0~sec, 4.0~sec, 5.0~sec and was lost 
  when it was put in the dark for 10.0~sec.  
  Note that when the same
  experiment was done with the DACs instead of batteries (and only the usual
  $1~k\Omega$ and $820~pF$ filtering) the maximum dark lifetime was 1~second.}

  \label{tab:dv16m:ionDarkTime}
\end{table}


\clearpage
\subsection{Motional heating}
\label{sec:dv16m:heating}
\begin{figure}[b] 
  \centering
  \includegraphics[width=0.9\textwidth]
  {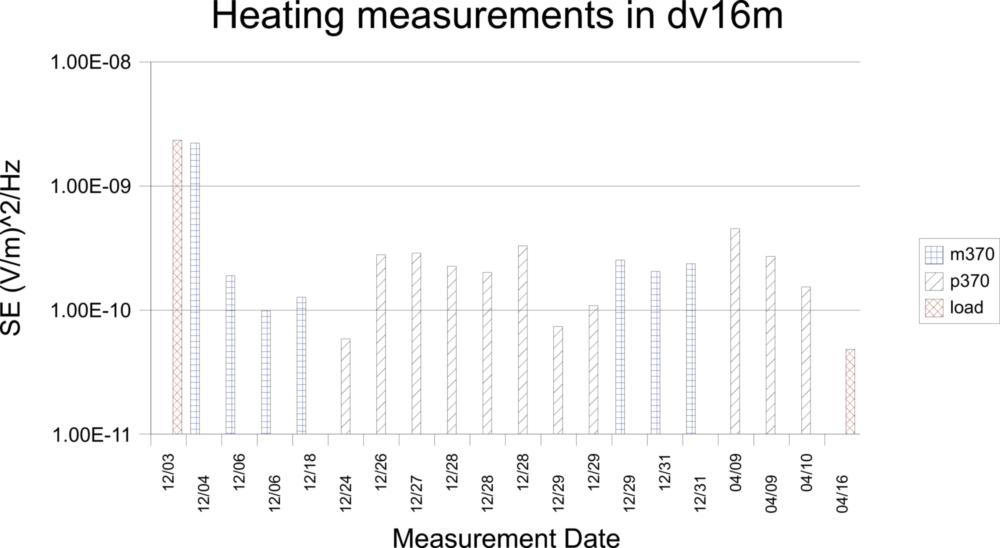}
  \caption[Motional heating measured in dv16m over time.]
  {Motional heating measured in dv16m over time. For each measurement the control
  potentials were supplied by batteries and the micromotion was nulled along the 
  cooling laser beam.}
  \label{fig:dv16m:motionalHeatingOverTime}
\end{figure}

In addition to trapping and transport, one of the main objectives for this trap was a
comparison of the motional heating for two material types: gold and bare boron
doped silicon.  It was anticipated that the heating for these two surfaces would
differ.  However, within the sensitivity of my experiments I observed no
difference.  Moreover, I saw that the electric field spectral density at the ion
in dv16m was about ten times higher than the best NIST SET~\cite{seidelin2006a}
(see Figure~\vref{ions:fig:ionHeatingScatterPlot}) and the heating rate varied by up
to an order of magnitude day to day (see
Figure~\vref{fig:dv16m:motionalHeatingOverTime}).  The large day to day variation
is not common for traps tested at NIST, however it is a common occurrence that
nearly identical traps have observed to have widely differing heating rates
(again, see Figure~\vref{ions:fig:ionHeatingScatterPlot}).  It was hypothesized that
these observations were due to an external source of electric field noise. This
section discusses several possible sources of heating and related experiments.

The following frequencies are known to couple to the motion of a single ion.
\begin{compactenum}
  \item $\omega_i$ (direct drive)
  \item $\Omega_{\rm RF}\pm n~\omega_i$ for $n\in {1,2,\ldots}$ 
     (only if micromotion along the $i^{th}$ trap axis)
  \item $\Omega_{\rm RF}\pm 2~n~\omega_i$ for $n\in {1,2,\ldots}$ 
    (only if no micromotion along the $i^{th}$ trap axis),
\end{compactenum}
where $\Omega_{\rm RF}$ is the trap drive frequency and $\omega_i$ is the
secular frequency along the $i^{th}$ principle axis $i\in\subset\{x,y,z\}$. 
The third coupling mechanism
is called parametric heating. Electric field noise at these frequencies may
originate from a variety of sources.  There are several sources of noise power
at these frequencies in the trap.  

\begin{compactitem}
  \item noise in potentials applied to trap control electrodes (e.g. DAC noise)
  \item noise in the RF drive potential (e.g. amplifier and RF oscillator noise)
  \item Johnson noise from finite electrode impedance and trap high pass RC
  filters.
  \item Johnson noise from the RF resonator's finite impedance
  \item ambient laboratory fields coupled to trap apparatus (e.g. poorly shielded
  cables) 
  \item field-emitters (due to sharp points on the electrodes)
  \item collisions with background atoms
  \item fluctuating patch potentials at the electrode surfaces
\end{compactitem}

The sections that follow discuss some of these noise sources and attempts to
mitigate them when possible. They are also discussed in the literature
\cite{turchette2000a,bible,deslauriers2006a,labaziewicz2008a,leibrandt2007a}.

\paragraph{noisy control potentials}
\label{sec:dv16m:noisyPotentialSources}

One source of electric field noise is potential noise on trap electrodes. The
resulting fields directly couple to an ion's motion via its charge and can cause
motional heating.  One way to characterize the heating is with $\Gamma_{0\to1}$,
the heating rate from the ground state to the first excited motional state in
units of quanta/ms.  To characterize the noise fields, the electric field noise
spectral density $S_E(\omega)$ is usually specified instead of $\Gamma_{0\to1}$. 
This is because $S_E(\omega)$ is useful when comparing traps; it normalizes
across ion species and trap frequencies. These quantities are connected by
\cite{turchette2000a},
\begin{equation*}
  \Gamma_{0\to 1}=\frac{q^2S_E(\omega )}{4 m \hbar  \omega }
\end{equation*}
where $q$ is the ion charge, $m$ is the ion mass, $\omega$ is the motional
secular frequency, and $S_E(\omega )$ is the electric field noise spectral 
density at $\omega$.

The potentials applied to trap control electrodes were generated by DC supplies,
DACs and batteries (see Table~\vref{tab:apparatus:controlElectrodeSupplies}). The
noise of these potential sources (at 0~V and 1~V outputs) when they were
generating a constant potential was measured at $\omega_z/2\pi=3$~MHz.  A
spectrum analyzer (SA) with a DC block capacitor and a
$50~\Omega$ input impedance was used to measure the potential noise just before
the RC filter boxes (see Figure~\vref{fig:trapTestingControl} to see how they are
situated with respect to the trap). The SA noise floor was -145~dBm/Hz =
$12\times 10^{-9}~V/\sqrt{Hz}$.

The trap RC filters have a 3~dB point of approximately 
200~kHz (C=820~pF, R=1000~$\Omega$). 
The noise potential attenuation factor for the RC at $\omega_z/2\pi=3$~MHz is 
240.

Table~\vref{tab:dv16m:controlElectrodeNoise} summarizes the observed noise 
and the calculated noise at the trap electrodes after attenuation by the RC filters.

\begin{SCtable}[5]
  \centering
  \begin{tabular}{l|ll}
      &$(V/\sqrt{Hz})$&$(V/\sqrt{Hz})$\tabularnewline
      & Noise at SA &Noise at trap \tabularnewline  
    \hline
    DAC & $125\times 10^{-9}$ &$523\times 10^{-12}$\tabularnewline
    DC supply & $125\times 10^{-9}$ &$523\times 10^{-12}$ \tabularnewline
    battery & $12\times 10^{-9}$ & $52\times 10^{-12} $\tabularnewline
    $50~\Omega$ &$12\times 10^{-9}$&\\
  \end{tabular}
  \caption[Control potential supply noise.]
  {Control potential supply noise at $\omega/2\pi=3$~MHz in a 1~Hz bandwidth.
  The second column is the expected
  noise at the trap electrodes after attenuation by the RC filters.  $50~\Omega$
  termination was used to measure the SA noise floor, which was far larger than the 
  Johnson noise of the resistor.}
\label{tab:dv16m:controlElectrodeNoise}
\end{SCtable}

\subparagraph{p371: test case heating}
For zone p371 (Section~\vref{sec:dv16m:p371CEi}) 
calculate ion heating due to potential noise at a trap electrode.  For
example, for $^{24}Mg^+$ an electric field spectral density $S_E(\omega)$ of
$1.23\times10^{-11}$ $\left(\frac{V}{m}\right)^2\frac{1}{\text{Hz}}$ in the
$z$-direction at $\omega_z/2\pi=3$~MHz in gives rise to a heating rate of
1~quanta/ms. A potential of
$7.7\times10^{-9}$~$\frac{V}{\sqrt{Hz}}$ at electrode DC72 produces such a
field. Note that it is the noise power spectral density that is typically
measured on a spectrum analyzer
$\frac{S_V(\omega)}{50~\Omega}=-149~\frac{\text{dBm}}{\text{Hz}}$.  By way of
comparison, the Johnson noise potential of a 50~$\Omega$ resistor at room
temperature is $0.9\times10^{-9}/\sqrt{Hz}$, 
which as a noise power is $-167.8~\frac{dBm}{Hz}$ (into $50~\Omega$).  

For all
heating measurements for dv16m, I used shielded batteries referenced 
to the trap ground.  Control electrode noise is therefore not expected
to make a large contribution to motional heating in my experiments.


%
%
%

\paragraph{RC filter Johnson noise}
\label{sec:jNoiseAndRCfilters}

Thermal electronic noise results from finite electrode resistance and from other
lossy circuit elements. Since the wavelength of the secular frequencies and RF
drive are large compared with the trap geometry, its acceptable to treat the
fundamentally microscopic thermal noise with lump-circuit models. 
In practice, the dominant
lossy element is the resistor forming part of the RC low pass filter attached to
each control electrode (see Figure~\vref{ions:fig:properRC})\cite{bible}.

Typically, $R=1~k\Omega$ and $C=820$ pF. The $1~k\Omega$ resistor in this filter
has an associated room temperature Johnson noise. Its voltage spectral density
$S_{V_n}$ is
\begin{align*}
  S_{V_n}&=4 k_BT~R\\
  	&=1.6\times10^{-17}~V^2/{\rm Hz}\\
  \sqrt{S_{V_n}}&= 4~nV/\sqrt{\rm Hz}.
\end{align*}
As a power spectral density (into $50~\Omega$) this is $-154$~dBm/Hz 
which is difficult to measure on many spectrum analyzers.

The noise at the trap electrodes is less than $S_{V_n}$ due to the control
electrode RC filter attenuation. The filters can be modeled as a voltage noise
source and noise free resistor. The voltage spectral density attenuation is
\begin{equation*}
  A_{LP}=\frac{1}{1+(\omega  R C)^2}
\end{equation*}
for the impedance divider. At $\omega/2\pi=3$~MHz and the RC values above, the
attenuation of $S_E(\omega)$ is $240$.

\begin{SCfigure}
  \centering
  \includegraphics[width=0.65\textwidth]{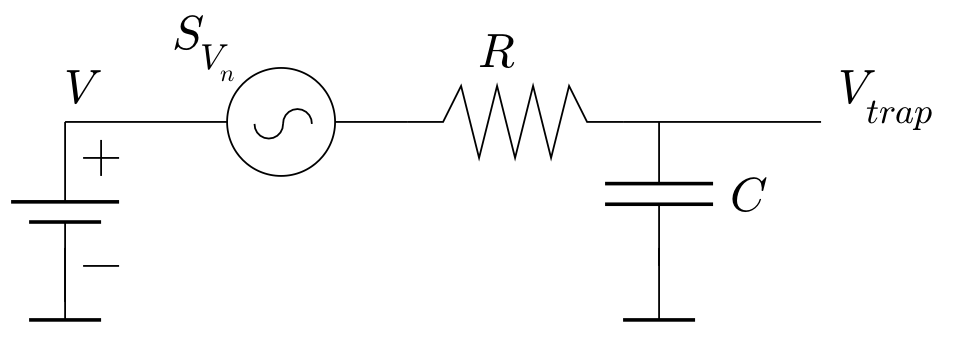} 
  \caption[Lumped circuit model for the control electrodes' low pass
  filters.]
  {Lumped circuit model for the control electrodes' low pass
  filters. The resistor is modeled as a perfect resistor in series with
  a Johnson noise source. }  
  \label{fig:resistorJohnsonNoise}
\end{SCfigure}

It is straightforward to calculate the expected heating rate as
\begin{align*}
  \Gamma_{0\to 1}&=4\times \frac{e^2}{4 m \hbar  \omega}
  \times S_{V_n} C_{E_z}^2\times
  A_{\text{LP}} \\
  	&=0.005~\text{quanta/ms}
\end{align*}
where $C_{E_z}$ is as for dv16m-p371 (see Equation~\vref{eq:dv16mp371z}) and
the factor of $4$ accounts for the four endcap control 
electrodes in this trap. This rate is
well below that observed experimentally. Therefore, the LP filter Johnson 
noise doesn't explain the observed ion heating.

\paragraph{RF resonator Johnson noise}
\label{sec:RFresonatorNoiseAndFiltering}

\begin{table}[b]
  \centering
  \begin{tabular}{l|l|ll|ll}
    &$\Omega_{\rm RF}$ & $\Omega_{\rm RF}-\omega_z$&$\Omega_{\rm
    RF}-\omega_x$&$\omega_x$&$\omega_z$ \\
    \hline
    $\lambda/2$ attn. (dB)&0&-5.9&-15.2&-30.7&-31.7\\
    noise power (dBm)&-139&-145&-154&-170&-171
  \end{tabular}
  \caption[Noise from RF resonator resistance.]{Noise from RF resonator
  resistance.  For, $\Omega_{\rm RF}/2\pi=71~\text{MHz}$, $Q_L=80$, $R=
  38~k\Omega$ and $T = 300$~K.  $\omega_z/2\pi=3$~MHz and
  $\omega_x/2\pi=10$~MHz are typical axial and radial secular frequencies,
  respectively.}
  \label{tab:L4attenuationTable}
\end{table}

The trap RF potential is derived from a voltage step-up transformer fed by
an RF oscillator. In the lab this takes the form of a $\lambda /4$ RF
cavity resonator. See Section~\vref{sec:cavityCoupling} for details on
resonant cavity oscillators modeled as lumped circuit RLCs and 
Section~\vref{sec:apparatus:trapRF} for details on the experimental apparatus.

On resonance, the cavity's impedance is resistive and is a source of Johnson
noise. The parallel resistance $R$ can be calculated and is typically around
$38~k\Omega$ (see Section~vref{sec:apparatus:trapQLoss}). 
The resonator line shape filters this and other RF noise
sources (e.g. amplifier noise), 
\begin{equation*}
  S_{V_n}(\omega) = 4 k_BT R \times \left(1+4~Q_L^2\left(
  \frac{\omega-\Omega_{\rm RF}}{\Omega_{\rm RF}} \right)^2\right)^{-1},
\end{equation*}
where $\Omega_{\rm RF}$ is the RF resonant frequency, $Q_L$ is the resonator
loaded quality factor, $T$
is the RF resonator temperature and $\omega$ is the observation frequency. 
Table~\vref{tab:L4attenuationTable} summarizes the amplitude of RF power due to
Johnson noise at various frequencies which can cause ion motional heating 
(Section~\vref{sec:RFAM}).

%
%
%
%

This noise power at $\omega_x$ and $\omega_z$ is far too small to cause noticable
heating in my traps (see Section~\vref{sec:dv16m:p371CEi}). As we will see in
Section~\vref{sec:RFAM} noise at secular difference frequencies may however play
a role in motional heating.


\paragraph{ambient laboratory fields}
\label{sec:heatingDueToLabFields}

Ion heating may also result from electronic noise originating outside the
trapping apparatus. These fields could couple to the trap structure via poorly
shielded/grounded cables. To check for this, ion heating was measured before and
after disconnecting the following apparatus connected to the vacuum system.
\begin{compactitem}
  \item ion gauge controller and cable
  \item Mg oven current supply and cable
  \item titanium sublimation pump controller and cable
\end{compactitem}
No change in heating was observed.

Ambient laboratory fields might also couple to the trap structure through optical
ports in the vacuum housing. I placed a pickup coil at the trap imaging port,
flush with the quartz window.  This port lies 3~cm above the trap.  The pickup
coil was attached to an operational amplifier (OpAmp) buffer and monitored on a
spectrum analyzer (SA).  The noise spectrum observed near  is in the top half of
Figure~\vref{fig:pickupCoilAmbientNoiseVsInjectedNoise}.

Since the environmental noise had narrow spectral features, it would be hit or miss
whether the noise had much power at the ion's axial secular frequency $\omega_z$.
It was not experimentally convenient to sweep $\omega_z$. Therefore, I injected
broad current noise on a second coupling loop concentric with the pickup loop. I
used an amplitude commensurate with the ambient noise as judged by the pickup
loop.

\begin{SCfigure}
  \centering
  \includegraphics[width=0.6\textwidth]{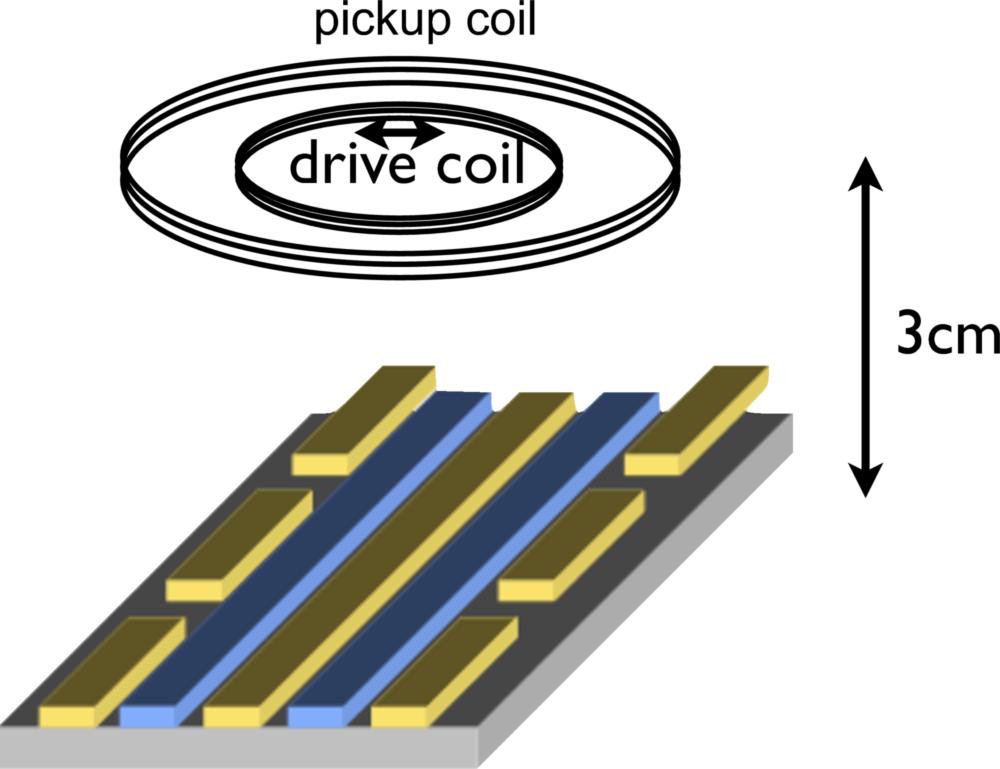}
  \caption[Coupling coils for tickle with external fields.]
  {Coupling coils for
  tickle with external fields. Small drive coil (2.75~cm by 11~turn)
  surrounded by a pickup coil (7~cm diameter by 9~turns). The self
  resonant frequency of the coils was $>100$~MHz. }
  \label{fig:couplingCoilFig}
\end{SCfigure}

\begin{figure}
  \centering
  \includegraphics[width=0.7\textwidth]
  {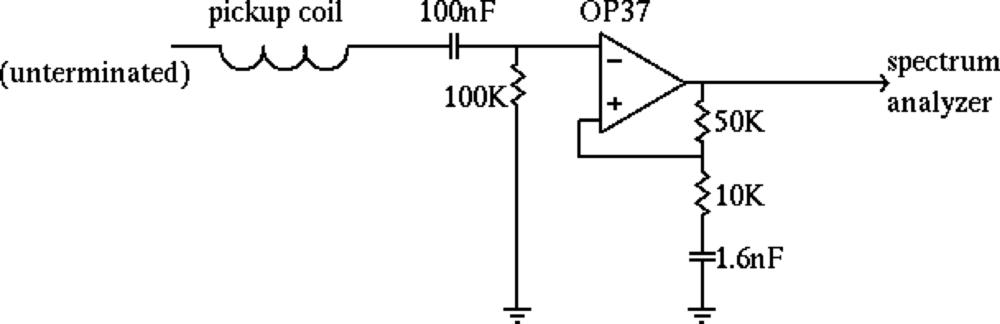}
  \caption[OpAmp driver for pickup coil.]
  {OpAmp driver for pickup coil. The voltage gain is 5~times and the
  low frequency 3~dB point is 10~kHz. The circuit is powered by a battery
  and is housed in a shielded enclosure.}
  \label{fig:couplingCoilOpAmpSchematic}
\end{figure}

Current noise was generated by frequency modulating a coherent RF source with
white noise generated by a SRS DS345. The RF center frequency coincided with the
ion axial frequency $\omega_{z}$. A $100$~kHz FM depth was used. With the
amplified pickup coil attached to a SA, the field from the drive coil could be
detected in addition to environmental noise.
Figure~\vref{fig:pickupCoilAmbientNoiseVsInjectedNoise} (bottom) shows injected
noise superimposed on laboratory noise spikes extending $10$~dB above the noise
floor. These spikes were not artifacts of the pickup coil or detector.

Ion heating was measured with various levels of current noise applied to the
drive coil. Table~\vref{tab:loopNoiseInjectionTable} shows ion heating and the
noise power detected by the pickup coil when the current noise was not attenuated
(0~dB), attenuated by 10~dB or disconnected ($\infty$~dB) from the drive coil.

\begin{figure}
  \centering 
  \includegraphics[width=0.9\textwidth]
  {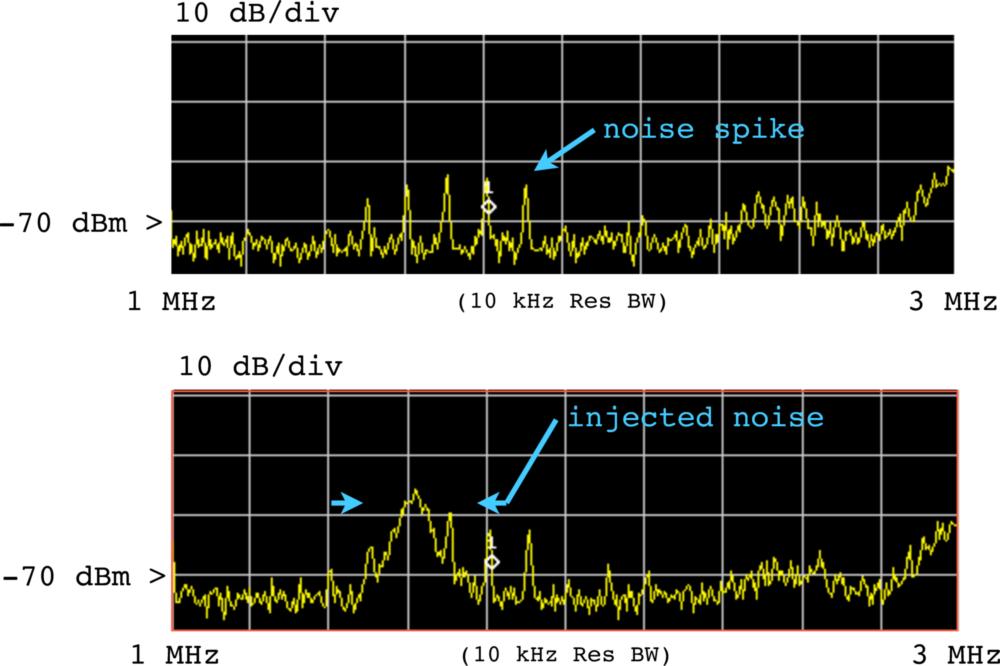}
  \caption[Pickup from coupling coil.]
  {Pickup from coupling coil. (top) Spectrum analyzer (SA)
  trace showing the ambient laboratory noise
  background. (bottom) SA trace
  showing ambient noise plus injected noise (20~dB attenuation). 
  The SA resolution bandwidth was 10~kHz.}
  \label{fig:pickupCoilAmbientNoiseVsInjectedNoise}
\end{figure}

\begin{SCtable}[5]
  \centering
  \begin{tabular}{l|ll}
    (dB)&($\frac{dBm}{10~kHz}$)&(quanta/ms)\\
    Noise Attn.&$P$		&$\Gamma_{0\to1}$\\
   	\hline
    $\infty$&	-75&	<10\\
    0&			-57&	790\\
    10&			-64&	413\\
    $\infty$&	-75&	<23\\
    10&			-63&	467\\
  \end{tabular}
  \caption[Ion heating with external field tickle.]
  {Ion heating ($\omega_z=1.6$~MHz) due to injected noise. 
  A sequence of 300~ms heating measurements were done with
  varying amounts of attenuation (0~dB, 10~dB, $\infty$~dB) for the injected noise. 
  $P$ is noise power in a 10~kHz bandwidth observed at 1.6~MHz using the pickup coil on
  a spectrum analyzer (SA).  The SA resolution bandwidth was 10~kHz.}
  \label{tab:loopNoiseInjectionTable}
\end{SCtable}

With 10~dB attenuation, the current noise caused considerable ion heating
and the signal from the pickup coil was difficult to discern from
background noise on the spectrum analyzer. This experiment suggests that
environmental noise could cause considerable heating if it were to overlap with
the ion's motional frequencies.

These observations suggests that ambient laboratory fields which may be difficult to
detect with laboratory electronic sensors may couple to
ion motion and cause heating. When striving to minimize ion motional heating it is
advisable to try several axial frequencies as there may be environmental noise.

%


\paragraph{micromotion compensation}
For quantum information applications it highly desirable that an ion have no
micromotion at the several experimental zones along the trap z-axis (e.g. at
$z=+370~\mu$m).  See Section~\vref{sec:micromotion:effectsof} for why
micromotion is problematic. If the ion is constrained to lie at a particular
point along the z-axis where an axial RF electric field exists, micromotion
results which can't be nulled. Conventional 2-layer ion traps (see
Figure~\vref{ions:fig:2layerTrap}) by symmetry have no axial electric field. In
this kind of trap there is no intrinsic axial micromotion and we need only to be
concerned with nulling micromotion in the radial directions. In contrast, surface
electrode ion traps (SETs) do not have this symmetry and gradients in the RF
potential along the trap axis are possible. Some SET electrode geometries result
in significant axial RF electric field and micromotion occurs that can't be
nulled with control potentials.

When I designed the dv16m electrode geometry I  did not take any measures to
minimize the axial RF electric field. Numerical simulations show that the dv16m
geometry had significant axial fields in all trapping zones except the load zone.
The tapered region was especially problematic in this respect. See for example
Figure~\vref{fig:dv16m:simulation:4zoneFieldPlot} which shows $\partial E_z^2(z)/
\partial z \propto \phi_{pp}$ along the trap axis.

\begin{SCfigure} 
  \centering
  \includegraphics[width=0.5\textwidth]
  {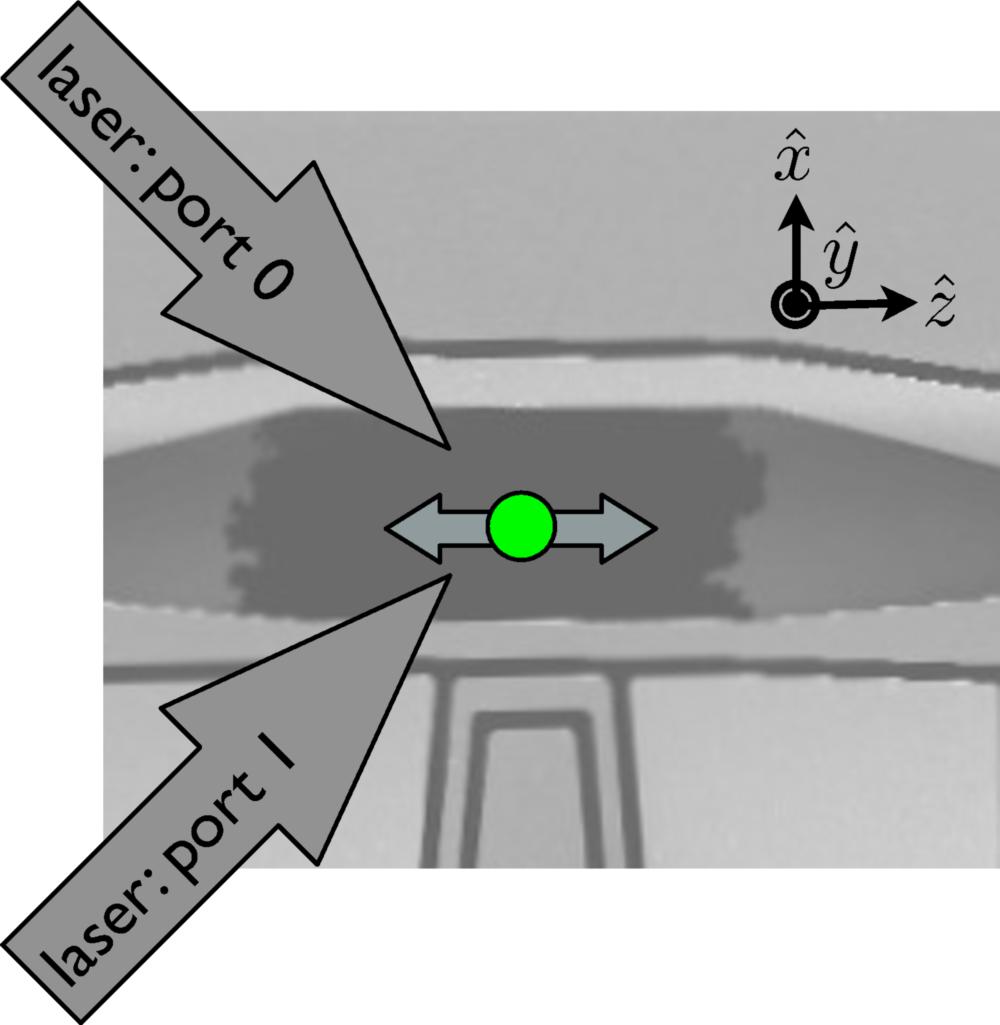}
  \caption[Schematic view of two Doppler cooling lasers in the plane of the
  surface electrode trap.]
  {Two Doppler cooling laser beams in the plane of the surface electrode trap. They are
  equally sensitive to the micromotion component along
  the trap axis (gray arrow). The ion location in this picture is the load trap
  ($0~\mu$m along the z-axis). }
  \label{fig:dopplerPairForMUMdet}
\end{SCfigure}

\subparagraph{load zone}
Simulations predicted a pseudopotential zero in the load zone (see
Figure~\vref{fig:dv16m:simulation:4zoneFieldPlot}).  To check this I applied a
pair of orthogonal Doppler cooling laser beams to an ion in the load zone
(illustrated in Figure~\vref{fig:dopplerPairForMUMdet}). Micromotion in a plane
parallel to the trap surface can be measured using this pair of beams which
propagate along the surface and have a large inclination with respect to the trap
z-axis. Motion along the z-direction gives a Doppler shift to both beams. z-axis
micromotion was minimized by minimizing the fluorescence on the
$n={1,2,\text{...}}$ RF sidebands on both beams. If sideband
amplitude remains and is symmetric for the two beams, then it is due to 
axial micromotion.

\begin{SCfigure}
  \centering
  \includegraphics[width=0.6\textwidth]
  {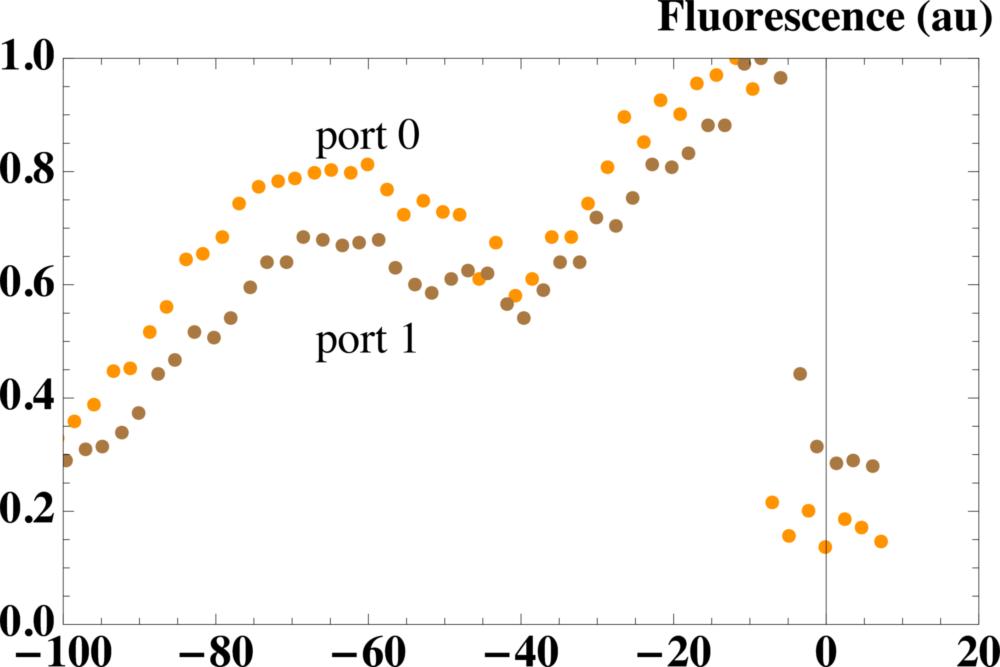}
  \caption[Fluorescence for the laser beams at port~0 and port~1 after following
  the micromotion nulling recipe.]
  {Fluorescence for the laser beams at port~0 and port~1 after following
  the micromotion nulling recipe. The residual axial micromotion corresponds to a
  modulation index of $\beta\sim 1.3$.}
  \label{fig:dopplerPairFluoresence}
\end{SCfigure}

This experiment was carried out in the load zone with all the endcaps shorted to ground.
In this configuration it is expected that the ion will sit at the axial field
minimum barring displacement due to stray electric fields; I saw no evidence of sizable
stray fields. From simulation, this is the only place in the trap structure
where there is zero axial micromotion. However, even at this location
I could not simultaneously null the micromotion sidebands for the pair of
beams.  That is, there was significant intrinsic axial micromotion. The plots in
Figure~\vref{fig:dopplerPairFluoresence} show symmetric sidebands that couldn't
be further minimized using control potentials.  This could indicate a nonzero
phase difference between the RF potentials on the RF electrodes as discussed in 
Section~\vref{sec:dv16k:causesofphiRFneqzero}.

\paragraph{ion heating due to noisy RF potentials and micromotion}

The RF pseudopotential which provides ion confinement can also cause motional
heating $\dot{\bar{n}}$ if the RF source is noisy 
and the ion has micromotion. See Section~\vref{sec:RFAMtheory} for details on 
this mechanism.

Micromotion in the radial direction arises due to uncompensated stray electric
fields; this micromotion can be nulled using control potentials if it can be
detected.  For a particular trap zone along the z-axis, 
micromotion in the axial direction arises due to an axial RF electric
field; this micromotion can't be nulled.  
Figure~\vref{fig:dv16m:simulation:4zoneFieldPlot} shows $E_z(z)$ (the electric
field along along the trap axis). Small micromotion amplitude does not adversely
impact the recooling heating measurement technique (discussed in
Section~\vref{sec:recooling}). However, in both cases noise on the RF potential
can cause heating of the axial and radial modes.

The importance of minimizing $E_z(z) \frac{\partial E_z(z)}{\partial z}>0$ 
in Equation~\vref{eq:rfam:axialheating:rate} was
not fully appreciated until after the trap was built.  In this section I calculate
the heating due RF noise and micromotion in dv16m zones m370 and p370.

I denote the noise power at frequency $\omega_n$ relative to the power at the 
carrier frequency $\Omega_{\rm RF}$ as
$\alpha^2 \equiv P_{noise}/ P = 2 S_{E_n}/E_0^2 = 2 S_{V_n}/V_0^2$. According to
the manufacturer's specification for the oscillator I used, an HP~8640B,
 $\alpha^2_{8640B}\cong -147$~dBc for noise
at ($\Omega_{\rm RF}-\omega_n)/2\pi=$~1~MHz 
(see Figure~\vref{fig:RFelectronics:8640Bnoise}).
There may also be additional noise added by
the RF amplifier, a 
\href{http://www.amplifiers.com}{Amplifier Research, Inc.} 1~W AR 1W1000A.
However, a noise figure is not available for this amplifier.  

\subparagraph{axial heating}
If $\left|E_z(z) \frac{\partial E_z(z)}{\partial z}\right|>0$ and there is noise on the
RF potential at the axial trap frequency $\omega_z$,  
as in Equation~\vref{eq:rfam:efieldnoise:EwithNoise}, ion heating 
$\dot{\bar{n}}$ results.  

Section~\vref{sec:rfam:axialheating}  derives the following expression for the 
axial heating rate,
\begin{align} 
  \dot{\bar{n}}(\omega_z)&=\frac{1}{4 m \hbar  \omega_z}\times 
  		\left[ 
			 \frac{q^2}{m\Omega^2_{\rm RF}}  
			E_0(z)\frac{\partial E_0(z)}{\partial z}
		\right]^2
		 \frac{S_{E_n}(\omega_n)}{E_0^2(\omega_n)},
\end{align}
in units of quanta per second. 

Consider zone p371 ($z=+371~\mu$m), discussed in
Section~\vref{sec:dv16m:p371CEi}. Typical trap parameters are $^{24}Mg^+$,
$\Omega_{\rm RF}/2\pi=70$~MHz, $\omega_z/2\pi=3$~MHz and $V_{\rm RF}$=50~V. 
Numerical simulation gives $E_z(z)=213~V/m$ and $\frac{\partial E_z(z)}{\partial
z}=23,500~V/m^2$ (Equations~\vref{eq:dv16mp371Ez} and~\vref{eq:dv16mp371dEz}). 
Accounting for 17~dB of attenuation due to the $\lambda/4$ resonator (see
Table~\vref{tab:apparatus:RF:attenuation}),
\begin{align}
	\frac{\dot{\bar{n}}(\omega_z)}{\alpha^2}&=1.3\times 10^{11}~quanta/ms.
\end{align}
If $\alpha^2=\alpha^2_{8640B}\cong -147$~dBc, then 
$\dot{\bar{n}}(\omega_z)=27\times10^{-6}$~quanta/ms.

This heating rate could be four orders of magnitude greater for locations 
just several 10's~$\mu$m away from position p370 
(see Figure~\vref{fig:dv16m:simulation:4zoneFieldPlot}) or if $\alpha^2$ were larger.   

\subparagraph{radial heating}
If an ion's mean position is displaced radially from the pseudopotential minimum
by a distance $x$, as from a stray static electric field, the ion undergoes
micromotion.  If the RF electric field is noisy, as in
Equation~\vref{eq:rfam:efieldnoise:EwithNoise}, ion heating results.

Section~\vref{sec:rfam:radialheating} derives the following expression for the 
the radial heating rate,
\begin{align} 
  \dot{\bar{n}}(\omega_x)
		 &=\frac{1}{4 m \hbar  \omega_x}\times 
  		\left[ 
			 2 m \omega_x^2 x
		\right]^2
		 \frac{S_{E_n}(\omega_x)}{E_0^2(x)}
\end{align}
in quanta per second and where $\omega_x$ is the frequency of radial motion.

Consider zone p371  ($z=+371~\mu$m), discussed in
Section~\vref{sec:dv16m:p371CEi}. Typical trap parameters are $^{24}Mg^+$,
$\Omega_{\rm RF}/2\pi=70$~MHz and $\omega_x/2\pi=10$~MHz.
Assume $\beta$=1.43, which gives a micromotion amplitude of
64~nm and a displacement of $x=1.05~\mu$m
(see Section~\vref{ions:sec:micromotion}). Accounting for 27~dB of attenuation
due to the $\lambda/4$ resonator (see Table~\vref{tab:apparatus:RF:attenuation}),
\begin{align}
	\frac{\dot{\bar{n}}(\omega_z)}{\alpha^2}&=2\times 10^{17}~quanta/ms.
\end{align}
If $\alpha^2=\alpha^2_{8640B}\cong -147$~dBc, then 
$\dot{\bar{n}}(\omega_z)=0.4$~quanta/ms.

This heating rate could greater if $\beta$ or  $\alpha^2$ were larger.   
 
\subsection{Trap environment and motional heating}

There are several key difference between the vacuum apparatus used in my
experiments and in Seidelin's at NIST where an order of magnitude lower motional
heating rate was observed.  I used an octagon style vacuum system discussed in
Section~\vref{sec:octagonVacSystem} for tests of dv16m.  Seidelin used a
$\lambda$/4 style vacuum discussed in
Section~\vref{sec:L4vacSystem}~\cite{seidelin2006a}. Some of the differences
between these systems could contribute to the difference in motional heating.
Further evidence for this possibility is suggested by results found in an ion
trap (Amini, NIST) very similar to the Seidelin design which had excellent
heating. This trap was tested in an octagon style vacuum system and also
exhibited an approximately order of magnitude higher heating rate (see
Figure~\vref{fig:ionHeatingScatterPlot}) than Seidelin's.

Following is a list of several differences between the octagon and $\lambda/2$
style vacuum systems which could be related to the electronic environment at the
ion.

\paragraph{RF lead length} The RF lead length in the vacuum systems differ. In
both it is a Kapton wrapped wire. It's length should be a minimum to prevent radiative
losses, inductive impedance and capacitive coupling to ground.
\begin{compactitem}
  \item \textit{octagon}: The RF lead length in the vacuum system is typically
  5~cm.
  \item \textit{$\lambda$/4}: The RF lead length beyond the end of the
  $\lambda$/4 center conductor is typically short, $<$ 2 cm. Moreover,
  the entire trap chip including the lead lie within the bore of the
  $\lambda$/4 resonator.
\end{compactitem}

\paragraph{UHV RF FT}The location of RF UHV feedthrough differs.
\begin{compactitem}
  \item \textit{octagon}: The UHV RF feedthrough lies at the tip of the
  resonator, near it's highest voltage point. It has a large (8-15~pF)
  capacitance to ground.
  \item \textit{$\lambda$/4}: The UHV RF feedthrough lies inside the
  $\lambda$/4 resonator at about its midpoint. It has a smaller (4.5~pF)
  capacitance to ground.
\end{compactitem}

\paragraph{Resonator geometry} The geometry of the RF resonators differ.
\begin{compactitem}
  \item \textit{octagon}: The resonator is a helical cavity resonator
  \cite{macalpine1959a}.
  \item \textit{$\lambda$/4}: The resonator is a coaxial cavity resonator
  \cite{jefferts1995a}.
\end{compactitem}

\paragraph{Loaded $Q_L$} The loaded $Q_L$ of the RF resonators differs with no
trap chip attached.
\begin{compactitem}
  \item \textit{octagon}: With the resonator attached to the vacuum system,
  with the RF routed to a pin on the chip carrier but with no trap chip
  installed, I observed $Q_{L}=225$ at $\Omega_{\text{RF}}=45~\text{MHz}$.
  \item \textit{$\lambda$/4}: With the resonator attached to the vacuum
  system but no trap chip attached to the center conductor of the resonator,
  typical values are $Q_{L}=800$ at $\Omega_{\text{RF}}=100~\text{MHz}$.
\end{compactitem}

\paragraph{Delivery of RF and control potentials to trap chip}
\begin{compactitem}
  \item \textit{octagon}: A cofired ceramic chip carrier was used (see
  Section~\vref{sec:chipCarrierSocket}) to supply the trap chip with both control
  and RF potentials.  These potentials were communicated to the chip carrier by
  spring-loaded socket-pin connectors. Also, cofired ceramic is a new material
  for ion trapping applications.
  \item \textit{$\lambda$/4}: An alumina or quartz substrate was used to route
  the control and RF potentials to the trap.  Wires supplying these potentials
  were attached directly to traces on the substrate by resistive welding.
\end{compactitem}

\paragraph{Shielding} The degree of shielding from ambient laboratory noise.
\begin{compactitem}
  \item \textit{octagon}: It's geometry is open with many view ports with
  line of sight to the trap chip.
  \item \textit{$\lambda$/4}: The entire trap chip including the lead
  lie within the bore of the $\lambda$/4 resonator. Openings for the
  lasers and imaging system are smaller than for the octagon.
\end{compactitem}

\paragraph{RC filter location} The location of the RC filters differ for the
two systems. These filters short to ground RF potential coupled to the control electrodes from
the RF electrodes by a small inter-electrode capacitance $C_{T}$.
In my traps $C_{T}\leq$0.1 pF. I used $C=820~\text{pF}$ and $R=1~\text{k$\Omega$}$.
A low impedance path between control electrodes and trap ground at
the trap RF frequency $\Omega_{\text{RF}}$ is desired.

\begin{compactitem}
  \label{par:octagonRCfiltersAndInductance}
  \item \textit{octagon}: The RC filters lie outside the vacuum system.
  The lead length from control electrode to RF shorting capacitor is
  7-10~cm. Along this length there are many metal-metal connections.
  Specifically, from control electrode to RC filter: 1~wire bond, 
  1~BeCu chip carrier spring connector, 2~crimps, 1~UHV D-Sub feedthrough,
  1~socket connection and several solder joints (on the RC filter PCB).\\
  The impedance of
  this lead  at $\left.\Omega_{\text{RF}}\right/2\pi=100$~MHz  
  can be significant. A lumped circuit model (vs transmission
  line model) is reasonable since $\lambda_{RF}>>10$~cm.
  A crude estimate of the lead's inductance may be calculated assuming
  it is a length ($l=$10~cm) of thin ($a=$0.5~mm) wire, near ($d=$1~cm) 
  a conducting plane \cite{terman1955a}.
  \begin{gather*}
 		L=\mu_{0}l(\ln(2d/a)+1/4)=495\text{ nH}\\
      	\begin{split}
    		Z_{L}&	=i~2\pi~\Omega_{\text{RF}}L\\
    		&		=i~2\pi~100\text{ MHz}\times495\text{nH}\\
    		&		=i~311~\Omega
    	\end{split}
  \end{gather*}
  In some traps there is a relatively large capacitance to ground $C'$
  at each control electrode. This capacitance coupled to L can be problematic.
  For example, if L=100~nH and $C'=30\text{pF}$ the resonant
  frequency of the resulting LC oscillator is
  \begin{equation*}
    f_{\rm LC'}=\omega_{\rm LC'}/2\pi=
      (2\pi\sqrt{\rm LC'})^{-1}=92~MHz
  \end{equation*}
  This frequency may lie near the trap drive frequency $\Omega_{\text{RF}}$.
  If this is the case resonant effects can suppress the shunting effect
  of the RC filters if the resonant Q is high. The result is a larger
  RF potential on the control electrodes. This potential may moreover
  acquire a phase shift (due to the LC's reactance) with respect to
  the RF on the RF electrodes, a situation which can cause intrinsic
  micromotion \cite{berkeland1998a}.
  
  \item \textit{$\lambda$/4}: So far, only simple traps have been used
  in $\lambda$/4 systems and the RC filters were located $<$ 1~cm
  from the trap chip inside the vacuum system. 
\end{compactitem}
\begin{figure}
  \centering
  \includegraphics[width=1\textwidth]
  {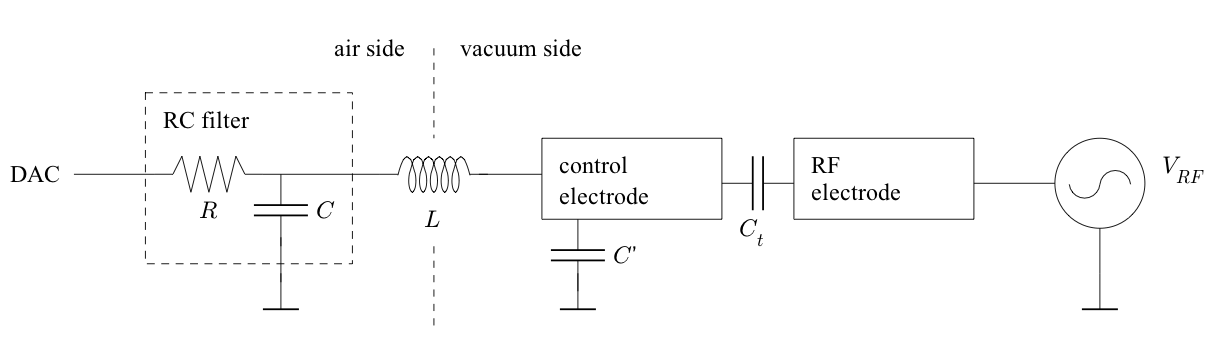}
  \caption[Schematic illustrating why it's better to put RC filters near the
  ion trap.]
  {Schematic illustrating why it is desirable to put RC filters near the
  ion trap (see text).}
  \label{fig:putRCnearTrap}
\end{figure}

\subsection{Conclusion}
The basic features of ion loading and transport between zones worked.  And,
predicted secular frequencies were verified by experiment. A heating comparison
between gold and bared doped silicon surfaces was inconclusive because ion heating
from externally injected noise could not be completely ruled out.

It may be that the difficulty I had nulling micromotion in this trap is due to a
phase difference between the RF electrodes (see Section~\vref{sec:micromotion}). 
Possible causes were discussed in the context of trap dv16k in
Section~\vref{sec:dv16k:causesofphiRFneqzero}.

The SOI wafer used in this experiment was obtained from a distributer that
resells surplus wafers. SOI with higher conductivity is available by special
order.  See Section~\vref{sec:SOI}.

    \chapter{Microtrap Testing Apparatus}
\label{sec:trapTestingApparatus}

The chapter provides an introduction to the apparatus used to house and test 
surface electrode ion traps.  Throughout this section dv16m, dv16k and dv10 
refer to specific ion trap geometries discussed in Chapter~\vref{sec:thetrapschapter}.

\section{Vacuum apparatus}

\label{sec:vacuumApparatus} An ultra high vacuum (UHV) environment
($\sim1\times10^{-11}$~Torr) is required for QIP with ion traps. At higher
pressures background collisions can cause ion heating and qubit decoherence.
Additionally, during some collisions there may be a chemical reaction
with the trapped ion, creating a different and undesired ion species.
This is especially problematic for $^{9}Be^{+}$ where ion lifetime
is fundamentally limited by collisions with $H_{2}$ \cite{molhave2000a,bible}.
I did not observe such problematic chemistry with $^{24}$Mg$^{+}$. 
However, the importance of good vacuum can't be underestimated since
$^{9}$Be$^{+}$ is emerging as a good candidate for QIP \cite{ozeri2007a}.

Obtaining UHV quality vacuum requires careful choice of trap materials,
meticulous cleaning and a long bake at high temperature. An excellent
guide to UHV processing is the 27~page appendix to Birnbaum's doctoral
thesis in the Kimble group at Caltech \cite{birnbaum2005a}. 

This section details distinguishing features of the two vacuum systems
I used in my experiments.

\subsection{Quarter-wave style vacuum system}

\label{sec:L4vacSystem} 
\begin{figure}
  \centering
  \includegraphics[width=0.75\textwidth]
  {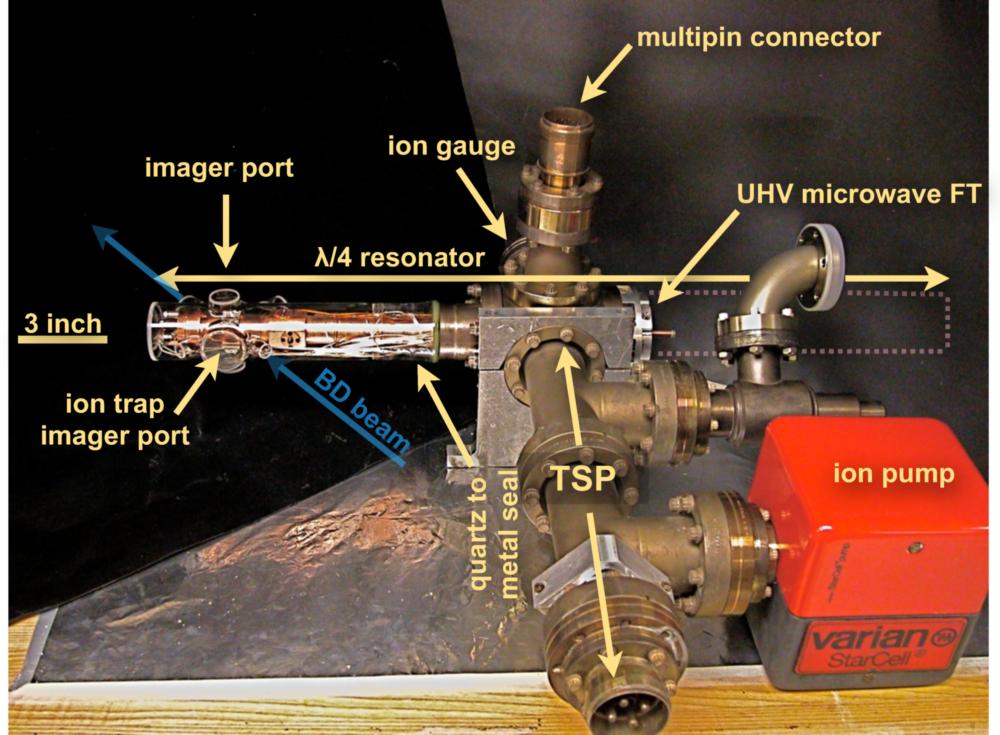}
  \caption[$\lambda/4$ style vacuum system.]
  {$\lambda/4$ style vacuum system.}
  \label{fig:L4vacuumSystemPhoto}
\end{figure}

The ion trap vacuum system design used in all the QIP
experiments at NIST to date is identifiable by the integration of a prominent
$\lambda$/4 coaxial RF resonator (see~\cite{jefferts1995a} for details). Such a
system was used for testing of early boron doped silicon ion traps 
(see trap ``dv10'' in Section~\vref{sec:dv10} and trap ``dv14'' in 
Section~\vref{sec:dv14}). It consists of a $\lambda$/4 copper
coaxial RF resonator surrounded by a quartz vacuum envelope with optical quality
quartz windows  for laser beams and imaging. The boundary conditions for the RF
cavity produce a voltage maximum at the tip of the resonator. The trap RF
electrodes are attached to this tip. The trap control electrodes are referenced
to the RF ground at the resonator sheath. See
Figure~\vref{fig:L4vacuumSystemPhoto} for a labeled photograph.

Many of the $\lambda$/4 vacuum system components were available off
the shelf from the following manufacturers. The stainless steel conflat
flange components were purchased from 
(\href{http://www.lesker.com}{Kurt J. Lesker, Inc.}) and 
(\href{http://www.mdc-vacuum.com}{MDC Inc.}). The two vacuum pumps visible
in the photo were used to maintain vacuum following the bake. 
They are a \href{http://www.varianinc.com}{Varian, Inc.}
titanium sublimation pump (TSP, p/n~916-0061) and 20~L/s StarCell
VacIon Plus ion pump (p/n~9191144) purchased from
\href{http://www.lesker.com}{Kurt J. Lesker, Inc}. The pumps were controlled
by a Varian MiniVac controller (p/n~9290191) and a Varian TSP controller
(p/n~9290022). The ion gauge is a \href{http://www.helixtechnology.com}{Granville-Phillips, Inc.} nude ion gauge (p/n~274022) and is controlled by
a \href{http://www.thinksrs.com}{SRS, Inc.} IGC100. A turbo pump backed
by a diaphragm pump, a large ion pump and a SRS RGA100 residual
gas analyzer were used during the bake. The multi-pin connector is
made by 
\href{http://www.insulatorseal.com}{Insulator Seal, Inc.}. The remaining
components were custom made.

Two-thirds of the $\lambda$/4 coaxial microwave resonator resides
in vacuum (shown in Figure~\vref{fig:L4vacuumSystemPhoto}), while
the remainder is in air (removed in the photo in the Figure, represented
by dashed lines). It is surrounded by a quartz envelope made by
\href{http://www.allenglass.com}{Allen Scientific Glass, Inc}. 
The optical quality quartz windows (\href{http://www.cvilaser.com}{CVI Laser
Optics, Inc.}) and quartz to metal seal
(\href{http://www.larsonelectronicglass.com}{Larson Electronic Glass, Inc.})
were attached to the envelope. 
The UHV microwave feedthrough is designed with a small capacitance to ground and to
minimize the impedance mismatch between the vacuum-side and air-side
of the $\lambda$/4 resonator. It was manufactured to
my specifications by Ceramaseal a division of
\href{http://www.ceramtec.com}{CeramTec, Inc}. 
\href{http://www.omley.com}{Omley Industries, Inc.} 
was unsuccessful at building the feedthrough. 
The vacuum-side copper resonator and air side aluminum
resonator (not shown) were made by the
\href{http://www.colorado.edu}{University of Colorado}
\href{http://cires.colorado.edu/iidf}{CIRES Integrated Instrument Development
Facility}. The copper is OFHC grade (oxygen free).

\paragraph{RC Filters}\label{par:L4RCfilters}
The trap chips were mounted on a ceramic filter board. Traces were
put on the filter board by silk screening with gold paste 
(see Section~\vref{sec:electricalInterconnect}).
Trap RF may be capacitively coupled to the control electrodes and
must be shunted to ground. This was done using an RC filter for each
control electrode mounted in the vacuum system on the filter board.
The capacitors and resistors were surface mount devices ($820$~pF, 
$1000~\Omega$) with a high breakdown voltage ($>100~V$). It is crucial to use
devices without any solder tinning which can spoil the UHV vacuum.

The 830~pF capacitors were made by \href{http://www.novacap.com}{Novacap, Inc.}
with the following characteristics. 

\begin{compactitem}
  \item size: 1.27 mm x 1.02 mm x 1.12 mm
  \item precision: +/-5$\%$
  \item maximum voltage: 300 V
  \item termination: palladium silver
  \item part number: 0504B821J301P
\end{compactitem}
The $1~k\Omega$ resistors were made by Amitron (formerly of North Andover,
MA). Amitron was purchased by \href{http://www.anaren.com}{Anaren, Inc}.

\begin{compactitem}
  \item size: 1.52 mm x 0.76 mm
  \item style: top contact
  \item base material: 0.010 inch alumina
  \item tolerance: $\pm 5 \% $
  \item termination: gold
  \item part number: R1A1508-100J5A0
\end{compactitem}

\subsection{Octagon vacuum system}

\label{sec:octagonVacSystem} 
\begin{figure}
  \centering
  \includegraphics[width=0.75\textwidth]
  {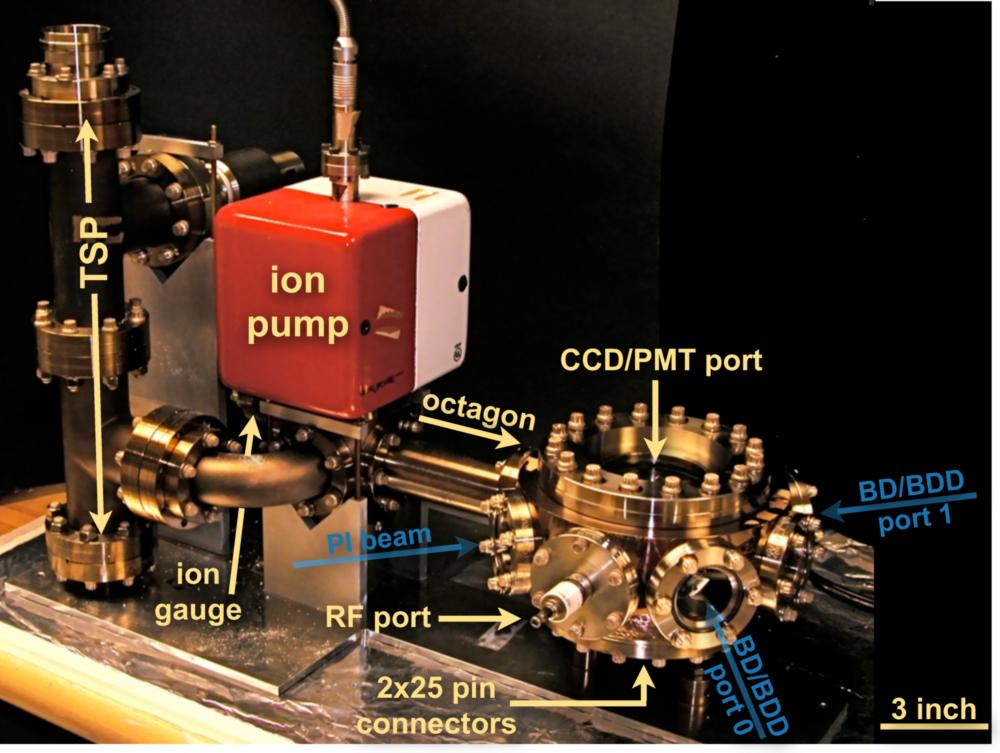}
  \caption[Octagon style vacuum system.]
  {Octagon style vacuum system.}
  \label{fig:octagonVacuumSystemPhoto}
\end{figure}

The vacuum system design used for recent NIST
surface electrode (SE) ion trap experiments has a prominent stainless steel
conflat flange (CF) enclosure with eight 3.75~inch CF ports. See
Figure~\vref{fig:octagonVacuumSystemPhoto}. It is an extension of the technology
used in the $\lambda$/4 style system (see Section \ref{sec:L4vacSystem}),
designed especially for many electrode SE ion traps. Its octagonal vacuum system
was favored for SE traps because it can be made of commercially available
components (see below) while meeting many SE trap needs.  My SOI SET 
 was tested in this vacuum system (see trap ``dv16m" in Section~\vref{sec:dv16m}).
\begin{compactitem}
\item 5 optical ports accommodate cooling and photoionization beams
that pass above the trap electrode surface.
\item Large top and bottom ports support F1 light collection optics and
accommodate high density UHV multi-pin electrical feedthroughs.
\item An open geometry and plug-and-play chip carrier make systematic testing of
multiple traps faster and less error prone (see \ref{sec:chipCarrierSocket}).
\item Since it is modular, electrical interconnect and vacuum
processing for a new ion trap chip can be completed in one week, at least a factor
of four improvement over the $\lambda$/4 system.
\end{compactitem}

I used this system because one of the surface electrode (SE) traps I 
tested had 43 control electrodes each with an independent control potential.
This necessitated an equal number of feedthroughs, a difficult proposition
with the $\lambda$/4 style vacuum system due to geometric and wiring
constraints. Routing of 43 (or more) wires to the trap electrodes
inside the vacuum system and outside to the DACs was a significant
engineering challenge. NIST collaborated with the University of Michigan
to develop a UHV compatible 100-pin socket for a chip carrier like
those used in the microelectronics industry \cite{stick2007a}.
A similar vacuum system was use to test traps produced by the Lucent
Inc. and Sandia National Labs ion trap foundries. See 
Figure~\vref{fig:octagonVacuumSystemPhoto}
for a labeled photograph.

Many of the octagon vacuum system components are available off the
shelf from the same manufactures as the $\lambda$/4 style trap (see
Section \ref{sec:L4vacSystem}). The RF feedthrough is a standard
\href{http://www.insulatorseal.com}{Insulator Seal, Inc.} part. 
The five quartz laser beam access windows are mounted on $2.75$ inch CF. 
They were made by \href{http://www.mpfi.com}{MPF Products, Inc.} (p/n
A0651-A-CF). The stainless parts were baked at $400^{\circ}$ c in air overnight
to outgas adsorbed hydrogen \cite{bernardini1998a}. 
The remaining components were custom made.

The RF resonator was of a helical variety and lies completely outside
the vacuum system \cite{cohen1965a}. Two 25-pin connectors (not
visible in the photo in Figure \ref{fig:octagonVacuumSystemPhoto}) 
were made by \href{http://www.insulatorseal.com}{Insulator Seal, Inc.}
and were welded to a CF by \href{http://www.accuglassproducts.com}{Accu-Glass
Products, Inc}. The CCD/PMT port is a large quartz
window attached to a a CF by a quartz to metal seal made by 
\href{http://www.larsonelectronicglass.com}{Larson Electronic Glass, Inc}. 
It uses a solder which melts at $300^{\circ}$ c; we follow Larson's
advice to bake at temperatures no higher than $200^{\circ}$ c.

\paragraph{RC Filters}
As in the $\lambda/4$ system RC filters were used for each control electrode
(see Section \ref{par:L4RCfilters}). Due to the large number of control
electrodes it was convenient to put the RC filters outside the vacuum
system. However, it is better to put them as
close as possible to the trap chip (inside the vacuum system). This
avoids the inductive impedance of the connecting wires and doesn't
rely on the control electrode UHV feedthroughs to be low impedance
at RF.  Possible ramifications of putting the RC filters outside the vacuum
system are discussed in Section \ref{par:octagonRCfiltersAndInductance}.

\subsection{Chip carrier and socket}
\label{sec:chipCarrierSocket}NIST collaborated with Dan Stick at
the University of Michigan to develop a UHV compatible-100 pin socket
for a chip carrier like those used in the microelectronics industry.
The chip carrier is a two-layer design made of Vespel and machined
with a grid of holes to house 50~gold-plated spring-loaded sockets
which interface with pins on a chip carrier. Each socket was connected
by a Kapton wrapped wire to a pin on one of the 25-pin UHV vacuum
feedthroughs. Both ends of the wire were crimped with a special crimp
tool. 

\href{http://www.dupont.com/kapton}{Kapton} is a
\href{http://www.dupont.com}{DuPont, Inc.} UHV compatible polyamide film.
\href{http://www.dupont.com/vespel/Vespel}{Vespel} is a 
\href{http://www.dupont.com}{DuPont, Inc.} UHV compatible
plastic that is easily machinable. The Kapton wrapped wire 
is $22$ AWG silver plated copper made by \href{http://www.accuglassproducts.com}{Accu-Glass
Products, Inc.} (p/n 100680).
See Stick's thesis for details including other part numbers \cite{stick2007a}.

The chip carrier was a UHV compatible 100-pin low-temperature co-fired
ceramic (LTCC) pin grid array (CPGA) package make by
\href{http://www.kyocera.com}{Kyocera, Inc.}. It can be purchased from
\href{http://www.globalchipmaterials.com}{Global chip Materials, Inc.} 
(p/n PGA10047002). Extraction of the chip carrier from the socket required care
to avoid bent pins. I used a chip extractor made by \href{
http://www.emulation.com}{Emulation Technologies, Inc.} (p/n TW2063,
S-PGA-13-101-A). It is helpful to conduct electrical 
tests on traps mounted on chip carriers outside of the vacuum system. 
A breakout printed circuit board (PCB) was built
for this purpose with a $169$-pin zero insertion force (ZIF) PGA socket
(\href{http://www.digikey.com}{Digi-Key, Inc.} p/n 169-PRS13001-12). See
Figure~\vref{fig:100pinNumbering} for the electrode numbering scheme I used.

A ceramic paste was used to attach silicon trap chips to the cPGA
chip carriers. It is made by \href{http://www.aremco.com}{Aremco, Inc.} with
part number Ceramabond $569$. The bulk of the paste is alumina. It is a very
poor adhesive in layers thicker than $3$ mil. It was cured overnight at
$200^{\circ}$ c. Be cautious to avoid application to surfaces near the wafer front
surface as it was once observed to shed small alumina flakes. Before
installing the trap chip a hole was machined in the back of the chip
carrier to permit back side loading.

Wire bonding with $0.001$ inch diameter gold was used to interconnect
the ion trap chip to the chip carrier. See Section~\vref{sec:electricalInterconnect}.
Before wire bonding the chip was attached with the ceramic paste.

\begin{SCfigure}
  \centering
  \includegraphics[width=0.6\textwidth]
  {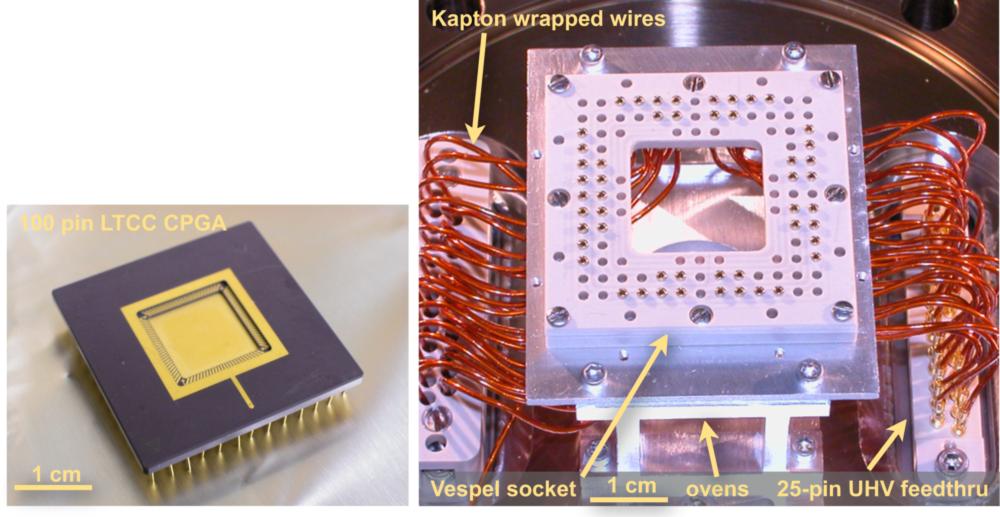}
  \caption[Photographs of 100-pin chip carrier and Vespel chip carrier socket.]
  {Photographs of 100-pin chip carrier and Vespel chip carrier socket.}
  \label{fig:100pinChipCarrier}
\end{SCfigure}
\begin{SCfigure}
  \centering
  \includegraphics[width=0.6\textwidth]
  {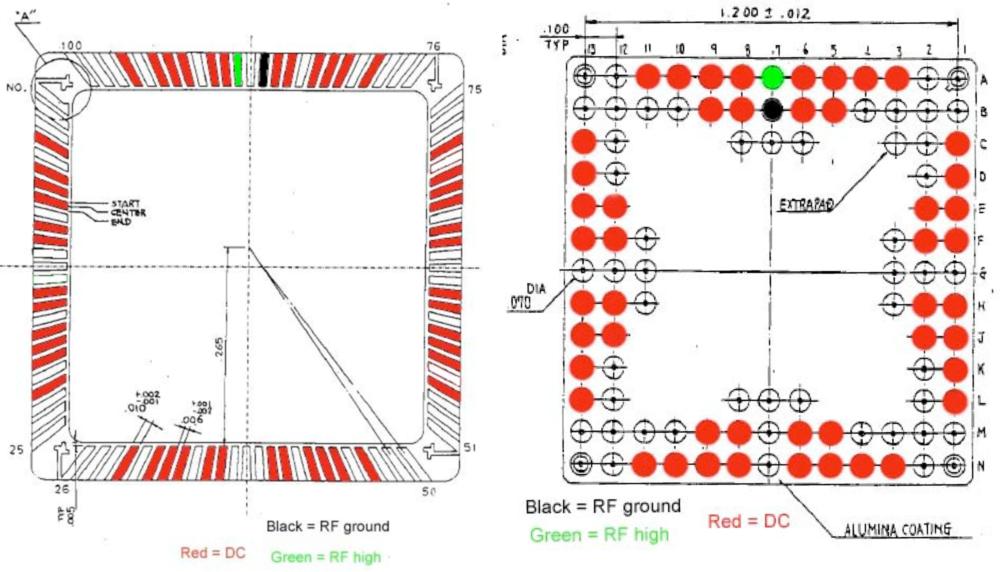}
  \caption[100 pin cPGA pin numbering.]
  {100 pin cPGA pin numbering.}
  \label{fig:100pinNumbering}
\end{SCfigure}
\begin{SCfigure}
  \centering
  \includegraphics[width=0.6\textwidth]
  {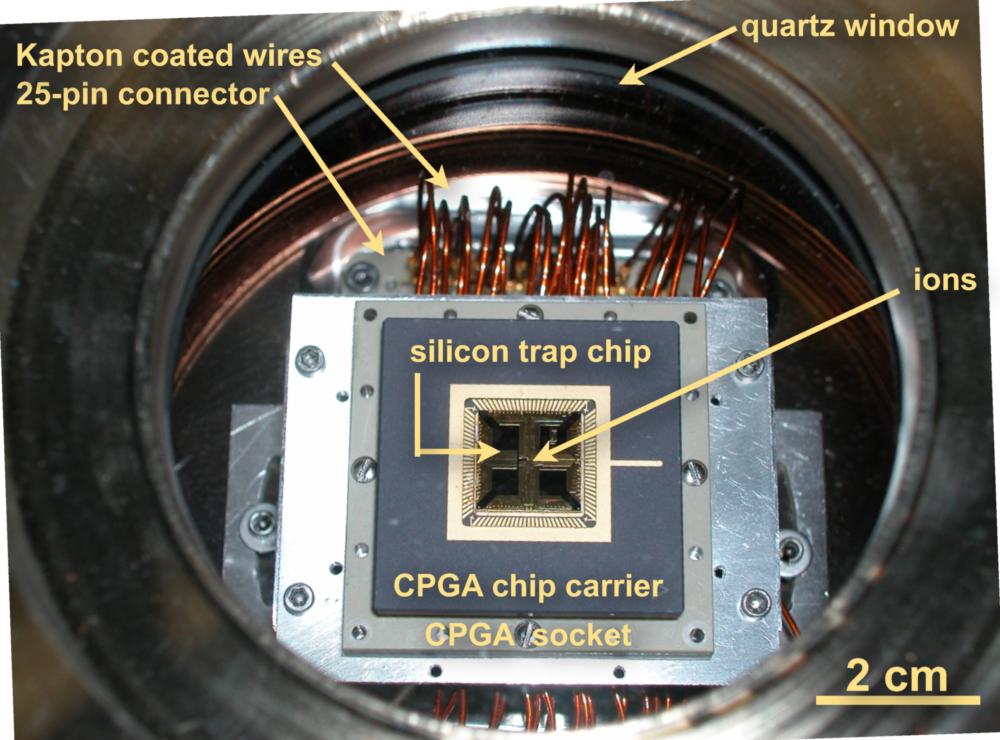}
  \caption[Trap chip in octagon vacuum system.]
  {Trap chip in octagon vacuum system.}
  \label{fig:trapChipInOctagonTopView}
\end{SCfigure}

\subsection{Prebake checklist}
\label{sec:prebakeChecklist}

Here's a list of some things to check when installing an ion trap
in a vacuum system prior to the bake.

\begin{compactitem}
  \item Don't let the imager have a direct line of sight to the electron filament
  because it will be blinded (and possible damaged) during loading.
  \item Have a direct view of ovens from outside vacuum system so you can
  look for it's glow. This is helpful to approximate their temperature.
  \item Have baffles for the ovens which can be viewed from outside the vacuum
  system. It's helpful to have an indicator of oven plating. Plating
  is difficult to observe with a flashlight; turn on the room lights.
  \item Don't operate Mg ovens in advance of bake. Magnesium oxide is helpful
  in that it will help the oven withstand higher temperature bakes.
  \item Check that the oven openings do not have a view of their own wiring,
  that of their neighbors, any trap traces, the imager window or the
  laser beam ports. Shorts or inconvenient obscuration may result.
  \item Make sure there is a direct line of sight from the electron filaments
  and ovens to the trap center.
  \item Test electron filaments prior to bake at low pressure ($<1\times
  10^{-6}$ Torr). Ramp up the current slowly. As the filaments are baked their
  emission current steadily improves. Don't walk away!
  \item Check for the presence of the RC filter for each electrode. Do this
  by looking on a scope for the filter response to a current limited
  square wave.
  \item Look for finite resistance paths ($<1~M\Omega$) between adjacent
  control electrodes (and to ground).
  \item Look for finite resistance paths ($<1~M\Omega$) between each control
  electrode and the RF electrodes.
  \item Laser beams should not clip any wires or trap features inside the
  vacuum system. check for clearance before and after bake.
  \item During the first half of the bake, run all the TSP filaments at $30$
  A for $2-3$ hours each. These filaments outgas a great deal and it is
  better if contaminants are pumped away and not deposited on the vacuum
  system walls. Note that the usual TSP operation cycle is $3$ min at
  $47$ A once per week (or as needed).
  \item Confirm that the RC filter ground is tied to the trap RF ground inside
  the vacuum system. It should be a good RF quality ground.
\end{compactitem}

\subsection{Bake procedure}

\label{sec:bakeChecklist} The vacuum systems were baked at $200^{\circ}$
C for about $5$ days. The bake was ended when at constant temperature
the observed pressure plateaued around $1\times 10^{-7}$ Torr. For advice on UHV
baking see \cite{birnbaum2005a}.

\subsection{Post bake checklist}

\label{sec:postBakeChecklist} Here's a list of tests to do (and not
do) to an ion trap after it's been baked but before you try to load
it.

\begin{compactitem}
\item Do not use HCl to clean the copper oxide off tarnished RF feedthroughs.
HCl attacks the braze joints. Use a Scotch-Brite pad instead.
\item Check for the presence of the RC filter for each electrode. Do this
by looking on a scope for the filter response to a current limited
square wave.
\item Look for finite resistance paths ($<1~M\Omega$) between adjacent
control electrodes (and to ground).
\item Look for finite resistance paths ($<1~M\Omega$) between each control
electrode and the RF electrodes.
\item Apply a little power to the RF resonator and tweak the coupling to
$>90~\% $. Slowly ramp up the power over several hours. Breakdown
may happen if there are field emitters present. In some cases they
can be burned off. Do this by applying $>\Gamma_{RF}/2$ frequency
modulation to the RF source while ramping up the power. $\Gamma_{RF}$
is the RF resonator linewidth, $\Gamma_{RF}=\Omega_{RF}/Q$. If breakdown
is observed (as a sharp discontinuity in reflected power) wait until
it goes away before applying more power. The trap temperature may change due to 
dissipation of RF in the trap structure.  In this case the RF capacitance of the trap
to ground may change causing a shift RF cavity resonant frequency and coupling.
\item Check for continuity between each trap electrode and external wiring.
It's possible that a wire has broken between the RC filter and the
trap electrode. There are several tricks of decreasing specificity.

\begin{compactenum}
\item Shine a UV laser beam on a trap electrode and look for the resulting
photocurrent. One observation in a functional trap with gold coated
boron doped silicon electrodes: $\sim 500~\mu$W, 280~nm, grazing incidence
on a control electrode produced $\times10^{-9}$ A to $0.1 \times10^{-9}$ A
induced current.
\item Turn on an electron filament with an emission current of $100~\mu$A and
a bias of minus $50-100$ V. There should be several $\mu$A collected on each
of the trap electrodes that have a line of sight of the filament (and
perhaps some others if the trap RF is turned on). One observation
in a functioning trap: filament $1.2 $V $830$ mA, $100~\mu$A emission, $-80$
V bias; $1-5$ mV on a Fluke 79III multimeter ($10~M\Omega$ input impedance,
$100-500$ nA); trap RF off.
\item Turn on the trap RF. Look for RF on the control electrodes (from capacitive
pickup). Note that it may be difficult to determine the RF coupling
path.
\end{compactenum}
\item Install a copper pinch-off on the vacuum system valve cF after the
bake is complete. This is a stopgap in the event of a valve failure.
A \href{http://www.varianinc.com}{Varian, Inc.} adapter is needed to connect
the turbo pump to the pinch-off tube (p/n FCP03750275).
\item Prior to loading with an electron gun, permit it to outgas for several
hours with an emission current of $\sim 100~\mu$A, $50-100$~V bias. The
RGA indicates that tungsten filaments are a source of hydrogen.
\end{compactitem}

\subsection{Mg ovens}

\label{sec:MgOvens} A vapor of Mg is produced by a metal sample in
a miniature oven with an aperture directed toward the trap center.
The tubes were heated by running current thru their bulk. The current
was supplied by thin wires with low thermal conductivity.
 
The ovens were thin walled stainless steel tubes packed with Mg fragments
and sealed at the ends by crimping. A small hole along the length
of the tube permitted Mg vapor to escape in a beam pointed at the
trap. The tubes were $5-8$ mm long by $1$ mm diameter with a wall thickness
of $100$ $\mu$m. They were made of $304$ stainless steel. The stainless
was baked at 400$^{\circ}$ c in air overnight to outgas adsorbed
hydrogen \cite{bernardini1998a}. The hole was made at the middle
of the tube by nicking it's circumference with a $150$ $\mu$m wide
saw blade. Mg fragments were inserted in the tub and the ends crimped
with pliers. Electrical connection to each end of the stainless oven
was made by gap welding a pair of thin ($75$ $\mu$m diameter) tantalum
wires. The tantalum was in turn gap welded to thick stainless wires
using Advance metal (see Section~\vref{sec:electricalInterconnect}).
The resistive losses in the stainless dominated so most of the power
was dissipated in the oven, not the leads. care must be taken to prevent
the metal vapor from contacting functional surfaces in the vacuum
system including electrical traces on the trap wafers and laser/imaging
ports. In the octagon trap this is accomplished with a glass tube
surrounding the ovens which abutted the back side of the ceramic chip
carrier.

The natural abundance of Mg is $79 \%$ for $\,^{24}$Mg, $10 \% $
for $\,^{25}$Mg and $11 \% $ for $\,^{26}$Mg \cite{crc2008a}. 
The isotopically enriched $\,^{24}$Mg
used in my experiments was obtained from \href{http://www.ornl.gov}{Oak Ridge
National Lab} (ORNL $54-0134$).

As the current to the Mg oven is increased its temperature rises.
An indication of the presence of a Mg vapor is metallization (plating)
of exposed trap surfaces. However, the onset of metallization is extremely
nonlinear in current. It is believed this is due to the natural oxide
(MgO) on the surface of Mg which forms in air. MgO has a
substantially higher vaporization temperature ($2830^\circ$ C)
than Mg ($650^\circ$ C)
(\href{http://www.webelements.com}{www.webelements.com}). It is observed that there is little vapor released by the oven until the oxide layer is
broken, then there is a lot of vapor. The abrupt increase in deposition
was accompanied by an sharp increase in vacuum system pressure (e.g.
from $1\times10^{-10}$ Torr to $>1\times10^{-9}$ Torr). It is advised to watch
for this pressure burst and quickly reduce the oven current by $50 \%$ to
prevent exhaustion of the oven Mg supply. Then, slowly ramp back up the oven
current.

In one of my ovens I observed the onset of plating at $380$~mW ($0.76$
A, $0.50$ V). At this point the oven glow was barely visible on an IR
viewer with the lab lights off and the ion gage filament off. In
another similar oven, plating was observed at $633$~mW ($1.02$ A, $0.65$ V).
Subsequent loading occurred at lower oven current. Note that some ion
trappers at NIST have consistently loaded Mg ions with oven currents
low enough that a pressure burst was not observed.

Note that the leads to the trap ovens are current loops which may
couple external electric fields into the vacuum system (and cause
ion heating). This can be minimized by sinking oven current to the
vacuum housing and using a stiff low pass filter on the oven current
supply.

\begin{figure}
  \centering
  \includegraphics[width=0.8\textwidth]
  {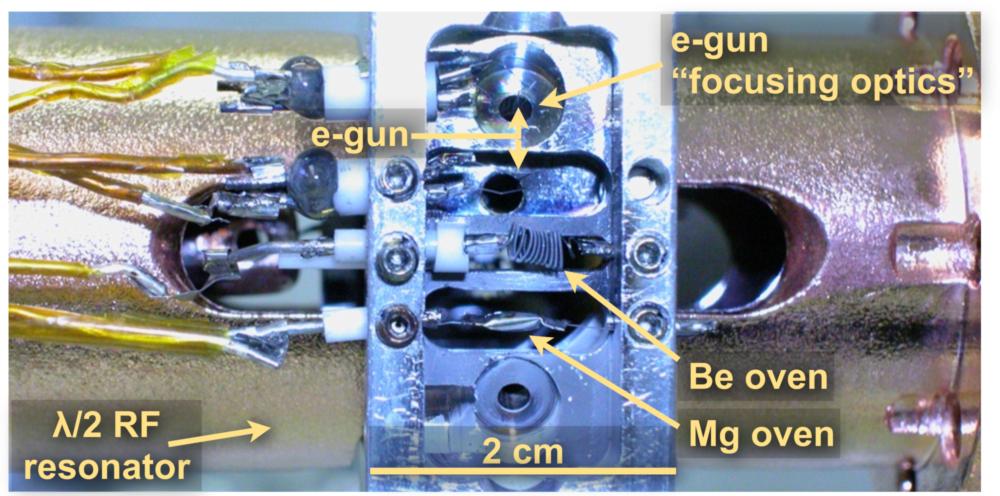}
  \caption[Photograph of a Mg oven installed in a $\lambda$/4 resonator.]
  {Photograph of electron filaments and a Mg oven attached to a $\lambda$/4
  resonator. They are mounted on an aluminum jig which orients the electron/neutral
  flux toward the ion trap center. The jig is attached to a quarter
  wave microwave resonator used to generate the trap RF potential. The
  ion trap is located at the center of the resonator. It is not visible
  since it lies behind the ovens in this picture. A Be oven is visible
  in this photograph but was not used in my experiments.}
  \label{fig:L2ovenMount}
\end{figure}

\begin{figure}
  \centering
  \includegraphics[width=0.7\textwidth]
  {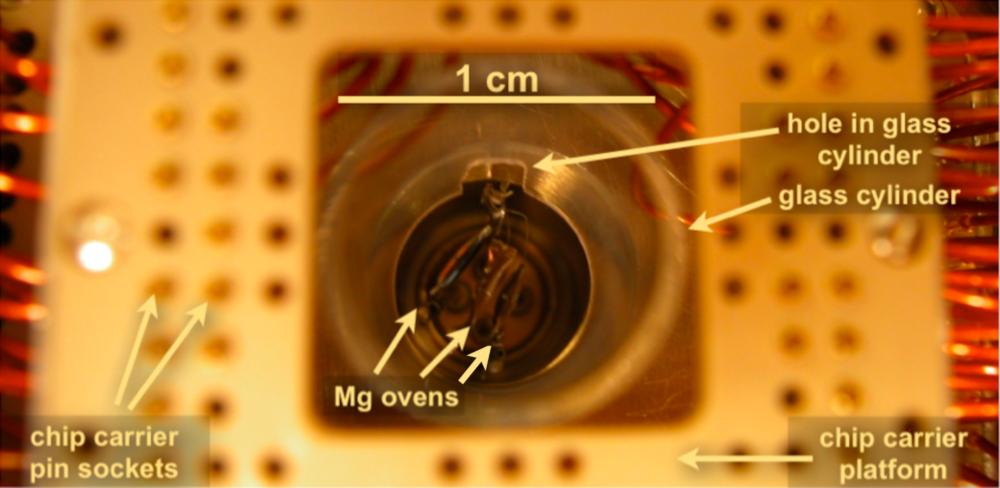}
  \caption[Photograph of Mg ovens.]
  {Photograph of three Mg ovens (for redundancy) installed in the octagon
  vacuum system. The Mg vapor flux is upward, out of the page. The ion
  trap chip is plugged into the pin sockets mounted in an array on the
  chip carrier platform. There is a slit in the trap wafer permitting
  Mg vapor to reach the trapping region (on the front side of the wafer).
  Surrounding the ovens is a glass cylinder which prevents deposition
  of Mg on the back side of the chip carrier platform and on the laser
  beam ports. A hole in the glass provides conductance of the cylinder
  to the ion pump (not visible).}
  \label{fig:OctOvens}
\end{figure}

\section{Ionization techniques}
\label{sec:ionizationTechniques}
\subsection{Electron impact ionization}

\label{sec:eGuns} Electron impact ionization is a standard technique
for generating single and multiple ionization states of atoms. It
is used at NIST for a variety of atomic species (Mg, Be, Hg, Al).
In 2005 a laser photoionization system was built for Mg (see Section~\vref{sec:MgPI}).
This section discusses electron impact ionization as used in my early
ion trap work.

Electrons were produced by thermionic emission from a hot ($\sim$1200$^{\circ}$
C) $75$ $\mu$m diameter tungsten filament. They were directed to the
trap center by crude electron optics biased to about $-100$ V with respect
to the trap RF. The electron energy is sufficient to overcome the
$7.65$ eV ionization threshold in Mg. Multiple ionization states of
Mg may have formed but were not trapped (\cite{jha2002a}).

Tungsten filaments work fine but are unpleasant in that they must
be very hot (causing pollution of the vacuum system), they burn out
immediately if exposed to air and become extremely fragile (brittle)
after first use. After my first trap I switched to 1$\%$ thoriated
tungsten which operates at lower temperatures and is less fragile
after first use. Yb-oxide and Th-oxide are also options. For a thoriated
filament, a $-80$~V bias and a filament current of $830$~mA, a $1.2$~V drop
across the $\sim 5$~mm long filament and $100~\mu A$ emission was observed.

Note that the (thermionic) work function for tungsten is $4.54$ eV while
for thoriated tungsten it can be as low as $2.96$~eV \cite{mason1990a}.

While a stalwart in ion trapping, electron impact ionization is problematic for
microtraps. The electron beam is poorly directed and is further deflected by the
high RF potentials on the trap electrodes ($\sim$100~V, $\sim$100~MHz). A low
electron density at the trap center and charging of dielectric surfaces may have
contributed to early unsuccessful attempts at loading my first surface electrode
trap in summer $2005$ (see Section~\vref{sec:dv14}).  Loading of this trap was
ultimately successful in 3/2006 using photoionization. The Seidelin trap was
initially loaded several times using electron impact ionization, but the loading
efficiency was low and ion lifetime extremely short due presumably to surface
charging \cite{seidelin2006a}.  To my knowledge no surface electrode microtrap
has been routinely loaded by electron impact.

\subsection{$^{24}$Mg photoionization}

\label{sec:MgPI}

Neutral $\,\text{Mg}$ can be photoionized (PI) with $1-10$~mW CW~$285$~nm laser
light (focused to a $40~\mu$m waist). Photoionization of $\,\text{Mg}$ was
first demonstrated experimentally by Madsen, Drewsen, \textit{et al.} in 2000
\cite{madsen2000a}.\footnote{Our labs also photoionize $^9$Be using a
frequency quadrupled pulsed $Ti:Sa$ \cite{deslauriers2006b}.}

\begin{SCfigure}
  \centering
  \includegraphics[width=0.5\textwidth]
  {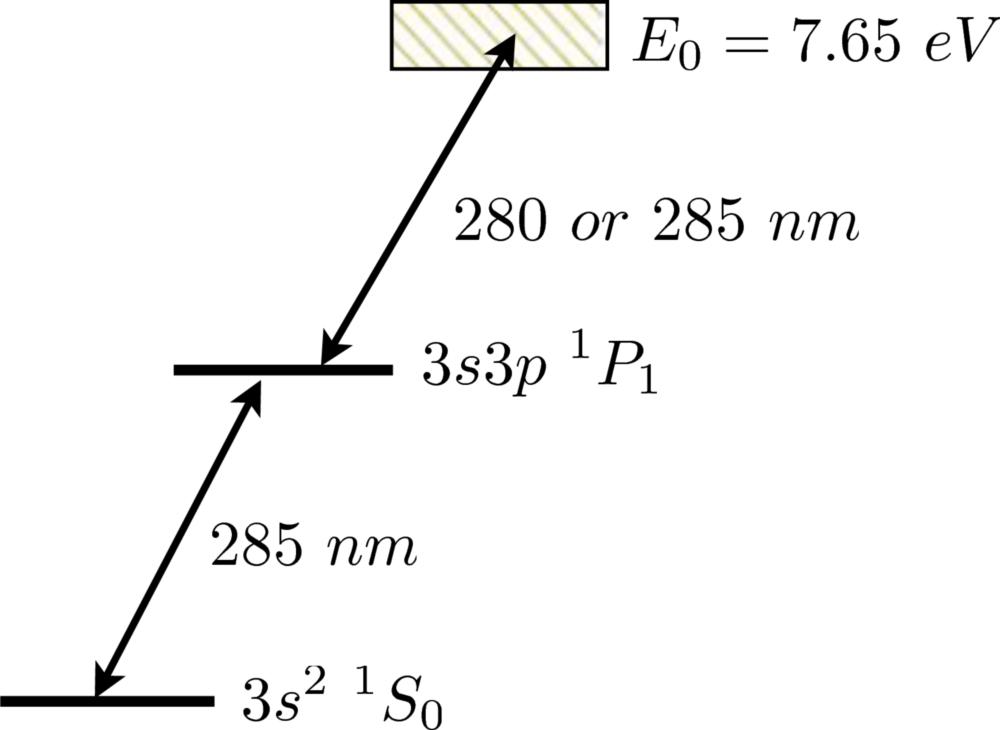}
  \caption[Two photon photoionization of $^{24}$Mg.]
  {Two photon photoionization of $^{24}$Mg. Two-photon ionization takes
  place thru the $3s^{2}\,^{1}S_{0}\to3\text{s3p}\,^{1}P_{1}$ transition.
  The first photon is resonant with the strong transition to the lowest
  lying P state. A second photon of the same frequency goes to the continuum.
  The second photon can also be a $280$~nm photon.}
  \label{fig:MgPIlevelDiagram}
\end{SCfigure}

There are several advantages of PI over electron impact ionization.
With the PI laser there is no need for a hot electron filament and
loading is possible with lower Mg oven currents. The result is less
contamination of the vacuum system during loading. The $285$~nm laser
beam's power is focused on the trap center on a trajectory which avoids
the trap electrodes. As as result there is less charging of trap dielectrics
near the ion. 

\begin{SCfigure}
  \includegraphics[width=0.7\textwidth]
  {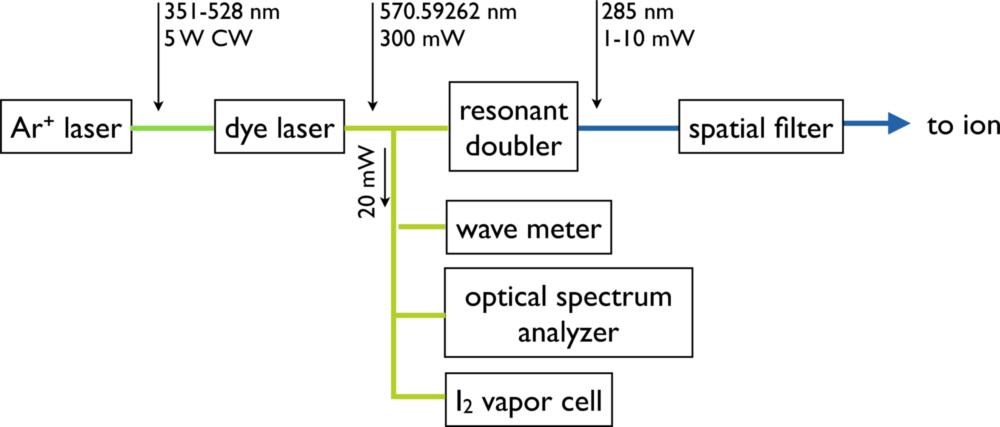}
  \caption[Schematic of laser setup for the 285~nm cW
  laser light used to photoionize
  $\,^{24}\text{Mg}$.]
  {Schematic of laser setup for the
  285~nm cW laser light used to photoionize
  $\,^{24}\text{Mg}$.}
  \label{fig:285optics}
\end{SCfigure}

The $285$~nm system is shown in Figure~\vref{fig:285optics}. The pump
is a \href{http://www.spectra-physics.com}{Spectra Physics, Inc.}
$Ar^+$ laser lasing multimode with $5$~W output power. This light pumps
a \href{http://www.coherent.com}{Coherent, Inc.}~$699$ ring dye laser stabilized
with a Hansch-Couillaud lock to a reference cavity 
\cite{hansch1980a}. The dye is \href{http://www.exciton.com}{Exciton,
Inc.} Rhodamine~$560$ in ethylene glycol ($1.6$ g per $400$ mL).
The green light pumps a Wavetrain resonant doubling cavity made by
\href{http://www.las.de}{Laser Analytical Systems, GmbH.} (Berlin). Second
harmonic generation occurs in an angle phase matched BBO crystal. The cavity
is stabilized by the Pound-Drever-Hall method \cite{drever1983a}.
The UV light beam had numerous lobes perpendicular
to the cavity plane.\footnote{These lobes lie along the matching axis of the
BBO and are due to beam walk off.  They are an intrinsic feature of angle phase
matched doublers.} The center lobe was highly astigmatic and required
correction by a pair of cylindrical lenses.  A spatial filter was used to select
the highest power mode and produce a clean beam.  

For $\,^{24}\text{Mg}$ photoionization the dye laser was tuned near the
$3s^{2}\,^{1}S_{0}\to3\text{s3p}\,^{1}P_{1}$ transition at 285.59262~nm using a
traveling cart wave meter \cite{hall1976a,fox1999a}. A Doppler broadened
transition in an $I_{2}$ vapor cell was used as a crude absolute frequency
reference at $570.6$~nm. I tuned approximately $300$~MHz to the red side of the
$1.2$~GHz Doppler broadened (at $300$ K) feature set at
$17,525.632$~$cm^-1$ (scan number $1749006$ in Volume $3$ of the JSPS iodine
atlas \cite{kato2000a}. Note that $1$~$cm^{-1}$ is approximately $30$~GHz. There
were no AOM's used in this setup. The laser stability and Doppler broadening of
the hot ($\sim400^\circ$ C) $\,^{24}$Mg vapor permitted efficient
photoionization with the laser system running open loop for hours at a time. Single mode
operation was confirmed with a Tropel L240 optical spectrum analyzer with a
$1.5$~GHz free spectral range in the visible ($3.0$~GHz in the UV).

\clearpage

\section{$^{24}$Mg$^{+}$ laser cooling}
\label{sec:dopplerLaserCooling}

The electronic structure of $\,^{24}\text{Mg}^{+}$ is simple. Like
an alkali neutral atoms it has only a single valence electron. And,
since its nuclear spin is zero it has no hyperfine structure. Laser
cooling is possible on the D1 and D2 transitions at 280~nm. Laser
cooling of $\,^{24}\text{Mg}^{+}$was first demonstrated by Wineland,
\textit{ et al.} in a Penning trap\textit{ }\cite{wineland1978a}\textit{.}

Ordinarily, laser cooling of many atomic ion species requires wavelengths
that can only be produced by dye lasers and frequency doubling. Recently,
however, commercial solid state fiber lasers have become
available at $1120$~nm, the fourth subharmonic of $280$~nm
\cite{friedenauer2006a}. I used a 
\href{http://www.koheras.com}{Koheras, Inc.} CW Boostik Y10 fiber laser
system is available in the range $1030-1121$~nm with $0.5$~nm tuning, 
a $70$~kHz line width
and $1-2$~W coherent emission ($\sim$1~W amplified spontaneous emission). 


The choice of $\,^{24}\text{Mg}^{+}$ for ion trap testing has several
motives. Beyond successful trapping, a critical measure of a trap's
promise for ion QIP is its electric field noise which can heat the
ions. Historically, this has been measured in ions with hyperfine
structure cooled to the motional ground state, an expensive and technically
challenging undertaking \cite{bible,monroe1995a}. In 2007 a measurement
technique relying on only a single Doppler cooling was demonstrated
for $\,^{24}\text{Mg}^{+}$ (see Section~\vref{sec:recooling} and 
\cite{wesenberg2007a,epstein2007b}). The absence of hyperfine structure
also obviates the need for a quantizing magnetic field, a repump laser
and careful laser beam polarization control. Also, Mg thermal ovens
are straightforward to build and operate (see Section~\vref{sec:MgOvens})
and Mg can be photoionized (see Section~\vref{sec:MgPI}).

\begin{figure}
  \includegraphics[width=1\textwidth]
  {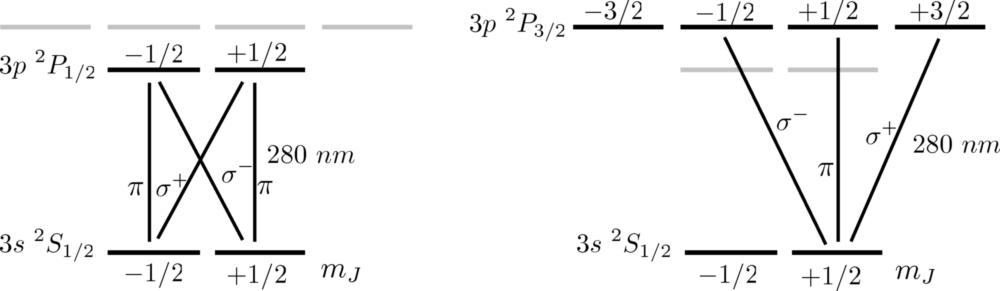}
  \caption[Atomic level diagram for $\,^{24}\text{Mg}^{+}$.]
  {Atomic level diagram for $\,^{24}\text{Mg}^{+}$ showing the low
  lying P states and two cooling transitions: (left) D1
  $3s^{2}\,^{2}S_{1/2}\to3p\,^{2}P_{1/2}$ and (right) D2
  $3s^{2}\,^{2}S_{1/2}\to3p\,^{2}P_{3/2}$.  In my experiments I used circularly
  polarized light ($\sigma^{+}$ or $\sigma^{-}$) on the D2 transition
  for laser cooling. In the $\sim 10$~Gauss ambient magnetic field the $m_{J}$
  levels were degenerate at the sensitivity of my experiments. }
  \label{fig:Mg24levelDiagram}
\end{figure}

To be conservative, the blue most $3s^{2}\,^{2}S_{1/2}\to3p\,^{2}P_{3/2}$
transition in $\,^{24}\text{Mg}^{+}$at $\lambda_{0}=279.636$~nm
was selected for laser cooling \cite{drullinger1980a}. This transition
has a linewidth of $\Gamma_{0}\simeq40\text{MHz}$. Note that the
fine-structure splitting between the $P_{1/2}$ and $P_{3/2}$ states
is about $0.7$~nm ($2.75$~THz) \cite{ozeri2007a}.  See also the
\href{http://physics.nist.gov/PhysRefData/ASD/lines_form.html}{NIST atomic
spectra database}.

The Boostik Y10 fiber laser I used had a factory
set center frequency of $1118.54$~nm. It did not meet many of the factory
specifications but worked well enough for my purposes.  It could be slowly
temperature tuned over a $0.5$~nm range centered at about $41.7^\circ$ c in
several minutes. Tuning is also possible with a piezoelectric element over a 
bandwidth of several GHz at a rate of $<$ 1~kHz, considerably less
than the $0.4$~nm and $100$~kHz, respectively, specified by Koheras.
The laser's output polarization was often stable, but some days drifted
at up to 1/2~radian per hour. Over the course of a typical day its
frequency drifted by $<$~1~GHz. Between power cycles the frequency
varied by $<$~10~GHz. The fiber laser is frequency stabilized by
locking it's frequency with a piezo tuning element to a saturated
absorption hyperfine feature in an $I_{2}$ vapor cell whose frequency
is nearly twice that of the cooling transition \cite{wieman1976a,preston1996a}.
Piezo tuning has considerable hysteresis but not enough to prevent
locking. The laser reliably produced $1.8$~W of coherent laser light
since $2006$.

The fiber laser light is doubled four-fold to $280$~nm by a pair of
resonant doubling cavities, crudely approximating the setup of the
Schätz group \cite{friedenauer2006a}. In the first cavity, visible
light at $560$~nm is generated by second harmonic generation in a noncritical
phase matched LBO crystal at $95.0^\circ$~C \cite{boyd1968a}.
The cavity is a folded ring cavity and is stabilized by the Hansch-Couillard
method \cite{hansch1980a}. Its output is largely $\text{TEM}_{00}$mode.
In the second cavity, UV light is produced by a Wavetrain resonant
doubling cavity made by \href{http://www.las.de}{Laser Analytical Systems,
GmbH}. Second harmonic generation occurs in a two-mirror Delta cavity with
a a critically phase matched BBO crystal (see \href{http://www.las.de}{LAS
website} for details). The cavity is stabilized by the Pound-Drever method
\cite{drever1983a}. The UV light beam was highly astigmatic and 
had numerous lobes perpendicular
to the cavity plane.\footnote{These lobes lie along the matching axis of the
BBO and are due to beam walk off.  They are an intrinsic feature of angle phase
matched doublers.} A spatial filter was used to select the middle
most mode and produce a clean beam. This light was then sent to a
pair of acoustooptic modulators (AOMs) before being directed to the
ion trap.

\begin{figure}
  \centering
  \includegraphics[width=1\textwidth]
  {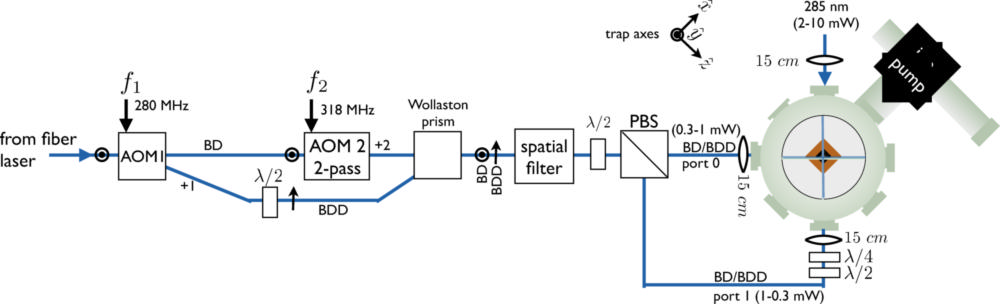}
  \caption[Schematic of $\,^{24}\text{Mg}^{+}$ Doppler cooling laser.]
  {Schematic of $\,^{24}\text{Mg}^{+}$ Doppler cooling laser.}
  \label{fig:Mg24dopplerCoolingLaserSetup}
\end{figure}

\begin{figure}
  \centering
  \includegraphics[width=1\textwidth]
  {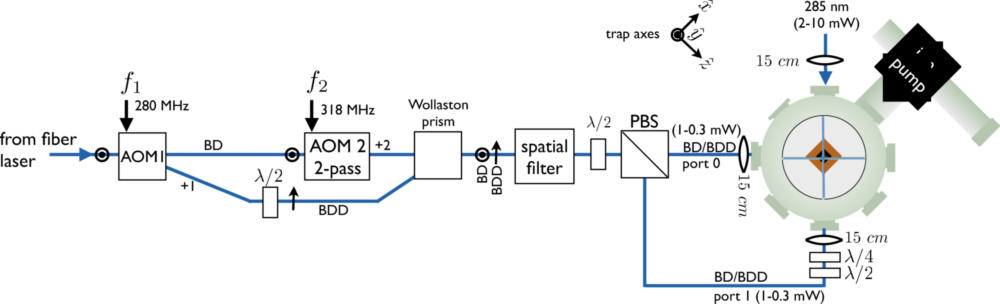}
  \caption[Schematic of AOMs and beam geometry.]
  {Schematic of AOMs and beam geometry.}
  \label{fig:testTrapBeamPath}
\end{figure}

A pair of acoustooptic modulators provides frequency shifts to the
Doppler laser setup which assist with cooling, micromotion compensation
and heating measurements. The first, AOM~1 in
Figure~\vref{fig:testTrapBeamPath}, splits the incoming 
light into two beams: a Doppler cooling laser
(BD) and a detuned Doppler cooling laser (BDD). AOM~2 varies the BD
detuning $\Delta_{\text{BD}}$ with respect to resonance at $f_{0}$,
$\Delta_{\text{BD}}=f_{\text{BD}}-f_{0}$. It is setup in a double
pass configuration making its deflection angle first order insensitive
to the RF drive frequency $f_{2}$. This AOM is used to measure the
atomic transition line shape and determine the transition's frequency
$\omega_{0}$ and linewidth $\Gamma_{0}$. These numbers are needed
because Doppler cooling efficiency varies with detuning $\Delta_{\text{BD}}$
and is an important parameter for estimating ion heating 
(see Section~\vref{sec:recooling}).

Ion fluorescence is sensitive to micromotion along the laser beam direction at
$\Delta=\omega_{0}-n~\Omega_{T}$, where $n\in\{1,2,\ldots\}$ and $\Omega_{T}$ is
the trap RF drive frequency (see Section~\vref{sec:umDetectionByFluoresence}).
When $f_{2}=318$~MHz, BD is on resonance $\Delta_{\text{BD}}=0$. Typically,
$f_{2}=313$~MHz giving $\Delta_{\text{BD}}=-\Gamma/4=-10$~MHz.

Frequency $f_{1}$ is applied to AOM 1 which puts power into a secondary, detuned
Doppler laser beam, BDD. When an ion's temperature is well above the Doppler
cooling limit, the near resonant BD beam has little cooling power due to the
Doppler broadening of hot ions. Additional cooling from BDD also prevents ion
loss. Typically, $f_{1}=280$~MHz giving $\Delta_{\text{BDD}}=-356$~MHz.

The AOMs are \href{http://www.intraaction.com}{IntraAction, Inc.} ASD-3102LA62-1
wide band modulators with a center frequency of 310~MHz and an efficiency of $>$
60$\%$ over a range of 100~MHz. The frequencies $f_{1}$ and $f_{2}$ are generated
by computer controlled direct digital synthesizers (DDS). The same computer
collects ion fluorescence with a PMT (see Section
\vref{sec:controlElectronics}).

It is desired that BD and BDD recombine after the AOM's and copropagate.
Combining beams on a 50/50 beam splitter is straightforward but wastes half the
power in each beam. Instead, the BDD beam's polarization was rotated by
90$^{\circ}$ with a wave plate and combined with the BD beam on a Wollaston
prism. While there is little power loss with this approach, the copropagating
beams have orthogonal polarizations. This is exploited to permit continuous
variation of the ratio of BD to BDD light incident on two ports of the vacuum
system using a wave plate. Note that the UV mirrors I used attenuate horizontally
polarized light more than vertically polarized light (for propagation in the
plane of the optics table). In some experiments the BDD beam was extinguished
completely by switching off power to AOM~1.

The vacuum system has several quartz optical ports (see
Figure~\vref{fig:testTrapBeamPath}). The beam at Port~0 was directed to the
trap's load zone by a 15~cm singlet lens. This beam was used principally for ion
loading. The ratio of BDD to BD power was~2-4. The beam at Port~1 was used to
cool ions in various experimental trap zones. Its polarization was controlled
(see below) and the ratio of BD to BDD power was~2-4. A 15~cm singlet was also
used for Port~1. The photoionization laser entered the vacuum system thru a port
counterpropagating with respect to the the Port~1 cooling beams. It was focused
to the position of the load trap by a 15~cm singlet lens. For $\lambda=280$~nm,
an initial beam diameter of $w=1\text{mm}$ and $f=15$~cm, the beam waist at
the trap is $\frac{\lambda f}{\pi w}\sim14~\mu$m.

The polarization of the BD and BDD beams for port $1$ is set by a pair
of adjustable wave plates ($\lambda$/2 followed by $\lambda$/4,
see Figure~\vref{fig:testTrapBeamPath}). They were set to maximize
ion fluorescence from the BD laser beam in Port $1$. This operation
compensates for birefringence in the quartz window mounted on the
vacuum system. Note that this compensation may not be complete in
the case of a spatial gradient to the birefringence across the laser
beam due for example to mechanical stress in the window. The window
is mounted on a vacuum conflat flange (CF). The range of fluorescence
observed for various laser beam polarizations was at most a factor
of two. Note that the ambient lab magnetic field was $\sim10$~Gauss so
the $m_{J}$ levels were degenerate at the sensitivity of my experiments.

It is important that the Doppler cooling laser overlap ($\geq$ 10$^{\circ}$)
with all trap principle axes. Otherwise, photon recoil can cause heating.
Meeting this requirement in surface electrode ion traps requires proper
choice of trap electrode geometry~\cite{chiaverini05b}. See 
Section~\vref{ions:sec:doppler}.

 \begin{figure}
    \centering
   \includegraphics[width=.90\textwidth]
   {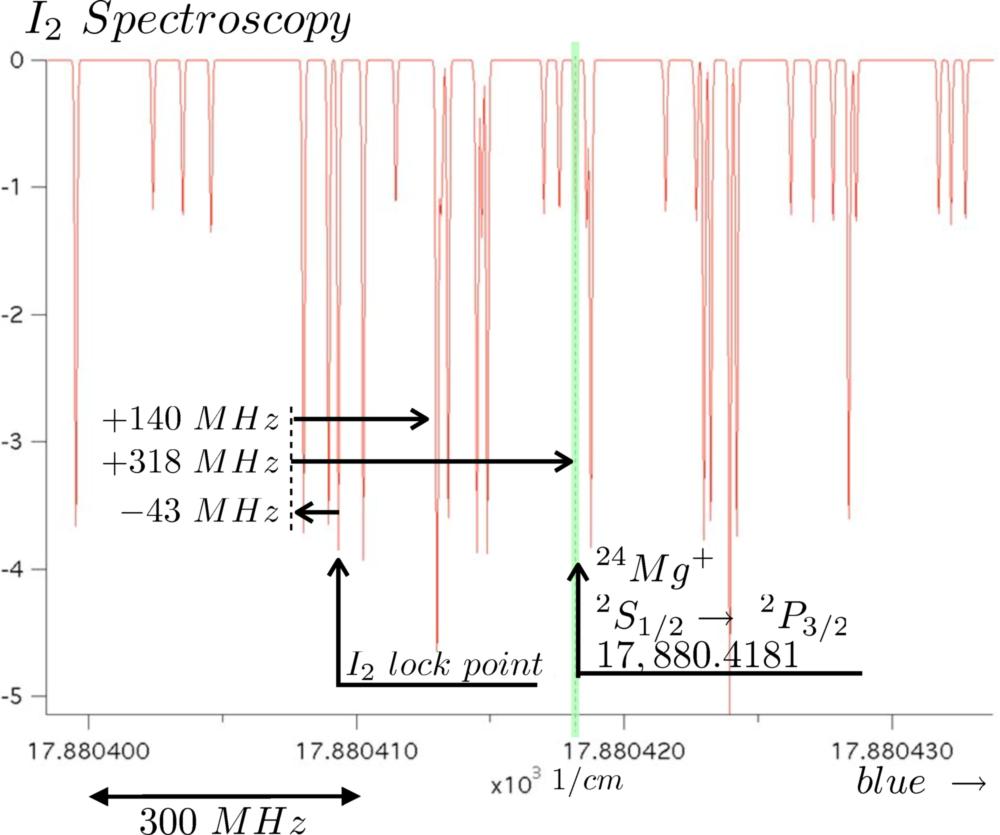}
   \caption[Iodine spectroscopy for the Doppler cooling transition 
   in $\,^{24}\text{Mg}^{+}$.]
   {Iodine spectroscopy for the Doppler cooling transition in $\,^{24}\text{Mg}^{+}$.
   Absorption features in iodine span much of the visible spectrum. In
   the figure is an absorption spectrum corresponding to the hyperfine
   spectra of a certain rho-vibrational transition. The same features
   are visible in my Doppler free spectroscopy setup and serve as an
   excellent absolute frequency reference. The green line marks the location
   in the iodine spectrum for the $\,^{2}S_{1/2}\to\,^{2}P_{3/2}$ transition
   in $\,^{24}\text{Mg}^{+}$ in the visible at 17,880.4181/cm (559.271039~nm)
   \cite{schaetz2006a}. See also \cite{herrmann2008a}.  The fiber laser is
   locked to the marked spectral feature using Doppler free spectroscopy which adds a -43~MHz shift. Laser
   light resonant at the atomic transition is obtained by doubling the 
   599~nm light and shifting it by $2\times318$~MHz
   in the UV with a double pass AOM. Nonresonant laser light detuned
   from resonance by -356~MHz is obtained with a 280~MHz AOM in the UV.
   See Figure~\vref{fig:testTrapBeamPath}. This $I_{2}$ spectrum was
   generated by the \href{http://www.toptica.com}{Toptica, GmbH.} Iodine Spec 
   software \cite{bodermann2002a,knockel2004a}.
   The actual signal in my setup has a dispersive line shape for easy
   locking (see Figure~\vref{fig:560I2setup}).}
   \label{fig:560I2spectrum}
 \end{figure}
 
\clearpage

\paragraph{$I_2$ spectroscopy}

\begin{figure}
  \centering
  \includegraphics[width=0.6\textwidth]
  {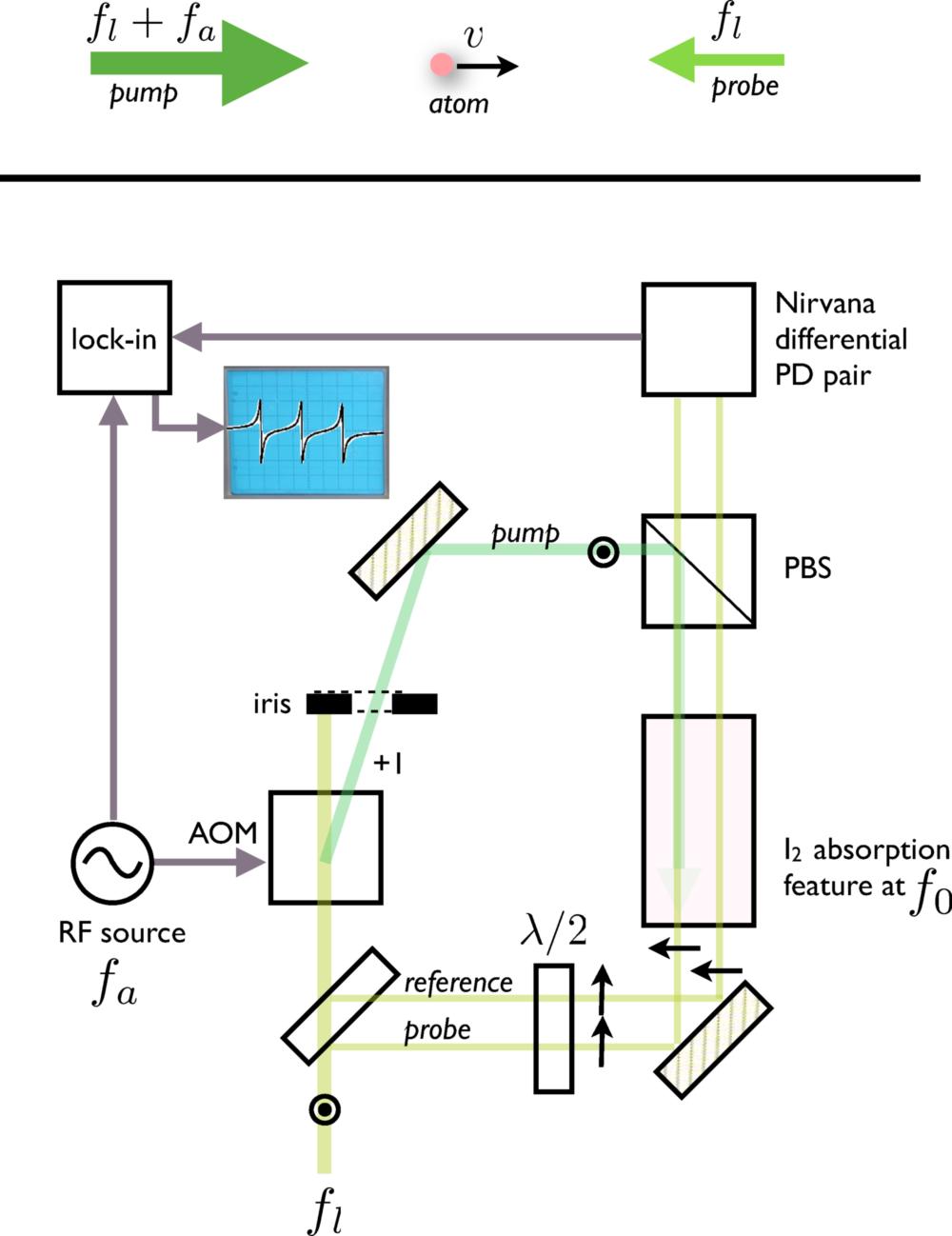}
  \caption[Schematic of saturated absorption spectroscopy setup for $I_{2}$
  using a single AOM.]
  {Schematic of saturated absorption spectroscopy setup for $I_{2}$
  using a single AOM. }
  \label{fig:560I2setup}
\end{figure}
 
At room temperature the hyperfine spectra plotted
  in Figure~\vref{fig:560I2spectrum} are each Doppler broadened (at room
  temperature) to about 1~GHz. Saturated absorption spectroscopy is a technique
  which uses spectral hole burning to resolve these features. The components I
  used to do this are shown in Figure~\vref{fig:560I2setup}.  A good pedagogical
  discussion is by D. Preston~\cite{preston1996a}.
  
  The $560$~nm laser beam at frequency $f_{l}$ has a linewidth ($\sim
  70$~kHz) that is narrow relative to the $I_{2}$ hyperfine feature
  spacing (several~MHz). The lock signal is offset from $f_{l}$ by
  $\left.-f_{a}\right/2$. This can be seen as follows. Suppose we wish
  to resolve an $I_{2}$ HF feature at $f_{0}$. Let $v$ be the velocity
  of atoms resonant with the pump at $f_{l}+f_{a}$, 
  \begin{equation*}
  	f_{0}=\left(f_{l}+f_{a}\right)(1-v/c). 
  \end{equation*}
  These atoms are pumped into the excited state, leaving a spectral
  hole. Let $v'$ be the velocity of atoms resonant with the probe at
  $f_{l}$,
  \begin{equation*}
  	f_{0}=f_{l}(1+v'/c). 
  \end{equation*}
  Note that the plus sign in this expression is due to counter propagation
  of the pump and probe beams (as compared to the previous expression
  where there is a minus sign and the beams copropagate). The probe
  is minimally absorbed when it is resonant with the population of pumped
  atoms moving at velocity $v$. That is, when $v=v'$. This happens
  when $f_{l}=f_{0}-\left.f_{a}\right/2$.
  
  A signal appropriate for the lock-in technique is generated by applying a
  100~kHz dither to $f_{a}$ and using a differential photodiode pair
  (\href{//http://www.newfocus.com}{New Focus, Inc.} Nirvana Detector). The
  signal from of the lock-in has a dispersive line shape symmetric about zero
  which can be easily used in a PID servo loop. In my setup
  $f_{a}=86.065\text{MHz}$. The iodine cell was purchased from
  \href{www.triadtechno.com}{Triad Technology, Inc.} (p/n~TT-I2-100-V-P).

\section{Ion imaging optics}

\label{sec:imagingOptics}
\begin{figure}
  \centering
  \includegraphics[width=1\textwidth]
  {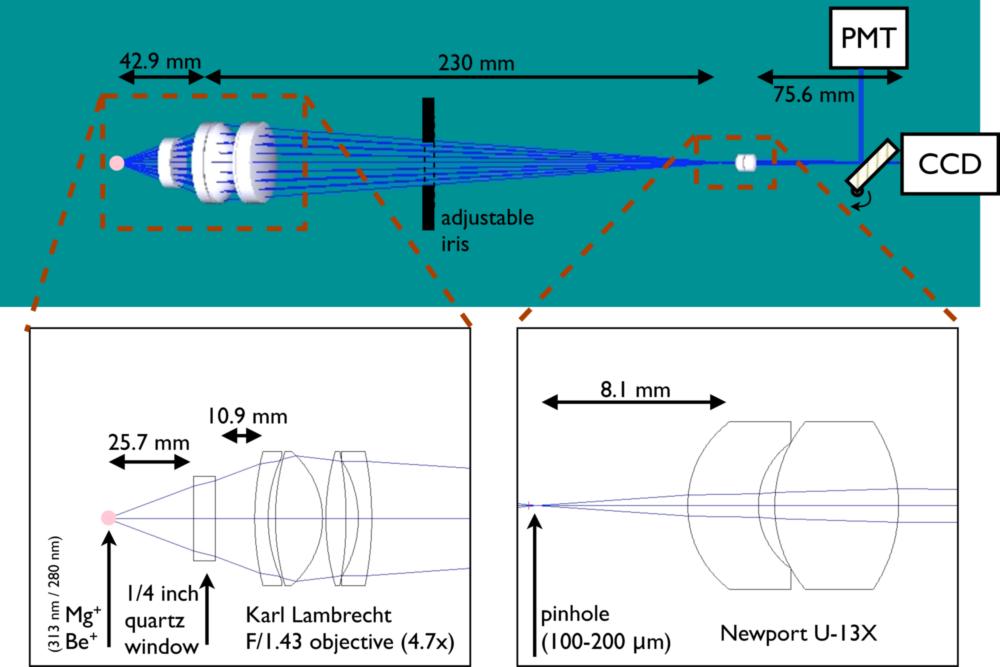}
  \caption[Schematic of ion imaging optics.]
  {Schematic of ion imaging optics. The ray trace diagrams are due to Pei Huang.}
  \label{fig:imagingOptics}
\end{figure}

The ion imaging system collects ion fluorescence
and directs it to an imaging camera or a photomultiplier tube. It
was designed for good collection efficiency and resolving power at
both 280~nm (for $\text{Mg}^{+}$) and 313~nm (for $\text{Be}^{+}$).
In particular it was desired that the system clearly resolve individual
ions in a linear crystal. For example, 5.5 $\mu$m is the inter-ion
spacing for a $\,^{24}\text{Mg}^{+}$ three ion crystal when $\left.\omega_{z}\right/2\pi=1.0\text{MHz}$.
$\omega_{z}$ is the frequency of the lowest (COM) vibrational mode
along the trap axis. The geometry is illustrated in Figure~\vref{fig:imagingOptics}.

The imaging system discussed below was designed for the ion trap testing setup.
It resides outside the vacuum system, oriented normal to the trap surface. A full
system simulation using \href{http://www.zemax.com}{ZEMAX} was done by Pei Huang
in 2006. For the geometry in Figure~\vref{fig:imagingOptics} he found the
resolution to be 4.5~$\mu$m at 280~nm for rays near the optical axis and
6.5~$\mu$m for rays at the edge of the lens.

The objective was a custom F/1.43, 4.7x, 4-element system made by
\href{http://www.klccgo.com}{Karl-Lambrecht, Inc}. On the way to the detectors
the ion's fluorescence passed thru a 0.25~inch (6.35~mm) thick quartz UHV
window. This window caused significant spherical aberration for far off axis
rays. The light then propagated in air to the objective. The objective was designed to
correct spherical aberration, but for a 2~mm window not a 6.35~mm one. An additional
refractive element could increase the lens's resolution (to 3.3~$\mu$m on axis)
by correcting this aberration. An adjustable iris was occasionally used to clip
these rays when resolution was valued above collection efficiency. Laser beam
light may be scattered into the objective from trap structures, a background that
degrades signal to noise. This stray light was reduced by placing a pinhole
at the primary's focus which clips rays not originating from the ion's focal depth.
A commercially available F/6.7, 13x UV microscope objective imaged the pinhole
and relayed light to the detectors (\href{http://www.newport.com}{Newport
Corporation Inc.} p/n U-13X). The system's design magnification was 30x but in 
practice was operated at around 50x. 

A flipper mirror directed the light to either a charge-coupled device (CCD)
camera or a photomultiplier tube (PMT). The entire optical train from the
objective to the detectors was in a light tight box (not illustrated in
Figure~\vref{fig:imagingOptics}). Micrometers translated the objective permitting
the 50~$\mu$m field of view to be panned across the trap surface. The focus was
set by a third micrometer.

The PMT was a Hamamatsu H7732P-11 designed for photon counting. The radiant
sensitivity of its cathode was $R_{\lambda}=$$\sim$55~mA/W at 280~nm, giving a QE
of 24$\%$. Note that 100$\%$ quantum efficiency at 280~nm is
 $\left(\frac{\hbar\omega}{e}\right)^{-1}=\left(\frac{hc}{\lambda
 e}\right)^{-1}$= (4.43~$W/A)^{-1}=$ 0.23~A/W. It's dark count at room
 temperature was specified to
be 80~$s^{-1}$. The PMT signal was processed by a discriminator and converted to
TTL for recording by an FPGA in the experiment control computer.

The camera was an \href{http://www.andor.com}{Andor, Inc.} iXon DV887 electron
multiplying CCD (EMCCD). Its $512$x$512$ pixel array was readout serially at
10~MHz, 26~ms/frame. In practice we used a longer integration time
($300\text{ms}/\text{frame}$) to improve signal to noise. The CCD pixel size was
16x16~$\mu$m and its quantum efficiency at 280~nm was $\sim$35$\%$. The
electron multiplication gain factor was variable: 1-1000x. The CCD saturated at
174,000 electrons/pixel/sec and could be exposed to room light. With minimum
gain, the total dark count rate was about 1~kHz when the CCD was cooled
(thermomelectrically) to -90$^{\circ}$~C.

The Andor replaced the imaging photon detector used in previous NIST ion trap
experiments, the \href{http://www.photek.com}{Photek, Inc.} IPD3 multichannel
plate (MCP) detector. Advantages of the EMCCD over the MCP include higher pixel
count, much higher saturation level, lower cost, higher quantum efficiency at
280~nm and more fully featured software. The MCP is superior for single photon
counting due to its lower dark count rate (64~Hz at 20$^{\circ}$ C) and when
excellent timing resolution (20~ns) is required.

\section{Experiment control electronics}
\label{sec:controlElectronics}
\subsection{Introduction}
A custom software/hardware platform
was developed at NIST to control the experiment apparatus used in
quantum logic and trap testing experiments. It's primary functions
are the following.

\begin{compactitem}
  \item define a syntax to specify controlled interactions between ions and
  trapping fields and modulated laser/RF source; the evolution of a
  particular experiment is specified in a digital control .dc file
  \item deterministically control experiment evolution
  \item measure and record ion qubit states by resonance fluorescence
  \item deterministically branch experiment evolution in minimal time based
  on measurement outcomes
  \item support calibration and diagnostic routines
\end{compactitem}
The architecture for this system was designed by Chris Langer and
is discussed in detail in his thesis \cite{langer2006a}. Since $2006$
the system was extended by many people at NIST. For example, the .dc file syntax 
was extended to control DACs and optics micropositioners.
The trap testing experiment utilized a subset of these capabilities.
Prior to Langer's platform, other tools were used. See for example
Section \vref{sec:earlyExpControl}.

\begin{figure}  
  \centering
  \includegraphics[width=0.9\textwidth]{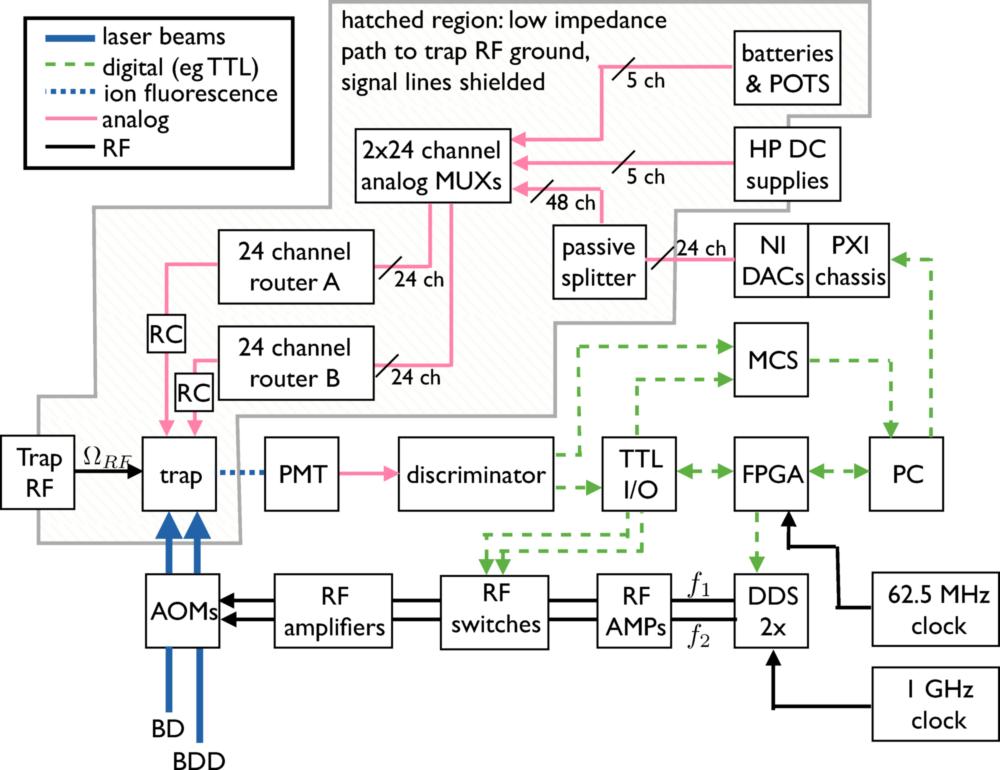}
  \caption[Functional schematic of the electronics controlling the trap testing
  experiment]{Functional schematic of the electronics controlling the trap testing
  experiment}
  \label{fig:trapTestingControl}
\end{figure}

Figure~\vref{fig:trapTestingControl} shows the experiment control
apparatus used for trap testing with $\,^{24}\text{Mg}^{+}$. The
sequence of operations comprising an experiment is written in a special
language and stored in a .dc file. This file is interpreted by software
on the PC which fills a table in a field programmable gate array (FPGA)
with the experiment steps. It also prepares control electrode potential
waveforms for issuance by the DACs. Our qubit evolves at a characteristic
frequency $\omega_{0}$ ($\sim$2 GHz) and some operations (eg laser
beam pulses) must be synchronous with its evolution. Such coherent
operations are are completely governed by the FPGA.

The heart of the system is an FPGA. An FPGA is a microelectronics
chip which implements a user defined logic. Traditionally this logic
was implemented using individual logic components (chips) like AND
gates, shift registers and memory interconnected on a printed circuit
board. An FPGA however accomplishes the same logic using a very large
(200,000) array of elementary logic elements whose interconnection
is determined by a reconfigurable bank of switches. Unlike microprocessors
which typically respond to external stimulus via interrupt requests
(IRQs) with $\mu$s delays, an FPGA is a state machine and responds
deterministically (with ns delay) to multiple simultaneous stimuli.
The logic it performs is specified in a language called VHDL and compiled
(and optionally simulated) for a specific FPGA chip. The FPGA logic
for ion trapping experiments provides a structure (a table) which
holds a sequence of operations (e.g. apply laser S to an ion or collect
photons from PMT X) and executes appropriate steps to implement each
operation in experiment hardware (e.g. turn on RF switch S or record
pulses from PMT X) at the right time.

The RF driving the acoustooptic modulators (AOMs) is generated by
direct digital synthesizers (DDS). Direct digital synthesis is a means
of generating arbitrary frequencies from a fixed frequency reference.
The RF is turned on and off by TTL controlled RF switches and amplified
by RF instrumentation amplifiers. The RF electronics responsible for
confining the ions is discussed in 
Section~\vref{sec:trapTestingRFelectronics}.

\subsection{Control systems built for this thesis work}
The remainder of this section discusses my contributions to the
experiment control system.

The FPGA communicates with the DDS modules and the TTL IO board over
a low voltage differential signal (LVDS) bus. LVDS is a digital signaling
standard for transmitting high speed data in noisy environments over
cheap twisted-pair cables. Multiple devices can tap the bus along
its length and all devices can transmit and receive. The ends of the
bus are terminated in $110$~$\Omega$. The performance of this bus
(lower bus error rate) was improved in recent years by reducing its
total length and using a custom rigid backplane instead of twisted-pair
cables.

Light scattered by the ion (and stray light) is detected by a photomultiplier
tube (PMT) (see Section ~\vref{sec:imagingOptics}). A discriminator
rejects low amplitude PMT pulses associated with dark noise and produces
TTL pulses for signals above a threshold. Two signal processing systems
exist which generate histograms of observations spanning many identical
experiments. In one system, the .dc file specifies the interval during
an experiment when the FPGA responds to PMT pulses. During this interval
the FPGA counts the number of PMT pulses it receives and adds the
count to a histogram on the FPGA (number of experiments vs. photon
count). Multiple distinct detection intervals (and resulting histograms)
can be defined within a single experiment. In the second system, PMT
pulses are sent to a multichannel scalar (MCS). The .dc file defines
a specific starting time when the MCS should begin to record PMT pulses.
Then for some period the MCS is programmed to record the arrival time
of each pulse it sees. These arrival times are recorded in histograms
on the MCS with a fixed bin duration (number of experiments vs. photon
arrival time).

\paragraph{control potentials}
Static control potentials applied to endcap electrodes provide axial
$\left(\hat{z}-\text{direction}\right)$ confinement. In ion traps
with a single trapping zone these potentials are sometimes derived
from voltage limited power supplies. The potentials are adjusted to
null micromotion in that single zone. Multiple zone (multi-zone) traps
require many more potential sources. Ion transport from zone to zone
and separation of ions requires time-varying potentials. Nulling micromotion
in multiple zones simultaneously requires position dependent shims
to the trapping potentials. In trap tests with a single ion, potential
sources can be shared by multiple electrodes in widely-separated zones
of the trap. However, in real QIP experiments with multiple ions in
different zones, source sharing may be less practical. This is because
ions' trajectory thru the trap structure are expected to differ,
hence requiring independent potentials for many of the control electrodes.

The control electrode potentials in my experiments were supplied by
several sources. DC supplies ($\pm$10~V) were used for initial trap
loading for convenience, DACs were used to test ion transport and
batteries ($\pm$18~V) were used for ion heating measurements (owing
to their low intrinsic noise). All supplies were located within 2~meters of the
trap and used shielded wiring schemes to bring the potentials to the 
vacuum system port. A pair of 24-channel analog multiplexers
could switch between the supplies without ion loss.

The DC supplies and battery box included a shielded routing box for
applying potentials to any of the 24~channels. The battery box contained
voltage dividers to permit continuous manual variation of its potentials.

The DACs were 8~channel, 16~bit, 750~kHz National Instruments 6733~DACs
residing in a PXI bus. DAC waveform data was held in PC 
memory and streamed real-time over a MXI-express bus (up to 100 MB/sec).
These 24~channels were shared by up to 48~trap electrodes as determined
by a pair of routing boxes (printed circuit boards with jumpers).
Different routing boxes were used for different trap layouts to accommodate
their specific wiring requirements. The cables were doubly shielded
24~pin D-Sub (parallel port) cables which mated to the vacuum feedthroughs.

Proper grounding and shielding of the apparatus supplying the control
electrode potentials is important to suppress injected noise which
can cause ion heating (see Section~\vref{ions:sec:heating}). A star-grounding
scheme referenced to the trap vacuum housing (continuous with the
RF ground) was used to suppress ground loops and pickup. A glaring
exception was the DACs. The 6733s' ground reference was hard wired
by National Instruments to the PXI chassis, continuous with PC ground.
That is, the DACs are single-ended and floating on the noisy PC ground,
a poor design choice. Better is a double-ended configuration where
each DAC output is referenced to a user supplied potential. Cost constraints
and a dearth of high channel count DAC options resigned me to using
this DAC product. See Figure~\vref{ions:fig:properRC} for how it should
be done.

Ground loops were avoided by using the trap vacuum housing as a ground
reference (DC supply and battery) or by breaking the ground shield
near the source (DACs). Note that the choice of breaking the DAC ground
reference is not ideal as it leaves the DACs floating on the PC ground.
However, this was done because the alternative results in a very large
ground loop through the ac line ground via the PC.

\begin{table}
  \centering
  \begin{tabular}{l|l}
    DAC  & National Instruments PCI-6733 card. \tabularnewline
    DC supply & Hewlett Packard 3620A (operated as a voltage source)\tabularnewline
    battery  & alkali batteries (in shielded box with a voltage divider) \tabularnewline
  \end{tabular}
  \caption[Supplies used for the control electrodes.]{Supplies used for the
  control electrodes.}
  \label{tab:apparatus:controlElectrodeSupplies}
\end{table}

\clearpage
\paragraph{apparatus list} Following is a list of apparatus mentioned in this
section.
\\
\begin{minipage}{1\textwidth}
  \singlespacing
  \begin{itemize}
    \item {\bf FPGA} \href{http://www.xilinx.com}{Xilinx, Inc} Xtreme DSP kit with
    Virtex IV FPGA and the \href{http://www.nallatech.com}{Nallatech, Inc.} FUSE
    library
    \item {\bf PCI~bus~interface} \href{http://www.nallatech.com}{Nallatech, Inc.}
    BenADDA V4-XC4VSX35-FF668
    \item {\bf DDS} \href{http://www.analog.com}{Analog Devices, Inc.} AD9858 PCB
    evaluation board with parallel interface
    \item {\bf LVDS} \href{http://www.ti.com}{Texas Instruments, Inc.} SN65LVDM1676
    (no termination) and SN65LVDM1677 (110 $\Omega$ termination)
    \item {\bf TTL~driver} \href{http://www.ti.com}{Texas Instruments, Inc.} SN64BCT25244
    \item {\bf DC~supplies} \href{http://www.agilent.com}{Agilent, Inc.} $\#$3610A and
    3620A
    \item {\bf mechanical shutters} \href{http://www.uniblitz.com}{Uniblitz, Inc.}
    D122 and VCM-D1 (not shown in schematics)
    \item {\bf PMT~discriminator} \href{http://www.paraelectronics.com}{PRA, Inc.}
    1762 (NIM module, 100~MHz)
    \item {\bf RF~amplifiers} \href{http://www.minicircuits.com}{Mini-Circuits, Inc.}
    ZHL-1000-3W, ZHL-03-05WF, ZFL-500HLN and
    \href{http://www.motorola.com}{Motorola, Inc.} CA2832C
    \item {\bf RF~switches} \href{http://www.minicircuits.com}{Mini-Circuits, Inc.}
    \item {\bf 1~GHz clock} \href{http://www.thinksrs.com}{Stanford Research Systems,
    Inc.} CG635
    \item {\bf RF counter} \href{http://www.mfjenterprises.com}{MFJ, Inc.} MFJ-888 (10
    Hz - 3 GHz) (not shown in schematics)
    \item {\bf RF~power~meter} \href{http://www.spectradynamics.com}{Spectra
    Dynamics, Inc.} RFPWR (-25 - 25 dBm, 1~MHz - 1 GHz) (not shown in schematics)
    \item {\bf DACS} \href{http://www.ni.com}{National Instruments, Inc.} PXI-6733 and
    PCI-6733; NI-DAQmx API
    \item {\bf MXI-express} \href{http://www.ni.com}{National Instruments, Inc.} PXI-8360 and NI PCIe-8361
    \item {\bf PXI chassis} \href{http://www.ni.com}{National Instruments, Inc.} PXI-1042Q
    \item {\bf MCS} \href{http://www.urscorp.com/EGG_Division}{EG$\&$G Ortec, Inc.}
    Turbo-MCS
    \item {\bf 24-channel~analog~multiplexer} parallel port 4-way switch box
  \end{itemize}
\end{minipage}

\subsection{Early experiment control}

\label{sec:earlyExpControl}I developed part of the software and hardware
platform used to control the quantum logic experiments conducted between
2002-2005. These packages addressed a practical need to automate and
simplify parts of our increasingly complex experiments. From 2005
onward a superior and extensible FPGA platform developed by Chris
Langer was used. His platform largely superseded many of these packages
though some were used thru 2007.

\paragraph{rdye.exe} is an application which permits remote control of an laser
servo lock over the internet. It was written in LabWindows CVI and
interfaced with \href{http://www.ni.com}{National Instruments, Inc.} (NI) analog
input/output hardware. The software included a signal processing algorithm
which could identify and lock to particular absorption features in
an iodine saturated absorption spectroscopy setup. rdye.exe was used
in a sympathetic cooling experiment \cite{barrett2003a}.

\paragraph{pmtreadout.dll} is a library which integrates with counter hardware
to permit multiple distinct measurement periods within a single experiment,
each with its own histogram. The counter hardware is a National Instruments
NI-6602 and is limited to a count rate of 5~MHz. The library was written
in C. The pmtreadout.dll platform was used in a number of experiments
\cite{chiaverini2005b,schaetz2005a,chiaverini2004a,barrett2004a,schaetz2004a}.

\paragraph{zaberc.exe} is an application which integrates PMT readings with
2-axis computer control of lenses to quickly locate and maximally overlap Raman
laser beams with ions. Lens motion was controlled by
\href{http://www.zaber.com}{Zaber, Inc.} linear actuators (T-LA28-S). In addition
to computer controlled position scans, it is helpful to walk beams manually.
However, using a computer mouse/keyboard to do this is not intuitive. Instead, a
natural, haptic control system was implemented using a pair of
\href{http://www.griffintechnology.com}{Griffin Technology, Inc.} PowerMate USB
digital encoders. The software was written in NI LabWindows cVI. zaberc.exe was
used in several experiments \cite{leibfried2005a,reichle2006a,knill2008a}.

\paragraph{6733wvf.exe} is a software and hardware system for applying
potentials to ion trap electrodes for ion transport, separation and logic gates.
It met a variety of complex needs including waveform modification
on the fly using experimental feedback (open loop), externally triggered
waveform production (eg for changing trap potentials synchronized
with laser pulses during phase gates) and features to diagnose ion
transport problems. The arbitrary waveform hardware was the National
Instruments 6733 and it was coded in Lab Windows CVI. The system was
developed over many years and was integral to several experiments
\cite{langer2006a,reichle2006a,leibfried2005a,langer2005a,
ozeri2005b,chiaverini2005b,schaetz2005a,chiaverini2004a,barrett2004a,schaetz2004a}.

The source code for these applications is available on the NIST Ion
Storage intranet.

\clearpage
\section{Trap RF}
\label{sec:apparatus:trapRF}
\label{sec:trapTestingRF}

\subsection{RF electronics}
\begin{SCfigure}[10][t]
	\centering
	\includegraphics[width=0.6\textwidth]{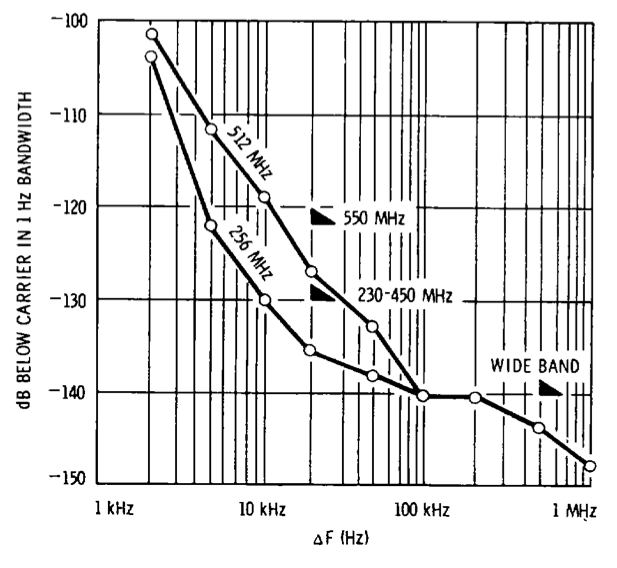}
	\caption[Noise in 8640B RF oscillator.]{Measured single sideband
	noise vs offset from carrier for HP 8640B RF oscillator.  The data
	was collected for two carrier frequencies 256~MHz and 512~MHz.  The 
	power is stated in a 1~Hz bandwidth.  Figure provided by the manufacturer.}
	\label{fig:RFelectronics:8640Bnoise}  
\end{SCfigure}
\label{sec:trapTestingRFelectronics}The trap RF is derived from a Hewlett Packard
8640B RF cavity oscillator stabilized to its internal temperature controlled
crystal oscillator (drift $<5\times10^{-8}$/hr). According to the manufacturer's manual
an 8640's phase noise at 1~MHz
from the carrier is -147~dBc (dB below the carrier) 
(see Figure~\vref{fig:RFelectronics:8640Bnoise}). Note that this
specification is important as noise can cause ion heating (see
Section~\vref{sec:RFAM}). The oscillator has amplitude and phase modulation inputs
and can be locked to an external 5~MHz reference. This source is amplified by an
\href{http://www.amplifiers.com}{Amplifier Research, Inc.} 1~W AR 1W1000A
amplifier. A \href{http://www.minicircuits.com}{Mini-Circuits, Inc.} XFDC-20-1H
directional coupler sends power reflected from the $\lambda$/4 resonator to a Agilent 8471D power
detector (diode rectifier, www.agilent.com). This reflected power is used to
measure coupling and loaded quality factor $Q_{L}$ for the $\lambda$/4 resonator
(see Section~\vref{sec:cavityCoupling}). Typical trap RF drive frequencies for
surface electrode traps are
$\Omega_{\rm RF}/2\pi=50-100$~MHz, coupling $>$ 95$\%$,
$Q_{L}=50-100$. A Minicircuits SHP-25 high pass filter attenuates RF at 10~MHz by
67~dBm (in addition to attenuation by the $\lambda$/4~resonator line shape).

A high~Q RF cavity also rejects noise from the RF source. That is,
its Lorentzian line shape attenuates off-resonant RF.
\begin{equation*}
  P/P_{max}\propto(V/V_{max})^{2}\propto\left(1+4Q_{L}^{2}
  \left(\frac{\Omega-\Omega_{0}}{\Omega_{0}}\right)^{2}\right)^{-1}
\end{equation*}
The power attenuation for several typical detunings is 
given in Table~\vref{tab:apparatus:RF:attenuation}.
\begin{table}
  \centering
  \begin{tabular}{c|ccccc}
      $\Omega$&$\Omega_{0}$&$\Omega_{0}-\omega_z$&$\Omega_{0}-\omega_x$&$\omega_x$&$\omega_z$\\
      \hline
      $P/P_{max}$&0&-16.8&-27.2&-42.7&-43.7\\
      (dBm)&&&&&
  \end{tabular}
  \caption[Attenuation of the $\lambda/4$ RF resonator for typical 
  parameters.]{Attenuation of the $\lambda/4$ RF resonator for typical 
  parameters: $Q_{L}=80$, $\Omega_{0}/2\pi=70$~MHz,  
  $\omega_x/2\pi=10$~MHz, $\omega_z/2\pi=3$~MHz.}
  \label{tab:apparatus:RF:attenuation}
\end{table}
\begin{figure}
  \centering
  \includegraphics[width=1\textwidth]
  {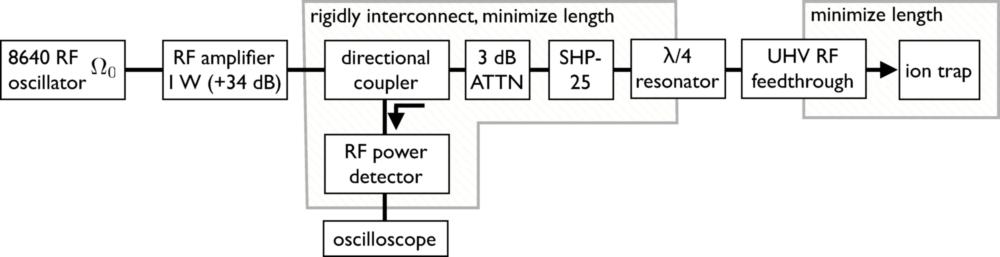}
  \caption[Schematic of trap RF electronics for the octagon style vacuum system.]
  {Schematic of trap RF electronics for the octagon style vacuum system.
  The left hatched region marks components which are rigidly interconnected
  with minimum length connectors. This and the 3 dB attenuator minimize
  RF line resonances. The right hatched region marks where the length
  of the RF wiring (a Kapton wrapped wire) should be a minimum to prevent
  radiative losses and capacitive coupling to ground and other trap
  features.}
  \label{fig:trapRFbasic}
\end{figure}

\subsection{$Q_{L}$ and ion trap chip losses}
\label{sec:apparatus:trapQL}
\label{sec:apparatus:trapQLoss}
\label{sec:QLchipLoss}The loaded quality factor $Q_{L}$ of the RF
resonator can be a good indicator of losses at the ion trap chip.
In the octagon
style vacuum housing RF is brought to the chip via a 
RF vacuum feedthrough (RF FT) which has a large capacitance to ground, 
unshielded wires, a chip carrier socket and a chip carrier.\\
\\
\begin{tabular}{l|l|p{0.4\textwidth}}
$Q_{L}$ & $\Omega_{\rm RF}/2\pi$ & RF components\tabularnewline
\hline
$\sim$300 & 156~MHz & resonator\tabularnewline
$\sim$270 & 63~MHz & ~~`` + RF FT \tabularnewline
$\sim$290 & 45~MHz & ~~`` + bare wire + chip carrier socket\tabularnewline
$\sim$170 & 38~MHz & ~~`` + empty Kyocera
chip carrier \tabularnewline
$\sim$80 & 43~MHz & ~~`` + dv16m trap chip (see Sec.~\ref{sec:dv16m:elecTests})
\end{tabular}\\
\\

The lowering of the resonator frequency as RF components are added is due to their
capacitance to ground.  Degradation in $Q_{L}$ can arise due to
losses in the trap chip. 

\begin{SCfigure}
  \centering
  \includegraphics[width=0.4  \textwidth]
  {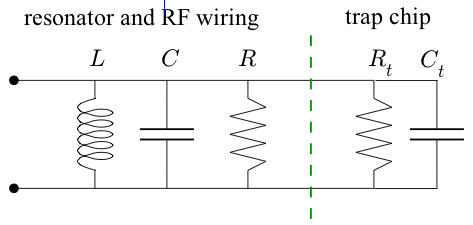}
  \caption[Parallel RLC model for ion trap chip losses.]
  {Parallel RLC model for ion trap chip losses. $R$ is due to the
  intrinsic losses in the bare resonator and other wiring. $R_{t}$ ($C_{t}$) accounts for the
  additional loss (capacitance) due to the trap chip.}
  \label{fig:trapChipLossesRLCmodel}
\end{SCfigure}

It is instructive to estimate what the losses might be in the trap chip. A
resonator can be modeled as a parallel RLC where,
\begin{equation*}
  \begin{aligned}
    \Omega_{\rm RF}^{2}=1/(LC)
  \end{aligned}
  \qquad\text{and}\qquad
  \begin{aligned}
    Q_{L}=R/|Z_{C}|=RC~\Omega_{\rm RF}.
  \end{aligned}
\end{equation*}
Note that this ignores the fact that $L$ and $C$ are distributed and not lumped
elements.  As an example of approximate experimental parameters, let  
$\Omega_{\rm RF}/2\pi=40~\text{MHz}$, $Q_{L}=170$ and 
$L=1~\mu$H then we
have $C=16~pF$ and $R=43~k\Omega$.  The resonator's inductance was modeled as a 1.9~meter
long vacuum coax with a ratio of outer to inner conductor of 12.  
Suppose that a trap chip is added to the circuit and the quality factor drops
to $Q_{Lt}=80$ (with no change in $\Omega_{\rm RF}$) as in 
Figure~\vref{fig:trapChipLossesRLCmodel}.  Then the new parallel 
resistance of $R$ and $R_t$ is $R_p=20~k\Omega$. And,  
\begin{equation*}
  R_t=\frac{R R_p}{R+R_p}=38~k\Omega.
\end{equation*}
If the RF potential at the trap is $V=50$~volts, then the power dissipated by
the trap is $\frac{1}{2}V^{2}/R_{t}=32$~mW.

\section{Time-resolved Doppler cooling}

\label{sec:recooling}A critical measure of an ion trap's utility in QIP is its
electric field noise which can heat the ions. Historically, this has been
measured in ions with hyperfine structure cooled to the motional ground state, an
expensive and technically challenging undertaking
\cite{caves1980a,diedrich1989a,monroe1995a,king1998a,turchette2000a,bible}. In
2007 a measurement technique relying on only a single Doppler cooling laser beam
was demonstrated experimentally for $\,^{24}\text{Mg}^{+}$ \cite{epstein2007b} and 
was explained theoretically \cite{wesenberg2007a}. $^{24}$Mg$^{+}$ has no
hyperfine structure; there is no need for a quantizing magnetic field, a repump laser or
careful laser beam polarization control~\cite{seidelin2006a,epstein2007b}. The
key physics for this method is that near resonance, an atom's fluorescence rate
is influenced by its motion due to the Doppler effect. Experimentally, this can
be exploited as follows.

\begin{compactenum}
  \item Cool a trapped ion to its Doppler limit.
  \item Let it remain in the dark for some time. Ambient electric fields couple
  to the ion's motion and heat it.
  \item Turn on the Doppler cooling laser. Measure the ion's time-resolved
  fluorescence.
  \item A fit from a theoretical model to ion fluorescence rate vs time 
  (during recooling) gives an estimate of the ion's temperature at 
  the end of step 2~\cite{wesenberg2007a}.
\end{compactenum}

\paragraph{recooling of hot ions}
The theory by Wesenberg, \textit{et al.} explored cooling of hot ions where the
Doppler shift is on the order of the cooling transition line width $\Gamma$. The
model is a one-dimensional semiclassical theory of Doppler cooling in the weak
binding limit where $\omega_{z}<<\Gamma$. It is assumed that hot ions undergo
harmonic oscillations with amplitudes corresponding to the Maxwell-Boltzman
energy distribution when averaged over many experiments~\cite{wesenberg2007a}.

As a one-dimensional (1D) model, only a single motional mode is assumed
to be hot. Since the electric field spectral density $S_{E}$ at the
ion is observed to scale approximately as $S_{E}\propto\omega^{-\alpha}$
where $\alpha=$ 1~to~1.5, the heating is effectively 1D if $\omega_{z}<<\omega_{x},\omega_{y}$
\cite{deslauriers2006a,turchette2000a}. This is also important experimentally
because efficient Doppler cooling requires laser beam overlap with
all modes simultaneously: a change in ion fluorescence can arise from
heating of any mode. 

Note that ion heating may result from effects
unrelated to electric field fluctuations, for example collisions with
ions or neutrals. Also, the model only applies to single trapped ions.
Other heating mechanisms are possible with multiple ions \cite{walther1993b}.
It also requires that the 3D micromotion be nulled as this opens the
door for other sources of heating (see Section~\vref{sec:RFAM}).

The model accuracy is related to three experimental parameters. They
should be optimized to the extent permitted by experimental constraints
like trap depth and the background collision rate.

\begin{compactenum}
  \item Doppler cooling laser detuning $\delta$ from resonance $\omega_{0}$
  should be small. While $\delta=-\Gamma/2$ is optimal for laser cooling
  of atoms near the Doppler limit, I used $\delta=\omega_{\text{laser}}-\omega_{0}=-\Gamma/4$.%
 	\footnote{This choice was made because -$\Gamma$/4 was better for the
	  	micromotion minimization experiment that frequently preceded heating
	  	experiments. It was experimentally convenient to not change the detuning.
	}
  \item The cooling laser beam intensity should be as low as is practical.
  I operated at a small saturation parameter: $s\sim1$. %
	\footnote{In the optical Bloch equation model for 2-level atomic systems, the
		saturation parameter $s$ relates laser intensity to the steady state
		excited population $\rho_{\text{ee}}$,
			\begin{equation*}
				\rho_{\text{ee}}=\frac{s/2}{1+s+(2\delta/\Gamma)^{2}}.
			\end{equation*}
	Assuming, $\delta=0$,
	\begin{compactitem}
		\item [{for~$s=1$,}] $\rho_{\text{ee}}=1/4$ and
		\item [{for~$s\to\infty$,}] $\rho_{ee}=1/2$.
	\end{compactitem}
	}
  \item The ion's initial temperature should be high enough so that its
    fluorescence is reduced considerably due to Doppler broadening. For my trap
    parameters this corresponded to about 1000 quanta.
\end{compactenum}
The saturation parameter for a particular laser beam intensity can
be estimated by recording ion fluorescence as a function of intensity
and doing a fit to this expression.

Measurement of an ion's temperature by time-resolved fluorescence was first
reported by Seidelin, \textit{ et al.} \cite{seidelin2006a}. The technique was
later confirmed experimentally by comparison with Raman resolved sideband cooling
(an established method, \cite{monroe1995a}. It was also confirmed that ion
heating rates measured at the $>1000$~quanta level can be extrapolated to the
single quantum level \cite{epstein2007b}. This is important in the context of ion
QIP where gate fidelity is impacted by very few motional quanta.

\begin{SCfigure}
  \centering
  \includegraphics[width=0.5\textwidth]
  {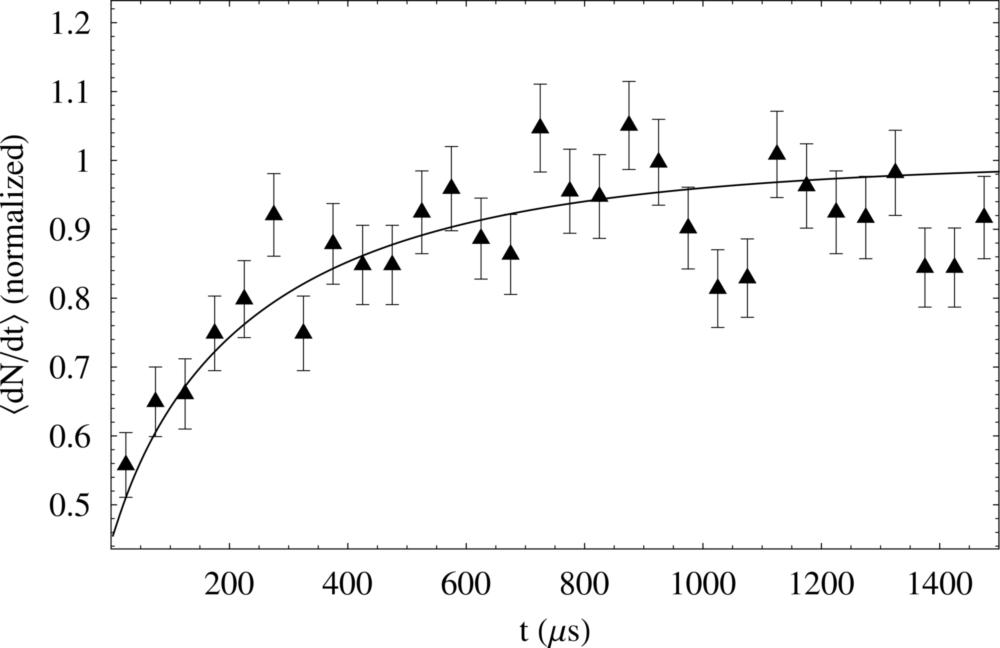}
  \caption[Plot showing normalized fluorescence rate dN/dt during Doppler cooling
  of a hot ion vs ion dark time.]
  {Plot showing normalized fluorescence rate $dN/dt$ during Doppler cooling
  of a hot ion vs ion dark time. The experimental data is fit to the
  1D model discussed in the text. The fit has a single free parameter:
  the ion's temperature at the outset of cooling (averaged over many 
  experiments). The error bars are
  based on counting statistics.}
  \label{fig:recoolingSampleData}
\end{SCfigure}

\paragraph{Experimental steps}

The apparatus used to control the Doppler cooling laser beam and collect
time-resolved fluorescence is discussed in Section~\vref{sec:controlElectronics}.
The experiment proceeded as follows.

\begin{compactenum}
\item Null ion micromotion.
\item Measure the axial trap frequency $\omega_{z}$.
\item Measure the ion saturation parameter $s$.
\item Doppler cool the ion.
\item Turn off the Doppler cooling laser beam for time $\tau$.
\item Turn on the Doppler cooling laser. Measure the ion's time-resolved
fluorescence.
\item Repeat steps $4-6$ and build up a histogram of fluorescence vs time.
\item Repeat steps $4-7$ for several periods $\tau$.
\end{compactenum}
Care must be taken owing to duty cycle related thermal effects in
AOMs. Some of the RF power delivered to an AOM is absorbed and causes
it to heat up. The deflection angle in the AOMs I used depends weakly
on temperature. During heating measurements, the RF to AOM~2 was switched
on/off in order to turn on/off the BD Doppler laser beam (see Figure~\vref{fig:testTrapBeamPath}).
The AOM temperature change during these on/off cycles was sufficient
to cause BD laser beam overlap with the ion to change. This is observed
as a change in ion fluorescence over several seconds after RF is applied
to a cool AOM~2. This effect was accounted for when measuring the saturation
parameter; saturation was measured with the AOMs operated on same
duty cycle as used in ion heating experiments.

Section~\vref{sec:recoolingDCfile} lists the .dc file used in this
experiment.

    \clearpage
\chapter{RF Cooling of a Micro Cantilever}
\label{sec:ccool}
In this chapter I discuss the demonstration of a method to cool a fundamantal
motional mode of a miniature cantilever by its capacitive coupling to a driven RF
resonant circuit. Cooling results from the RF capacitive force, which is phase
shifted relative to the cantilever motion. The technique was demonstrated by
cooling a 7~kHz cantilever from room temperature to $45$~K, obtaining reasonable
agreement with a model for the cooling, damping, and frequency shift. Extending
the method to higher frequencies in a cryogenic system could enable ground state
cooling and may prove simpler than related optical experiments in a low
temperature apparatus. 

Precise control of quantum systems occupies the efforts of many laboratories; an
important recent application of such control is in quantum information
processing.  Some of this work is devoted to quantum-limited measurement and
control of the motion of mechanical oscillators. This has already been
accomplished in a ``bottom-up'' approach where a single atom is confined in a
harmonic well and cooled to the ground state of its quantum mechanical
motion~\cite{diedrich1989a}. From this starting point physicists have made
nonclassical mechanical oscillator states such as squeezed, Fock and
Schrodinger-cat states~\cite{meekhof1996a,monroe1996a, leibfried2005a,
ben-kish2003a}.  Quantum control of motional states has also emerged as powerful
tool for spectroscopy and coupling of quantum
systems~\cite{leibfried2005a,haffner2005a,schmidt2005a,blatt2008a}.

For various applications, there is also interest in a ``top-down'' strategy,
which has approached the quantum limit by using ever smaller mechanical
resonators~\cite{braginskii1970a,braginskii1977a}. In this case, small ($\sim
1~\mu m$) mechanical resonators, having fundamental motional frequencies of
10-100~MHz, can approach the quantum regime at low temperature ($<1$~K);
mean thermal occupation numbers of 25 have been achieved~\cite{naik2006a}.  For a
review of this approach see~\cite{schwab2005a}.

Cooling of macroscopic mechanical oscillators also has been achieved with optical
forces and active external electronics to control the applied force
\cite{cohadon1999a,arcizet2006a,weld2006a,kleckner2006a,poggio2007a}. Passive
feedback optical cooling has also been realized in which a mirror attached to a
mechanical oscillator forms an optical cavity with another stationary mirror. For
appropriate tuning of radiation incident on the cavity, a delay in the optical
force on the oscillator as it moves gives cooling. This delay can result from a
photothermal effect~\cite{metzger2004a,harris2007a} or from the stored energy
response time of the cavity~\cite{arcizet2006a,gigan2006a,corbitt2007a}. Closely
related passive cooling has been reported in~\cite{naik2006a,schliesser2006a}.

We demonstrate a passive cooling mechanism where the damping force is the
electric force between capacitor plates that here contribute to a resonant RF
circuit~\cite{braginskii1977a,wineland2006c}. This approach has potential
practical advantages over optical schemes; the elimination of optical components
simplifies fabrication and integration into a cryogenic system, and the RF
circuit could be incorporated on-chip with the mechanical oscillator.

This work was reported in Physical Review Letters in 2007 in
an article titled ``Passive cooling of a micromechanical oscillator with a
resonant electric circuit"~\cite{brown2007a}.  The theory describing the
cantilever-resonator coupling and cooling mechanism appeared in a
2006 Archive paper titled ``Cantilever cooling with radio frequency
circuits"~\cite{wineland2006c}.  

\begin{SCfigure}
\centering \includegraphics[width=0.7\textwidth]{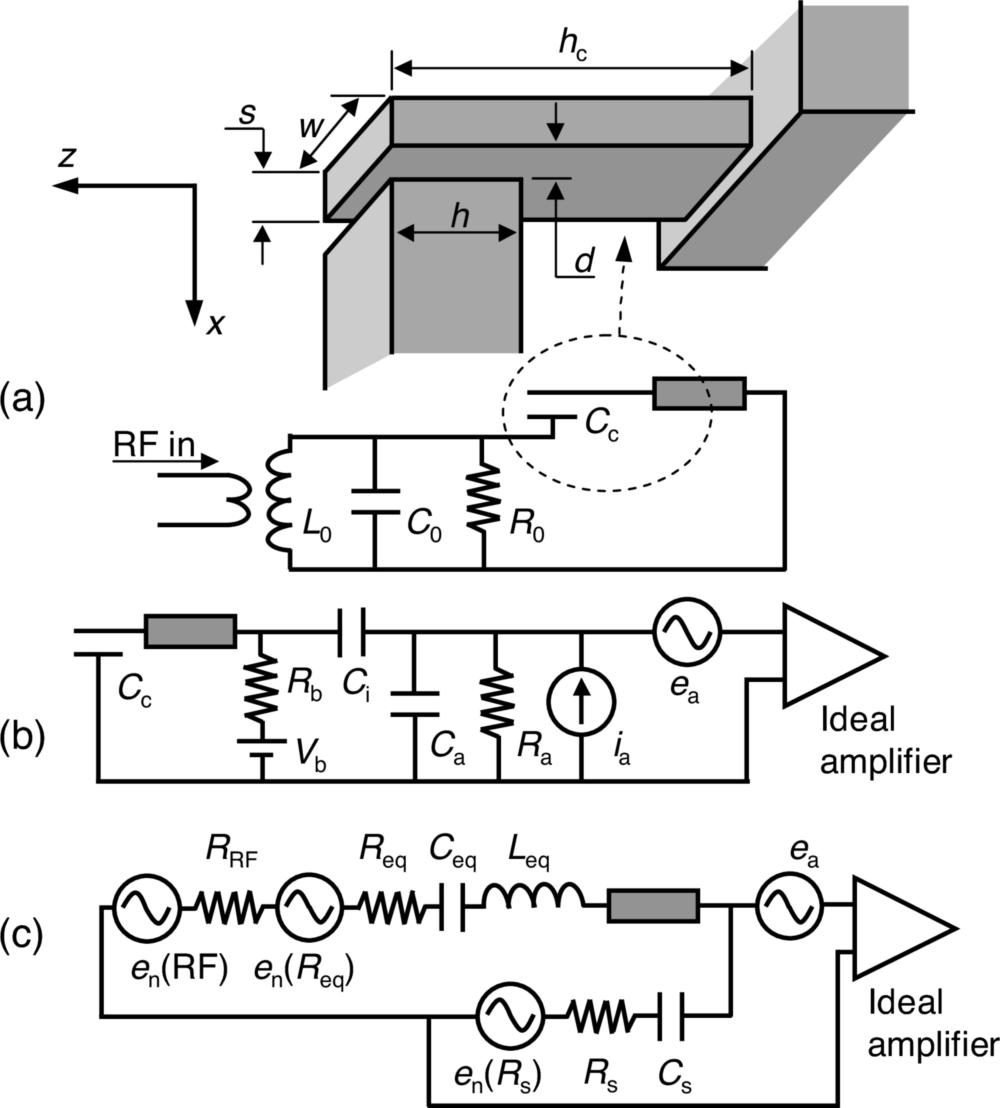} 
\caption[Schematics of the cantilever cooling and detection
  electronics.]{Schematics of the cantilever cooling and detection
  electronics. (a) Cantilever and associated RF circuitry. (b)
  Motional detection electronics. Near $\omega_\mathrm{c}$ the RF
  circuit looks like a short to ground as shown. (c) Equivalent
  circuit for the cantilever and detection electronics near $\omega
  \approx \omega_\mathrm{c}$.} 
\label{fig:schematic}
\end{SCfigure}

A conducting cantilever of mass density $\rho$ is fixed at one end
(Fig.~\ref{fig:schematic}(a)).
One face is placed a distance $d$ from a rigidly mounted plate of area
$w \times h$, forming a parallel-plate capacitor $C_\mathrm{c} =
\epsilon_0wh/d$, where $\epsilon_0$ is the vacuum dielectric constant.
An inductor $L_0$ and capacitor $C_0$ in parallel with $C_c$ form a
resonant RF circuit with frequency $\Omega_0 = 1/\sqrt{L_0(C_0 +
  C_c)}$ and with losses represented by resistance $R_0$. We assume
$Q_\mathrm{RF} \gg 1$, where $Q_\mathrm{RF} = \Omega_0 L_0/R_0 =
\Omega_0/\gamma$ and $\gamma$ is the damping rate.

We consider the lowest-order flexural mode of the cantilever, where
the free end oscillates in the $\hat{x}$ direction (vertical in
Fig.~\ref{fig:schematic}(a)) with angular frequency $\omega_c \ll
\gamma$. We take $x$ to be the displacement at the end of the
cantilever, so the displacement as a function of (horizontal) position
$z$ along the length of the cantilever is given by $x(z) = f(z)x$,
where $f(z)$ is the mode function (see, e.g.,~\cite{butt1995a}). Small
displacements due to a force $F$ can be described by the equation of
motion,
\begin{equation}
m\ddot{x} + m \Gamma \dot{x} + m \omega_c^2 x = F,
\label{equation_of_motion}
\end{equation}
where $\Gamma$ is the cantilever damping rate and $m$ its effective
mass, given by $\rho \xi_\mathrm{c}'' w h_\mathrm{c} s$, where
$\xi_\mathrm{c}'' \equiv \frac{1}{h_\mathrm{c}} \int_{h_\mathrm{c}}
f(z)^2 \, \mathrm{d}z = 0.250$ for a rectangular beam.

For simplicity, first assume $h \ll h_\mathrm{c}$, so that the force
is concentrated at the end of the cantilever. If a potential $V$ is
applied across $C_\mathrm{c}$, the capacitor plates experience a
mutual attractive force $F = \epsilon_0 w h V^2/(2 d^2) = C_\mathrm{c}
V^2/(2 d)$.  Consider that $V$ is an applied RF potential
$V_\mathrm{RF} \cos (\Omega_\mathrm{RF} t)$ with $\Omega_\mathrm{RF}
\approx \Omega_0$. Because $\omega_\mathrm{c} \ll \Omega_0$, the force
for frequencies near $\omega_\mathrm{c}$ can be approximated by the
time-averaged RF force
\begin{equation}
  F_\mathrm{RF} = \frac{C_\mathrm{c} \langle V^2 \rangle}{2 d} = 
  \frac{C_\mathrm{c} V_\mathrm{RF}^2}{4 d} = \frac{\epsilon_0
  w h V_\mathrm{RF}^2}{4 d^2},
\label{capacitor_force}
\end{equation}
where, for a fixed input RF power, $V_\mathrm{RF}$ will depend on
$\Delta \Omega \equiv \Omega_0 - \Omega_{\mathrm{RF}}$, according to
\begin{equation}
\frac{V_\mathrm{RF}^2}{V_\mathrm{max}^2} = \frac{1}{1 + \bigl[2
Q_{\mathrm{RF}} \Delta \Omega/\Omega_0 \bigr]^2} \equiv
\mathcal{L}(\Delta \Omega)\ .
\label{eq:L}
\end{equation}
As the cantilever oscillates, its motion modulates the capacitance of
the RF circuit
thereby modulating
$\Omega_0$. As $\Omega_0$ is modulated relative to
$\Omega_{\mathrm{RF}}$, so too is the RF potential across the
capacitance, according to Eq.~(\ref{eq:L}). The associated modulated
force shifts the cantilever's resonant frequency. Due to the finite
response time of the RF circuit, there is a phase lag in the force
relative to the motion.  For $\Delta \Omega > 0$ the phase lag leads
to a force component that opposes the cantilever velocity, leading to
damping. If this damping is achieved without adding too much force
noise then it cools the cantilever.

The average force due to applied potentials displaces the equilibrium position
$d_0$ of the cantilever. We assume this displacement is small and is absorbed
into the definition of $d_0$, writing $d \equiv d_0 - x$, where $x \ll
d_0$\footnote{This expression for $d$ neglects the curvature of the flexural
  mode and is strictly true only for $h \ll h_\mathrm{c}$ and $d_0 \ll w,\ h$.}.~
Following~\cite{braginskii1977a} or~\cite{wineland2006c} we find $\omega_\mathrm{c}^2
\rightarrow \omega_\mathrm{c}^2 (1 - \kappa)$ and $\Gamma \rightarrow \Gamma +
\Gamma'$, with
\begin{equation}
  \kappa \equiv \frac{C_\mathrm{c} V_\mathrm{max}^2 \mathcal{L}(\Delta \Omega)}
  {2 m \omega_\mathrm{c}^2 d_0^2} \biggl[ \xi'' + \frac{2 (\xi')^2 Q_{\mathrm{RF}} 
    \Delta \Omega \mathcal{L}(\Delta \Omega)}{\gamma}
  \frac{C_\mathrm{c}}{C_\mathrm{c} + C_0} \biggr],
\label{kappa}
\end{equation}
\begin{equation}
\Gamma' \equiv  \frac{Q_{\mathrm{RF}} V_\mathrm{max}^2 C_\mathrm{c}^2}{m
\omega_\mathrm{c} d_0^2 (C_\mathrm{c} + C_0)} \frac{(\xi')^2 \Delta
\Omega \mathcal{L}(\Delta\Omega)^2}{\gamma} \sin{\phi},
\label{gammaprime} 
\end{equation}
where $\xi' \equiv \frac{1}{h} \int_h f(z)\,\mathrm{d}z$ and $\xi''
\equiv \frac{1}{h} \int_h f(z)^2\,\mathrm{d}z$ are geometrical factors
required when $h \ll h_\mathrm{c}$ is not satisfied.  The phase $\phi$
is equal to $\omega_\mathrm{c} \tau$, where $\tau =
4\mathcal{L}(\Delta \Omega)/\gamma$ is the response time of the RF
circuit~\cite{marquardt2007a}. For $\Delta \Omega > 0$, $\Gamma'$ gives
increased damping.  For $\Delta \Omega = \gamma/2$ and $h \ll
h_\mathrm{c}$ ($\xi' \approx \xi'' \approx 1)$, we obtain the
expressions of~\cite{wineland2006c}.  For our experiment, $h \approx
h_\mathrm{c}$, $\xi' = 0.392$, and $\xi'' = \xi_\mathrm{c}'' = 0.250$.

We detect the cantilever's motion by biasing it with a static
potential $V_\mathrm{b}$ through resistor $R_\mathrm{b}$ as shown in
Fig.~\ref{fig:schematic}(b), where $R_\mathrm{a}$, $C_\mathrm{a}$,
$i_\mathrm{a}$, and $e_\mathrm{a}$ represent the equivalent input
resistance, capacitance, current noise, and voltage noise,
respectively, of the detection amplifier. We make $R_\mathrm{a}$ and
$R_\mathrm{b}$ large to minimize their contribution to the current
noise $i_\mathrm{a}$. We assume $C_\mathrm{i} \gg (C_\mathrm{c} +
C_\mathrm{a})$ and $\omega_\mathrm{c} R (C_\mathrm{c} + C_\mathrm{a})
\gg 1$, where $1/R \equiv 1/R_\mathrm{b} + 1/R_\mathrm{a}$. As the
cantilever moves, thereby changing $C_\mathrm{c}$, it creates a
varying potential that is detected by the amplifier.

The (charged) cantilever can be represented by the series electrical
circuit in Fig.~\ref{fig:schematic}(c). From
Eq.~(\ref{capacitor_force}) and following~\cite{wineland1975a}, the 
equivalent inductance is given by $L_\mathrm{eq} = m
d_0^2/(q_\mathrm{c}\xi')^2$, where $q_\mathrm{c}$ is the average
charge on the cantilever. From $L_\mathrm{eq}$, $\omega_\mathrm{c}$,
and $\Gamma$, we can then determine $C_\mathrm{eq} =
1/(\omega_\mathrm{c}^2 L_\mathrm{eq})$ and $R_\mathrm{eq} =
L_\mathrm{eq} \Gamma$. Additional damping due to the RF force is
represented by $R_\mathrm{RF} = L_\mathrm{eq} \Gamma'$. For
frequencies $\omega \approx \omega_\mathrm{c}$, the parallel
combination of $R_\mathrm{b}$, $C_\mathrm{a}$, and $R_\mathrm{a}$ can
also be expressed instead as the Th\'{e}venin equivalent
$R_\mathrm{s}$-$C_\mathrm{s}$ circuit in
Fig.~\ref{fig:schematic}(c). The amplifier's current noise
$i_\mathrm{a}$ is now represented as $e_\mathrm{n}(R_\mathrm{s})$. The
intrinsic thermal noise of the cantilever is characterized by a noise
voltage $e_\mathrm{n}(R_\mathrm{eq})$ having spectral density $4
k_\mathrm{B} T_\mathrm{c} R_\mathrm{eq}$, where $k_\mathrm{B}$ is
Boltzmann's constant and $T_\mathrm{c}$ is the cantilever temperature.
 
We must also consider noise from the RF circuit. In
Eq.~(\ref{capacitor_force}), we replace $V_\mathrm{RF}$ with
$V_\mathrm{RF}+ v_\mathrm{n}(\mathrm{RF})$, where
$v_\mathrm{n}(\mathrm{RF})$ is the noise across the cantilever
capacitance $C_\mathrm{c}$ due to resistance in the RF circuit and
noise injected from the RF source. The cantilever is affected by
amplitude noise $v_\mathrm{n}(\mathrm{RF})$ at frequencies near
$\Omega_{\mathrm{RF}} \pm \omega_\mathrm{c}$, because cross terms in
Eq.~(\ref{capacitor_force}) give rise to random forces at the
cantilever frequency.  This force noise can be represented by
$e_\mathrm{n}(\mathrm{RF})$ in the equivalent circuit. The noise terms
sum to $e_\mathrm{n}^2 = e_\mathrm{n}^2(R_\mathrm{eq}) +
e_\mathrm{n}^2(R_\mathrm{s}) + e_\mathrm{n}^2(\mathrm{RF})$
($e_\mathrm{a}$ does not drive the cantilever), which gives a
cantilever effective temperature
\begin{equation} 
T_\mathrm{eff}=\frac{e_\mathrm{n}^2
  }{4k_\mathrm{B}(R_\mathrm{eq} + R_\mathrm{RF} + R_\mathrm{s})}.
\label{effective_temperature}
\end{equation}

Our cantilever has nominal dimensions $h_ \mathrm{c} \approx 1.5$~mm, $s \approx
14$~$\mu$m, and $w \approx 200$~$\mu$m, created by etching through a p++-doped
($\sim 0.001$ $\Omega$ cm), 200~$\mu$m thick silicon wafer with a standard Bosch
reactive-ion-etching process. Its resonant frequency and quality factor are
$\omega_\mathrm{c}/(2\pi) \approx 7$~kHz and $Q \approx 20,000$. The cantilever
is separated by $d_0 \approx 16$~$\mu$m from a nearby doped silicon RF electrode,
forming capacitance $C_\mathrm{c}$\footnote{The uncertainties for these numbers
are a few micrometers,
  stemming from spatial nonuniformity in our etch process.}. The sample is
  enclosed in a vacuum chamber with pressure less than $10^{-5}$~Pa. The RF
  electrode is connected
via a vacuum feedthrough to a quarter-wave resonant cavity with $L_0 =
330(30)$~nH and with loaded quality factor $Q_\mathrm{RF}=234(8)$ at
$\Omega_\mathrm{RF}/(2 \pi)=100$~MHz when impedance matched to the source. The
cantilever is connected by a short length of coaxial cable and blocking capacitor
$C_\mathrm{i}=4$~nF to a low-noise JFET amplifier (see
Fig.~\ref{fig:schematic}(b)). We have $C_\mathrm{a}=48(1)$~pF, with
$R_\mathrm{a}=R_\mathrm{b}=1$~G$\Omega$. We use $V_\mathrm{b} = -50$~V, which
gives a measured 2.5~$\mu$m static deflection at the cantilever end.

We temporarily lowered $R_\mathrm{a}$ to 600~k$\Omega \approx
1/(\omega_\mathrm{c}C_\mathrm{a})$, in which case the cantilever noise
spectrum strongly distorts from a Lorentzian lineshape (not shown),
and it becomes straightforward to extract the equivalent circuit
parameters of Fig.~\ref{fig:schematic}(c). We find $L_\mathrm{eq} =
27,000(600)$~H. To lowest order in RF power this equivalent inductance
remains constant, so we assume this value for $L_\mathrm{eq}$ in
subsequent fits to the thermal spectra, while $R_\mathrm{RF}$ is
allowed to vary to account for RF power induced changes in the
cantilever damping.

\begin{SCfigure}
\centering \includegraphics[width=0.6\textwidth]{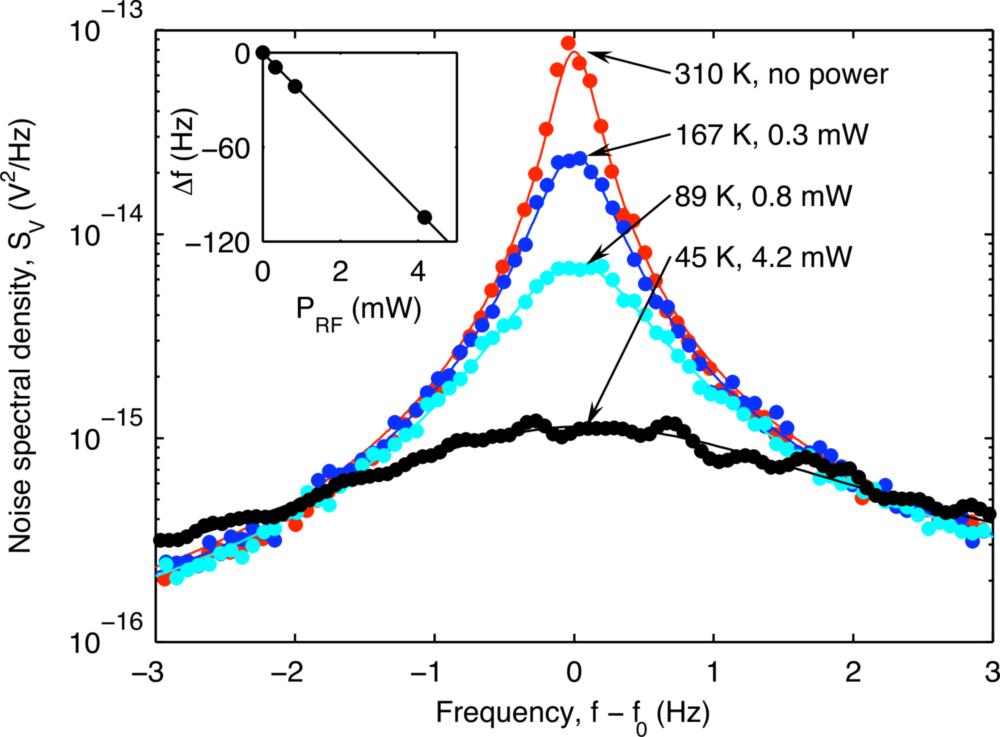} 
\caption[Cantilever thermal spectra for four values of
  RF power.]{Cantilever thermal spectra for four values of
  RF power. The $x$-axis for each spectrum has been shifted to align
  the maxima of the three datasets. $S_\mathrm{v}$ is the measured
  noise referred to the input of the amplifier. Solid lines are fits
  to the model in Fig.~\ref{fig:schematic}(c), giving the temperatures
  indicated by arrows. (Inset) Cantilever frequency shift $\Delta f$
  versus RF power.}
\label{fig:thermal_spectra}
\end{SCfigure}

For $R_\mathrm{a} = 1$~G$\Omega$ we measure
$e_\mathrm{a}=1.5$~nV/$\sqrt{\mathrm{Hz}}$ and
$i_\mathrm{a}=16$~fA/$\sqrt{\mathrm{Hz}}$.
Figure~\ref{fig:thermal_spectra} shows a series of thermal spectra
acquired with this value of $R_\mathrm{a}$ at different values of RF
power $P_\mathrm{RF}$ but at constant detuning $\Delta\Omega = 2 \pi
\times 90$~kHz $= 0.21 \gamma$.
Both the lowering and the broadening of the spectra with increasing
$P_\mathrm{RF}$ are evident, in accordance with
Eq.~(\ref{gammaprime}). Here, the effective temperature is very nearly
proportional to the area under the curves, although there is a slight
asymmetric distortion from a Lorentzian lineshape, fully accounted for
by the equivalent circuit model. The center frequency of each spectrum
also shifts to lower frequencies for increasing $P_\mathrm{RF}$, as
predicted by Eq.~(\ref{kappa}) and the definition of $\kappa$ in terms
of $\omega_\mathrm{c}$. After calibrating the gain of the amplifier,
we extract $e_\mathrm{n}^2$ for each spectrum from a fit to the model
of Fig.~\ref{fig:schematic}(c). The absolute effective temperature is
then given by Eq.~(\ref{effective_temperature}).

\begin{SCfigure}
\centering \includegraphics[width=0.7\textwidth]{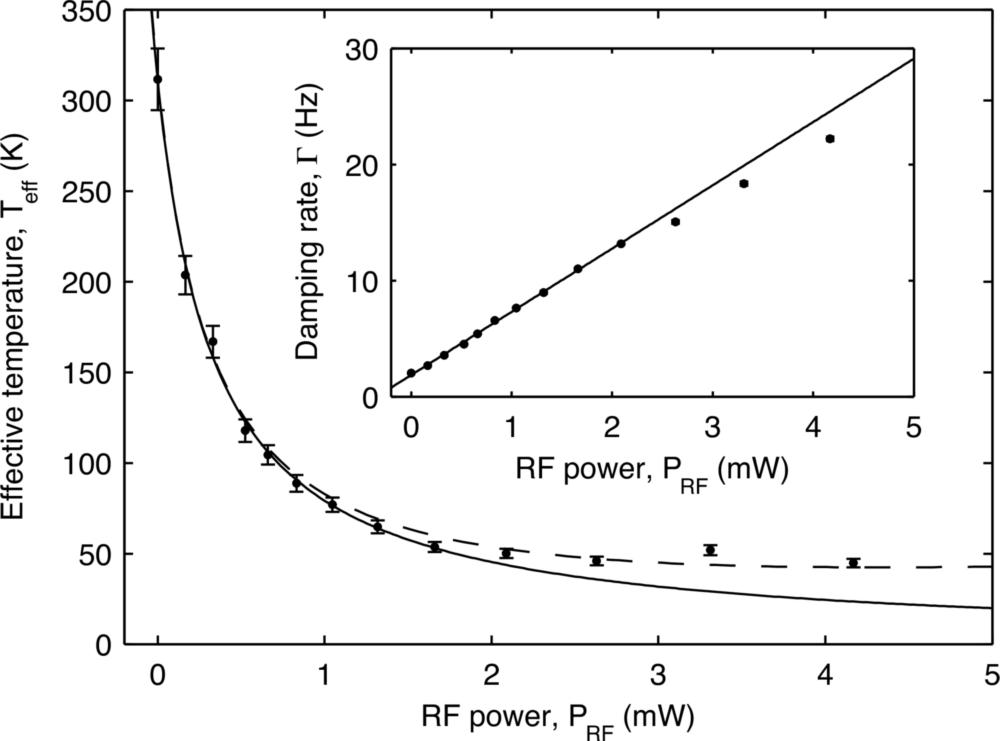}
\caption[$T_\mathrm{eff}$ as a function of RF power.]{$T_\mathrm{eff}$ as a
function of RF power. The solid line is the temperature predicted by Eq.~(\ref{effective_temperature})
  using $\Gamma$ from the fit in the inset and $e_\mathrm{n} =
  e_\mathrm{n}(R_\mathrm{eq})$, while the dashed line takes into
  account additional noise due to the RF source. (Inset) Damping rate
  versus RF power. The solid line is a linear fit to the first ten
  points.}
\label{fig:effective temperature}
\end{SCfigure}

Equations~(\ref{gammaprime}) and (\ref{effective_temperature}) predict
that the cantilever's effective temperature should fall with
increasing $P_\mathrm{RF}$, as demonstrated by the data in
Fig.~\ref{fig:effective temperature} for low power.
With no RF applied we find $T_\mathrm{eff} = 310(20)$~K. The coldest
spectrum corresponds to a temperature of 45(2)~K, a factor of 6.9
reduction. The minimum temperature appears to be limited by AM noise
from our RF source. This noise power at $\Omega_\mathrm{RF} \pm
\omega_\mathrm{c}$ is constant relative to the carrier, leading to a
noise power at $\omega_\mathrm{c}$ given by
$e_\mathrm{n}(\mathrm{RF})^2 \propto P_\mathrm{RF}^2$. We fit the
residual noise $e_\mathrm{n}(\mathrm{RF})^2$ to a quadratic in
$P_\mathrm{RF}$, giving the dashed line temperature prediction in
Fig.~\ref{fig:effective temperature}. From this fit we determine that
the AM noise of our source is -170~dBc/Hz, reasonably consistent with
the value (-167~dBc/Hz) measured by spectrum analysis.

The inset shows the cantilever damping rate $\Gamma$ versus
$P_\mathrm{RF}$. The slope is $\Gamma'/P_\mathrm{RF} =
5,450(70)$~Hz/W, slightly higher than the value 3,970~Hz/W calculated
from Eq.~(\ref{gammaprime}) and the nominal cantilever dimensions. The
nonlinearity in $\Gamma'/P_\mathrm{RF}$ at higher powers is consistent
with the cantilever being pulled toward the RF electrode. We have
numerically simulated this effect and find reasonable agreement. The
variation of $\kappa$ with $P_\mathrm{RF}$ (not shown) is also linear,
with a slope $\kappa/P_\mathrm{RF} = 7.64(8)$~$\mathrm{W}^{-1}$,
compared with the value 3.45~$\mathrm{W}^{-1}$ calculated from
Eq.~(\ref{kappa}).

Although $\Gamma'/P_\mathrm{RF}$ and $\kappa/P_\mathrm{RF}$ differ
from their predicted values, this disagreement is not unexpected
considering the relatively large variations in dimensions $d_0$ and
$s$\footnote{The uncertainties for these numbers are a few micrometers,
  stemming from spatial nonuniformity in our etch process.}. Another indication of these uncertainties is
that optical measurements of the static deflection of the cantilever
along its length disagree with predictions based on a constant
cantilever cross section.  This will lead to deviations from our
calculated values of $\xi'$, $\xi''$, and $\xi_\mathrm{c}''$.
However, we stress that these deviations should not give significant
errors in our measured values of $L_\mathrm{eq}$, $R_\mathrm{eq}$, and
therefore our determination of $T_\mathrm{eff}$.

\begin{SCfigure}
\centering \includegraphics[width=0.7\textwidth]{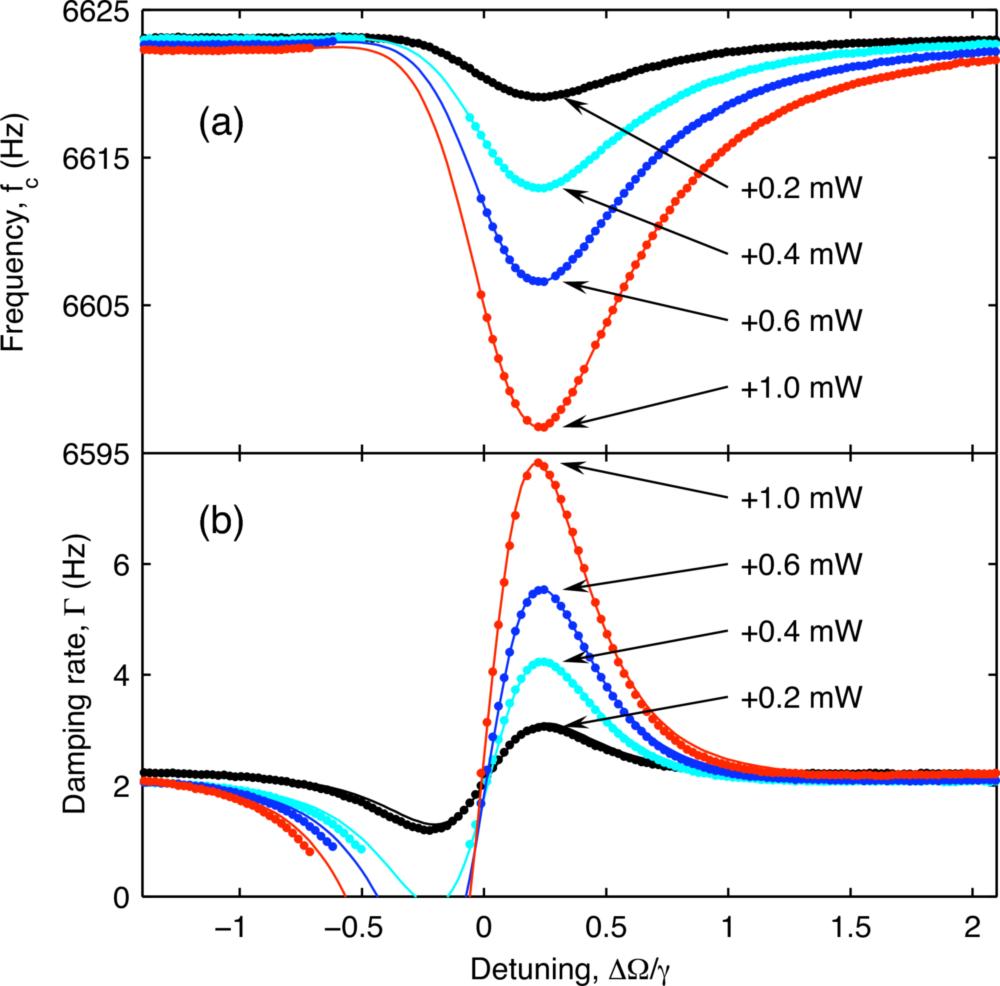}
\caption[Variation of the cantilever resonance with
  respect to RF frequency.]{Variation of the cantilever resonance with
  respect to RF frequency. (a) Cantilever frequency
  $f_\mathrm{c}$ and (b) damping rate $\Gamma$ versus
  normalized RF detuning $\Delta\Omega/\gamma$ for several values of
  $P_\mathrm{RF}$. The missing points between $-$0.7~MHz and 0~MHz
  correspond to a region of instability where $\Gamma$ becomes
  negative. Solid lines are fits to Eqs.~(\ref{kappa}) and
  (\ref{gammaprime}).}
\label{fig:ringdown}
\end{SCfigure}

To further test the model, we examine $\Gamma$ and $\omega_\mathrm{c}$
as a function of $\Delta\Omega$ (Fig.~\ref{fig:ringdown}).
For large detunings $\Delta\Omega$, $\Gamma$ asymptotically approaches
the value obtained in Fig.~\ref{fig:effective temperature} for
$P_\mathrm{RF}=0$. Near $\Delta\Omega=0$, $f_\mathrm{c}$ is generally
shifted to a lower value, while $\Gamma$ is either enhanced or
suppressed, according to the sign of $\Delta \Omega$. Data cannot be
obtained for $\Delta\Omega < 0$ when the RF power level is sufficient
to drive the cantilever into instability ($\Gamma<0$). The solid line
fits show good agreement with the predicted behavior. From these fits
we extract $C_\mathrm{c} = 0.09$~pF, lower than the value $0.17$~pF
obtained from the physical dimensions. This disagreement is not
surprising for the reasons mentioned above.

Some experiments using optical forces have observed strong effects
from the laser power absorbed in the cantilever mirror. A conservative
estimate of the RF power dissipated in our cantilever gives a
temperature rise of less than 1~K at the highest power, so these
effects should not be significant.

Although rather modest cooling is obtained here, the basic method
could eventually provide ground state cooling. For this we must
achieve the resolved sideband limit, where $\omega_\mathrm{c} >
\gamma$ ~\cite{marquardt2007a,wilson-rae2007a}, and the cooling would be very
similar to the atomic case~\cite{diedrich1989a,monroe1995a}. To insure
a mean quantum number $n$ less than one, the heating rate from the
ground state $\dot{n}_\mathrm{heat} = \Gamma k_\mathrm{B}
T_\mathrm{c}/(\hbar \omega_\mathrm{c})$ must be less than the cooling
rate for $n = 1 \rightarrow 0$. The cooling rate
$\dot{n}_\mathrm{cool}$ can be estimated by noting that each absorbed
photon on the lower sideband (at the applied RF frequency $\Omega_0 -
\omega_\mathrm{c}$) is accompanied by re-radiation on the RF
``carrier'' at $\Omega_0$. If we assume the lower sideband is
saturated for $n \approx 1$, $\dot{n}_\mathrm{cool} \approx
\gamma/2$. Hence we require $R \equiv
\dot{n}_\mathrm{heat}/\dot{n}_\mathrm{cool} \approx 2 k_\mathrm{B}
T_\mathrm{c} Q_\mathrm{RF}/(\hbar \Omega_0 Q_\mathrm{c}) \ll 1$. For
example, if $T_\mathrm{c} = 0.1$~K, $\Omega_0/(2 \pi) = 20$~GHz,
$Q_\mathrm{RF} = 5,000$ (e.g., a stripline), and $Q_\mathrm{c} =
20,000$ we have $R \approx 0.05$. For resolved sidebands, we require
$\omega_\mathrm{c}/(2 \pi) > 4$~MHz.

    \clearpage
\chapter{Semiclassical Quantum Fourier Transform}
\label{sec:qft}
In this chapter I summarize an implementation of the semiclassical quantum Fourier
transform (QFT) in a system of three beryllium ion qubits (two-level quantum
systems) confined in a segmented multi-zone trap.  The quantum Fourier transform
is a crucial step in Shor's algorithm, a quantum algorithm for integer
factorization which is superpolynomially faster than the fastest known classical
factoring algorithm. The QFT acts on a register of qubits to determine the
periodicity of the input states' amplitudes. With coworkers at NIST I applied the
transform to several input states of different periodicities; the results enable
the location of peaks corresponding to the original periods.

Among quantum algorithms discovered up to this time, Shor's method for factoring
large composite numbers~\cite{shor1994a} is arguably large-scale quantum
information processing's most prominent application; efficient factoring would
render current cryptographic techniques based on large composite-number keys
vulnerable to attack. The key component of this algorithm is an order-finding
subroutine that requires application of the quantum discrete Fourier transform
to determine the period of a set of quantum
amplitudes~\cite{shor1994a,coppersmith1994a,ekert1996a,nielsen2000a}. In addition
to this application, the QFT is also an essential part of quantum algorithms for
phase estimation and the discrete logarithm~\cite{nielsen2000a}.  In fact, the
polynomial-time QFT is responsible for most of the known instances of exponential
speedup over classical algorithms.

Relative phase information of the output state from the QFT is not required when
applied in any of the algorithms mentioned above; only the measured probability
amplitudes of each state are used. This allows the replacement of the fully
coherent QFT with the semiclassical, or ``measured'' QFT~\cite{griffiths1996a},
in which each qubit is measured in turn, and the prescribed controlled phase
rotations on the other qubits are conditioned on the classical measurement
outcomes. This eliminates the need for entangling gates in the QFT protocol,
which for an ion implementation considerably relaxes the required control of the
ions' motional states.  In addition, the semiclassical version is quadratically
more efficient in the number of quantum gates when compared with the fully
coherent version.  That is, for $n$ qubits the required number of quantum gates
is $O(n)$ instead of $O(n^2)$. 

Prior implementations of the QFT were in nuclear magnetic resonance (NMR) systems
and used the coherent version of the
protocol~\cite{vandersypen2000a,weinstein2001a,
vandersypen2001a,lee2002a,weinstein2004a}. This implementation was distinguished
in that it used the more efficient measured-QFT and that ions are a scalable
system for quantum information processing, while NMR is not~\cite{cory2000a}.
Extension of our implementation to larger quantum registers requires a linear
increase in the number of qubits and qubit
operations~\cite{bible,kielpinski2002a}.

The details of the QFT as realized in the lab were published in a paper which
appeared in Science in 2005 titled ``Implementation of the semiclasscial
quantum Fourier transform in a scalable system"~\cite{chiaverini2005b}. 
Details on the apparatus, experimental techniques and 
principles of quantum control of beryllium ion qubits are 
discussed in the 
literature~\cite{bible,kielpinski2001a,barrett2004a,langer2006a}.


The QFT is a basis transformation in an $N$-state space that
transforms the state $|k\rangle$ ($k$ is an integer ranging from 0
to $N-1$) according to

\begin{equation}
|k\rangle \rightarrow {1 \over \sqrt{N}} \sum_{j=0}^{N-1} e^{i2\pi j
k/N} |j\rangle.
\end{equation}

\noindent The action on an arbitrary superposition of states may be
written as

\begin{equation}
\sum_{k=0}^{N-1} x_k |k\rangle \rightarrow \sum_{j=0}^{N-1} y_j
|j\rangle,
\end{equation}

\noindent where the complex amplitudes~$y_j$ are the discrete
Fourier transform~\cite{arfken1985a} of the complex amplitudes~$x_k$. For
three qubits, switching to binary notation, where~$k_1$, $k_2$,
and~$k_3$ are the most to least significant bits, respectively, in
the label for the state $|k_1 k_2 k_3\rangle = |k_1\rangle \otimes
|k_2\rangle \otimes |k_3\rangle$ ($k_i \in \{0,1\}$), the transform
can be written as~\cite{nielsen2000a}

\begin{equation}
|k_1 k_2 k_3\rangle \rightarrow {1 \over \sqrt{8}} \left( |0\rangle
+ e^{i2\pi \left[0.k_3\right]} |1\rangle \right)  \otimes \left(
|0\rangle + e^{i2\pi \left[0.k_2 k_3\right]} |1\rangle \right)
\otimes \left( |0\rangle + e^{i2\pi \left[0.k_1 k_2 k_3\right]}
|1\rangle \right), \label{prodrep}
\end{equation}

\noindent where $\left[0.q_1 q_2 \cdots q_n\right]$ denotes the
binary fraction $q_1/2 + q_2/4 + \cdots + q_n/2^n$. When written in
this form, it can be seen that the QFT is the application to each
qubit of a Hadamard transformation $\left[ |0\rangle\rightarrow {1
\over \sqrt{2}}\left( |0\rangle + |1\rangle \right)\ {\rm and}\
|1\rangle\rightarrow {1 \over \sqrt{2}}\left( |0\rangle -
|1\rangle\right)\right]$ and a $z$~rotation conditioned on each of
the less significant qubits, with a phase of decreasing binary
significance due to each subsequent qubit, all followed by a
bit-order reversal~\cite{nielsen2000a}. The three-qubit quantum
circuit, without the bit-order reversal, is shown in Fig.~1a. The
simplified circuit for the measured QFT is shown in Fig.~1b.

\begin{figure}
\begin{center}
\hbox{a) \hfill}
\includegraphics[width=0.90 \columnwidth]{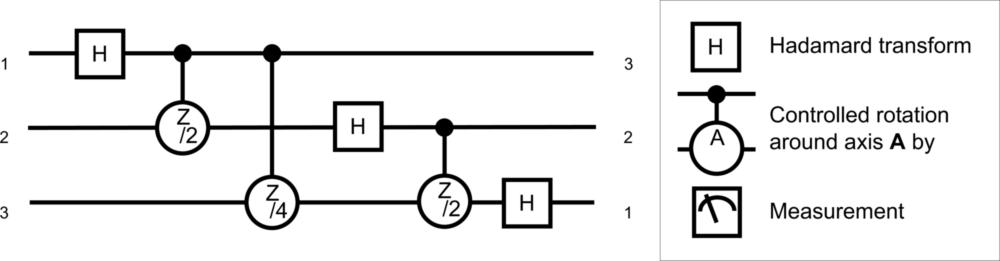}\\
\hbox{b) \hfill}
\includegraphics[width=0.75 \columnwidth]{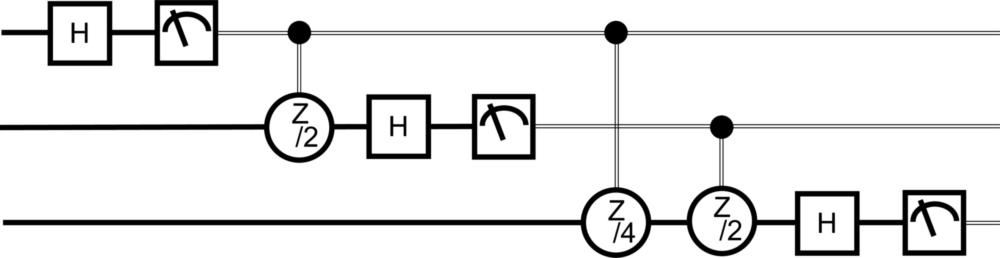}\\
\caption[Circuits for the quantum Fourier transform (QFT) of three
qubits.]{Circuits for the quantum Fourier transform (QFT) of three
qubits. (a) The QFT as composed of Hadamard transforms and two-qubit
conditional phase gates~\cite{nielsen2000a}. The gate labeling scheme
denotes the axis about which the conditional rotation takes place
and, below the axis label, the angle of that rotation.  The
$|\varphi_i\rangle$ and $|\chi_i\rangle$ are the input and output
states, respectively, of qubit~$i$. The most-significant qubit
corresponds to $i=1$. This circuit produces the QFT in reverse bit
order, so in practice, the qubits are simply read out in reverse
order~\cite{nielsen2000a}. (b) The semiclassical (or ``measured'')
QFT~\cite{griffiths1996a}. The double lines denote classical
information.  This circuit can be implemented by means of a single
classically-controlled quantum operation on each qubit.  The
protocol is preceded by state preparation (not shown in figure) of
the quantum state to be transformed.} \label{circuits}
\end{center}
\end{figure}

In the experiment, $z$~rotations are transformed into $x$~rotations,
which are more straightforward to implement in our system, and
rotations are redistributed to accommodate required spin-echo
refocussing pulses ($\pi$ rotations) which reduce dephasing due to
fluctuating magnetic fields~\cite{hahn1950a,allen1975a,rowe2002a}, but
this does not change the basic protocol. Because of the substitution
of $\pi/2$ rotations for Hadamard operations and our choice of
conditional-rotation direction, the coherent QFT corresponding to
our measured QFT is described by

\begin{equation}
|k_1 k_2 k_3\rangle \rightarrow {1 \over \sqrt{8}} \left( |0\rangle
- e^{-i2\pi \left[0.k_3\right]} |1\rangle \right)  \otimes \left(
|0\rangle - e^{-i2\pi \left[0.k_2 k_3\right]} |1\rangle \right)
\otimes \left( |0\rangle - e^{-i2\pi \left[0.k_1 k_2 k_3\right]}
|1\rangle \right).
\end{equation}

\noindent The sign differences from Eq.~3 are unimportant, because
only the probability amplitudes of the output state are measured;
the relative phases of the output basis states are arbitrary. We
have applied this three-qubit QFT to input states of several
different periodicities.

The qubits comprise two states of the ground-state hyperfine
manifold of $^9$Be$^+$: the state $|F=1, m_F=-1\rangle$, labeled
$|0\rangle$, and the state $|F=2, m_F=-2\rangle$, labeled
$|1\rangle$.  These states are separated in frequency by 1.28~GHz.
Rotations

\begin{equation}
R(\theta,\phi)= \cos {\theta \over 2} \:I - i \sin {\theta \over 2}
\cos \phi \:\sigma_x - i \sin  {\theta \over 2} \sin \phi \:\sigma_y
\end{equation}

\noindent are performed by means of two-photon stimulated-Raman
transitions~\cite{monroe1995a,wineland2003a}.  Here $\theta$ is the
rotation angle, $\phi$ is the angle of the rotation axis from the
$x$~axis in the $xy$~plane of the Bloch sphere~\cite{allen1975a},
$I$ is the identity operator, and $\sigma_x$ and $\sigma_y$ are the
usual Pauli spin operators. The beryllium ions are confined in a
linear radio-frequency Paul trap~\cite{barrett2004a} similar to
that described in~\cite{rowe2002a}. This trap contains six zones, and
the ions can be moved between these zones, together or separately,
by means of synchronized variation of the potentials applied to the
trap's control electrodes. State determination is made by projection
of the qubit state with the use of resonance
fluorescence~\cite{monroe1995a} (an ion in the $|1\rangle$ state
fluoresces, whereas an ion in the $|0\rangle$ state does not).
Measurement results were recorded, and laser pulses were applied by
means of classical logic to implement conditional operations.  The
QFT protocol proceeded as depicted in Fig.~2a, with ions located in
the multizone trap as shown in Fig.~2b.

\begin{figure}
\begin{center}
\hbox{a) \hfill}
\includegraphics[width=0.85 \columnwidth]{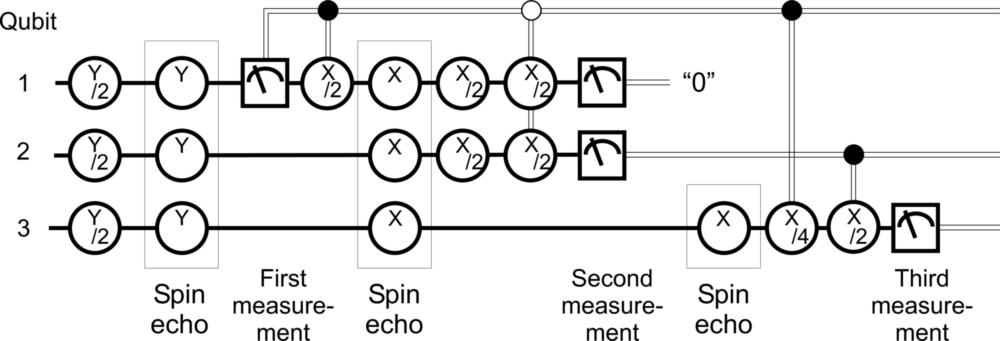}\\
\hbox{b) \hfill}
\includegraphics[width=0.66 \columnwidth]{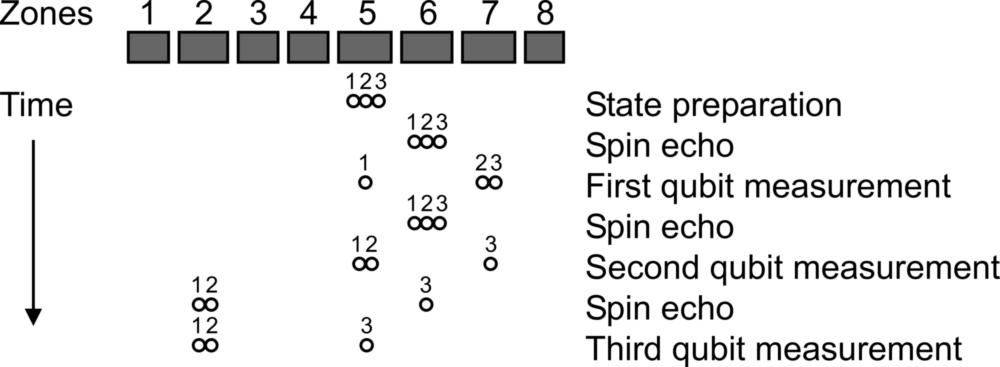}\\
\caption[Circuit for the QFT and locations of the ions in the
multizone trap during protocol execution.]{Circuit for the QFT and locations of the ions in the
multizone trap during protocol execution.  (a) The semiclassical
QFT~\cite{griffiths1996a} as implemented in this Report. The double
lines denote classical information.  The closed circles on control
lines denote rotation conditional on ``1;'' the open circles denote
rotation conditional on ``0.''  The initial conditional rotation of
qubit~1 ensures that it is in the non-fluorescing state when the
second ion is measured (the second ion is measured in the presence
of the first ion, which contributes negligibly to the fluorescence
signal during the second measurement~\cite{schaetz2004a}; refer to
``Second qubit measurement'' (b)).  This circuit, up to some
irrelevant phases, can be obtained from that in Fig.~1b through
conjugation of rotations and reordering of some operations. (b) The
locations of the ions in the multizone trap structure during the QFT
protocol as a function of time. Separation of ions and refocussing
operations are performed in zone~6, and all other qubit operations
are performed in zone~5.} \label{circuits2}
\end{center}
\end{figure}

Five different states were prepared to test the QFT protocol
(Table~1). These states have periods of~1, 2, 4, 8, and
approximately~3. The three-qubit state space consists of eight
states, labeled $|000\rangle$, $|001\rangle$, $\dots$, $|111\rangle$
in binary notation and ordered lexicographically.  The periodicity
is derived from the recurrence of the quantum amplitudes in a
superposition of these eight states.

The period~1 state was generated by applying the rotation
$R(\pi/2,-\pi/2)$ to all three ions in the initial state
$|111\rangle$.  The period~2 state was generated from $|111\rangle$
by physically separating ion~3 from ions~1 and~2, applying a
rotation $R(\pi/2,-\pi/2)$ to ions~1 and~2, and then bringing all
three ions back together. Similarly, the period~4 state was created
by applying the rotation $R(\pi/2,-\pi/2)$ to only the first ion
after separating it from ions~2 and~3.  The period~8 state was
simply the state of the ions after initialization,~$|111\rangle$.

The most obvious (approximate) period~3 state in this eight-state
space is $|000\rangle + |011\rangle + |110\rangle$ (here and in the
following, we omit normalization factors). Becasue this state's
periodicity is not commensurate with the state space, the addition
of the next (in a sequence of three) basis state $|001\rangle$ to
this superposition also results in an approximate period~3 state,
$|\psi_{3}'\rangle=|000\rangle + |011\rangle + |110\rangle
+|001\rangle$.  We used a cyclic permutation of $|\psi_{3}'\rangle$;
in particular, adding 3~(mod 8) to each state produces
$|\psi_3\rangle = |011\rangle + |110\rangle + |001\rangle +
|100\rangle$. This state is the tensor product of $|01\rangle_{1,3}
+ |10\rangle_{1,3}$ (ions~1 and~3) with $|0\rangle_2 + |1\rangle_2$
(ion~2).  Starting from $|111\rangle$, this state can be prepared by
entangling the outer two ions (ions~1 and~3) with a geometric phase
gate embedded between two rotations--- $R(\pi/2,\pi/4)$ and
$R(3\pi/2,\pi/4)$ ---applied to all three
ions~\cite{leibfried2003b,barrett2004a}. This was followed by a rotation
$R(\pi/2,-\pi/2)$ to all three ions. This state was produced with a
fidelity of approximately~0.90 (resulting from the reduced fidelity
inherent in multi-qubit entangling operations compared with
single-qubit rotations).

\begin{table}

\begin{center}
\caption{Periodic states prepared to test the semiclassical QFT
protocol}
\smallskip
\begin{tabular*} {1.0 \columnwidth} {@{\extracolsep{\fill}} c  l  r }
\hline \hline
Periodicity     & State (normalization omitted) & Preparation fidelity\\

\hline
1&          $|\psi_1\rangle=|000 \rangle + |001\rangle + |010\rangle + \cdots + |111\rangle$ & $0.98(1)$              \\
2&          $|\psi_2\rangle=|001 \rangle + |011\rangle + |101\rangle + |111\rangle $ & $0.98(1)$      \\
$\sim3$&  $|\psi_3\rangle=|001 \rangle + |011\rangle + |100\rangle + |110\rangle $ & $0.90(2)$       \\
4&          $|\psi_4\rangle=|011 \rangle + |111\rangle $ & $0.98(1)$ \\
8&          $|\psi_8\rangle=|111\rangle$   &  $> 0.99(1)$\\

\hline \hline
\end{tabular*}

\label{preptable}
\end{center}

\end{table}

Each experiment began with Doppler cooling and Raman-sideband
cooling to bring the ions to the ground state of all three axial
vibrational modes of the trapping potential and optical pumping to
prepare the three ions in the internal state
$|111\rangle$~\cite{bible,king1998a}. The aforementioned input
states were then prepared as described above. For each input state,
several thousand implementations of the QFT were performed, each
involving: ({\it i})~rotation of ion~1, ({\it ii})~measurement of
ion~1, ({\it iii})~rotation of ion~2 conditional on the measurement
of ion~1, ({\it iv})~measurement of ion~2, ({\it v})~rotation of
ion~3 conditional on the first two measurements, and ({\it
vi})~measurement of ion~3.  Each experiment required
approximately~4~ms after initial cooling, optical pumping, and state
preparation.

The measured output state probabilities after application of the QFT
algorithm are shown in Fig.~3 along with the theoretically expected
probabilities for the five different input states.  The data
generally agree with the theoretical predictions, although the
deviations from the predicted values are larger than can be
explained statistically, and are due to systematic errors in the
experiment. These systematic errors are associated with the state
preparation (not associated with the QFT protocol) as well as with
the separate detections and conditional rotations of the three ions
(intrinsic to the QFT protocol). The first, second, and third ions
were measured approximately 1.2~ms, 2.4~ms, and 3.5~ms after the
beginning of the algorithm. Dephasing due to slow local
magnetic-field fluctuations, though mitigated by the refocussing
(spin-echo) operations, grows as a function of time during each
experiment; the chance that an error occurs because of dephasing
grows from approximately~5\% for the first ion to approximately~13\%
for the third ion.

\begin{figure}
\begin{center}
\hbox{a) \hspace{.48 \columnwidth} b)}
\includegraphics[width=0.38 \columnwidth]{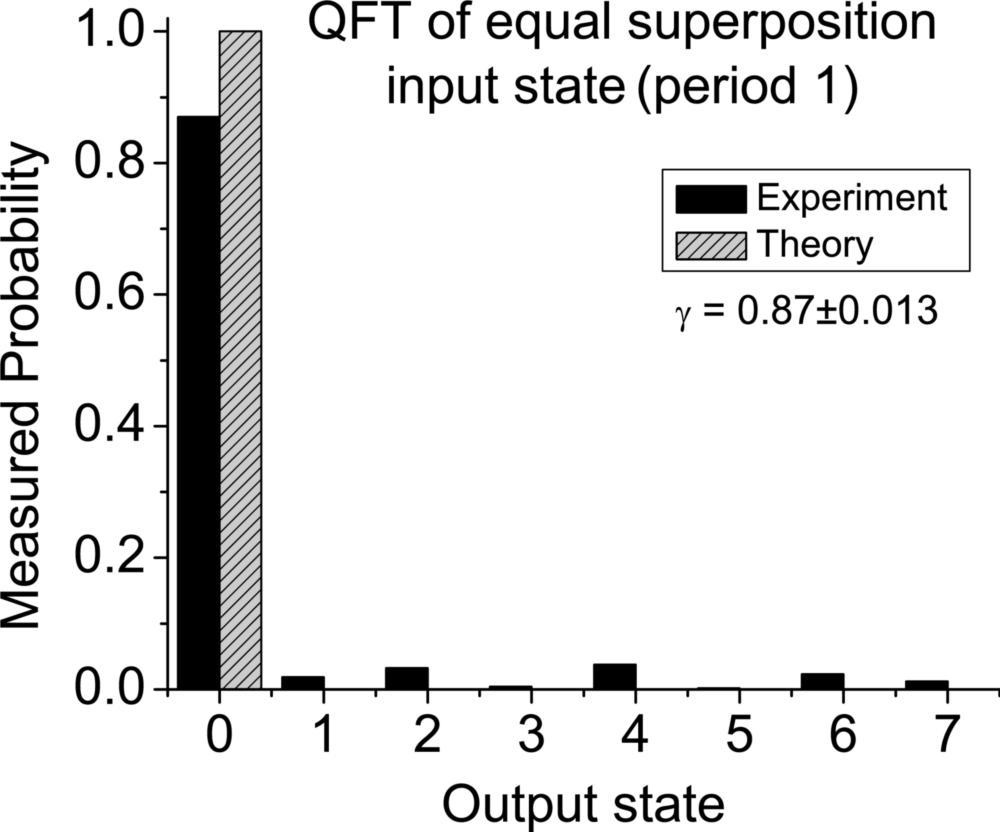} \hspace{0.12
\columnwidth} \includegraphics[width=0.38 \columnwidth]{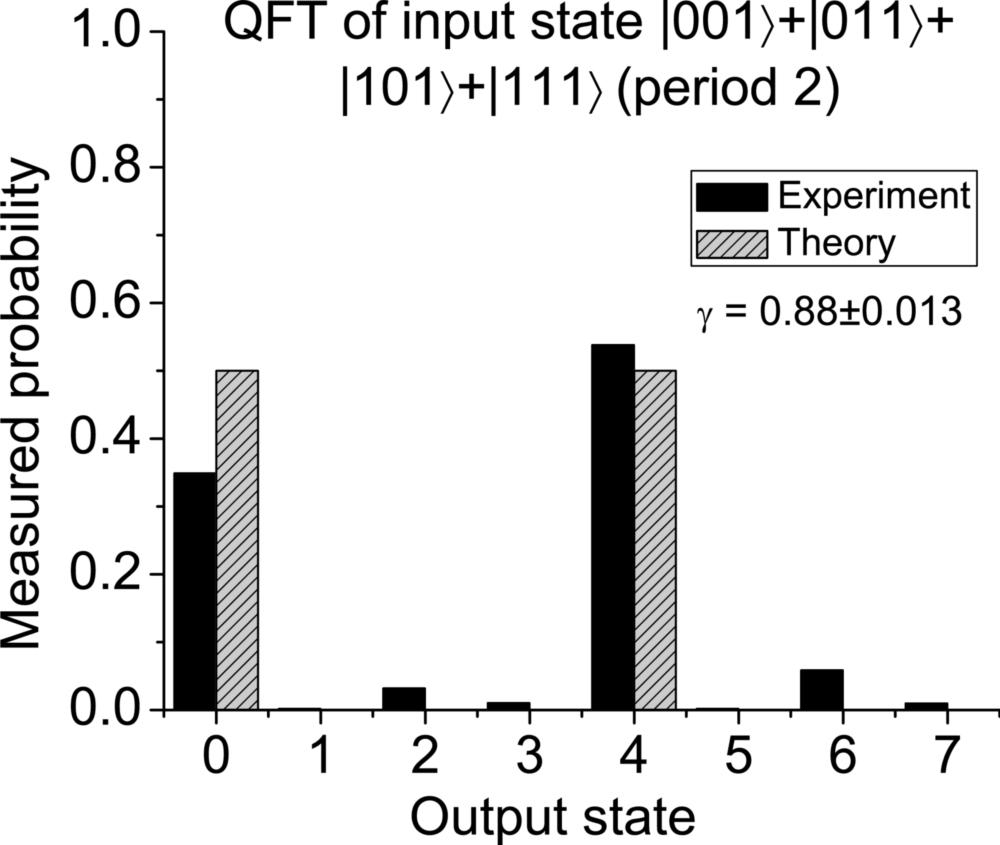}
\\
\hbox{c) \hspace{.48 \columnwidth} d)}
\includegraphics[width=0.38 \columnwidth]{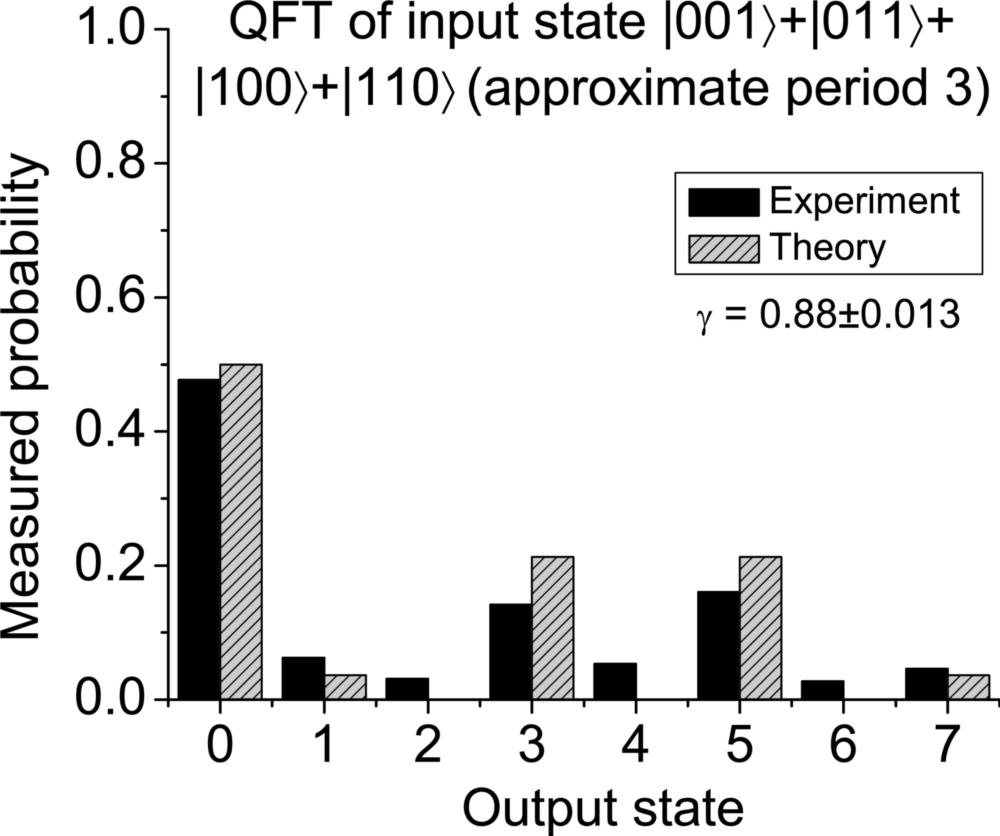} \hspace{0.12
\columnwidth} \includegraphics[width=0.38
\columnwidth]{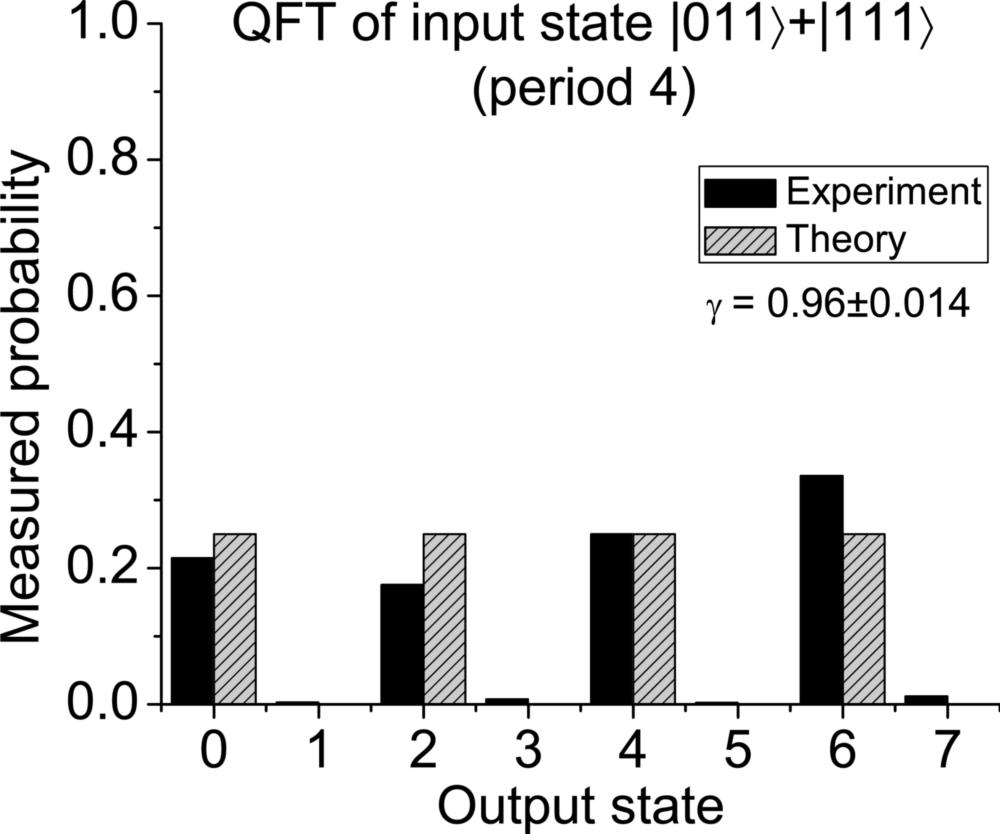}\\
\hbox{e) \hfill}
\includegraphics[width=0.38 \columnwidth]{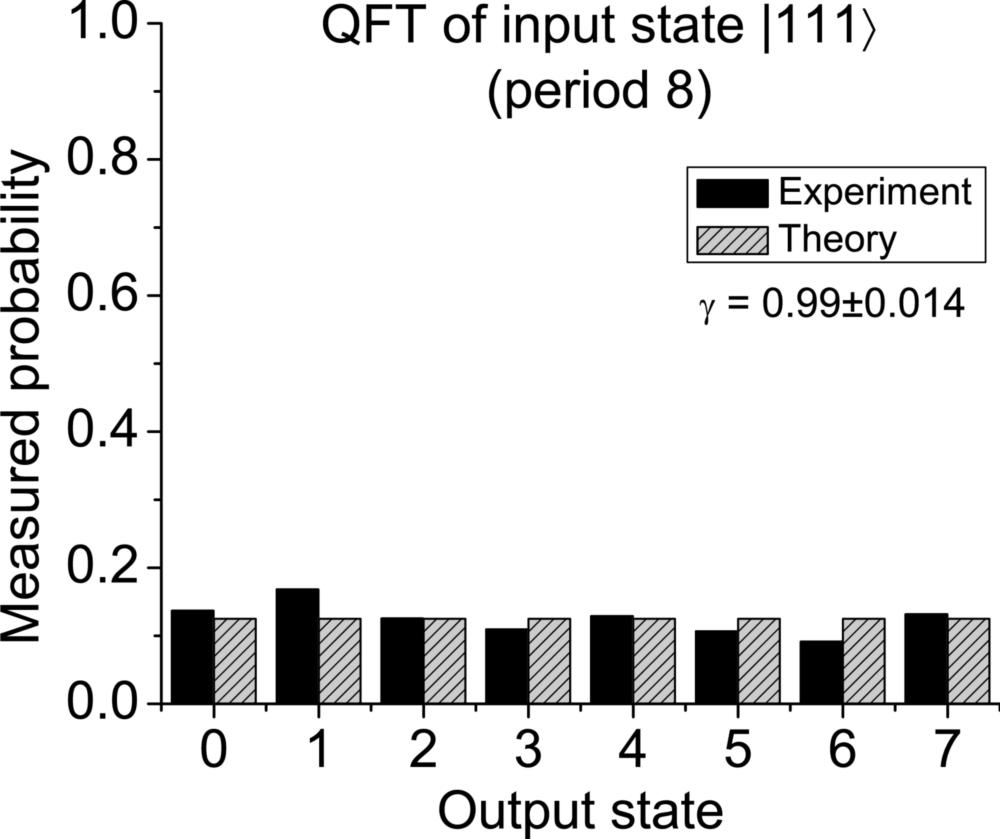}
\caption[Results of the semiclassical QFT.]{Results of the semiclassical QFT.
Measured probability of each output state occurring after the application of the protocol is
shown along with the expected transform output.  Each plot contains
data from 5000 experiments. The~SSO~$\gamma$ is a measure of
transform accuracy. Uncertainties quoted for the SSO are statistical
and do not include systematic errors. (a) to (e) are the QFTs for
$|\psi_1\rangle$, $|\psi_2\rangle$, $|\psi_3\rangle$,
$|\psi_4\rangle$, and $|\psi_8\rangle$, respectively.} \label{data1}
\end{center}
\end{figure}

Even with these systematic errors, the results compare well with
theory, as can be shown by examining the squared statistical
overlap~(SSO) (derived from the statistical overlap
of~\cite{fuchs1996a}) of each set of data with the associated
predictions. Here, we define the SSO as $\gamma = \left(
\sum_{j=0}^{7} m_j^{1 \over 2} e_j^{1 \over 2} \right)^2$, where
$m_j$ and $e_j$ are the measured and expected output-state
probabilities of state~$|j\rangle$, respectively. This is an
effective measure of fidelity without regard for relative output
phases.  The lowest measured SSO for the five prepared states is
0.87, suggesting that peaks can be reliably located to determine
periodicities as required for Shor's factorization algorithm.  To
verify the reliability of the protocol for this task, one should
compare the experimental and theoretical values of the measurement
probabilities of the output states where peaks are located.  For the
period~2 state, the measured probability for the output
state~$|4\rangle= |100\rangle$ (the measurement outcome sufficient
to determine the periodicity as~2), was 0.538, an 8\% difference
from the expected value of 0.5. For the period~3 state, the sum of
the measured probabilities for output state~$|3\rangle=|011\rangle$
and state~$|5\rangle=|101\rangle$ (the states corresponding to the
most correct periodicity) was 0.301, a 29\% difference from the
expected value of 0.426. Notably, the preparation fidelity of this
state was not as high as for the others.

One other set of input states was created to demonstrate that the
semiclassical QFT protocol is sensitive to relative input phases.
All the states of Table~1 had amplitudes (of the basis states in the
superpositions) with the same phase.  We also prepared a period~3
state with a relative phase between some states in the
superposition.  By incrementing the phase of the three uniform
rotations used in the creation of the period~3 state with respect to
the operations in the QFT protocol by a phase~$\phi_R$ (that is,
$R(\theta,\phi) \rightarrow R(\theta,\phi+\phi_R)$), we can create
the state

\begin{equation}
|\psi_3(\phi_R)\rangle =  |001\rangle + e^{i \phi_R} |011\rangle +
|100\rangle + e^{i \phi_R} |110\rangle. \label{phir}
\end{equation}

\noindent  The relative phase between pairs of basis states in this
superposition leads to a Fourier transform that depends on $\phi_R$.
The measured probabilities of the eight output states are plotted in
Fig.~4a along with the theoretical values in Fig.~4b. The level of
agreement can be seen in Fig.~4c, a plot of the SSO as a function of
preparation phase.

\begin{figure}
\begin{center}
\hbox{a) \hspace{.31 \columnwidth} b) \hspace{.31 \columnwidth} c)}
\hbox{\includegraphics[width=0.33 \columnwidth]{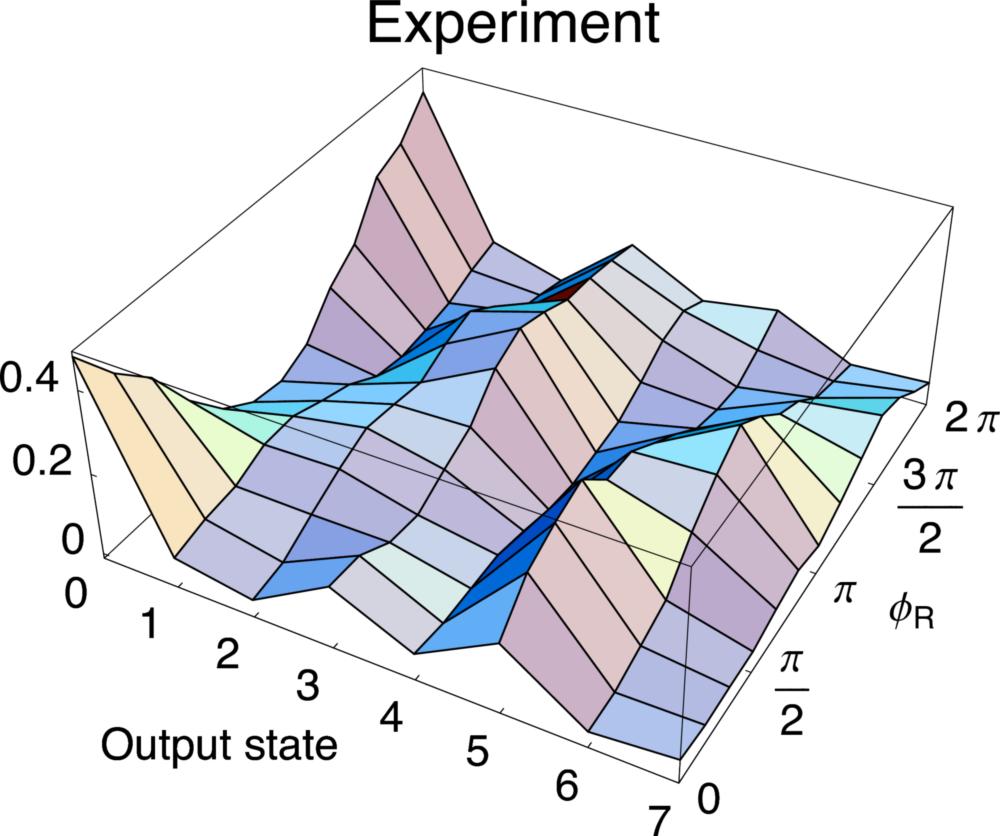}
\includegraphics[width=0.33 \columnwidth]{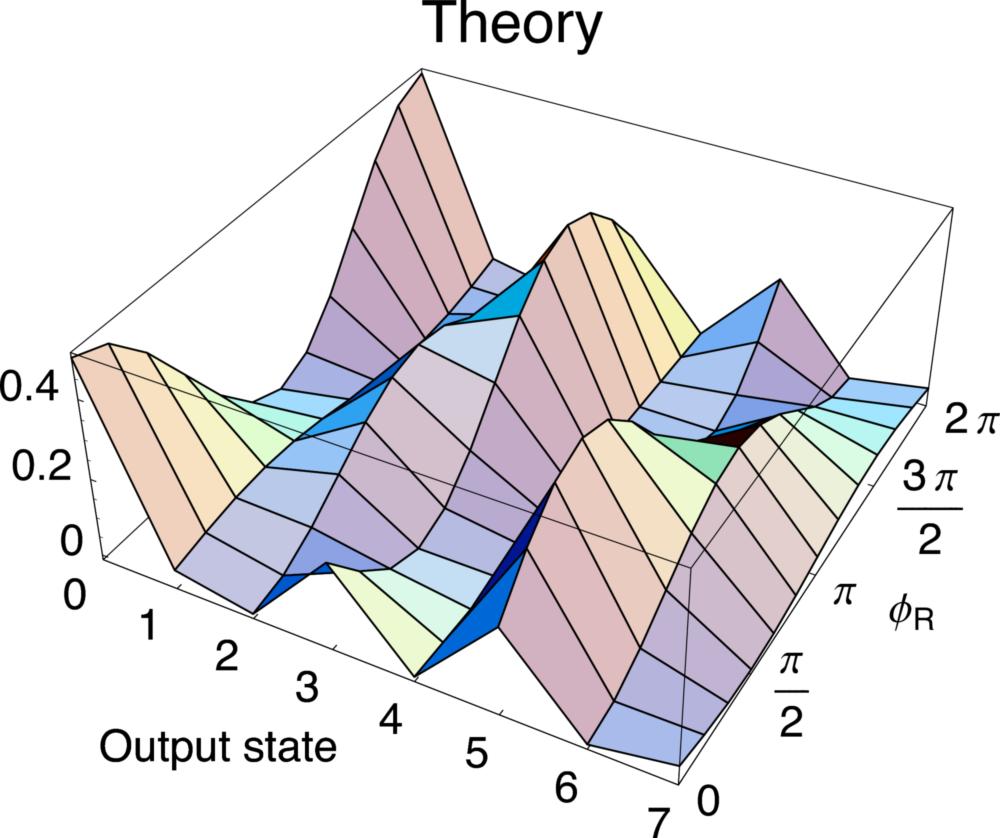}
\includegraphics[width=0.33 \columnwidth]{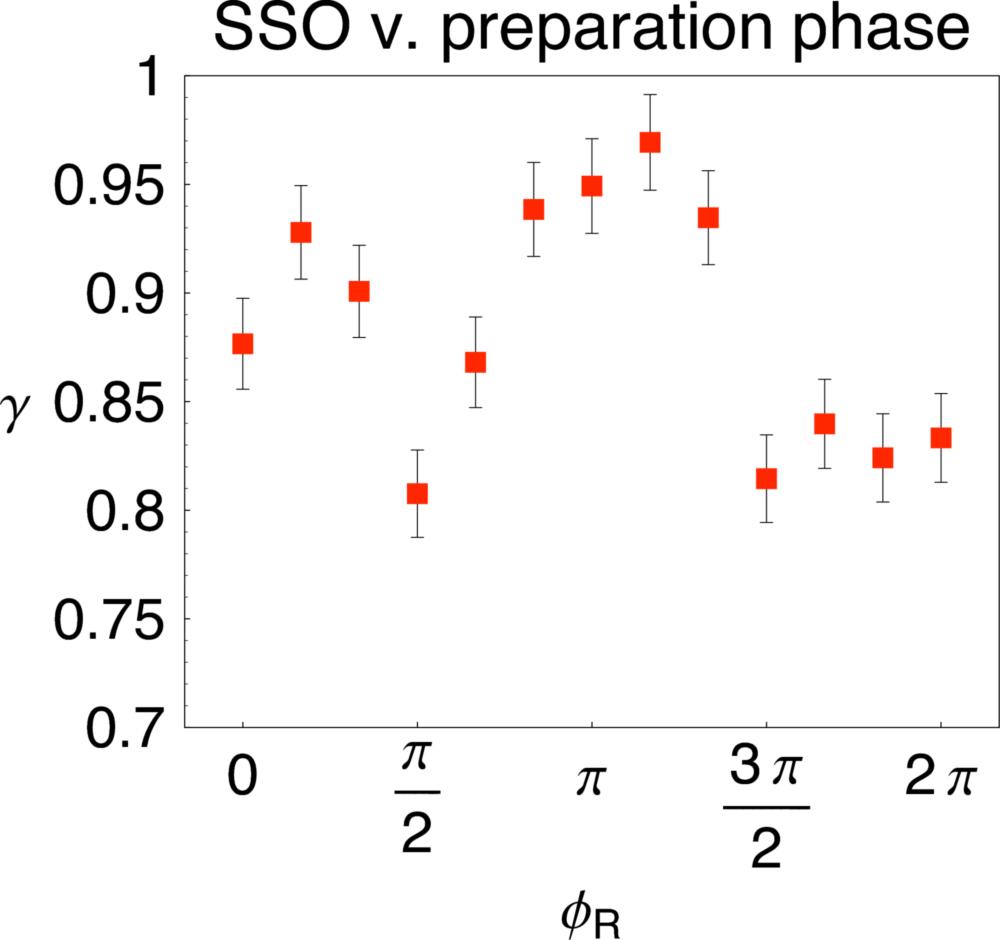}}
\caption[Semiclassical QFT of nominal period~3 state as a function
of preparation phase $\phi_R$.]{Semiclassical QFT of nominal period~3 state as a function
of preparation phase $\phi_R$ (Eq.~6). (a) The measured
probabilities are plotted as a function of the output state and the
phase of the state preparation after application of the QFT
protocol. The QFT at each phase is based on~2000 experiments. (b)
The expected probability plotted in the same manner. (c) The
SSO~$\gamma$ as a function of preparation phase. Error bars
represent 1$\sigma$~statistical uncertainties only.} \label{data2}
\end{center}
\end{figure}

These results demonstrate that for small state-spaces, the QFT can
be performed semiclassically with a signal-to-noise level sufficient
for period-finding in quantum algorithms by means of a system of
trapped-ion qubits. Even with input state infidelities as large
as~0.10, as in the period~3 state created here, the measured QFT had
substantial SSO with the theoretical prediction for the correct
input state. Furthermore, the effect of the incommensurability of
state periodicity with state space is diminished as the size of the
state space increases.  Peaks due to similar incommensurate
periodicities will be easier to locate in larger state spaces
(compared to the case of a period~3 state in an eight-state space).
Extension of the technique described here to larger quantum
registers~\cite{bible,kielpinski2002a} is a function only of trap-array size
and involves a linear overhead in ion separation and movement.  The
main source of intrinsic error in our implementation was qubit
dephasing resulting from magnetic-field fluctuations.  Use of
first-order magnetic-field-independent qubit
transitions~\cite{bible} can mitigate this problem and lead to a
high-fidelity method for implementation of the QFT, a necessary step
toward large-number-factorization applications of quantum computing.

    \chapter{Conclusion}

This thesis work investigated new ways of building ion traps for use in quantum
information applications. Several traps made of boron doped silicon were
demonstrated with ion-electrode separations down to $40~\mu$m. Improvements in
trap fabrication and testing reduced the time from concept to trapping by an
order of magnitude compared to previous demonstrations. In my traps upper bounds
on ion heating rates were determined; however, heating due to externally injected noise could not be
completely ruled out.  Noise on the RF potential responsible for providing
confinement was identified as one source of injected noise.

Future directions for this work include further studies of the sources of ion
heating.  A next step for boron doped traps is to positively identify
sources of injected noise, with special attention to the possible role of
sideband noise on the trap RF potential source. In addition, systematic
investigation of ion heating as a function of trap
temperature \cite{labaziewicz2008a}, electrode metal type and electrode surface
preparation is needed. Trapping with ion-electrode separations as small as
$10~\mu$m is interesting not just for QIP purposes, but also because ions are
sensitive probes of near near field electric and magnetic
fields~\cite{maiwald2008a}.  In several surface electrode traps it is
observed that in the absence of laser cooling ions depart from the trap more
quickly than can be accounted for by background collisions.  This may be due 
to parametric heating or inelastic collisions with the trap ponderomotive
confining potential. For scaling to the number of trapping zones needed for
large-scale QIP, integration with CMOS electronics, MEMS optics and optical
fibers may be fruitful.

Using the microfabrication technology developed for ion traps, I made a
cantilevered micromechanical oscillator and with coworkers demonstrated a method
to reduce the kinetic energy of its lowest order mechanical mode via its capacitive coupling to
a driven RF resonant circuit. Cooling results from a RF capacitive force, which
is phase shifted relative to the cantilever motion. The technique was
demonstrated by cooling a 7~kHz fundamental mode from room temperature to
$45$~K. Ground state cooling of the mechanical modes of motion of harmonically trapped
ions is routine; equivalent cooling of a macroscopic harmonic oscillator has not
yet been demonstrated. Extension of this method to devices with higher motional
frequencies in a cryogenic system, could enable ground state cooling and may prove
simpler than related optical experiments.  

I discussed an implementation of the semiclassical quantum Fourier transform
(QFT) using three beryllium ion qubits.  The QFT is a crucial step in a number of
quantum algorithms including Shor's algorithm, a quantum approach to integer
factorization which is exponentially faster than the fastest known classical
factoring algorithm. This demonstration incorporated the key elements of a
scalable ion-trap architecture for QIP.  Future work could improve the 
fidelity of the QFT, by for example using magnetic field
independent qubit transitions~\cite{langer2005a}, and incorporate it as a
subroutine in more complex algorithms.  In order for it to be considered a true subroutine
experimental control apparatus is needed which automates the many calibration
steps involved in operating the QFT.

    \clearpage
\chapter{Appendix}
\label{sec:appendix}
 
%

\section{Mathieu equation of motion}
\label{sec:mathieuEquation}
If potential $V_0\cos(\Omega_{\rm RF} t)+V_{\rm DC}$ is applied to the RF
electrodes with the others grounded ($V_1=V_2=0$), as in
Figure~\vref{ions:fig:LinEle}, the potential near the geometric center of the
trap takes the form
\begin{equation}\label{sec:mathieuEquation:fullPotential}
  \Phi\approx\frac{1}{2}\left(V_0\cos(\Omega_{\rm RF}
  t)+V_{\rm DC}\right)(1+\frac{x^2-y^2}{R^2}),
\end{equation}
where $R$ is a distance scale that is approximately the distance from the trap
center to the surface of the electrodes
\cite{bible}.

The classical solutions to the equations of motion for a particle of mass $m$ and
charge $q$ for the potential in Equation~\vref{sec:mathieuEquation:fullPotential}
were first solved by Paul, Osbergahaus, and Fischer \cite{paul1958a}.  Here, only
the x-direction will be solved since the solutions are decoupled in the spatial
coordinates.  The x coordinate equation of motion is
\begin{equation}
  \ddot{x} = - \frac{q}{m} \frac{\partial \Phi}{\partial x} = 
  -\frac{q}{m}\left( V_0 \cos(\Omega_{\rm RF} t)+V_{\rm DC}\right)\frac{x}{R^2}.
\end{equation}
With the following substitutions, 
\begin{equation}\label{eqq1}
  \xi = \frac{\Omega_{\rm RF} t}{2}, \\
  a_{x}=\frac{4 q V_{\rm DC}}{m \Omega_{\rm RF}^{2}R^2}, \\
  q_{x}=\frac{2 q V_0}{m \Omega_{\rm RF}^{2}R^2},
\end{equation}
the equation of motion can be cast in the form of the Mathieu differential equation,
\begin{equation}
  \ddot{x}+\left[a_{x}-2q_{x}\cos(2 \xi)\right]x=0.
\end{equation}
The solution to this equation to first order in $a_x$ and second order 
in $q_x$ is~\cite{ghosh1995a},
\begin{align}
	u_x(t)&=A_x\{
	\cos(\omega_x+\phi_x)
	\left[1+\frac{q_x}{2}\cos(\Omega_{\rm RF}t)+
	\frac{q_x^2}{32}\cos(2 \Omega_{\rm RF})\right]\\
	&+\beta_x \frac{q_x}{2}\sin(\omega_x t+\phi_x)\sin( \Omega_{\rm RF} t ) \},
\end{align}
where $A_x$  depends on initial conditions, $\omega_x=\beta_x\frac{\Omega_{\rm RF}}{2}$
and $\beta_x\sim\sqrt{a_x+q_x^2/2}$.  The oscillation at $\omega_x$ is called \emph{secular motion}
while that at $\Omega_{\rm RF}$ and $2 \Omega_{\rm RF}$ is called \emph{micromotion}.  When
$a_x<<q_x^2<<1$ we can ignore the micromotion terms and the ion behaves as if it were
trapped radially by a harmonic potential $\Phi_{pp}$ called the \emph{pseudopotential},
\begin{align}
	q\Phi_{pp}=\frac{1}{2}m \omega_x^2 x^2
\end{align}
where $\omega_x=\frac{qV_0}{\sqrt{2}\Omega_{\rm RF}m R^2}=\frac{q_x\Omega_{\rm RF}}{2\sqrt{2}}$ is 
the radial secular frequency.

A quantum-mechanical solution to the problem is possible
\cite{glauber92,leibfried2003a} and it is found to match the classical one for
the range of parameters commonly encountered in my experiments.

Books on ion trapping include those by Gosh~\cite{ghosh1995a} and Major, \emph{et
al.}~\cite{major2005a}. There is a review by Paul~\cite{paul1990a} and a study of
the effects of higher order terms in the trapping potential by
Home and Steane~\cite{home2006a}.

\section{Micromotion}
This section discusses in more detail micromotion, which was introduced in
Section~\vref{ions:sec:micromotion}.
\label{sec:micromotion}


The first-order solution to the Mathieu equation (see
Section~\vref{sec:mathieuEquation}) in the limit of small trap parameters
($\left|q_i\right|<<1$ and $\left|a_i\right|<<1$) is
\begin{equation*}
	u_i(t)\approx u_{1i}\cos \left(\omega
	_it\right)\left(1+\frac{q_i}{2}\cos\left(\Omega_{\rm RF} t\right)\right)
\end{equation*}

where $\omega_i\cong \frac{1}{2}\Omega_{\rm RF}\sqrt{a_i+\frac{1}{2}q_i{}^2}$,
$i\in \{\text{{x},y,\text{z}}\}$. The secular harmonic motion $u_i(t)$ is at
frequency $\omega_i$ with amplitude $u_{1i}$. This motion carries the ion back
and forth thru the RF pseudopotential minimum. The coherently driven motion at
$\Omega_{\rm RF}$ is called \emph{micromotion}. It can't be cooled since 
it's driven motion but it's amplitude can be minimized by reducing the secular
amplitude $u_{1i}$, that is cooling the $i^{\text{th}}$ mode at $\omega_i$.  
Intrinsic micromotion also arises from a nonzero phase difference 
$\phi_{\rm RF}$ between the potentials applied to the RF trap electrodes.  This
can happen due to a path length difference leading to the electrodes or
differential coupling to their environment.

Additional micromotion can arise if the ion's mean position doesn't
lie at a zero in the pseudopotential.  This
can happen if a static electric field $\pmb{E}_{\text{DC}}$ pushes the 
ion radially off axis.  The resulting motion is called excess micromotion can
can be nulled by applying a compensating field. The ion position along the x (or
y) radial direction may be calculated \cite{berkeland1998a}, 

\begin{equation*}
  u_x(t)\cong \left(u_{0x}+u_{1x}\cos \left(\omega
  _xt\right)\right)\left(1+\frac{1}{2}q_x\cos \left(\Omega
  _T\right)\right)-\frac{1}{4}q_xR
  \alpha  \phi _{\text{ac}}\sin \left(\Omega _Tt\right).
\end{equation*}

The x motion due to $\pmb{E}_{\text{DC}}$ is has amplitude

\begin{equation*}
u_{0x}\cong \frac{4 Q \pmb{E}_{\text{DC}}\cdot \overset{\wedge }{u_x}}{m \omega
_x{}^2}.
\end{equation*}

Excess micromotion disappears when $\pmb{E}_{DC}=0$. 
$\pmb{E}_{DC}=0$ can achieved by applying feedback to the trap control
electrode potentials, pushing the ion back to the pseudopotential minimum. 
Axial RF fields can result from asymmetric trap geometries like tapers or gaps
between electrodes. Careful simulation can minimize these fields in trap regions where it
may be problematic.

Micromotion can also result when there is a phase difference between the RF 
potentials applied to the trap RF electrodes.  This is discussed in 
Section~\vref{sec:dv16k:causesofphiRFneqzero}.  

\paragraph{effects of micromotion}
\label{sec:micromotion:effectsof}
Micromotion causes a first-order Doppler shift which can alter the fluorescence
spectrum of an ion.  This can result in suppressed fluorescence during state
readout, decresed Doppler cooling efficiency and ion heating from the Doppler
cooling laser beam. Micromotion may lead to ion heating if there is noise on the
RF potential at a motional difference frequency (see~\vref{sec:RFAMtheory})
\cite{devoe1989a}. Other effects include a second-order Doppler shift and an AC
Stark shift. Micromotion energy can be parametrically exchanged between secular
modes (see~\cite{bible}~~\cite{hunt2008a}~4.1.5). Minimization of excess micromotion is
therefore desirable.

\subsection{Micromotion detection}
\label{sec:micromotion:detection}
There are a variety of approaches to detect the presence of micromotion. One
technique is amplitude modulation of the trap RF drive at a secular difference
frequency $\Omega_{\rm RF}\pm\omega_i$, see Section~\vref{sec:RFAM}. This
approach can be used to null the 3D micromotion. Another approach is measurement
of the ion fluorescence spectrum vs the detuning of the Doppler cooling laser.
Finally, a time-resolved fluorescence measurement correlated with the ac drive
produces a signal indicative of micromotion~\cite{berkeland1998a}. The latter two
approaches only provide information about micromotion along the direction of the
cooling laser.
 
\paragraph{micromotion detection by Doppler fluorescence}
\label{sec:umDetectionByFluoresence}

A straight forward means of measuring ion micromotion is observing the
fluorescence spectrum from the ion as the frequency of the Doppler cooling laser
is varied. An ion with excess micromotion of amplitude $\pmb{u}'$ experiences a
modulated laser field. For a laser of frequency $\omega_{laser}$, amplitude
$\pmb{E}_0$ and wave vector $\pmb{k}$,

\begin{align*}
  \pmb{E}(t)=&\pmb{E}_0 \exp^{i\pmb{k\cdot u}-i\omega_{laser}t} \\
  &\cong \pmb{E}_0\exp^{i\pmb{k\cdot(u_0+u')}- i\omega_{laser}t}.
\end{align*}

It is assumed that the micromotion amplitude is small relative to the secular
amplitude. The modulation amplitude is

\begin{equation*}
  \pmb{k}\pmb{\cdot }\pmb{u}'= \beta  \cos \left(\Omega_{\rm RF} t\right)
\end{equation*}

where the modulation index $\beta $ is
\begin{equation*}
  \beta =\frac{1}{2}\sum _{i=x,y} k_iu_{0i}q_i.
\end{equation*}

Let $\rho_{22}$ be the excited state population of the cooling transition. Let
$E_0$ be the amplitude of the micromotion modulated laser electric
field at the ion. $\rho _{22}$ can be calculated the low intensity limit using
the steady-state solutions of the optical Bloch equation 

\begin{equation*}
  \rho _{22}=\left(\frac{\mu  E_0}{2\hbar }\right){}^2\sum _{n=-\infty }^{\infty
  } \frac{J_n{}^2(\beta )}{\left(\Delta +n \Omega
  _T\right){}^2+\left(\frac{1}{2}\gamma \right)^2},
\end{equation*}

where $\Delta=\omega_{atom}-\omega_{laser}$, $\gamma$ is the
transition line width and $J_n^2(\beta)$ are $n^{th}$ order Bessel functions
and $\mu$ is the transition electric dipole moment. $\gamma~\rho_{22}$ is the
ion fluorescence rate \cite{allen1975a,devoe1989a}.

\begin{SCfigure}
  \centering
  \includegraphics[width=0.5\textwidth]{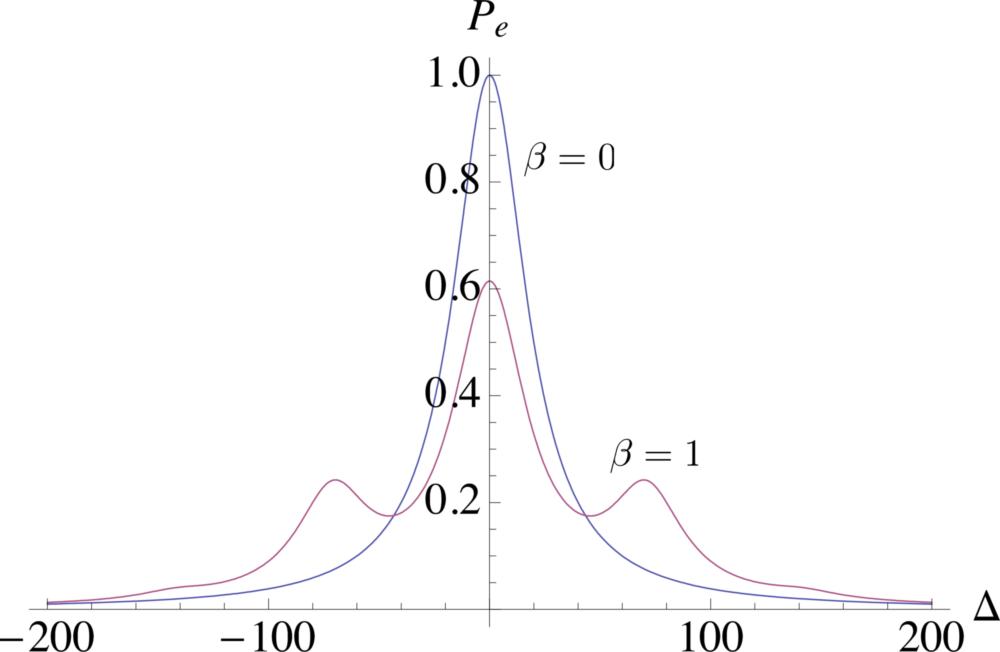}
  \caption[Plot of normalized ion fluorescence vs cooling laser detuning.]
  {Plot of normalized ion fluorescence vs cooling laser detuning.
  $\Omega_{\rm RF}/2\pi=71$ ~MHz, $\gamma =40$~MHz and $\beta ={0,1}$. Note
  that for $\beta =1$ there is a drop in carrier fluorescence and sidebands 
  appear at $\Delta =\pm n~\Omega_{\rm RF}$ for  $n\in
  {0,1,2,\ldots}$. If $\beta >0$ there is micromotion along the
  direction of the beam.}
  \label{fig:fluoresenceDuringRecooling}
\end{SCfigure}

Micromotion is considered significant when it impacts ion fluorescence. Consider
for example the case where the amplitude of the carrier and first micromotion
sideband are equal. That is, $\frac{J_0{}^2(\beta )}{J_1{}^2(\beta )}=1$ which
happens at $\beta =1.43$.

In practice ion micromotion is minimized by tuning the cooling laser so that
$\Delta =- n \Omega_{\rm RF}$ for $n\in \{1,2,3,\ldots\}$ and minimizing
the observed ion fluorescence. 

In a 3D Paul trap with an open geometry it is possible to direct three laser
beams at an ion at large relative angles thereby spanning 3-space. Each laser is
then sensitive to motion along it's direction of propagation and full 3D nulling
is possible.

Surface electrode traps pose special challenges for micromotion nulling due to
their planar geometry. In demonstrated microscale surface electrode ion traps the
ion lies about 40~$\mu$m above the surface and photons are collected from
overhead. In such traps it is not practical to use a vertical beam because light
scattered off the surface can enter the camera. Vertical micromotion therefore
can't be detected in this fashion and other techniques must be used~\cite{berkeland1998a}.

\begin{SCfigure}
  \centering
  \includegraphics[width=0.6\textwidth]
  {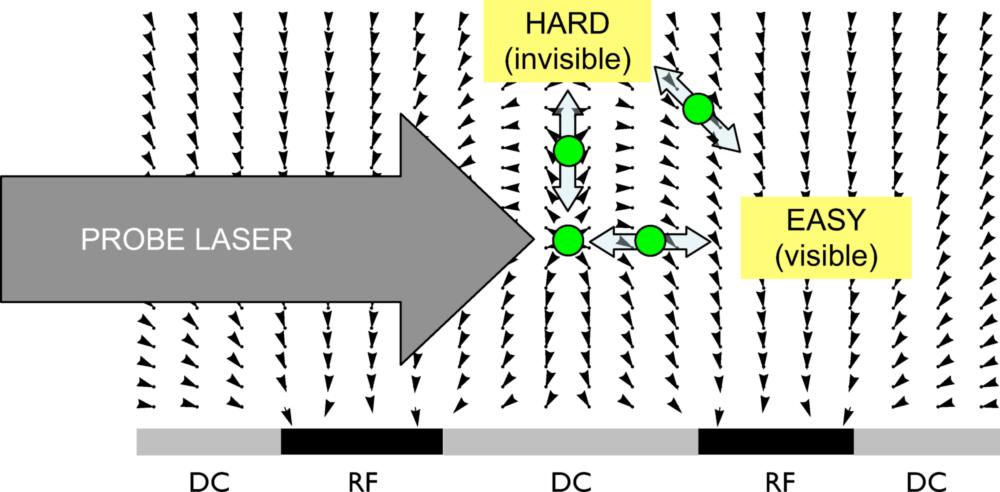}
  \caption[Drawing showing electric field vectors 
  for a symmetric 5-wire surface-electrode trap.]
  {Drawing showing electric field vectors for a symmetric
  5-wire surface-electrode trap. The ion's micromotion motion 
  lies along the field lines at the ion's position. The probe 
  laser propagating parallel to the surface is sensitive only to 
  the micromotion resulting from horizontal
  displacements of the ion from the RF null.}
  \label{fig:fiveWireTrapEfieldAndTrapAxes}
\end{SCfigure}

\clearpage
\section{Ion heating due to noisy RF potentials}
\label{sec:rfam}
\label{sec:RFAM}
\label{sec:RFAMtheory}
This section discusses a way in which noise on the trap RF potential can couple
to ion motion and cause heating $\dot{\bar{n}}$~\cite{bible}.  The heating rate
for a secular mode of frequency $\omega_z$ due to force noise at
$\omega_n=\omega_z$ is related to the force noise spectral density
$S_{F_n}(\omega)$ by,
\begin{align} 
  \dot{\bar{n}}(\omega_z)&=\frac{S_{F_n}(\omega_z)}
  {4 m \hbar  \omega_z}\label{eq:rfam:dndt},
\end{align}
in quanta/second \cite{turchette2000a} \footnote{
\begin{align}
  S_{F_n}(\omega)&\equiv 2 \int^{\infty}_{-\infty}\exp^{i \omega\tau} 
  \left<F(t) F(t+\tau)\right>d\tau
\end{align}
is the spectral density of force fluctuations in units of $N^2 Hz^{-1}$.}.

\subsection{Ion motion in an oscillating electric field}
\label{sec:rfam:inhomoEfieldForce}
This section starts with a derivation of the equation of motion for a charged
particle in a sinusoidally oscillating electric field.  For simplicity, all
calculations are one-dimensional.

\begin{SCfigure}[10]
  \centering
  \includegraphics[width=0.5\textwidth]
  {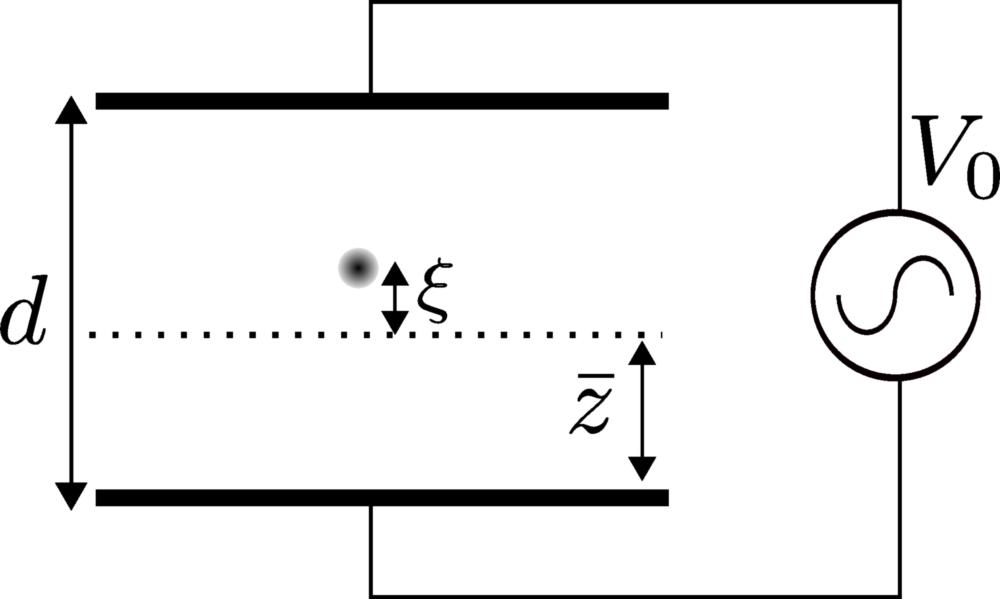}
  \caption[Ion in a homogeneous RF field as from a parallel 
  plate capacitor.]
  {Ion in a homogeneous RF electric field of amplitude $V_0/d$ from a parallel 
  plate capacitor.  
  The guiding center for the ion's motion is $\bar{z}$ with
  micromotion excursions of amplitude $\xi$.  Homogeneity of the electric 
  field requires that the capacitor lateral dimenstions be much larger than $d$.}
  \label{fig:rfam:ionInCap}
\end{SCfigure}

First, consider the motion of an ion of mass $m$ and charge $q$ in a homogeneous RF
electric field.  Such a field is found inside a parallel plate capacitor of
separation $d$ as in Figure~\vref{fig:rfam:ionInCap}.  Let $\bar{z}$ be the
initial position of the ion.  The ion's equation of motion is,
\begin{align}
	m \ddot{z} &= F_z = q E_0(\bar{z})\cos(\Omega_{\rm RF} t),
\end{align}
where $E_0(\bar{z})=V_0/d$. 

The resulting ion motion is,
\begin{align}
	z(t)&=\bar{z}+\xi(t)\\
	\xi(t)&=\xi_0 \cos(\Omega_{\rm RF}t) \text{, where }
	\xi_0=-\frac{q E_0(\bar{z})}{m \Omega^2_{\rm RF}},
\end{align}
where the ion's initial position $\bar{z}$ is assumed to be constant 
about which there is 
small  micromotion $\xi(t)$ (of amplitude $\xi_0$) at frequency $\Omega_{\rm RF}$.  
The time average of the force on the ion due to the homogeneous electric field is,
\begin{equation}
  \left< F_z(t) \right>=0.
\end{equation}

Suppose the electric field acquires a spatial dependence $E_0'(z)$, but with the
same value at $\bar{z}$: $E_0'(\bar{z})=E_0(\bar{z})$~\cite{dehmelt1967a}. 
I will show
that this inhomogeneity in the electric field gives a finite time-average force
in the direction of weaker RF field. 
This new field perturbs ion motion (now $z(t)=\bar{z}+\xi'(t)$) and the force
acting on it (now $F_z'(t)$),  
\begin{align}
	m \ddot{z} &= F_z' = q E_0'(z)\cos(\Omega_{\rm RF} t) \label{eq:rfam:inhomoEfield}\\
	\xi'(t)&=\xi_0 \cos(\Omega_{\rm RF}t) \text{, where }
	\xi_0=-\frac{q E_0'(\bar{z})}{m \Omega^2_{\rm RF}}.
\end{align}

Expanding the field about position $\bar{z}$,
\begin{align}
	E_0'(z)&=E_0'(\bar{z})+
		\frac{\partial E_0'(\bar{z})}{\partial z}(z-\bar{z})
		+ \ldots,
\end{align}  
where $z-\bar{z}=\xi'$.  Assume that the inhomogeneity is small,
\begin{align}
	\left| E_0'(\bar{z})\right| &>> 
	\left| \frac{\partial E_0'(\bar{z})}{\partial z}\right|\xi_0.
\end{align}
The time average of the new force $F'(z)$ is,
\begin{align}
	\left< F'(z) \right>&=
	\left< q E_0'(\bar{z})\cos(\Omega_{\rm RF}t)\right>+
	\left< q \frac{\partial E_0'(\bar{z})}{\partial z}\xi' \cos(\Omega_{\rm RF}t)\right>\\
		&\text{(to good approximation, assume $\xi'=\xi$~\cite{dehmelt1967a})}\\
	&\cong-\left< q \frac{\partial E_0'(\bar{z})}{\partial z}\xi_0 \cos^2(\Omega_{\rm RF} t)\right>\\
	&=-\frac{q}{2}\xi_0 \frac{\partial E_0'(\bar{z})}{\partial z}\\
	&=-\frac{q^2}{2 m \Omega^2_{\rm RF}} E_0'(\bar{z}) \frac{\partial }{\partial z}E_0'(\bar{z})\\
	&=-\frac{q^2}{4 m \Omega^2_{\rm RF}}\frac{\partial}{\partial z}E_0'^2(\bar{z}) 
		\label{eq:RFAM:inhomoEfieldForce}.
\end{align}
The force always points in the direction of weaker RF fields independent of the sign
of the ion charge.

\subsection{RF electric field noise and associated force noise}
\label{sec:rfam:EfieldNoise}
Suppose the electric field has a small noise term at $\Omega_{\rm RF}+\omega_n$.
The noise is a perturbation to
the drive at $\Omega_{\rm RF}$, 
\begin{equation}
  E(z,t) = E_0(z)  
  	\left[ 
  		\cos(\Omega_{\rm RF}t) +\alpha \cos(\Omega_{\rm RF}+\omega_n)t 
  	\right]\label{eq:rfam:efieldnoise:EwithNoise},
\end{equation}  
where $\Omega_{\rm RF}>>\omega_n$ and $\alpha<<1$. 
Following similar steps as those leading to 
Equation~\vref{eq:RFAM:inhomoEfieldForce} for the resulting force noise,
\begin{align}
	\left<F(z) \right>&=
	-\frac{q^2}{4 m \Omega^2_{\rm RF}}
		\frac{\partial}{\partial z}\left<E^2(z,t)\right>.
\end{align}

Calculating $\left<E^2(z,t)\right>$ over period $\tau=2\pi/\Omega_{\rm RF}$,
\begin{align}
	\left<E(z,t)^2 \right>
		&=E_0^2(z)  
  	\left<\left[ 
  		\cos(\Omega_{\rm RF}t) +\alpha \cos(\Omega_{\rm RF}+\omega_n)t 
  	\right]^2\right>\\
		&=E_0^2(z) \left<\cos^2(\Omega_{\rm RF}t)\right>
				+ E_0^2(z) \alpha^2 \left< \cos^2(\Omega_{\rm RF}+\omega_n)t\right>\\
				&+ 2 E_0^2(z) \alpha \left< \cos(\Omega_{\rm RF}+\omega_n)t \times \cos(\Omega_{\rm RF} t)\right>]\\
		&\cong\frac{1}{2}E_0^2(z)
			\left[
				1
				+ 2\alpha \left<\cos(2 \Omega_{\rm RF}+\omega_n)t\right>
				+ 2\alpha \left<\cos(\omega_n t)\right>
			\right]\\
		&=\frac{1}{2}E_0^2(z)\left[
			 1
			+ 2 \alpha \cos(\omega_n t)
		\right],
\end{align}
where a noise term in the force equation at $\omega_n$ has emerged from the cross term.  
The time-average force 
(Equation~\vref{eq:RFAM:inhomoEfieldForce}) over period $\tau$
depends on $\left<E_0(z,t)^2\right>$.
The force noise is then
\begin{align}
	 F_z(z) &=  \frac{q^2}{m\Omega^2_{\rm RF}}  
		E_0(z) \frac{\partial E_0(z)}{\partial z} \times \frac{E_n(\omega_n)}{E_0(z)},
\end{align}
where $E_n=\alpha E_0(z) \cos(\omega_n t)$.
Noise is often expressed in terms of a spectral density function.     
\begin{align}
	S_{F_n}(\omega_n) &= 
		\left[ 
			 \frac{q^2}{m\Omega^2_{\rm RF}}  
			E_0(z)\frac{\partial E_0(z)}{\partial z}
		\right]^2
		\times \frac{S_{E_n}(\omega_n)}{E_0^2(z)}\label{eq:rfam:noiseSFn}.	
\end{align}
Note that $S_{V_n}/V_0^2=S_{E_n}/E_0^2$ can be measured on a spectrum
analyzer.
\subsection{Motional heating due to RF electric field noise}
\noindent\paragraph{axial heating} 
\label{sec:rfam:axialheating} 
Consider an ion trapped in a potential with an
axial trapping frequency of $\omega_z$. Suppose there is a time-varying RF
field at $\Omega_{\rm RF}$ along the axial direction   
with noise at $\Omega_{\rm RF}+\omega_n$ as in 
Equation~\vref{eq:rfam:efieldnoise:EwithNoise}.
This gives rise to force noise which causes ion heating.  
Using Equation~\vref{eq:rfam:dndt},
\cite{turchette2000a},
\begin{align} 
  \dot{\bar{n}}(\omega_z)&=\frac{1}{4 m \hbar  \omega_z}\times 
  		\left[ 
			 \frac{q^2}{m\Omega^2_{\rm RF}}  
			E_0(z)\frac{\partial E_0(z)}{\partial z}
		\right]^2
		 \frac{S_{E_n}(\omega_z)}{E_0^2(z)}\label{eq:rfam:axialheating:rate},
\end{align}
in quanta/second.   

Therefore, noise at $\Omega_{\rm RF}+\omega_z$ can cause axial heating if there is 
a nonzero gradient to the RF electric field along the trap axis.

\noindent\paragraph{radial heating}
\label{sec:rfam:radialheating}
Consider now a specific spatial dependence of the electric field. In the radial
direction near the trap axis, the electric potential is given by
\begin{align}
	\phi(x,y,t)&\cong\frac{1}{2}V_0 
	\cos(\Omega_{\rm RF}t)\left(1+\frac{x^2-y^2}{R^2}\right)
	\label{eq:rfam:radialheating:pot},
\end{align}
where $R$ is determined by the electrode geometry.
The electric field in the x direction due to $\phi$ is,
\begin{align}
  E(x,t)&= -\frac{\partial }{\partial_{x}}\phi(x,t)\\
  	&=E_0(x)\times \cos(\Omega_{\rm RF}t) 
  	\text{, where } E_0(x)=-V_0 \frac{x}{R^2}\label{eq:rfam:radialheating:t1}.
\end{align} 
Assume
that the time variation of $E_0(x)$ is small over $\tau=2\pi/\Omega_{\rm RF}$.  That is,
$E_0(x,t) \cong E_0(x,t+\tau)$.  Noticing that the force on an ion from 
this field is harmonic with frequency $\omega_x$, $E_0(x)$ can be rewritten as follows
\footnote{
Equation~\vref{eq:rfam:radialheating:t1} is
of the form of \vref{eq:rfam:inhomoEfield} and so the time-average
force is given by \vref{eq:RFAM:inhomoEfieldForce}. For this potential, 
over period $\tau$,
\begin{align}
	\left<F_x(x)\right>&=
		-\frac{q^2}{4m\Omega^2_{\rm RF}}
		\frac{\partial}{\partial x}\left<E_0(x)^2\right>\\
	&=-\frac{q^2 V_0^2}{2 m \Omega_{\rm RF}^2 R^4} \times x\\
	&=-m \omega_x^2 \times x,
\end{align}
the equation of motion for harmonic motion of mass at frequency
$\omega_x=\frac{q V_0}{\sqrt{2} m \Omega_{\rm RF}R^2}$.  
For this potential there is slow
harmonic motion at $\omega_x$ with superimposed, low amplitude harmonic motion
at $\Omega_{\rm RF}$ (see Mathieu equation solution, 
Section~\vref{sec:mathieuEquation}).  
This holds only when $\omega_x<<\Omega_{\rm RF}$.
}
, 
\begin{align}
	E_0(x)=-\sqrt{2}\frac{m \Omega_{\rm RF}\omega_x}{q}x.
\end{align}

Suppose there is noise on the RF at $\Omega_{\rm RF}+\omega_n$ as in 
Equation~\vref{eq:rfam:efieldnoise:EwithNoise}.
This gives rise to force noise which causes ion heating.  Using Equations~\vref{eq:rfam:dndt} 
and~\vref{eq:rfam:noiseSFn},
\begin{align} 
  \dot{\bar{n}}(\omega_x)
		 &=\frac{1}{4 m \hbar  \omega_x}\times 
  		\left[ 
			 2 m \omega_x^2 x
		\right]^2
		 \frac{S_{E_n}(\omega_x)}{E_0^2(x)}
\end{align}
in quanta/second.   

Therefore, noise at $\Omega_{\rm RF}+\omega_x$ can cause radial heating if there is 
a stray electric field displaces the mean position of the ion to position $x$. 
If the ion lies at $x=0$ there is no heating due to this mechanism.


\section{Secular frequency measurement}
\label{sec:secularFreqMeasurement}

I used several techniques to measure the axial and radial potential curvatures
in my traps.  The corresponding experimental parameter is the secular
oscillation frequencies of an ion.   

\paragraph{endcap tickle}
\label{sec:secularFreqMeasurement:endcapTickle}
Due to the Doppler effect, an ion's fluorescence can change if a secular mode is
heated.  An electric field oscillating at an ion's secular mode frequency (and
with a component along that principle axis) causes heating. This causes a Doppler
shift of the Doppler cooling laser beam and can cause a change in ion
fluorescence. This resonance effect can be used to measure an ion's secular
frequencies.

For example, the axial oscillation frequency $\omega_{z}/2\pi$ was measured
experimentally by adding an oscillating potential at frequency $\omega$ to
an endcap electrode's static potential (e.g. any one of the electrodes marked
$V_1$ in Figure~\vref{ions:fig:LinEle}).
\begin{equation*}     
  	V_{\rm endcap}= V_{\rm dc}(1+\epsilon \cos(\omega t) ) 
\end{equation*} 
Ion fluorescence was observed while sweeping $\omega$. When
$\omega=\omega_{z}$,
the ion(s)'s motion is excited and the observed fluorescence decreases (see
Section~\vref{sec:micromotion:detection}).  Radial secular modes can be
measured by tickling using an electrode at the center of a trap zone.

\begin{figure}
  \centering \includegraphics[width=0.8\textwidth]
  {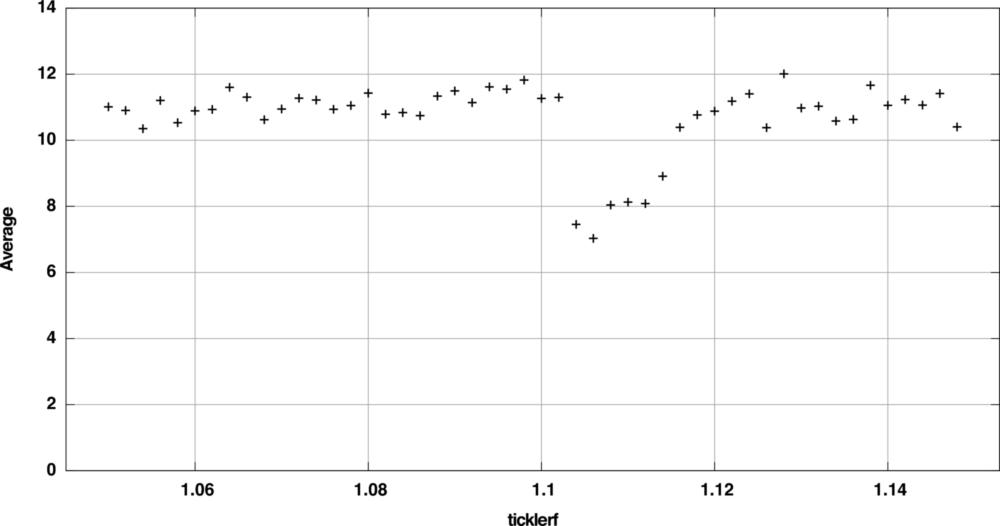} \caption[Plot showing
  the axial secular mode identified by the control electrode tickle technique.]
  {Plot showing the axial secular mode identified by the control electrode tickle
  technique.  Each point in the plot is an average of about 100 experiments. 
  Each experiment consisted of applying an oscillating potential to an endcap 
  control electrode and counting the number of photons observed on the PMT in a
  $400~\mu$s interval. The laser beam was continuously applied to the ion
  for these experiments.  The vertical axis is the average photon number.  A drop
  in ion fluorescence is observed at the secular frequency
  $\omega_z/2\pi\cong1.11$~MHz.}
  \label{fig:secularFreqMeasurement:EndcapTickleExampleSweep}
\end{figure}

Sample data are in Figure~\vref{fig:secularFreqMeasurement:EndcapTickleExampleSweep} for a
sweep of $\omega$ in search of the radial secular frequencies in trap dv16m.

\paragraph{RF amplitude modulation}
\label{sec:secularFreqMeasurement:RFAM}

Due to the Doppler effect, an ion's fluoresence can change if a secular mode is
heated. In this case intentional heating can be caused if the ion has excess
micromotion and the RF~drive is amplitude modulated (RF~AM) at a secular
difference frequencies $\Omega_{\rm RF}\pm n \omega $. This can be used to measure the
frequency of the secular modes.  This phenomena is discussed in more detail in
Section~\vref{sec:RFAMtheory}.

\begin{figure}
  \centering
  \includegraphics[width=0.8\textwidth]
  {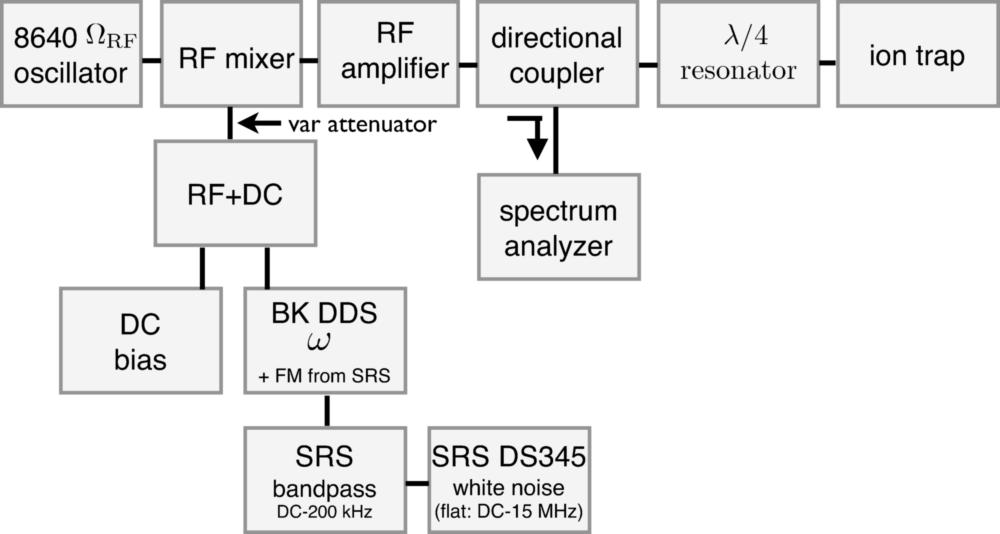}
  \caption[Schematic of RF electronics used for the RF AM tickle.]
  {Schematic of RF
  electronics used for the RF~AM tickle.  The 8640 is a Hewlett Packard cavity oscillator.  The BK DDS is a 
  BK Precision direct digital synthesizer.  SRS is Stanford Research Systems. }
  \label{fig:secularFreqMeasurement:RFAMcircuitDiagram}
\end{figure}

The experimental apparatus used to apply RF~AM is drawn in
Figure~\vref{fig:secularFreqMeasurement:RFAMcircuitDiagram}. The trap drive RF at
$\Omega_{\rm RF}$ was amplitude modulated by a spectrally broadened, dense noise
source centered at $\omega$. The AM was accomplished using an RF mixer fed by a
DC~biased noise source.  The noise source started as a continuous wave source
centered at $\omega_{motion}$ which was then frequency modulated by white noise
generated by a SRS~DS345.  A $100$~kHz FM~modulation depth was used. The DC~bias
was a heavily filtered ($\tau_{RC}\sim 100$~ms) power supply.

\begin{figure}
  \centering \includegraphics[width=0.8\textwidth]
  {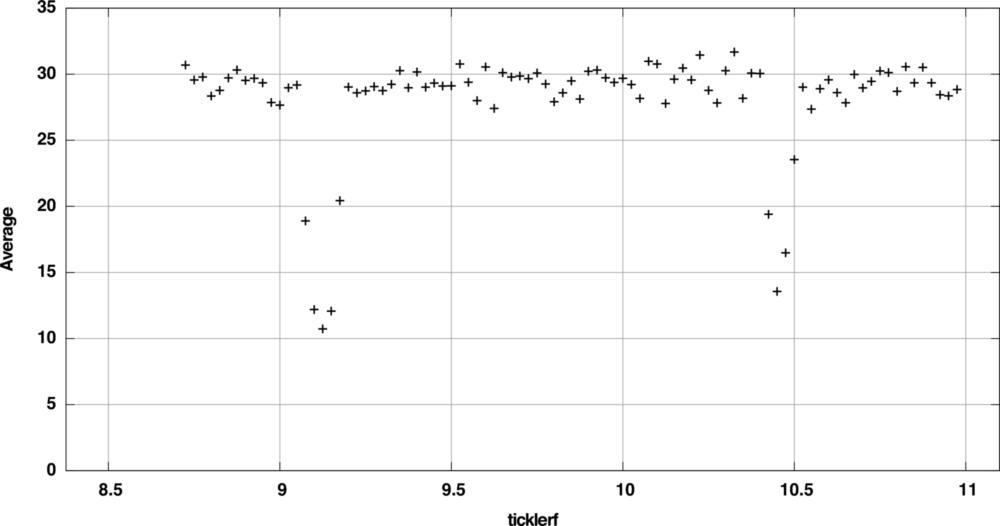} \caption[Plot showing radial
  secular modes identified by RF~AM.] {Plot showing radial secular modes
  identified by RF~AM.  Each point in the plot is an average of about 100
  experiments.  Each experiment consisted of applying AM to the trap drive RF 
  and counting the number of photons observed on
  the PMT in a $400~\mu$s interval. The laser beam was continuously applied
  to the ion for these experiments.   The vertical axis is the average photon
  number. A drop in ion fluorescence is observed at the two secular frequencies
  $\omega_x/2\pi\cong=9.2$~MHz and $\omega_y/2\pi\cong=10.45$~MHz.}
  \label{fig:secularFreqMeasurement:RFAMexampleSweep}
\end{figure}

Sample data are in Figure~\vref{fig:secularFreqMeasurement:RFAMexampleSweep} for
a sweep of $\omega$ in search of the axial secular frequency for a single ion in
trap dv16m.

Another, simpler appratus that also worked was use of a directional coupler to add noise power 
directly at $\Omega_{\rm RF}\pm n \omega $.

\paragraph{ion spacing}
\label{sec:secularFreqMeasurement:ionSpacing}
The axial secular frequency $\omega_{z}/2\pi$ can be estimated by observing
the inter ion spacing for a linear crystal.  The frequency follows 
assuming equilibrium between the ions' mutual Coulomb repulsion and the restoring force of the axial
trapping field.  This follows assuming equilibrium
between the ions' mutual Coulomb repulsion and the restoring force of the axial
trapping field. For a three ion crystal with inter-ion spacing $s_{3}$,
$\omega_{z}=\frac{q}{2}\left(\pi\epsilon_{0}ms^{3}\right)^{-1/2}$
where $s_{3}=(5/4)^{1/3}s$~\cite{bible}. In some traps this measurement was
made for a three-ion crystal to determine $\omega_{z}/2\pi$.

\section{RF cavity coupling}
\label{sec:RFelectronics}
\label{sec:cavityCoupling}

In ion trap systems we want to communicate maximum power to the
$\lambda$/4 resonator which generates the RF trapping potential. Under what
condition does this happen?

\paragraph{coupling to a series $RLC$ load}
\label{sec:rf:seriesRLC}
Consider the case where the load is a series $RLC$ oscillator. Such a model
can represent the distributed resistance, inductance and capacitance of a
transmission line or tank resonator. 


\begin{SCfigure}[5]
  \includegraphics[width=0.1\textwidth]{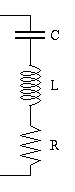}
  \caption[Series RLC circuit.]{Series RLC circuit.}
  \label{fig:seriesRLCcircuit}
\end{SCfigure}

When the RLC is not coupled to an external circuit it's said to be
\emph{unloaded} and its resonant frequency $\omega_0$ and unloaded $Q_0$ are related as
\begin{align*}
	\omega_0 & = 1/\sqrt{L C}\\
	Q_0 & = Z_L/R=\omega_0~L/R \\
	Q_0 & = Z_C/R = \frac{1}{\omega_0 C~R}
\end{align*}

In practice it may not be possible to directly measure these circuit parameters
if they are distributed as in a transmission line. Also,
we must be aware of how the measurement apparatus couples to the circuit and
modifies its quality factor and resonant frequency. 
Such an apparatus is said to \emph{load} the
resonator.

\begin{SCfigure}
  \centering 
  \includegraphics[width=0.6\textwidth]{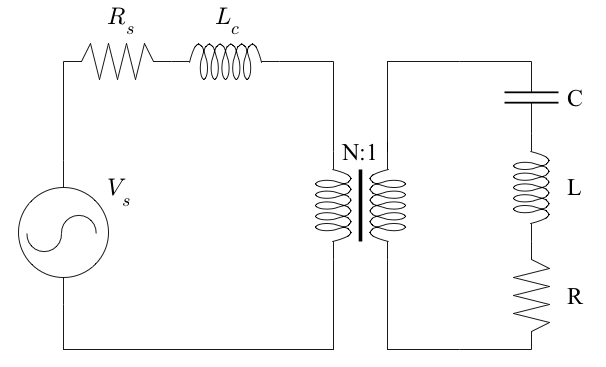} 
  \caption[Lumped circuit representation of cavity impedance matching.]
  {Lumped circuit representation of cavity 
  impedance matching. The cavity is modeled as a series RLC oscillator
  inductively coupled to a source with resistance $R_s$ by an ideal N-turn transformer
  with a series inductance $L_c$.}
  \label{fig:lumpCircuitCavityZmatching}
\end{SCfigure}

To couple maximum power to the resonator, its on resonance impedance $R$ must
equal the transformed source impedance. Figure~\vref{fig:lumpCircuitCavityZmatching}
is circuit showing a RLC inductively coupled to
a source with output impedance $R_s$ by a N-turn ideal transformer with series
inductance $L_c$. 

\begin{SCfigure}[20]
  \centering
  \includegraphics[width=0.5\textwidth]
  {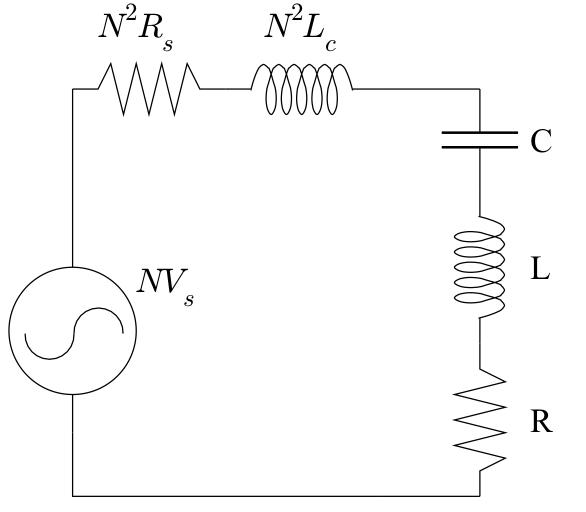} 
  \caption[Simplified lumped circuit representation of cavity impedance
  matching.] {Simplified lumped circuit representation of cavity impedance matching.}
	\label{fig:simplifiedLumpCircuitCavityZmatching}
\end{SCfigure}  

Following Liao, a simplified equivalent circuit is drawn in
Figure~\vref{fig:simplifiedLumpCircuitCavityZmatching} \cite{liao1980a}. As seen
by the RLC, the transformed source impedance is now reactive $N^2\left(i \omega L_c+R_s\right)$ and
its voltage is $N V_s$, where N is the transformer turns ratio. When the coupling
impedance $N^2L_c$ is small compared to L, the Q factor is shifted. 
\begin{align*}
Q_0&=\left.Z_L\right/R\\
&=\omega_0 L/R
\end{align*}
to
\begin{align*}
	Q_L&=\frac{\omega_0L}{R+N^2R_s}\\
	&=\frac{1}{1+N^2\left.R_s\right/R},
\end{align*}
where $Q_L$ is called the \emph{loaded Q}.  It is assume that the resonant
frequency shift due to the coupling is negligible. As in the coupled resistors
example a coupling coefficient may be defined.

\begin{equation*}
  \kappa=\frac{N^2R_s}{R}.
\end{equation*}

In terms of $\kappa $ the loaded $Q_L$ becomes

\begin{equation*}
	Q_L=\frac{Q_0}{1+\kappa}.
\end{equation*}

\paragraph{how to measure $Q_L$}
When the source and load impedances in an RF circuit are perfectly
matched ($\kappa = 1$), no power is reflected by the load and $Q_0=2~Q_L$.  

The loaded $Q_L$ may be determined by measuring the power reflected from the
cavity with a directional coupler,
\begin{equation*}
	Q_L= \frac{\omega _0}{\Delta \omega },
\end{equation*}
where $\Delta \omega$ is the full width at half max.  That is, $\Delta \omega
=\left|\omega_1-\omega_2\right|$ where $\omega_1$ and $\omega_2$ are the 
frequencies where the reflected power is half the maximum reflected power.

For more on microwave electronics see \cite{liao1980a, montgomery1948a,
kajfez1999a,terman1943a,purington1930a}.

\clearpage

\section{Time-resolved Doppler laser cooling .dc file}

\label{sec:recoolingDCfile}The detailed experiment evolution for
the ion heating rate measurement technique discussed in Section \ref{sec:recooling}is
specified in the computer file below. The syntax is defined in Chris
Langer's thesis \cite{langer2006a}. 
\singlespacing
\begin{verbatim}
/* 	
		- cool to Doppler limit
		- reheat for heattime_in_ms in the dark 
		- trigger multichannel scaler to integrate
		- detect ion for 50 us to see whether it is still around 
		
		MCS: pass length: 200
		MCS: dwell time: 50us		
*/

uvar interpn=10;
name det_pmt1_bd = pmt(1),bd;
name det_pmt2_bd = pmt(2),bd;
tvar heattime_in_ms = 11 dlb:11 dub:2000; //never below 3
fvar null=0 dlb:0 dub:1;

////////////////////////
///// START PULSES /////
////////////////////////

pulse bdd 10; //first pulse

//block to reset DDS 0 to 3 because they tend to conk out at random times
pulse wait minsetselt setsel(dds0);
pulse wait 10;
pulse wait minsetrft setfreq(313.0);
pulse wait 10;
pulse wait minsetselt setsel(dds1);
pulse wait 10;
pulse wait minsetrft setfreq(280);
pulse wait 10;

//initial cool of ion
pulse bdd 30e3;

//measurement into PMT 1 for monitoring ion. */
pulse det_pmt1_bd 400;
if(pmt(1) >= 3)
lbrace
//time delay for reheating (in ms)
pulse shutter_bd on;
pulse wait heattime_in_ms*1000-10001;
pulse shutter_bd off;
pulse wait 10000;

//trigger pulse for multichannel device
pulse exptrig 2.0; //Trigger, for programs/hardware that require it.
pulse mult_ch 2;

//extra multch_delay waittime after trigger
pulse wait 250;

//pulse with BD and RD to observe change in fluorescence on multichannel scaler
//dwell time = 50us; pass length=
pulse bd 20e3; //equal to MCS passlength*dwelltime
rbrace
if(pmt(1)<3)
lbrace
 pulse bdd 100e3;
rbrace

pulse bdd 10;
\end{verbatim}
\doublespacing
The following file is prepended to the .dc file above. We use a graphical
user interface to adjust experiment parameters like the potential
on control electrodes (as when nulling micromotion). This file specifies
the relationship between an experimental parameter and a computer
user interface element. 
\singlespacing
\begin{verbatim}
//Below are some macros for simplifying the waveform
//syntax. In particular it compartmentalizes explicitly 
//defined waveforms listed in series in a .dc file and 
//the automatic interpolation (if any) between them. 

@include initial_defs.dch //type definitions
@include stdnames.dch //eg name shutter =5;
@include IonProperties_inc.dc

//NOTE: Files that include this file need to be passed thru GNU cpp
//c:\cygwin\bin\cpp.exe -x c -P -CC filename.dc filename_cpp.dc

//define a two point waveform for joining 
wvfdef(interp) // Exactly two points only!
0,0,0,0,0,0,0,0,0,0,0,0,0,0,0,0,0,0,0,0,0,0,0,0;
0,0,0,0,0,0,0,0,0,0,0,0,0,0,0,0,0,0,0,0,0,0,0,0;
wvfend

//define some macros that properly implement waveforms
//A0 should be the first in a series of waveforms
//B for subsquent waveforms
//A1 should be the last in a series of waveforms
//
//PWVF is the waveform name
//PINTERPPTS is the number of points to use to interpolate
//PARGS are whatever other arguments should be passed in the body of wvf()
//POFFSETS is a list of offsets to append

//Variables to control from pulser_ui
pvar dwell=500e3 dlb:500e3 dub:1e7;
pvar load_cennear=ip_load_cennear dlb:-5 dub:5;
pvar load_cenfar=ip_load_cenfar dlb:-2 dub:2;
pvar load_middle=ip_load_middle dlb:-1 dub:1;
pvar load_endcaps=ip_load_endcaps dlb:-1 dub:1;
pvar load_endcapsd=ip_load_endcapsd dlb:-1 dub:1;
pvar load_shimAll=ip_load_shimAll dlb:-1 dub:1;
pvar load_shimRadial=0 dlb:-1 dub:1;

//include as Ion Properties
pvar m370_97=ip_m370_97 dlb:-1 dub:1;
pvar m370_35=ip_m370_35 dlb:-1 dub:1; //-0.9
pvar m370_middle=ip_m370_middle dlb:-1 dub:1;
pvar m370_endcaps=ip_m370_endcaps dlb:-2 dub:2;
pvar m370B_97=ip_m370B_97 dlb:-1 dub:1;
pvar m370B_35=ip_m370B_35 dlb:-1 dub:1; //-0.9
pvar m370B_middle=ip_m370B_middle dlb:-1 dub:1;
pvar m370B_endcaps=ip_m370B_endcaps dlb:-2 dub:2;
pvar m370B_95=0 dlb:-1 dub:1;
pvar m370B_4=0 dlb:-1 dub:1;
pvar m370B_36=0 dlb:-1 dub:1;
pvar m370B_33=0 dlb:-1 dub:1;
pvar m370C_97=ip_m370C_97 dlb:-1 dub:1;
pvar m370C_35=ip_m370C_35 dlb:-1 dub:1; //-0.9
pvar m370C_middle=ip_m370C_middle dlb:-1 dub:1;
pvar m370C_endcaps=ip_m370C_endcaps dlb:-2 dub:2;
pvar p370_79=ip_p370_79 dlb:-1 dub:1;
pvar p370_41=ip_p370_41 dlb:-1 dub:1; //-0.9
pvar p370_p79m41=0 dlb:-1 dub:1;
pvar p370_middle=ip_p370_middle dlb:-1 dub:1;
pvar p370_endcaps=ip_p370_endcaps dlb:-2 dub:2;
pvar p370_endLnear=0 dlb:-1 dub:1;
pvar p370_endLfar=0 dlb:-1 dub:1;
pvar p370_endRnear=0 dlb:-1 dub:1;
pvar p370_endRfar=0 dlb:-1 dub:1;
pvar p370_vertPush=0 dlb:-1 dub:1;
pvar p370_twist=0 dlb:-2 dub:2;
//differential shim on endcaps to translate ion axially
pvar p370_endcapsDif=0 dlb:-1 dub:1; 
pvar shim_dac1=0 dlb:-5 dub:5;
\end{verbatim}
\doublespacing
\clearpage
\section{Microfabrication Techniques}
\label{sec:fab} 
\label{sec:microfabTechniques}A variety of microfabrication tools were used to
build the ion traps and cantilevers discussed in this thesis. The techniques of
patterning and machining micro scale devices in materials like silicon fall under
the umbrella of micro-electromechanical systems (MEMS). The NIST cleanroom is
operated by the NIST Cryoelectronics Division which focuses on Josephson junction
devices. Therefore, many common MEMS techniques were not routine in our facility
(circa~2002) and had to be developed. They are discussed in this section.
Techniques commonly exercised in the NIST cleanroom (and many other facilities)
like photolithography are not discussed. The NIST
\href{http://grumpy.boulder.nist.gov/qffwiki}{Cryoelectronics} and
\href{htts://847wiki2.bw.nist.gov/}{Ion Storage} wikis are good
resources for such details for NIST staff.

This section is written with the physics AMO community as the target
audience but it is not meant to be comprehensive introduction to MEMS.
For a good introduction to MEMS see Liu's Foundations of MEMS \cite{liu2006a}.
A handy reference on a wide variety of MEMS topics is Mandous Fundamentals
of Microfabrication \cite{madou2002a}.
 
\subsection{Silicon deep etch}

\label{sec:DRIE} Plasma etching of silicon permits high aspect ratio
features with good repeatability and excellent mask selectivity. The
NIST etcher utilizes a variant of the BOSCH etch process optimized
for silicon deep etch ($>50\mu m$) \cite{larmer1992a}. This process
forms high aspect ratio features in silicon by interleaving etch and
surface passivation steps. The etch step is a chemically active RF
plasma ($SF_{6}$ and $O_{2}$) inductively coupled to the wafer surface.
The charged component of the plasma is accelerated normal to the surface,
enhancing its etch rate in the vertical direction. The passivation
step ($C_{4}F_{8}$) coats all exposed surfaces including sidewalls
with fluorocarbon polymer. Etch and passivation cycles (typically
12 and 8 seconds respectively) are balanced so that sidewalls are
protected from over/under etching as material is removed thru the
full wafer thickness. The cycled etching causes the sidewalls to have
a microscopic scalloped appearance with an amplitude of 100-500~nm
and a period of 200-1000~nm.

\begin{figure}
  \includegraphics[width=1\textwidth]{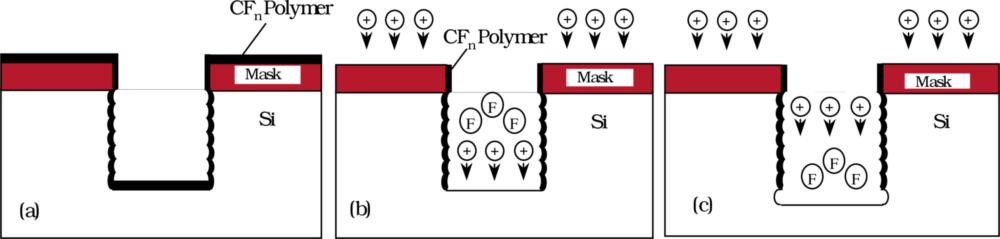}
  \caption[An illustration of the Bosch silicon deep etch mechanism.]
  {An illustration of the Bosch silicon deep etch mechanism. a) During
  passivation a fluorocarbon polymer uniformly coats all exposed surfaces.
  b) During etch the surfaces oriented normal to the plasma flux (vertical)
  degrade fastest. c) The sidewalls are protected. The figure is from
  \cite{mcauley2001a}.}
  \label{fig:DIRE:deepEtchCycleIllustration}
\end{figure}

The details of the etch chemistry for the STS etcher are proprietary.
However, the general topic of fluorine reactive ion etch is discussed
in the literature \cite{flamm1990a,madou2002a}. For anisotropic processes
common etch gas combinations include $CF_{4}/O_{2}$ and $SF_{6}/O_{2}$.
Other chemistries are possible but these are popular in commercial
systems due to the convenience and safety of the reactants and their
selectivity of silicon over photoresist and silicon oxide.

\begin{SCfigure}
  \centering
  \includegraphics[width=0.60\textwidth]{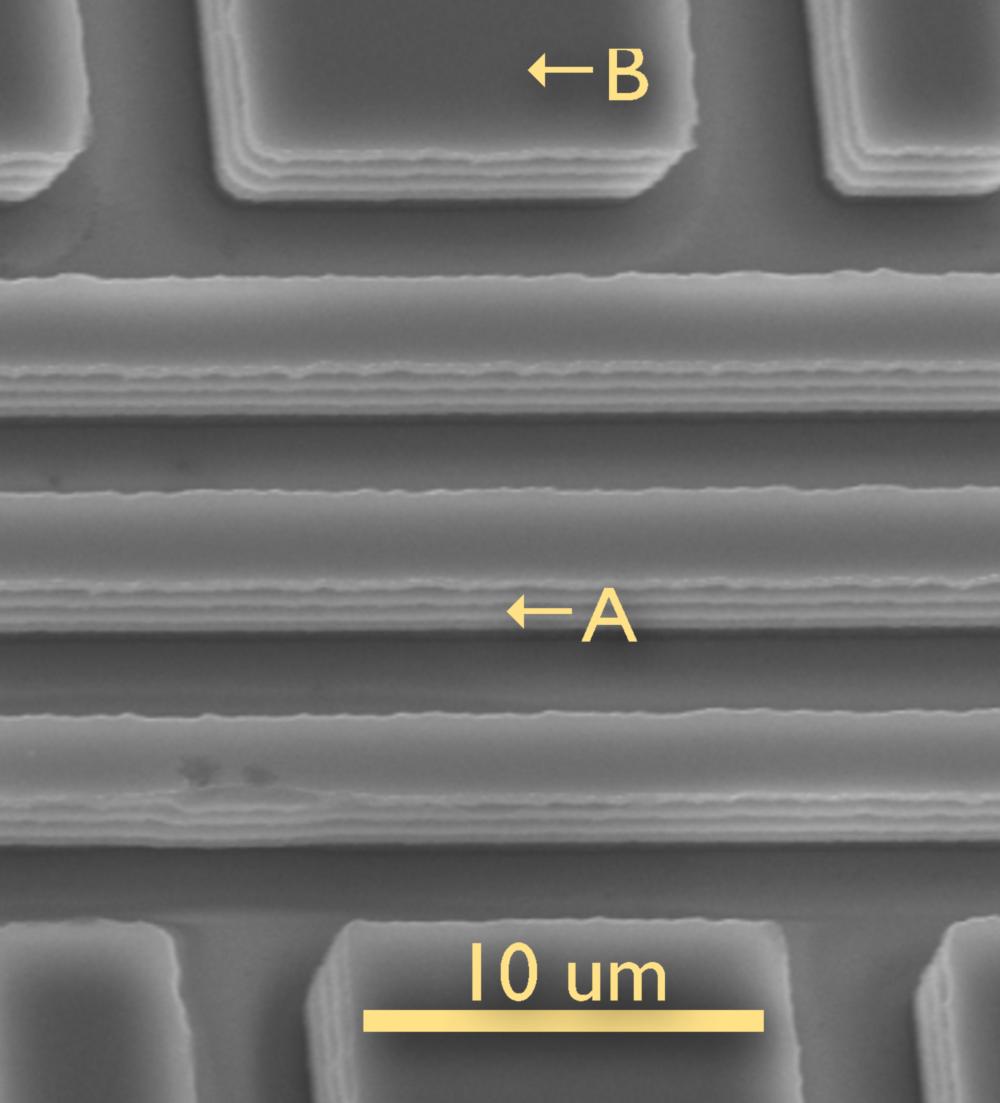}
  \caption[Micrograph showing scalloping along the edges of a 3~$\mu m$ deep silicon
  etch performed on the NIST STS etcher.]
  {Micrograph showing scalloping along the edges of a $3~\mu m$ deep silicon
  etch performed on the NIST STS etcher. Arrow A points to sidewall
  scalloping due to process cycling. Arrow B points to the top wafer
  surface that was protected during etching by photoresist (removed
  for this micrograph). This etch was performed using recipe SPECB which
  is optimized for etches $>200\mu m$ deep. Somewhat smoother sidewalls
  can be obtained using recipe SPECA at the price of a slower etch rate.}
  \label{fig:DIRE:etchScallopingSEM}
\end{SCfigure}

Low pressure ($<$10~mTorr)\footnote{1~Torr is 133~Pa} and high ion density
($>10^{11}~\text{cm}^{-3}$) are best for silicon deep etch. In the STS the plasma
is confined by an ac axial magnetic field at $\Omega=13.56$~MHz that causes an
azimuthal electric field (from $dB/dt = \nabla\times B$), confining the plasma
current (see Figure~\vref{fig:DRIE:STSschematic} and~\cite{bhardwaj1995a}). Phase
locked to axial field source is an ac bias voltage at $\Omega$ applied to the
platen (wafer chuck) which independently controls the ion energy. A frequency of
13.56~MHz selected because of sanction as an industrial, scientific, and medical
band by the \href{http://www.fcc.gov}{FCC} \cite{madou2002a}.  See the
\href{http://www.ntia.doc.gov}{National Telecommunications and Information
Administration's} \href{http://www.ntia.doc.gov/osmhome/allochrt.html}{United
States Frequency Allocation Chart}.


\begin{SCfigure}
  \centering
  \includegraphics[width=0.50\textwidth]{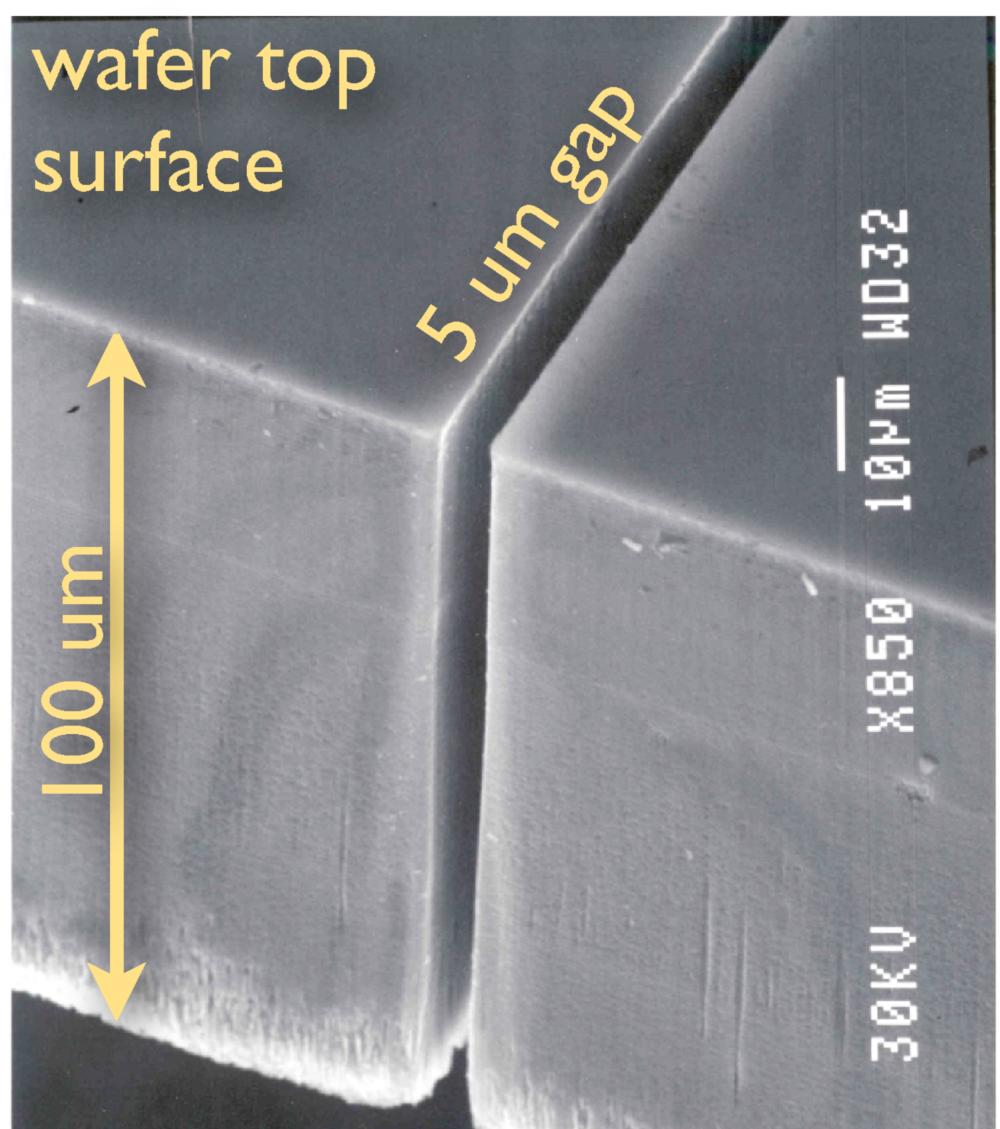}
  \caption[Micrograph of silicon etched by the NIST STS DRIE.]
  {Micrograph of silicon etched by the NIST STS DRIE.}
  \label{fig:DRIE:deepEtchedSiliconSampleSEM}
\end{SCfigure}

The primary etch mechanism for $SF_{6}/O_{2}$ is due to chemical
reactions between neutral atomic fluorine formed in the plasma and
silicon at the surface. The fluorine is adsorbed by the surface before
or during the chemical reaction. Ion bombardment can assist with the
adsorption, reaction and ultimate evolution of $SiF_{4}$ as a gaseous
product. The details are complex but occur roughly stepwise as follows.

\begin{eqnarray*}
	\mbox{etchant formation} & e+F_{2}\mbox{to}2F+e & \mbox{(in plasma)}\\
	\mbox{chemical reaction } & Si+4F\mbox{ to }SiF_{4} & \mbox{(at surface)}
\end{eqnarray*}

Oxygen acts to degrade the fluorocarbon polymer $CF_{x}$ to form
gaseous products.
\begin{equation*}
	O_{2},O+CF_{x}\mbox{ to }COF_{2}\mbox{, }CO\mbox{, }CO_{2}+F
\end{equation*}
The gaseous products are pumped away and do not redeposit on the process
wafer. Following a deep etch there may be residual fluorocarbon polymer
on the wafer surface. Use of a short oxygen plasma cycle (DESCUM recipe,
see below) appears to remove it.


As one of the first NIST users of the STS etcher (2002) I developed
several nonstandard techniques that were new at NIST. These are documented
in the following sections.

\begin{figure}
  \includegraphics[width=1\textwidth]{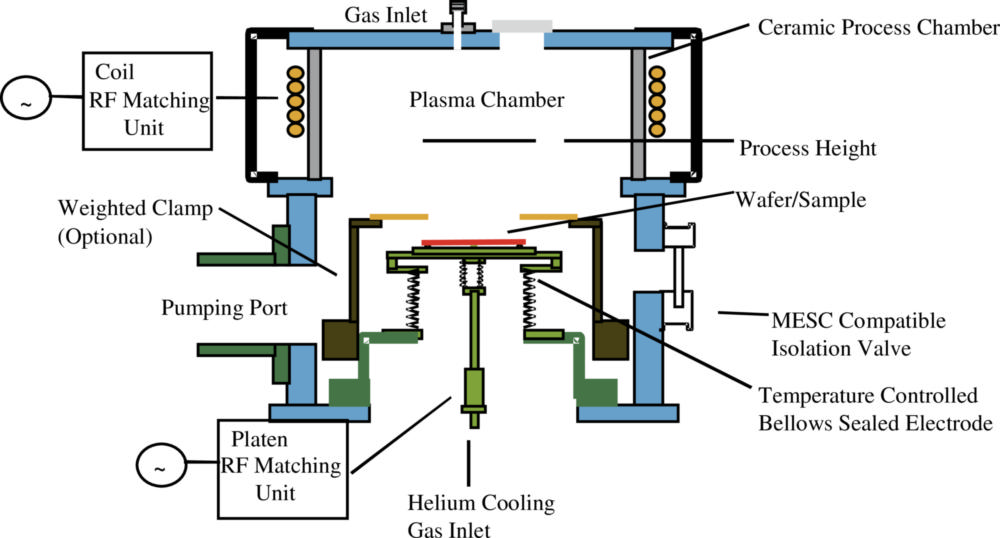}
  \caption[STS silicon deep etch tool schematic.]
  {STS silicon deep etch tool schematic. The figure is from \cite{mcauley2001a}.}
  \label{fig:DRIE:STSschematic}
\end{figure}

\begin{table}
  \centering
  \begin{tabular}{l|lp{3cm}ll}
    \textbf{Recipe} & \rotatebox{45}{\textbf{Step}} & \rotatebox{45}{\textbf{Gas
    Flow}} & \rotatebox{45}{\textbf{Coil Power}} &
    \rotatebox{45}{\textbf{Platen Power}}\tabularnewline
    \hline
    SPECA  & etch & \raggedright 130 sccm $SF_{6}$\\ and 13 sccm $O_{2}$ & 600 W &
    600 W\tabularnewline
    & passivation & 120 sccm $C_{4}F_{8}$ & 12 W & 0W\tabularnewline
    SPECB & etch & \raggedright 130 sccm $SF_{6}$\\ and 13 sccm $O_{2}$ & 600 W & 12
    W\tabularnewline
    & passivation & 85 sccm $F_{4}F_{8}$ & 600 W & 0 W\tabularnewline
    DESCUM & etch & 6 sccm $SF_{6}$ & 0 W  & 0 W\tabularnewline
  \end{tabular}
  \caption[STS etcher recipe details.]{STS etcher recipe details. In all several dozen parameters fully
  specify this recipe. They are proprietary to STS but provided to owners
  of STS etch systems. SPECA: This recipe produces smoother sidewalls
  than SPECB and is appropriate for etches $<$50~$\mu m$ in depth. The etch
  rate is about 1.5~$\mu m$/min. SPECB: This recipe is appropriate for etches
  $>$200~$\mu m$ in depth. The etch rate is about 3~$\mu m$/min. sccm: standard
  cubic centimeter per minute, but there is little agreement as to what
  constitutes standard temperature and pressure (STP) (see Wikipedia)}
  \label{tab:DRIE:stsEtchRecipes}
\end{table}


\subsubsection{deep etch ($>50\mu m$)}
\label{sec:DRIE:deepEtchPR}STS etch patterns are defined by photolithography of a
mask layer. Typically, this mask is 1 $\mu m$ thick Shipley 510L-A photoresist.
The selectivity of a mask is the rate at which it is degraded relative to the
desired etch process. For the 3~$\mu m$/min silicon etch recipe SPECB, 510L-A has
a selectivity of 1:35. Very deep etches require higher selectivity and a thicker
mask. Alternatives include silicon oxide (1:100) and silicon nitride (1:75-125),
but using standard growth techniques low pressure chemical vapor deposition
(LPCVD\footnote{LPCVD: low pressure chemical vapor deposition}) and thermal
oxidation (\cite{wiki2008thermox, tystar2008}) films thicker than 2 $\mu m$ are
impractical. Moreover, dielectric mask patterning requires additional etch steps.

It was found that for deep etches ($>50\mu m$) Shipley SPR-220 photoresist
applied in $3\,\mu m$ or $7\,\mu m$ films and cured at elevated
temperatures provides greater selectivity to the STS etch (1:75).

\paragraph{$3~\mu m$ SPR-220 recipe ($<200~\mu m$ deep etches)}
\begin{compactenum}
  \item clean: on spinner rinse with acetone then isopropanol
  \item prime with HMDS then spin (4 krpm, 35 sec)
  \footnote{Hexamethyldisilazane (HMDS) is a photoresist-silicon adhesion
  promoter \cite{wiki2008hmds}.}
  \item apply photoresist: SPR-220 $3.0~\mu m$ on wafer with dropper
  \item spin (3.0 krpm 40 sec, yields 3 $\mu m$)
  \item prebake: 90$^{\circ}$ C 60 sec, 115$^{\circ}$ C 60 sec (hot plate)
  \item expose photoresist: 2:00 min
  \item NIST Karl Suss exposure tool parameters: no UV300, soda lime mask,
  CH2, 300 mW
  \item hold wafer in air for 10 min
  \item post exposure bake: 90 sec on 115$^{\circ}$ C hot plate
  \item develop with Solitec: 40-60 sec, 45 rpm (also use low rpm for spin
  dry)
  \item inspect
  \item harden: 110$^{\circ}$ C 30 min (Blue M oven)
\end{compactenum}
\paragraph{$7~\mu m$ SPR-220 recipe ($<$500 $\mu m$ deep etches)}
\begin{compactenum}
  \item clean: on spinner rinse with acetone then isopropanol
  \item prime with HMDS then spin (4 krpm, 35 sec)
  \item apply photoresist: SPR-220 7.0~$\mu m$ on wafer with dropper
  \item spin (3.5 krpm 35sec -$>$ yields 7.0 $\mu m$)
  \item prebake: 90$^{\circ}$ C 100 sec, 110$^{\circ}$ C 100 sec (hot plate)
  \item expose photoresist: 2:30 min
  \item NIST Karl Suss exposure tool parameters: no UV300, soda lime mask,
  CH2, 300 mW
  \item hold wafer in air for 30 min
  \item develop with Solitec: 120-180 sec, 45 rpm (also use low rpm for spin
  dry)
  \item inspect
  \item bake: 90 min at $80{}^{\circ}-90^{\circ}$ C (BlueM oven)
  \item bake: 30 min at $110^{\circ}$ C (BlueM oven; skip if not doing deep
  etch or if rounded edges are not desired)
\end{compactenum}
\paragraph{Process tips for SPR-220.}
\begin{compactitem}
  \item Store unused photoresist in a refrigerator.
  \item Only spin photoresist that has warmed to room temperature.
  \item Note that in SPR-200 the chemical process initiated by exposure to
  UV light requires water to come to completion (see product data sheet).
  This is the reason to leave the wafer exposed to air at room temperature
  prior to the post exposure bake.
\end{compactitem}

\subsubsection{thin wafer etching}

\label{sec:DRIE:deepEtchThin} Etching a wafer $<$ 500 $\mu m$ thick
requires care. Inside the STS etcher the process wafer sits atop a
pedestal in the plasma stream (see Figure~\vref{fig:DRIE:STSschematic}).
Chilled helium gas flowed across the backside of the wafer to control
its temperature during etching. 
The helium gas is excluded from the process
environment by a Viton o-ring seal on a stage beneath the wafer. The seal integrity is enhanced
by application of force to the wafer circumference. The force is communicated
by several thin ceramic fingers situated around the wafers circumference
(the Weighted Plate in Figure~\vref{fig:DRIE:STSschematic}). The fingers
are attached to a 5 kg weight. The spatial nonuniformity of this force
atop the elastic o-ring can cause the wafer to flex and crack.

\begin{SCfigure}
  \includegraphics[width=0.50\textwidth]{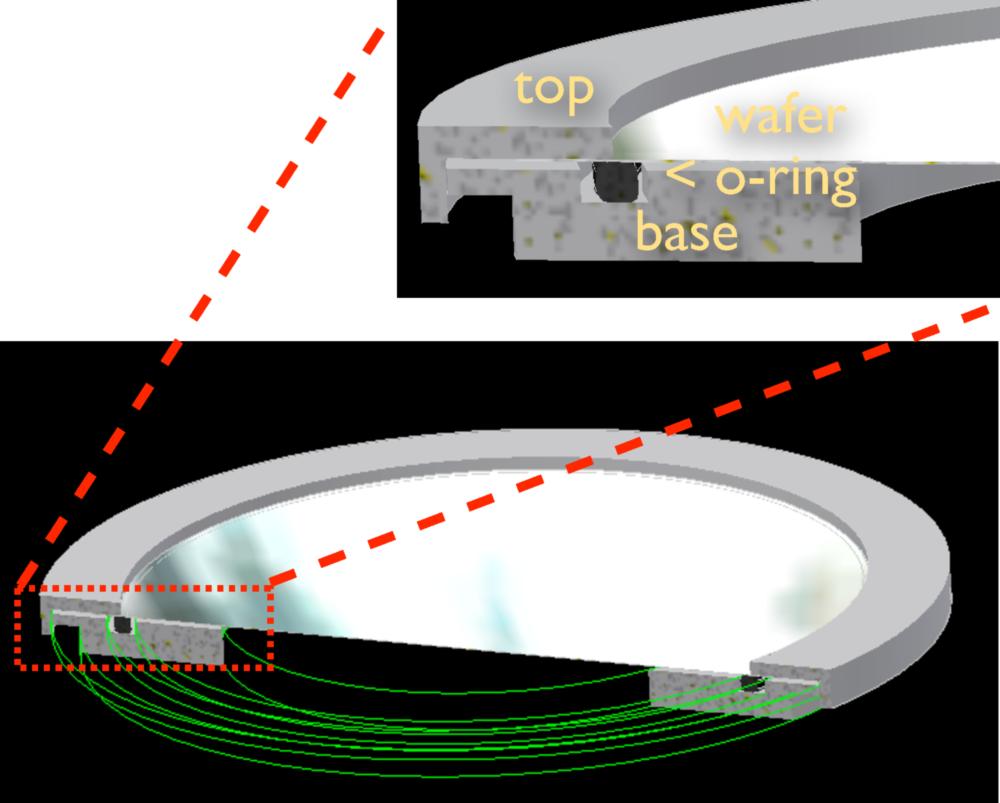}
  \caption[STS wafer carrier schematic.]
  {STS wafer carrier schematic. This aluminum wafer carrier housed the
  silicon wafer during processing. An o-ring between the top and base
  sealed the wafer to the carrier. The STS stage o-ring sealed to the
  carrier base to the STS etchers chilled stage. See Figure~\vref{fig:DRIE:STSschematic}.}
  \label{fig:DRIE:STSwaferCarrier}
\end{SCfigure}

A special wafer carrier was built to more uniformly distribute the
fingers force around the wafer circumference. Use of this carrier
prevents wafer cracking. The carrier is made of aluminum, the only
metal compatible with the silicon STS etch process. It consists of
three parts: a top aluminum retaining ring, a Viton o-ring, and a
sturdy aluminum base. See Figure~\vref{fig:DRIE:STSwaferCarrier} for a
schematic of the carrier.

\subsubsection{thru wafer etching}

\label{sec:DRIE:thruWaferAdvice}Etches that penetrate the full wafer thickness
require use of a backing wafer to prevent helium leakage, process
contamination and loss of wafer backside chilling. The backing wafer
is thermally contacted to the process wafer by a thin adhesion layer
of Crystalbond 509 wax (Structure Probe Inc., www.2spi.com). It flows
easily at 125$^{\circ}$ C and is soluble in acetone. Sapphire was
used as a backing wafer because it is transparent in the visible and
permits inspection of the wafer back side to check for proper etch
termination. With a sapphire backing wafer and a thin layer of wax
thermal conductivity is reduced. Therefore decrease the helium backside
chiller temperature to around $10^{\circ}$ C to keep the process
wafer cool enough.

STS suggests use of Cool Grease (AI Technology, www.aitechnology.com)
as a thermal contacting layer to backing wafers. Cool Grease is a
suspension of alumina powder. I had trouble completely removing it
after processing and do not recommend it.

If using a transparent backing wafer like sapphire the STS load lock
may not see your wafer. It detects wafers using a LED mounted beneath
the wafer loading arm. Attach opaque tape to the wafer backside opposite
the wafer flat. Take care to not compromise the o-ring seal region.
Remove the tape before use of a solvent which can transport adhesive
to the process surface.

\begin{SCfigure}
  \includegraphics[width=0.60\textwidth]{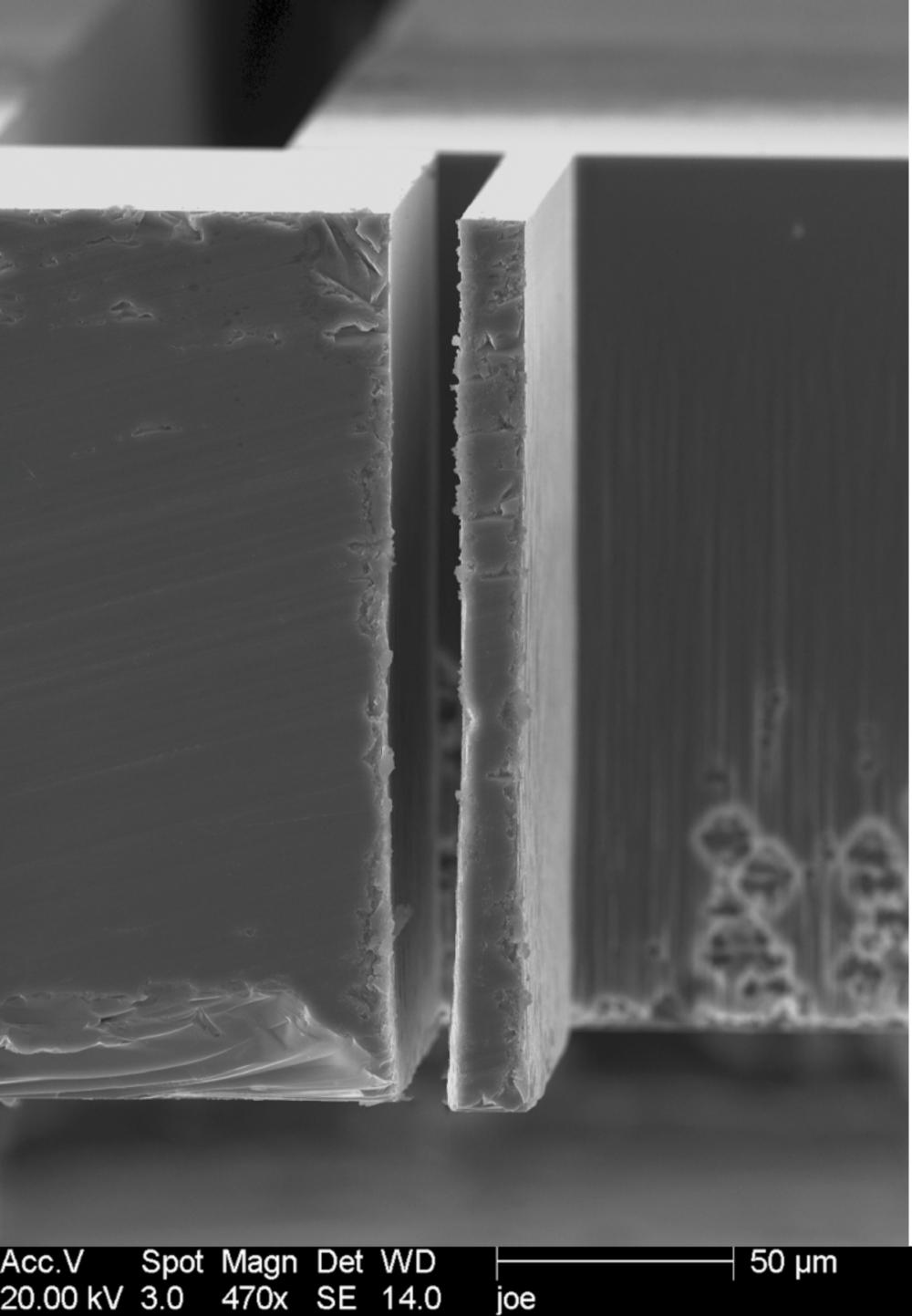}
  \caption[SEM showing the etch profile for a deep etch.]
  {Scanning electron micrograph showing the etch profile for a 10~$\mu m$
  wide, 200~$\mu m$ deep trench cut using recipe SPECB.The foreground surfaces
  were exposed by a dicing saw cut and are not representative of the
  deep etching process surface quality. Observe that the 10~$\mu m$
  wide trench tapers to about 8~$\mu m$ wide near the back side. In
  the photograph the photoresist (removed for the micrograph) protected
  the top surface from etch. A backing wafer (also removed) was used
  for this thru-wafer etch.}
  \label{fig:DRIE:deepEtchProfile}
\end{SCfigure}

\subsubsection{etch termination on dielectric}
\label{sec:DRIE:dielectric}

\begin{SCfigure}
  \includegraphics[width=0.50\textwidth]{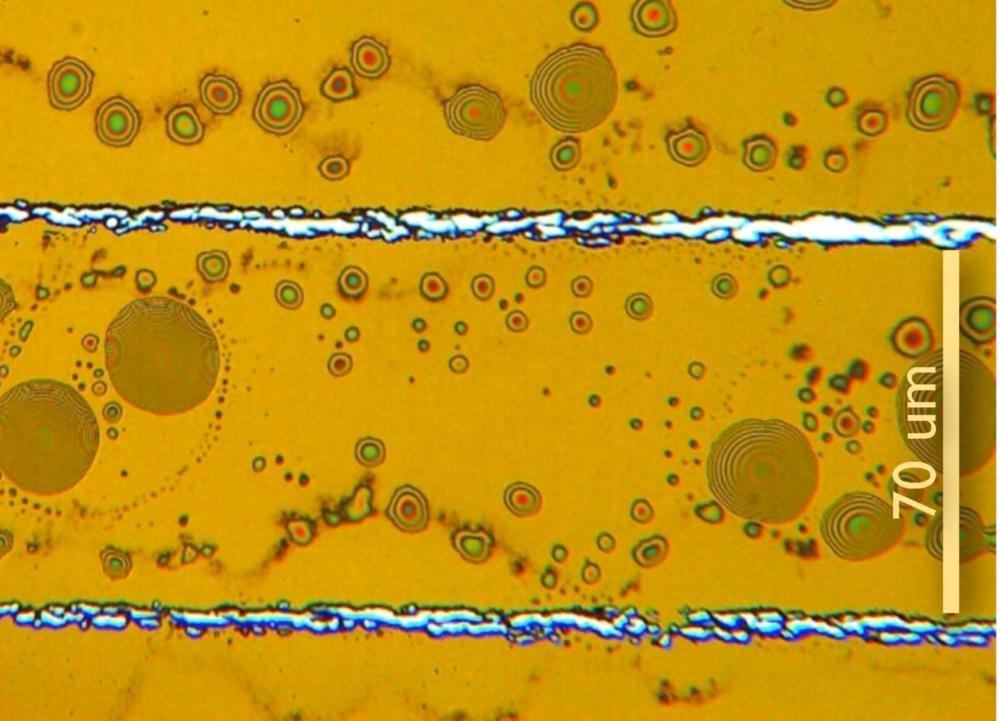}
  \caption[Optical micrograph showing an etch improperly terminated on a dielectric
  layer.]
  {Optical micrograph showing an etch improperly terminated on a dielectric
  layer. Two $10~\mu m$ wide trenches (separated by $70~\mu m$)
  were etched thru a silicon wafer using recipe SPECB. Prior to etching
  the silicon wafer was bonded to a sapphire backing wafer with wax.
  The trenches are backlit and the view is looking through the sapphire
  at the silicon-wax interface. The ragged appearance of the bottom
  side of the trenches is due to dielectric charging. Remedy: use Recipe~4 for
  the final $10~\mu m$ of the etch.}
  \label{fig:DRIE:badEtchToDielectric}
\end{SCfigure}

Special care is required when an etch is terminated by a dielectric layer. The
problem is that charges from the plasma stick to the dielectric and cause
vertical or lateral deflection of the plasma depending on the geometry of the
etch feature. This results in incomplete etching (swiss-cheese pattern) or severe
undercutting on the wafer back side respectively. A special etcher configuration
(STS SOI Upgrade Kit) and STS supplied recipe (eg NIST standard Recipe~4;
NIST readers, see~\href{https://847wiki2.bw.nist.gov}{Ion Storage Wiki})
interleave periodic low frequency (380~kHz) RF pulses that sweep away stray charges between etch/passivation cycles. To use this recipe,
first process the wafer as usual stopping 10~$\mu m$ before encountering the
dielectric layer. Then use Recipe~4. Note that the etch rate of bulk silicon is
reduced below that of SPECB for this recipe.

\begin{SCfigure}
  \includegraphics[width=0.50\textwidth]{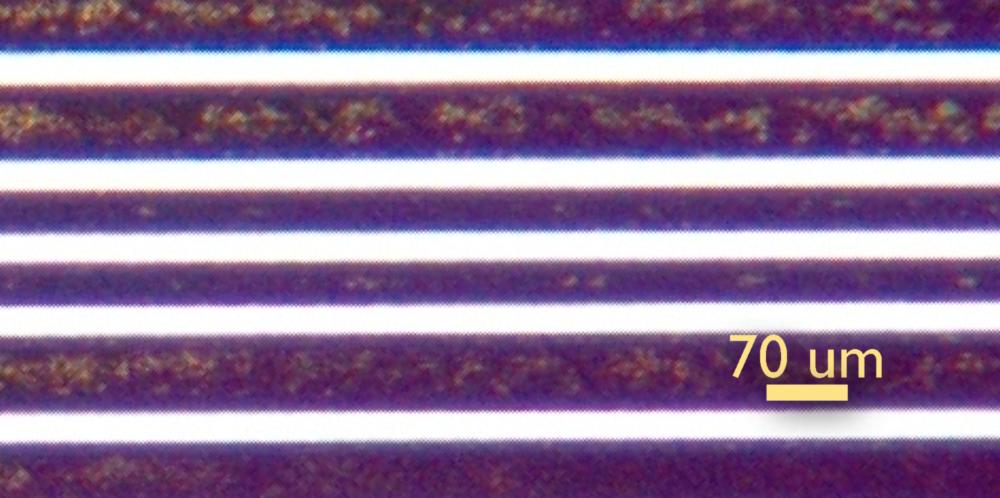}
  \caption[Optical micrograph showing an etch properly terminated on a dielectric
  layer.]{Optical micrograph showing an etch properly terminated on a dielectric
  layer.}
  \label{fig:DRIE:goodEtchToDielectric}
\end{SCfigure}

Common wafers with a dielectric etch termination layer include the
following.

\begin{compactitem}
  \item silicon on oxide (SOI) wafers
  \item silicon wafers coated with thin film of silicon oxide or silicon nitride
  \item silicon wafers adhered to a backing wafer by wax or Cool Grease
\end{compactitem}

\subsubsection{deep etch pitfalls}
\label{sec:DRIE:pitfalls} 

This section consists of micrographs illustrating
several of the most common fabrication errors associated with deep
silicon etch. Remediary actions are suggested for each.

\begin{figure}
  \centering
  \includegraphics[width=0.70\textwidth]{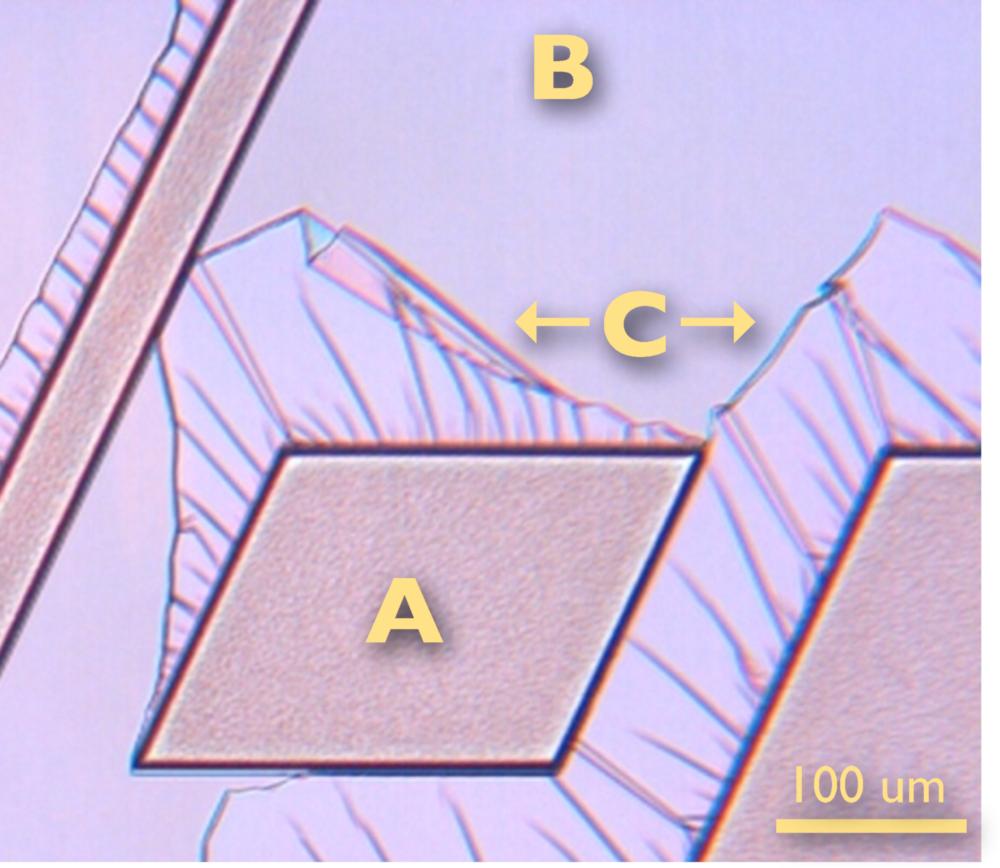}
  \caption[Optical micrograph of baked-on photoresist.]
  {Optical micrograph of baked-on photoresist.If the process wafer temperature
  becomes too high during deep etching photoresist will burn on to the
  wafer. While acetone is an excellent solvent for photoresist, when
  burned is insoluble in acetone. The micrograph shows deep etched silicon
  (A), the unetched silicon surface (B), covered with photoresist during
  etching) and remnant snake skin like sheets photoresist after washing
  in acetone (C). Use of the ultrasonic bath with acetone almost always
  removes baked-on photoresist but can damage delicate structures. Shipley
  SVC-150 photoresist stripper at 80$^{\circ}$ C can often remove it
  without use of ultrasound as can piranha etch (see~\vref{sec:RCA}).
  Resolution: lower the wafer temperature during processing by improving
  thermal contact to the backing wafer or lowering the backside chiller
  temperature set point.}
  \label{fig:DRIE:snakeSkinPhoto}
  \label{fig:DRIE:snakeSkin}
\end{figure}

\begin{figure}
  \centering
  \includegraphics[width=0.90\textwidth]{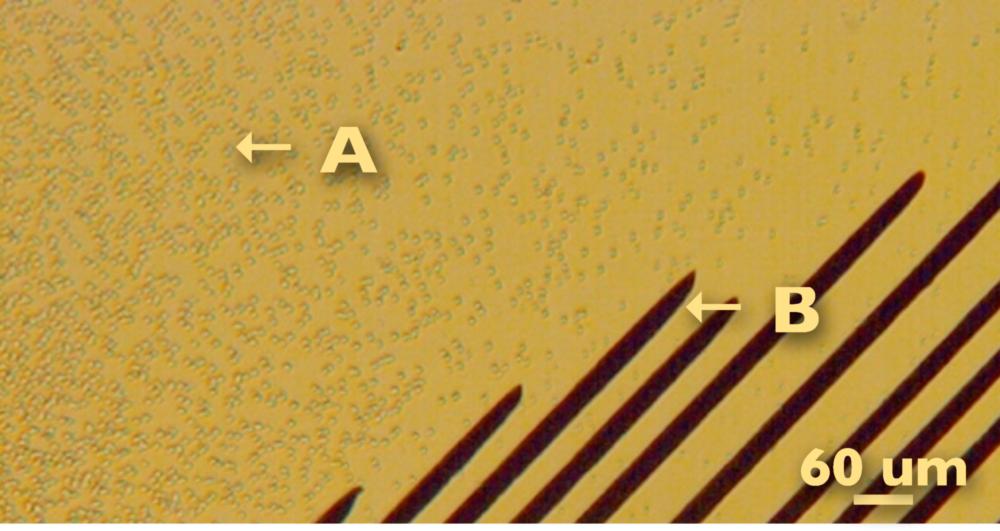}
  \caption[Optical micrograph showing excess fluorocarbon polymer.]
  {Optical micrograph showing excess fluorocarbon polymer. During a
  normal etch cycle the fluorocarbon polymer is fully etched away from
  surfaces normal to the plasma flux while persisting on channel sidewalls.
  If the process wafer surface is too cool during etching the polymer
  will not be fully etched and will accumulate on the wafer. In the
  micrograph the translucent $\sim 1~\mu m$  sized dots (A) are excess polymer
  in early stages of accumulation. The dark stripes (B) are deep etched
  silicon channels. If accumulation persists a nearly homogeneous layer
  will develop. The presence of this layer can also be identified by
  its extreme hydrophobicity. Resolution: raise the backside chiller
  temperature set point. }
  \label{fig:DRIE:fluorocarbonScumPhoto}
\end{figure}

\label{sec:DRIE:needles}
\begin{figure}
  \centering
  \includegraphics[width=0.80\textwidth]{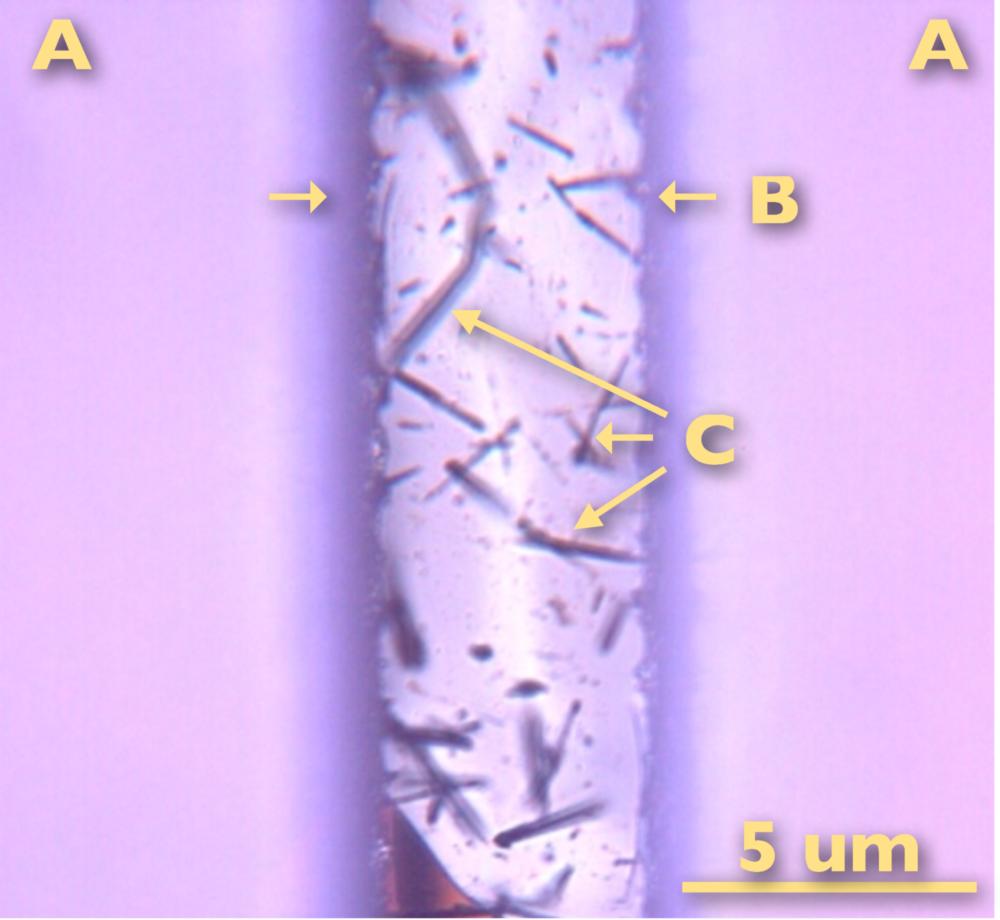}
  \caption[Optical micrograph showing silicon grass.]
  {Depicted is a trench~(B) etched into the surface~(A) of a silicon
  wafer but not fully through to the back side. The camera focus is
  several $\mu m$s above the bottom of the trench. There are many needle
  shaped features (aka grass) tens of $\mu m$s in length, arrows~C.
  They arise when moisture and sputtered contamination produce microscopic
  masks at various levels during etching. These masked areas are shielded
  from etching and form tall needles, some of which collapse laterally.
  These etch artifacts can appear when the process wafer surface temperature
  is too high. Resolution: lower the wafer temperature during processing
  by improving thermal contact to the backing wafer or lowering the
  backside chiller temperature set point. Also, dry the process wafer
  on a hot plate at 90$^{\circ}$~C for 10~minutes before etching (see
  STS documentation, \cite{flamm1990a}).}
  \label{fig:DRIE:needles}
\end{figure}

\begin{SCfigure}
  \centering
  \includegraphics[width=0.60\textwidth]{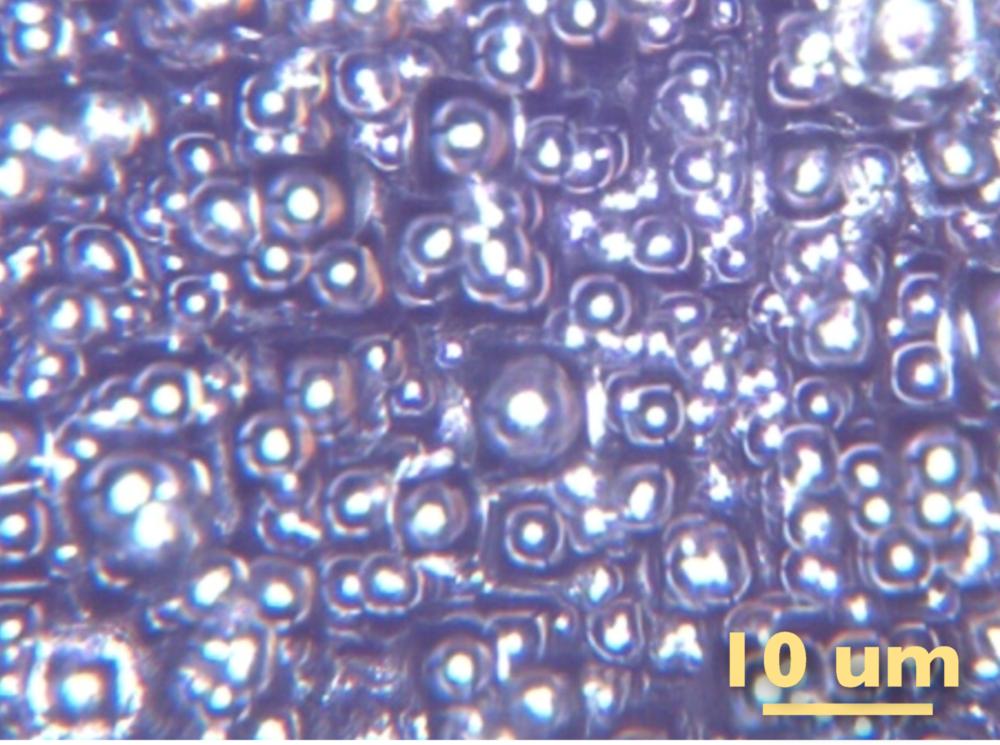}
  \caption[Photo of deep etched silicon surface morphology.]
  {Photo of deep etched silicon surface morphology. This optical micrograph
  shows silicon surfaces when etch parameters are properly adjusted.
  Depicted in the top image are trenches~(A) etched into the surface
  of a silicon wafer but not fully thru to the back side. The focus
  in micrograph is at the bottom of the trenches -- photoresist coated
  silicon ridges~(B) protrude upward beyond the field of view. The bottom
  image shows how properly etched surfaces appear: shiny with bubbly
  looking surface topology.}
  \label{fig:DRIE:siSurfaceMorphologyDRIE}
\end{SCfigure}

\clearpage
\subsection{Silicon oxide etching}
\label{sec:oxideEtch} Silicon oxide etch is a common MEMS process.
Two techniques are common: a wet etch with a solution containing HF
and a dry etch with a fluorine plasma. In this thesis both methods
were used to remove bulk structural $\text{SiO}_{2}$ (eg in SOI wafer
processing), to strip away native $\text{SiO}_{2}$ on doped silicon
ion trap surfaces and to pattern oxide layers which served as the
dielectric in MEMS capacitors. Neither method etches bulk silicon.

\begin{SCfigure}
  \centering
  \includegraphics[width=0.50\textwidth]{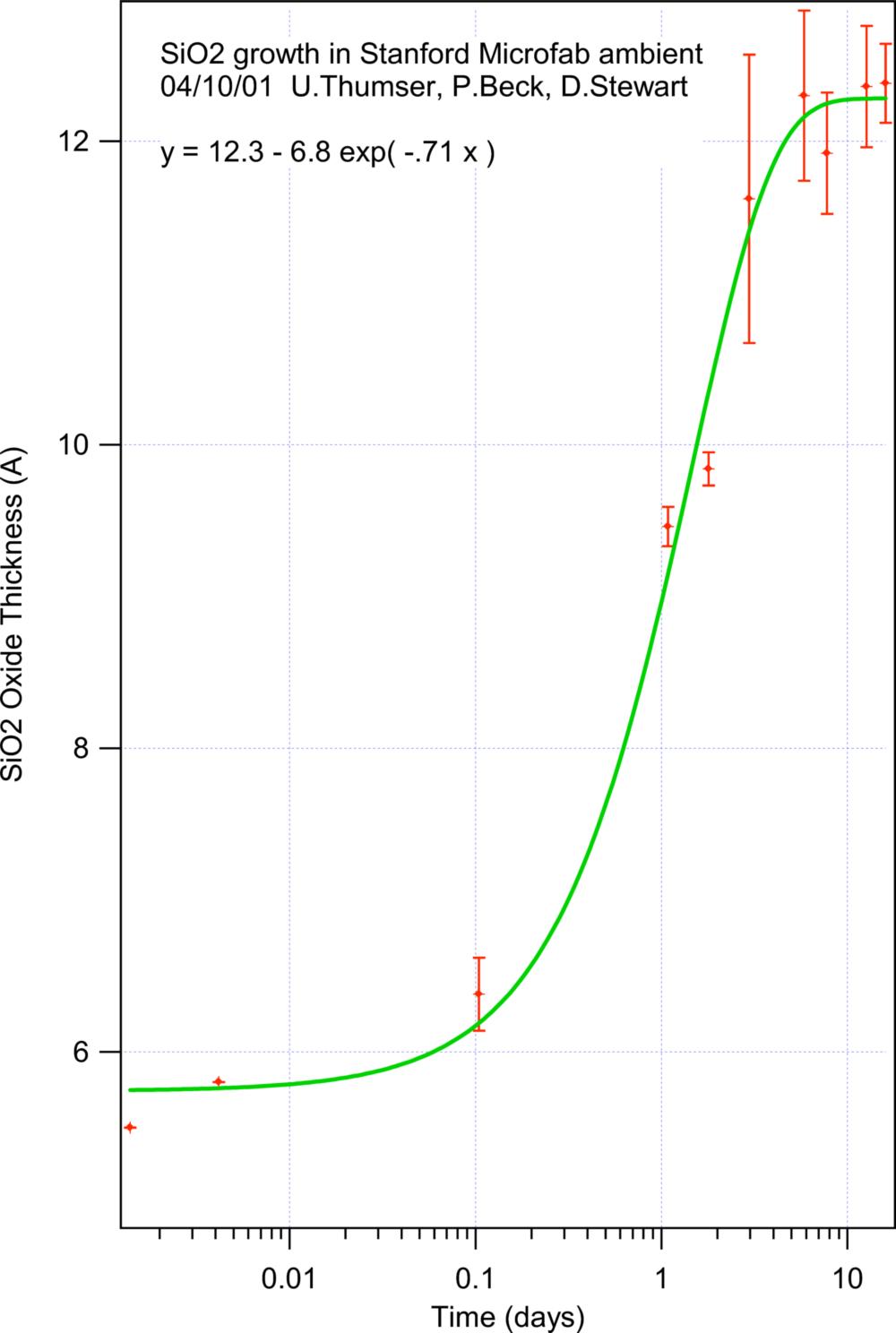}
  \caption[Native $SiO_{2}$ thickness on bare silicon.]
  {Native $SiO_{2}$ thickness on bare silicon. A bare silicon wafer
  quickly oxidizes at room temperature in air. The plot shows the oxide
  thickness over time for a wafer stripped of oxide by immersion in
  a 50:1 water:HF solution then left exposed to air. The thickness was
  measured using spectrometric reflectometry (Nanospec~250). It is expected
  that the oxide thickness is $\ll0.6~nm$ at $t=0$. The measure film
  thickness doesn't approach zero at zero time due to limitations of the
  measurement technique.
  \href{http://snf.stanford.edu/Process/Characterization/NativeOx.html}{Data}
  from the
  \href{http://snf.stanford.edu/}{Stanford Nanofabrication Facility} (M. Tang
  and U. Thumser.}
  \label{fig:SNFthermalOxide}
\end{SCfigure}

A 1-2~nm oxide forms on bare silicon in air at room temperature (see
Figure~\vref{fig:SNFthermalOxide} or~\cite{madou2002a}). Exposed
dielectrics are known to be problematic to ion traps because they
can support stray charges and it is unclear if such a very thin oxide
layer might sustain surface charges. As a precaution, the silicon
ion traps discussed in this thesis were processed to strip away this
native oxide within 20~minutes of installation into a vacuum system.
Both wet and dry etches were used in this application.

\paragraph{dry etch}
\label{sec:oxideEtch:plasma}
The dry plasma etching was performed in a Plasma Ops III etcher. In this device
$\text{CHF}_{3}$ is the process gas. A plasma containing atomic fluorine is
produced above wafer. Complex surface chemistry occurs at the surface which is
responsible for etching; primary reaction products are $\text{SiF}_{4}$,
$\text{CO}$ and $\text{CO}_{2}$. Oxygen may be optionally introduced into the
plasma to slowly degrade photoresist masks resulting in tapered (concave) etch
sidewalls. Photoresist is the usual etch mask. The key process parameters are in
Figure~\vref{fig:$SiO_2$plasmaEtch}.

\begin{SCfigure}
  \includegraphics[width=0.60\textwidth]{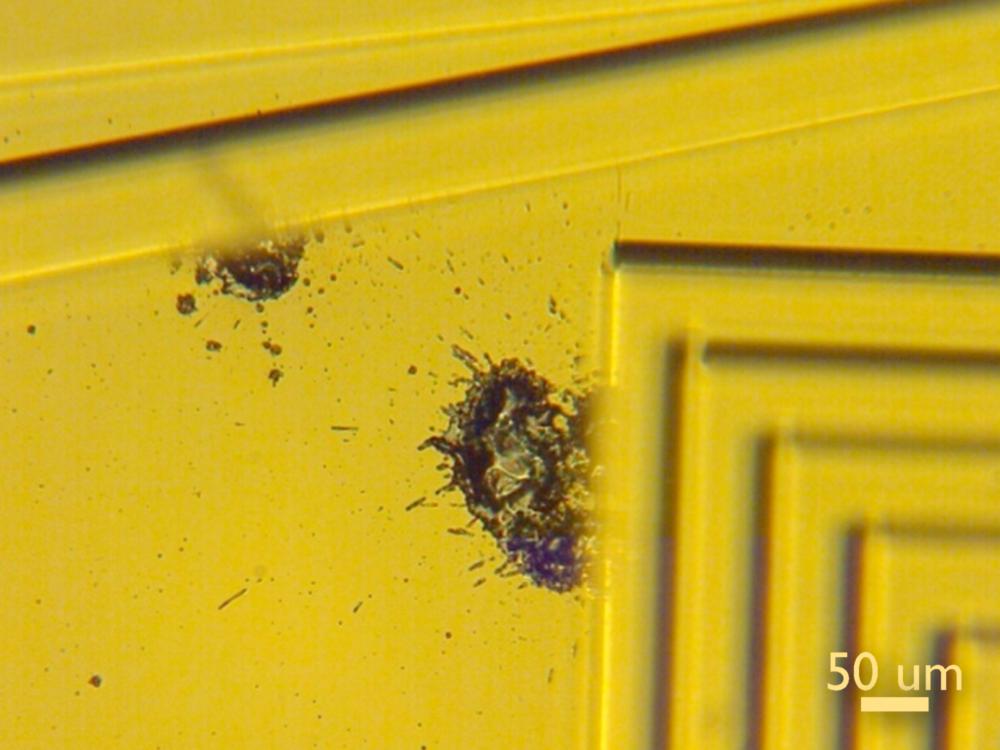}
  \caption[Breakdown on an SOI structure due to RF potential in plasma etcher.]
  {The micrograph shows evidence of breakdown on a SOI structures due
  to charging in the Plasma Ops III oxide etcher.}
  \label{fig:SOIbreakdown}
\end{SCfigure}

\begin{table}
  \centering
  \begin{tabular}{l|lllll}
    \textbf{Name} & \begin{sideways}
		\textbf{$O_{2}$ (sccm)}%
		\end{sideways} & \begin{sideways}
		\textbf{$CHF_{3}$ (sccm) }%
		\end{sideways} & \begin{sideways}
		\textbf{RF power (W) }%
		\end{sideways} & \begin{sideways}
		\textbf{RF DC BIAS (volts) }%
		\end{sideways} & \begin{sideways}
		\textbf{Pressure (torr)}%
		\end{sideways}\tabularnewline
		\hline
		$SiO_{2}$.RCP & 25 & 350 & 120  & -260 & 100e-3 \tabularnewline
		$SiO_{2}$FAST.RCP & 0 & 30 & 200 & -350 & 40e-3\tabularnewline
  \end{tabular}
  \caption[Typical $SiO_{2}$ etch recipe parameters.]
  {Chart showing typical $SiO_{2}$ etch recipe parameters for the Plasma
  Ops III etcher. Recipe $SiO_{2}$FAST.RCP etches at a rate of about
  25~nm/min for a fused quartz $\text{SiO}_{2}$ wafer. Recipe $SiO_{2}$.RCP
  etches at a rate of 12~nm/min for a fused quartz wafer. It produces
  $\sim$45$^{\circ}$ tapered side walls when used in conjunction with
  510LA photoresist. Such a taper is necessary if continuity is desired
  for subsequently deposited metallization as when making capacitors.
  Note that 1~Torr is 133~Pa.}
  \label{fig:$SiO_2$plasmaEtch}
\end{table}
High energy ions in the plasma can charge surfaces on the process
wafer to many hundreds of volts. In particular, catastrophic breakdown
was observed across $3~\mu m$ vacuum gaps during processing of SOI wafers
(see Figure~\vref{fig:SOIbreakdown}). The solution was to short all
isolated conducting surfaces to ground prior to processing in plasma
etcher.

\begin{SCfigure}
  \centering
  \includegraphics[width=0.50\textwidth]{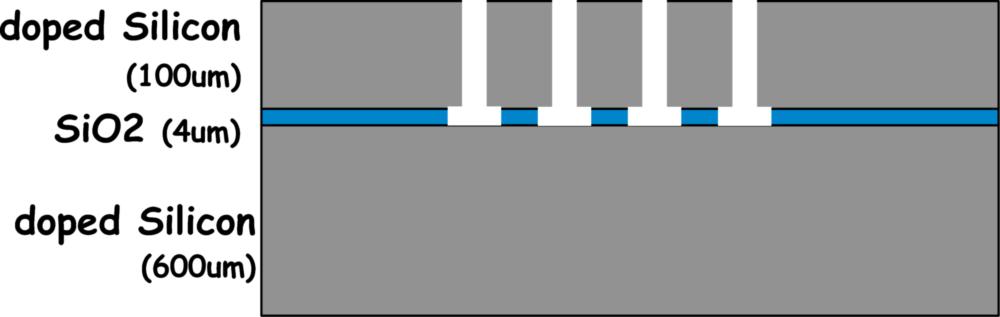}
  \caption[Aspect ratio of SOI oxide exposed to BOE during wet etch.]
  {Figure showing aspect ratio of SOI oxide exposed to BOE wet etch.Using
  BOE with agitation, 4~$\mu m$ thick $SiO_{2}$ was fully removed
  at the base of trenches 4-5~$\mu m$ wide 100~$\mu m$ deep in 35-40 minutes
  (100-114~nm/min). This technique was used in dv16m and dv16k series SOI ion
  traps. TODO ref to SOI trap section}
  \label{fig:SOIBOE}
\end{SCfigure}

\paragraph{wet etch}
\label{sec:oxideEtch:wet}
The wet etch of $SiO_{2}$ made use of industry standard buffered oxide etch
(BOE). BOE consists of 6~parts 40$\%$ ammonium fluoride, 1~part 49$\%$
hydrofluoric acid. The manufacturer specified etch rate at 16$^{\circ}$~C is
87~nm/min. BOE like most wet etches is isotropic and so results in undercutting
(see Figure \vref{fig:SOIBOE}). When etching $SiO_{2}$ at the bottom of deep
crevices (as in SOI processing) the solution was agitated on an orbital shaker.

Some precautions are in order for BOE. Titanium, a common thin film
adhesion layer, is rapidly etched by BOE \cite{williams2003a}. In
the course of this thesis work it was observed that even buried titanium
can be degraded by a BOE wet etch. This may be mitigated by thorough
rinsing (with agitation) using deionized water ($\sim$10 min). Glass
too is etched by BOE (eg 43~nm/min for Corning Pyrex 7740) and so
Teflon containers should be used \cite{williams2003a}. Note that
HF is extremely dangerous. Read the MSDS.

Thin film thickness (including $\text{SiO}_{2}$) can be estimated
by its color or ellipsometry \footnote{Optical fringes are visible in the sub
wavelength thickness air-film-substrate cavity.  There is a color vs
thickness chart
\href{http://www.ee.byu.edu/cleanroom/color_chart.parts/sio2_chart.jpg}{online}.}.
Note too that oxide surfaces are hydrophobic while bare silicon surfaces are hydrophilic.

\clearpage
\subsection{Doped silicon backside alignment}

\label{sec:backsideAlignment}Some devices require features on the
front and back side of a wafer. At the outset of wafer processing
it is common to etch registered alignment marks into the wafer surfaces.
Registration and pattern transfer can be done with several standard
techniques.

\begin{figure}
  \centering
  \includegraphics[width=0.70\textwidth]{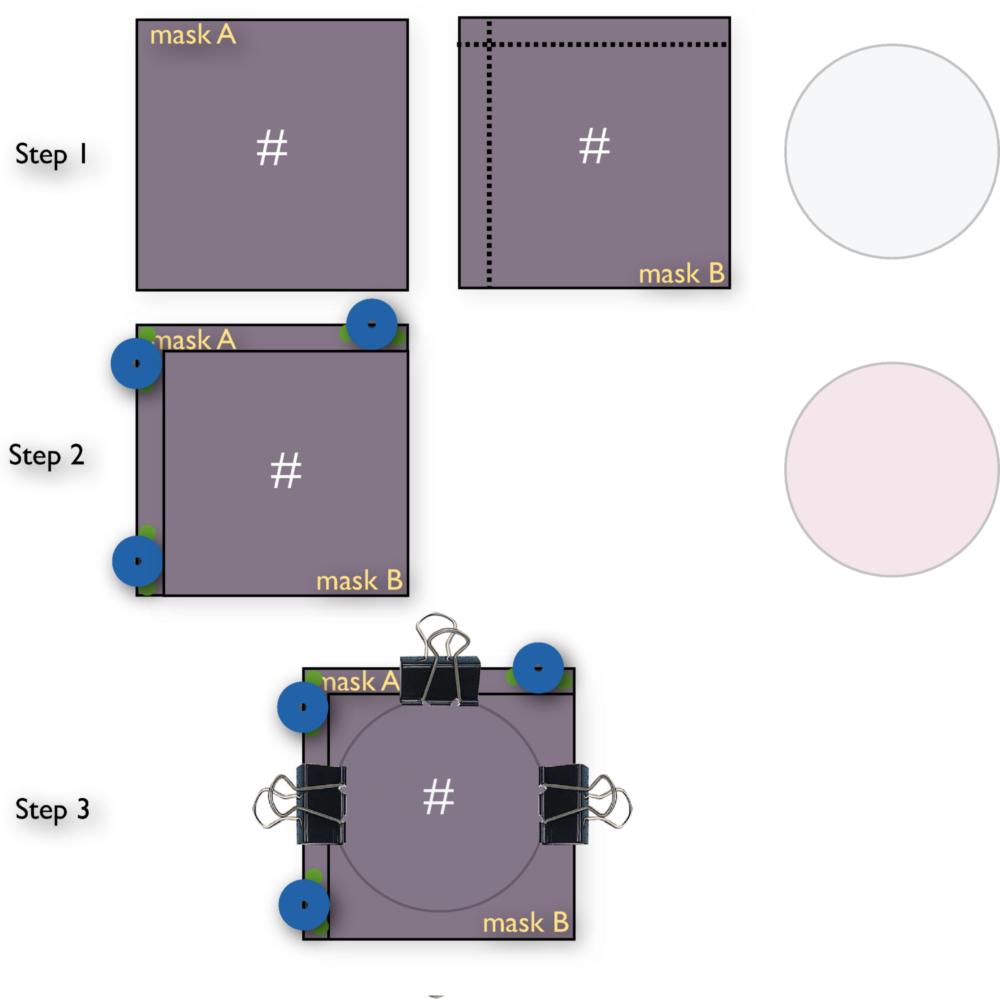}
  \caption[Kinematic backside alignment technique.]
  {Kinematic backside alignment technique.
  \textbf{Step 1}: Masks~A and B are
  4~inch glass-chromium photolithographic plates patterned with identical
  alignment marks~$\#$. A glass cutting saw was used to cut fully thru
  mask~B along the dotted lines, reducing its size by 0.5~inch on two
  sides. \textbf{Step 2}:  Put Mask~A on the stage of a microscope. Under an
  optical microscope slide mask~B across the top of mask~A, chromium
  sides touching. Apply a tiny bit of the epoxy to the edges of the
  masks so that it slightly wicks between the plates and provides viscous
  resistance to relative motion. Position mask B so its alignment pattern
  overlaps with that on mask~A. Let the glue cure. Position three 0.25~inch
  thick, 1.0~inch diameter stainless steel posts atop mask A and flush against
  the edge of mask~B as in the Figure. Apply epoxy to the stainless 
  posts fixing them to mask A. Do not put any epoxy near
  the intersections between mask~B and the stainless posts. Let the
  glue cure. Crack the tiny bit of glue fixing mask~A to mask~B and
  remove this glue with solvent. Apply photoresist to both sides of
  a 3~inch wafer. \textbf{Step 3}: Slide the wafer between the masks. Press
  mask~B flush with the stainless posts fixed to mask~A. Use binder 
  clamps to hold the stack together. The masks remain well aligned even after
  sliding the silicon wafer between them. Expose as usual on the Karl
  Suss UV exposure tool. Use an etch technique (eg silicon etch for
  a silicon wafer) to transfer the photoresist pattern to each side
  of the wafer.
  }
  \label{fig:kinematicAlignment}
\end{figure}

Ordinary (undoped) silicon wafers are transparent in the infrared
(IR). This is exploited by the NIST Karl Suss contact exposure tool
to image etch/metallization patterns on the wafer backside when viewing
from the front through the wafer bulk. This IR image is used to align
a mask to the front side and expose a front side photoresist pattern.
With this technique alignment mark registration to better than 5~$\mu m$
is possible. Highly doped silicon is however highly absorbing in the
IR and this technique isn't practical.

Other techniques rely on more expensive tools. For example, some wafer
suppliers offer laser etching of front-back alignment marks to $0.5~\mu m$
resolution. This is only available for standard size wafers. A common 
in-house tool is commercial wafer alignment machines which solve this problem 
using computerized alignment platforms and front/back
digital microscopes. Their cost at $\$$0.5~million is however prohibitive.

\begin{SCfigure}
  \centering
  \includegraphics[width=0.60\textwidth]{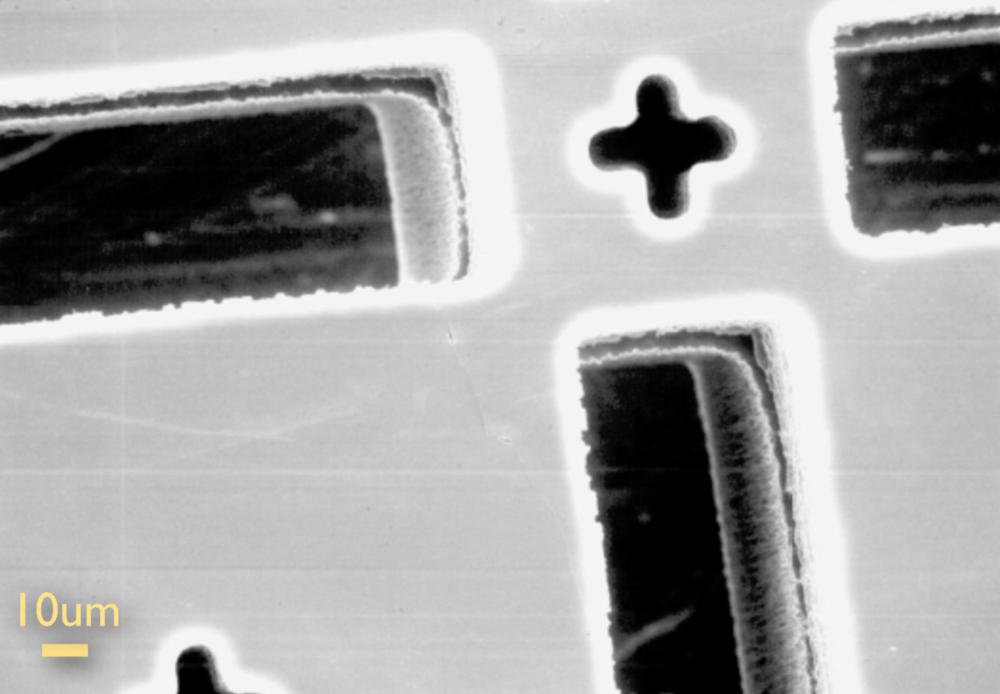}
  \caption[SEM micrograph of a silicon wafer showing front side to backside registration
  of an alignment pattern.]
  {SEM micrograph of a silicon wafer showing front side to backside
  registration of an alignment pattern. The kinematic jig described
  in Figure~\vref{fig:kinematicAlignment} was used to transfer an alignment
  pattern to thick photoresist on both the front and back side of a
  200~$\mu m$ thick silicon wafer. Deep etches from both sides meet
  and reveal an alignment error of about $5~\mu m$.}
  \label{fig:kinematicAlignmentDemoPhoto}
\end{SCfigure}

Inexpensive, repeatable backside alignment to about $5\mu m$,
$0.01^\circ$ accuracy was done in the NIST cleanroom using a simple kinematic approach
discussed in Figure~\vref{fig:kinematicAlignment}. It makes use of
a NIST in-house tool for easily fabricating high resolution photolithography
masks.
\clearpage

\subsection{Metallization of silicon}
\label{sec:AuMetallization} Deposition of gold, copper, titanium and
aluminum thin films are done in an e-beam evaporator. In a low pressure
($<1\times 10^{-6}$~Torr)  environment a high energy
(10~kV) electron beam (0-20~mA) is directed at one of several water cooled
tungsten crucibles containing the process metals. The electron energy
is deposited directly in the process metal causing it to melt. The
metal vapor plume which results intersects the process wafer about
0.5~m away. The deposition rate (0.1-2~nm/sec) is controlled by the
e-gun current. The chuck holding the process wafer is water cooled
and acts as an RF anode for \textit{in situ} plasma cleaning. The
NIST deposition system used is named HTS1.

An argon plasma can be used to perform \textit{ in situ} cleaning
of a wafer surface in advance of metal deposition. This is necessary
in many cases to ensure good adhesion of the metal to the silicon.
It slightly etches exposed surfaces.

\begin{table}
  \centering
  \begin{tabular}{l|lllll}
    \textbf{Name} & \begin{sideways}
		\textbf{Argon (mT) }%
		\end{sideways} & \begin{sideways}
		\textbf{Oxygen (mT) }%
		\end{sideways} & \begin{sideways}
		\textbf{Time (min) }%
		\end{sideways} & \begin{sideways}
		\textbf{DC Bias (V)}%
		\end{sideways} & \begin{sideways}\textbf{RF (W)}\end{sideways}\tabularnewline
		\hline
		metal atop $SiO_{2}$ & 15 & 0  & 3  & -350 & 25\tabularnewline
		standard clean 15-30 & 0 & 2 & -400 to -350  & 50 & \tabularnewline
  \end{tabular}
  \caption[Table of standard argon plasma etch recipes in the HTS1 e-beam evaporator.]
  {Table of standard argon plasma etch recipes in the HTS1 e-beam evaporator.}
  \label{tab:hts1PlasmaCleanRecipes}
\end{table}
An oxygen plasma etch may be used to strip off residual photoresist
if its believed to be causing adhesion problems.

Gold is commonly used in ion traps owing to its high conductivity and the absence
of a surface oxide. It does not stick to silicon oxide but Ti and Cr do
\cite{williams2003a}. Hence, these metals and are commonly used as an adhesion
layer (10-50~nm) for gold. Titanium is not magnetic and is convenient to deposit
on our e-beam evaporator, but requires care in process design as it is rapidly
etched by HF (which in turn is used as an silicon oxide etchant (see
Section~\vref{sec:RCA}). Titanium was used exclusively as a adhesion layer in the
devices discussed in this thesis.

Chromium is attractive as an adhesion layer for Au because it is not etched by
HF, however it is antiferromagnetic. That is, below the Neel temperature (for
bulk Cr, $T_N=311^\circ$~K \cite{zabel1999a}) it is favorable for
adjacent spins to antialign. The Neel temperature is analogous to the Curie
temperature for ferromagnetic materials. That is, it is the temperature at which
thermal energy becomes large enough to destroy microscopic ordering in the
material, rendering it paramagnetic. Thin films of Cr are especially complex and
exhibit a property called spin density wave (SDW) magnetism
\cite{fawcett1988a,zabel1999a}. Cr has not been used at NIST as an adhesion
layer for Au in ion traps. In a neutral atom experiment a Cr adhesion 
layer was used and the resulting magnetic fields found to be negligible \cite{lev2006a}.
\footnote{In his thesis work Ben Lev used Cr as a sticking layer for
electroplated gold wires used to make mots in his thesis. Cr was also used as a sticking layer in the
commercial hard disk platter he etched with a fine grating for the purposes of
making a magnetic mirror for neutral atoms \cite{lev2006a}.}

\clearpage
\subsubsection{deposition of thick gold films}
\label{sec:AuMetalThickFilms} 
\begin{SCfigure}
  \centering
  \includegraphics[width=0.60\textwidth]{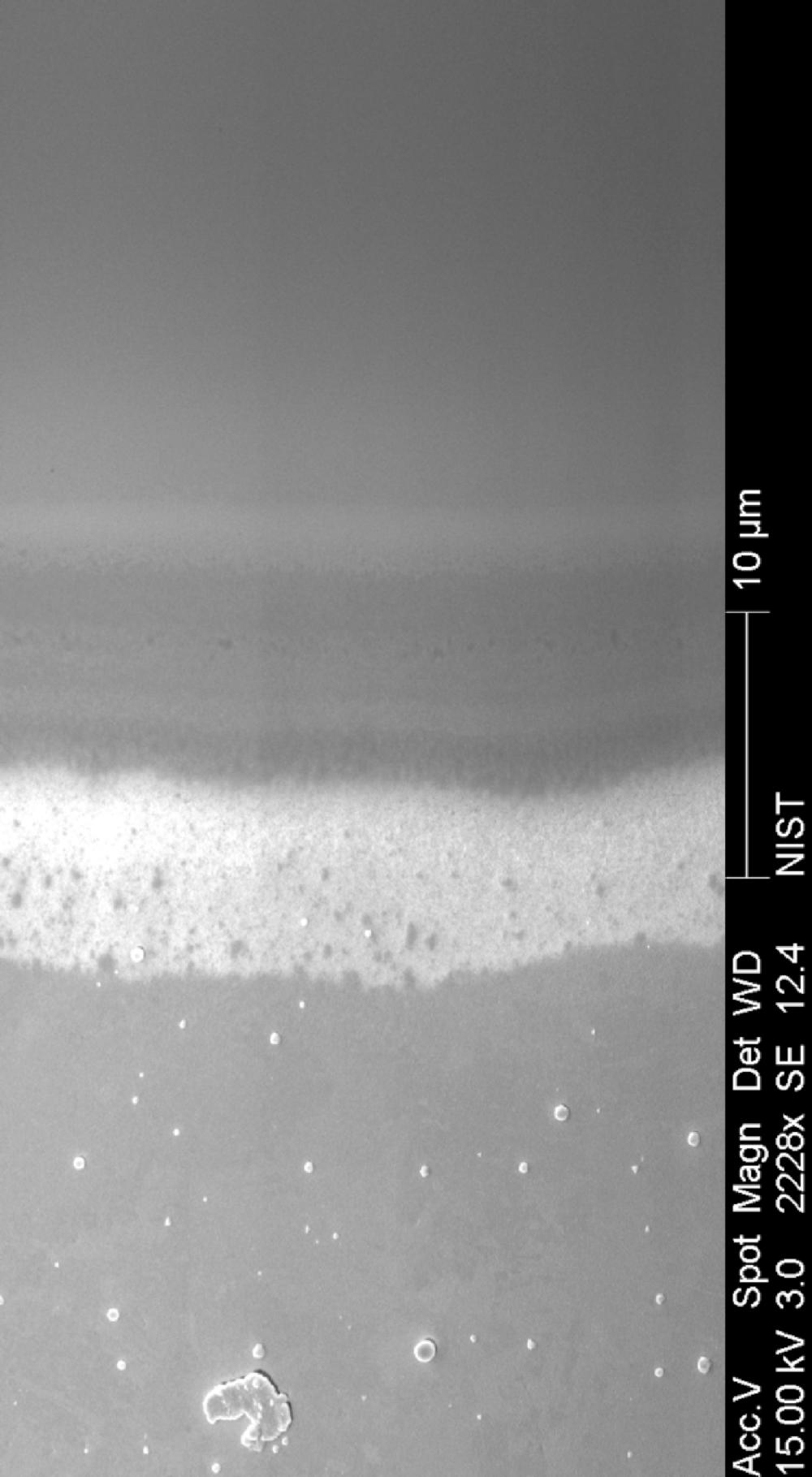}
  \caption[SEM showing contamination of a gold film deposited by e-beam evaporation.]
  {SEM showing contamination of a gold film deposited by e-beam evaporation.
  During deposition a stainless steel shadow mask 
  (see Section~\vref{sec:shadowMask})
  was positioned over the wafer which shielded the right hand side from
  gold deposition. The left side is a $\sim 1\mu m$ gold film using the
  FASTAU deposition recipe. The right hand side is bare silicon. The
  whitish strip in the middle is a transition between the two regions
  corresponding to the edge of the mask. The blobs on the left side
  are $200-500$~nm in height as measured by a stylus profilometer. Debris
  of this size are not expected to cause trouble for the ion traps discussed
  in this thesis if they are made of gold. However, since it couldn't
  be ruled out that they might instead be dielectric care was taken
  to avoid them (e.g. recipe SLOWAU).}
  \label{fig:semGoldBlobs}
\end{SCfigure}

Gold films $1$~nm to $1~\mu m$ are routine. For
depositions $>200$~nm in the NIST e-beam evaporator some care must be taken to
prevent contamination by debris (see Figure~\vref{fig:semGoldBlobs}). Films much
thicker than $1~\mu m$ suffer from rough surfaces ($> 500$~nm~rms) for reasons
that are not well understood. The precautions in recipe SLOWAU permitted growth
of up to $1.5~\mu m$ gold films with good surface quality. Unfortunately, this
recipe did not always reproduce debris free films.

\begin{SCtable}[40]
   \centering
   \begin{tabular*}{0.5\textwidth}{l|lll}
   		\textbf{Name} &
		\begin{sideways}\textbf{Pressure (Torr)} \end{sideways}&
		\begin{sideways}\textbf{Al (nm/s, nm) } \end{sideways}&
		\begin{sideways}\textbf{Ti (nm/s, nm) }\end{sideways}\tabularnewline
		\hline
		FASTAU &  & 1, 10 & 1, 10\tabularnewline
		SLOWAU &  & 0.3, 10 & 0.3, 10\tabularnewline
	\end{tabular*}
	\caption[Table of gold deposition recipes for the HTS1 e-beam evaporator.]
		{Table of gold deposition recipes for the HTS1 e-beam evaporator.
		Note that the aluminum deposition step is only necessary if making
		Ohmic contacts to doped, bare silicon wafers.}
	\label{tab:HTS1goldDepositionRecipes}
\end{SCtable}
%

The FASTAU recipe can be considered the baseline recipe for deposition
in HTS1. It consists of putting a wafer into the system and depositing
as rapidly as possible. The SLOWAU recipe includes the following additional
precautions beyond slow deposition rates.

\begin{compactitem}
  \item Don't use the communal gold crucible in HTS1 as it is used by many
  cleanroom users and may be contaminated. Instead, use a new crucible
  and gold pellets cleaned in solvents in an ultrasonic bath. Use a
  tungsten crucible, not a carbon crucible which may shed carbon flecks.\\
  
  \item In addition to deposition on the wafer surface, the interior of the
  HTS1 vacuum system is coated with metal during depositions. Some of
  these metals oxidize (eg Cu) when the vacuum is broken during wafer
  loading. Flakes from interior surfaces may fall into the process crucible
  during deposition. To reduce this likelihood check that the system
  has been recently cleaned. In addition, with the crucible removed
  use a clean stainless knife to scrape away the thick metal crust which
  forms at the rotating hearth aperture. Use the HEPA vacuum cleaner
  to remove flakes \vref{fn:Note-NIST-HTS1}.
  \item Inspect by eye the metal surfaces for all the crucibles to be used
  in a deposition. Look for discoloration, carbon deposits or debris.
  \item Attach the process wafer to the water cooled chuck. Use all the screws
  on the metal retaining ring to achieve good thermal contact. The process
  wafer heat load is lower for lower deposition rates. Cool surfaces
  result in smaller gold grain sizes (personal correspondence (2007)
  Ron Folman of Ben-Gurion University, Israel).
  \item HTS1 has a mechanical shutter that optionally shields the process
  wafer from the metal plume. Keep the shutter closed while heating
  up a crucible using the e-beam. Adjust the e-gun current so metal
  deposits on the acoustic film thickness detector at a rate of 1~nm/sec.
  For the next 3-4~minutes proceed as follows. Dither the e-beam across
  the metal surface in the crucible \vref{fn:Note-NIST-HTS1}. Inspect
  the molten metal by eye looking for islands of floating scum. Chase
  them with the e-beam until they are blasted away. You may see occasional
  sparks jumping from the crucible surface. Then, look by eye perpendicular
  to the metal flux for more sparks. If you don't see any sparks over
  a minute or so, lower the e-gun current to get the desired deposition
  rate, open the mechanical shutter and proceed with deposition.
\end{compactitem}

\footnote{\label{fn:Note-NIST-HTS1}Note NIST~HTS1 users, these procedures can
(easily) cause expensive damage to the e-beam evaporator. Talk to
an expert first.}

\begin{SCfigure}
  \centering
  \includegraphics[width=0.60\textwidth]{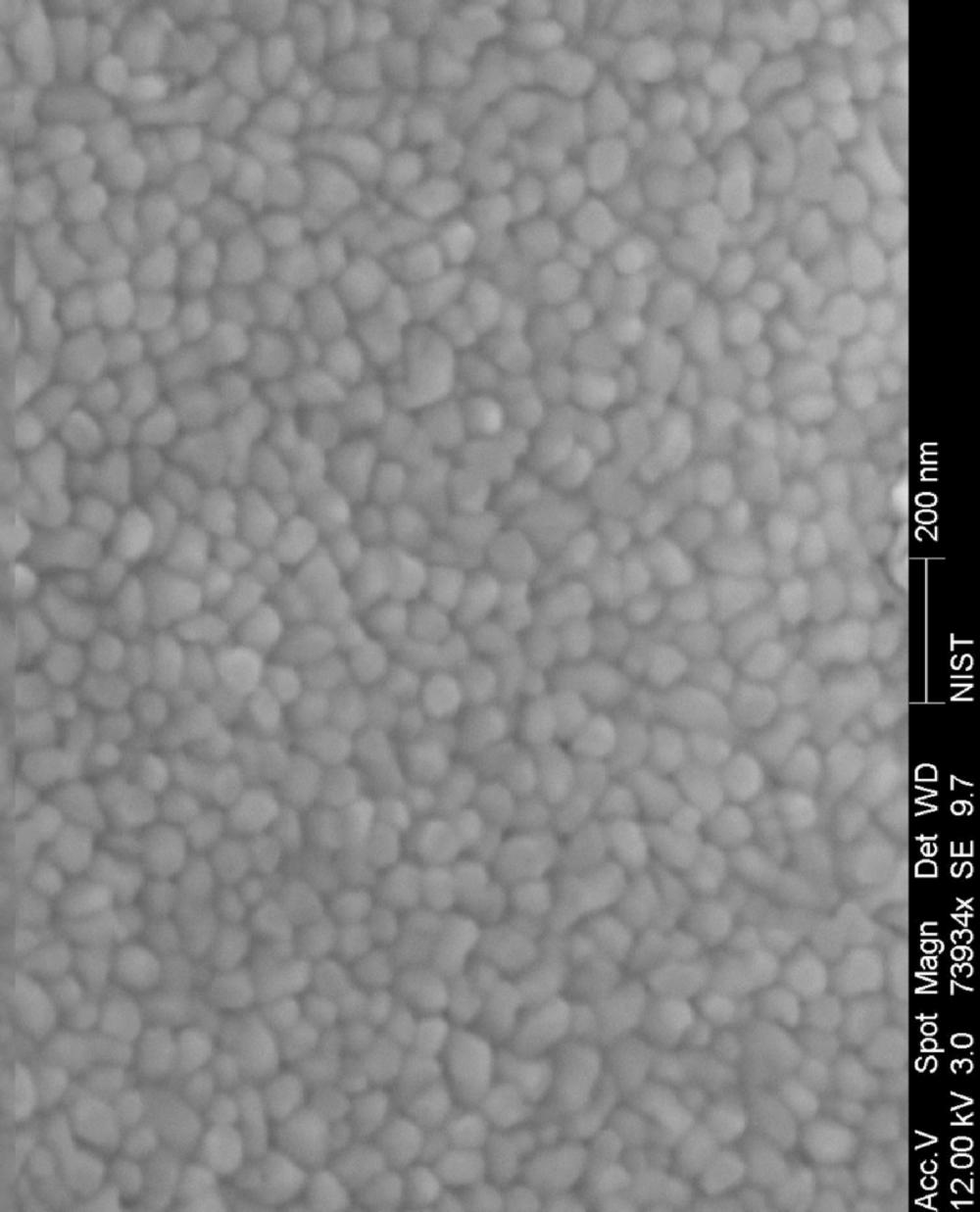}
  \caption[SLOWAU recipe performance: SEM showing 30-50~nm gold grains.]
  {SLOWAU recipe performance: SEM showing 30-50~nm gold grains. This
  is the intrinsic surface roughness of a gold film $1~\mu m$ thick deposited
  in the NIST e-beam evaporator at room temperature for the SLOWAU recipe. }
  \label{fig:slowauGoldGrainSize}
\end{SCfigure}

\begin{SCfigure}
  \includegraphics[width=0.60\textwidth]{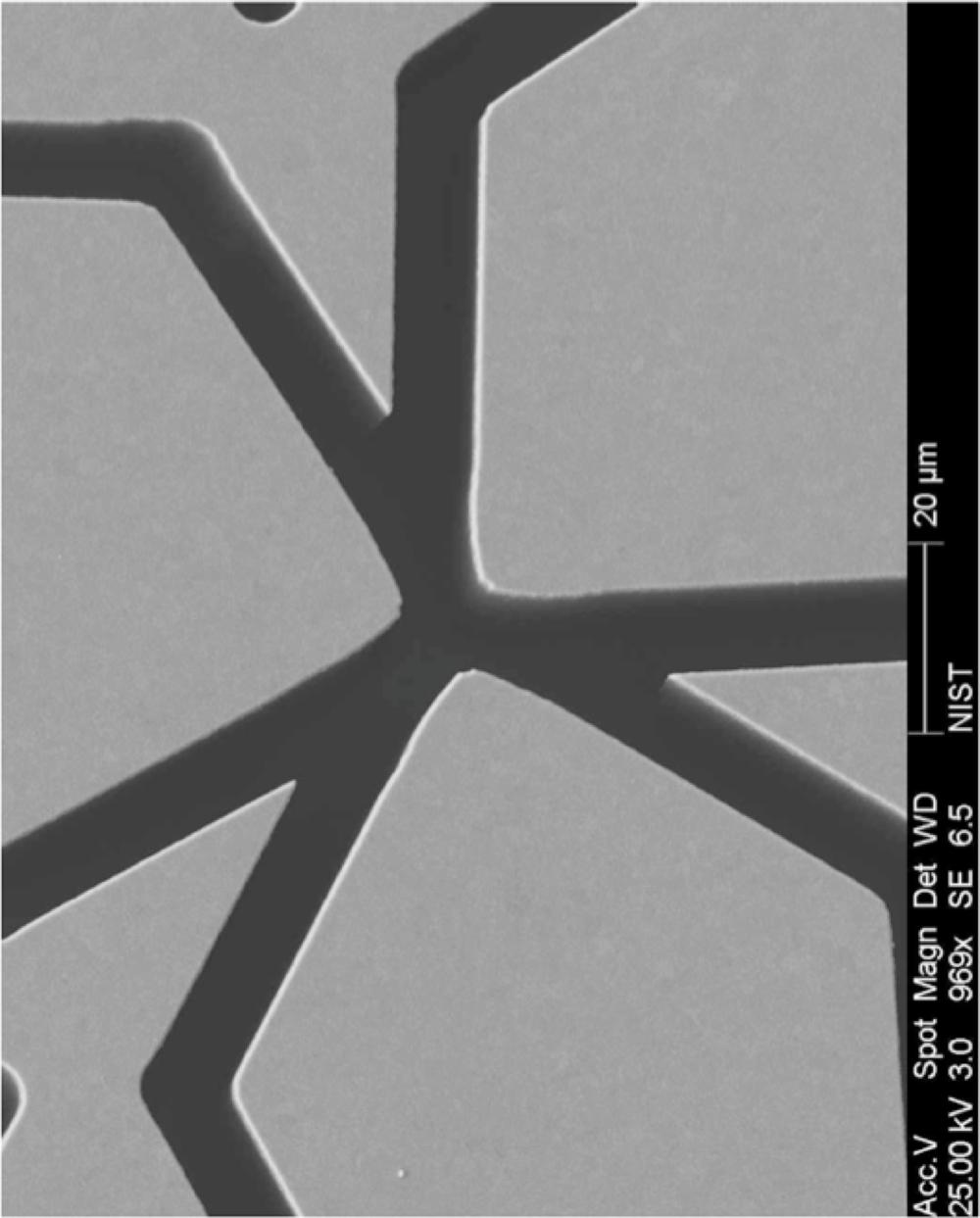}
  \caption[SLOWAU recipe performance: SEM showing a wide field of view of a
  1~$\mu m$ gold deposition.]{SLOWAU recipe performance: SEM showing a wide 
  field of view of a 1~$\mu m$ gold deposition. This illustrates the degree to
  which the SLOWAU recipe can reduce the prevalence of debris. The dark channels
  are trenches ($>100~\mu m$ deep) cut into the silicon.}
  \label{fig:slowauGoldWideView}
\end{SCfigure}

\clearpage

\subsubsection{shadow mask deposition}
\label{sec:shadowMask}
\label{sec:shadowMaskRecipe}A shadow mask 
is a thin sheet of metal with apertures. It masks off parts of a wafer
during metallization in an e-beam evaporator. Use the following recipe
to prepare a silicon chip for metalization using a shadow mask.

\begin{compactenum}
  \item Thoroughly clean a sacrificial silicon backing wafer (front and back
  side).
  \item On a hotplate at 130$^{\circ}$~C bond one or two chips to the backing
  wafer using wax. Use wax sparringly. No excess wax should protrude
  from under the chip at its edges.
  \item Strip the native oxide off the wafer by doing a 30~sec BOE dip (see
  Section~\vref{sec:RCA}) or 1~min in the Axic etcher with recipe SIO2\_FAST.RCP
  (see Section~\vref{sec:oxideEtch}). Complete the following steps
  in $<$ 10 minutes to prevent reoxidization of the silicon surface.
  \item Align the shadow masks to the chips under a stereoscope. Use Kapton
  tape to hold the masks in place.
  \item Attach the backing wafer to the water cooled HTS1 wafer chuck.
  \item Proceed with the FASTAU or SLOWAU deposition recipes in Section \vref{sec:AuMetalThickFilms}.
  Use an Ar plasma clean if necessary.
  \item Release the chips from the backing wafer using step IV in Section~\vref{sec:RCA}.
\end{compactenum}
Stainless shadow masks are laser machined by Solder Mask Inc. (www.soldermask.com).
The cost per mask was $\$$100 in lots of~1, same day shipping. The
GERBER format is used to specify the mask pattern. Typical specifications:
\begin{compactitem}
  \item material: stainless steel
  \item thickness: $250~\mu m$
  \item resolution: $25~\mu m$
\end{compactitem}

\clearpage

\subsubsection{Ohmic contacts in heavily doped silicon}
\label{sec:dopedSilicon} Care must be taken when making electrical contact to
semiconductors. This section provides a brief overview of how doping influences
semiconductor conductivity and what happens at metal-semiconductor interfaces.

In semiconductors there is band gap, a range of electron energies for which there
are no available states in the crystal. States lying below (above) this gap are
called the valence (conduction) band. The Fermi distribution predicts the
temperature dependent statistical distribution of electrons among available
states. Electrons populating the mostly empty conduction band can conduct
electricity, as can holes (missing electrons) in the valence band. Undoped
silicon at room temperature is a poor conductor because nearly all the available
electron states in the valence band are occupied and there is negligible
population in the conduction band.

Silicon is a semiconductor with a band gap of 1.1~eV and four valence electrons
per lattice site \cite{sze1981a,kittel1996a}. If a trivalent atom like B, Al or
Ga is substituted for silicon it steals an electron from the valence band in
order to satisfy the covalent bond with its neighbors. If many such substitutions
are made, the silicon is then called p-type and has holes in the valence band.
These holes permit electrical conduction. Conductivity results without additional
electrons being excited to the conduction band.

The doped silicon used in this thesis work is p-type due to doping with Boron.
The doping level is $0.1-1\%$ resulting in an electrical conductivity of
$0.0005-0.001~\Omega-cm$ (B*Si) \cite{sze1981a}). The wafers were obtained
from \href{http://www.virginiasemi.com}{}{Virginia Semiconductor, Inc.}

\begin{figure}
  \centering
  \begin{minipage}[b]{0.35\linewidth}
    \begin{tabular}{l|ll}
      	\textbf{$Metal$} & \textbf{$T$}& \textbf{$\sigma$} \tabularnewline
		\hline
		B*Si & 300 & 500-1000\tabularnewline
		Al & 300 & 2.73\tabularnewline
		Au & 300 & 2.27\tabularnewline
		Ag & 300 & 1.63\tabularnewline
		Cu & 300 & 1.73\tabularnewline
		Ti & 273 & 40\tabularnewline
    \end{tabular}
    \caption[Resistivity of common metals compared with the boron doped silicon
    (B*Si) used in this thesis.]{Resistivity of common metals compared with the boron doped silicon
    (B*Si) used in this thesis \cite{crc2008a}. The units for temperature $T$
    is $^\circ~K$ and for resistivity $\sigma$ is $\times 10^-6 \Omega-cm$.}
    \label{tab:metalConductivity}
  \end{minipage} 
  \hspace{0.05\linewidth}
  \begin{minipage}[b]{0.55\linewidth}
    \includegraphics[width=1\textwidth]{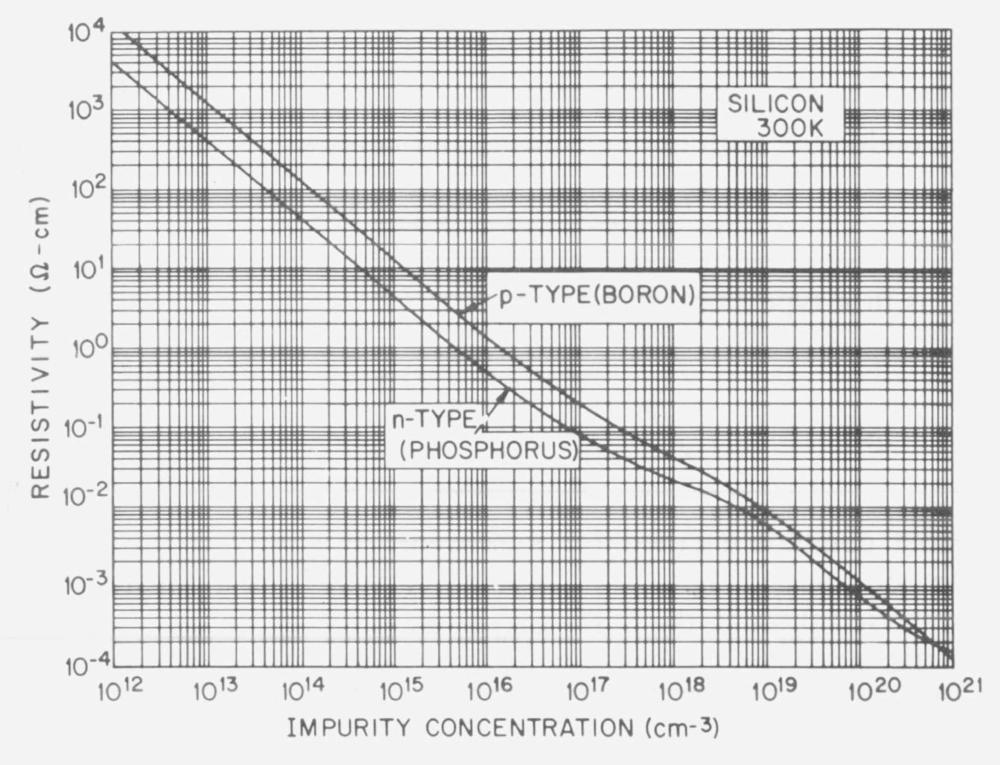}
    \caption[Resistivity vs impurity (dopant) concentration for silicon at 300$^{\circ}$
    K.]{Resistivity vs impurity (dopant) concentration for silicon at 300$^{\circ}$
    K \cite{sze1981a}.}
    \label{tab:szeResistivityVsDoping}
  \end{minipage}
\end{figure}

The fabrication of reliable contacts between metals and semiconductors is crucial
for integrated circuits. Usually, a Schottky barrier occurs at a
metal-semiconductor interface which has nonlinear, rectifying characteristics
\cite{sze1981a}. When a semiconductor is brought into contact with a metal,
current flows so that their respective chemical potentials are equilibrated at
the interface. A potential drop across the interface results along with a local
depletion of charge carriers called a depletion layer. In lightly doped
semiconductors this depletion region is wide and presents a barrier to
conduction. However, if a sufficient DC electronic bias is present electrons have
enough energy to exceed the barrier. That is, the current-voltage (I-V) response
of the junction is nonlinear.

An ohmic contact is a metal-semiconductor interface with a negligible contact
resistance relative to that of the bulk semiconductor. During device operation
the voltage drop across an ohmic contact should be small relative to the drop
across the whole device \cite{sze1981a}. That is, the current-voltage (I-V)
response of the junction is linear and symmetric. In highly doped semiconductors
there are so many carriers that the depletion region is very narrow and
conduction at room temperature is sometimes possible. Standard practice is to use
aluminum as the conductor and to anneal at $>350^\circ$~C \cite{card1976a}.
The exact conditions at semiconductor-metal interfaces which result in ohmic
contacts are not fully understood by engineers. Considerations include metal
type, the presence of oxide and water vapor at the interface and interdiffusion
due to annealing \cite{card1976a}.

I made ohmic contacts between doped silicon and gold as follows. First, the
native silicon oxide was removed using a 90~second plasma etch
\vref{sec:oxideEtch}. Second, with only 1-2~minute exposure to air, the wafer was
transferred to the e-beam evaporator. There, a 10~nm layer of Al was deposited on
the silicon, then 10~nm Ti (as an adhesion layer) and finally 1000~nm Au. The
quality of the junction was tested on a home built I-V measurement apparatus. It
was found to be ohmic without annealing probably due to heavy doping. Vacuum
processing for ion traps is a 3-4~day $200^\circ$~C bake and it is expected to
improve the junction conductivity. Gold-aluminum intermetallic compounds (aka
purple plague) can degrade the contact but were not observed, probably due to the
intervening Ti layer.

\clearpage
\subsection{Wafer bonding}

\label{sec:waferBonding} Semiconductor wafer bonding is a MEMS technique for
adhering wafers together without the use of glue \cite{wallis1969a},
\cite{tong1999a}). Bondable materials include silicon, oxidized silicon, glass
and fused silica. The resulting bonds are robust and hermetic. They find
application in MEMS device packaging and fabrication of multilayer
semiconductor-insulator structures. The process is simple. Two highly polished,
flat, hydrophilic surfaces are brought into contact and over several seconds
their surfaces bond. The resulting bond is usually weak at room temperatures but
can be made stronger with greater pressure and higher temperatures. The physics
behind bonding is part electrostatic, part chemical.

For traps with few electrodes and a large ion-electrode separation ($>200~\mu
m$), the traditional wafer bonding used mechanical screws. However, smaller traps
with many trapping zones requires higher alignment precision than is practical
using screws. A variety of alternative wafer bonding techniques were explored in
the context of ion trap microfabrication. The required UHV pressures
($<1\times10^{-11}$~Torr) and processing temperatures ($~200^\circ$~C), excludes
many common approaches including ordinary epoxy glue and solder due to
outgassing. Technologies investigated include thermocompression bonding, anodic
bonding and commercial silicon on insulator (SOI). Only the latter two are
discussed in this thesis.

Wafer bonding was first demonstrated in 1969~by Wallis \cite{wallis1969a}) for
silicon and glass. Wallis bonded the borosilicate glass Corning 7740~Pyrex to
silicon at $400^\circ$~C with a 300~V~DC bias in 1~min. His explanation of
the attraction between silicon and glass is purely electrostatic, with long term
adhesion at room temperature attributed to unspecified covalent chemical bonds at
the wafer interface. Subsequently water adsorption at the wafer surfaces and
resulting hydrogen bonds were implicated in both silicon-glass and
silicon-silicon bonding.

Both chemical and electrostatic processes contribute to wafer bonding. This
section begins with a practical guide to anodic wafer bonding and then discusses
these bonding processes in detail. It concludes with comments on commercially
available silicon on insulator (SOI) wafers.

\subsubsection{waferbonding practice}

\label{sec:bondingInPractice} Commercial bonding apparatus is available that
simultaneously accommodates anodic, thermocompression and silicon direct bonding
techniques and provides wafer-wafer alignment to better than $1~\mu m$. However,
at $>\$$0.5m dollars the cost is prohibitive. It turns out that its possible to
build a simple anodic bonding setup for small scale bonding.

Early in my thesis work initial attempts at waferbonding 1~$\text{cm}^{2}$
silicon to 7070 glass ($100~\mu m$ thick) were unsuccessful. Bonding was done in
air at 400$^{\circ}$~C and electrical breakdown occurred at 300-400~V, too low
for bonding. Breakdown occurs when the electron mean free path is just sufficient
to allow build up of enough energy to ionize a gas molecule upon subsequent
impact. The dielectric strength of air is $3\times10^6$ V/m, which is exceeded at
300~V for a $100~\mu m$ gap. Bonding at low pressure ($1\times 10^{-3}$~Torr;
1~Torr is 133~Pa) and in 1~atmosphere of the electronegative gas $\text{SF}_{6}$
(whose breakdown is three times that of air) were of middling success
\cite{madou2002a,thomson1928a,paschen1889a}. Ultimately, it was found that
bonding could be reliably accomplished with the following parameters and the
apparatus in Figure~\vref{fig:bondingPlatformVacuumStage}.

\begin{compactitem}
  \item $150~\mu m$ glass
  \item 400$^{\circ}$~C hot plate
  \item 500~V
  \item 30~min piranha dip within 6~hr of bonding (see Section~\vref{sec:RCA})
  \item 3~minutes per bond interface
\end{compactitem}

Other bonding tips include the following. At least 15~minutes before bonding use
a dry nitrogen jet to knock any dust off the bonding apparatus. Preheat the
bonding apparatus in advance of bonding. Always, inspect the chip surfaces for
debris or scratches. As a gauge of wafer flatness, when chips are stacked look
for $<6$~Newtons fringes per cm across their surface.

Its not certain if success was due to cleaner surfaces after the piranha etch or
if there is some favorable surface chemistry at work. I observe that wafers not
cleaned in piranha would bond successfully only to fall apart after cooling to
room temperature. Conversely, wafers recently washed in piranha formed bonds more
quickly and they proved much more robust at room temperature. Compatible with
this observation is observation by Seu, \emph{et al.} that piranha makes glass
more hydrophilic \cite{seu2007a}). It is worth noting also that surface treatment with
sodium silicate and other silicates is reported to permit bonding at temperatures
as low as 150$^{\circ}$~C \cite{puers1997a}. The efficacy of sodium silicate was
not investigated experimentally. At any rate, the importance of surface chemistry
was insufficiently appreciated early in my thesis work and should not be
overlooked.

Bonding of wafer fragments 1~$cm^{2}$ in size can be initiated from a single
needle point contact with the top wafer surface. This was helpful as final wafer
alignment could then be confirmed with a microscope prior to bonding (see
Figure~\vref{fig:bondingPlatformVacuumStage}). Larger areas wafers require the use of
electrodes commensurate with the wafer area. Silicon-7070-silicon structures as
large as 3~inches in diameter were successfully bonded.

Wafers suitable for bonding were obtained from several manufacturers (see
Section~\vref{sec:waferManufacturers}). As a crude gauge of surface roughness and
flatness, wafers that bonded easily had $<6$~Newtons fringes per cm across their
surface when placed on an optically flat glass plate.

\begin{figure}
  \centering
  \includegraphics[width=0.75\textwidth]{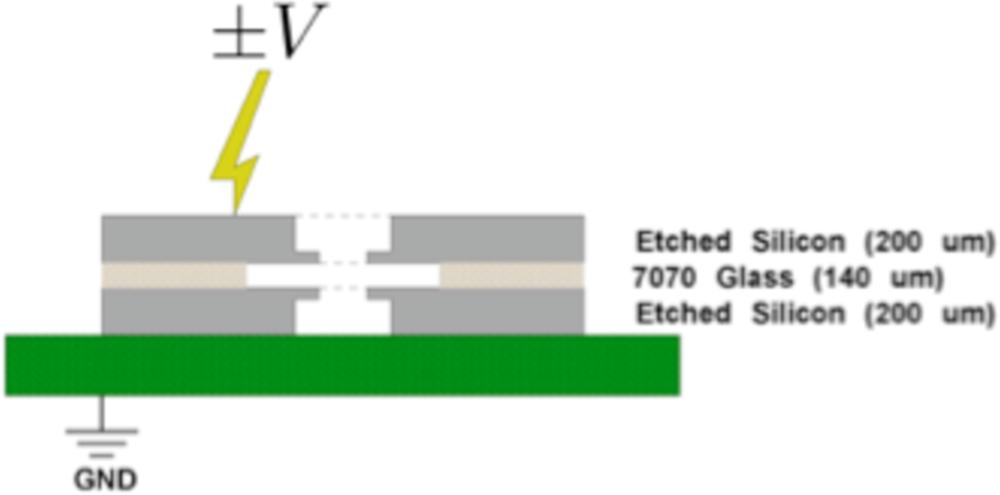}
  \caption[Side-on schematic view of two-layer silicon ion trap on the anodic
  bonding apparatus stage.]
  {Side-on schematic view of two-layer silicon ion trap on the anodic
  bonding apparatus stage. A glass wafer sandwiched between a pair of
  patterned (etched) silicon wafers sits atop a grounded aluminum base
  plate. Bonding to the bottom (top) silicon layer takes place by applying
  potential -V (+V) for several minutes each polarity.}
  \label{fig:twoLayerBonding}
\end{figure}

\begin{figure}
  \centering
  \includegraphics[width=0.60\textwidth]{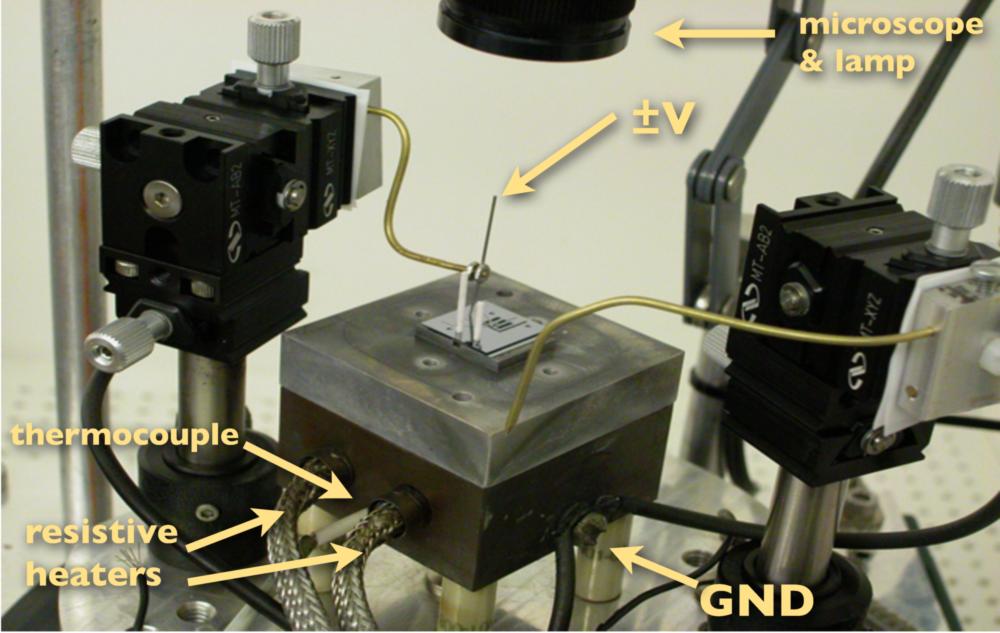}
  \caption[Anodic bonding apparatus for 1-3~$\text{cm}^{2}$ trap chips.]{Anodic
  bonding apparatus for 1-3 $\text{cm}^{2}$ trap chips. The bonding platform is heated with resistive heaters regulated by a PID
  controller attached to a thermocouple. The platform itself is at ground
  and the high voltage potential is applied by a pair of needles on
  translation stages. Two alumina pins protrude from the stage surface.
  These pins pass thru alignment holes in the silicon and glass wafers.
  Wafer alignment to about 25~$\mu m$ is possible using this approach.
  Not visible is a bell jar, vacuum apparatus and needle valve for bonding
  at low pressure or in a $\text{SF}_{6}$ environment. Also not visible,
  is an ammeter to measure the current traveling to the cathode in
 Figure~\vref{fig:anodicBondingElectrostaticPhysics}.
  The bonding takes place in a clean room hood (class 100) on a stage
  (see Figure~\vref{fig:bondingPlatformVacuumStage}).}
  \label{fig:bondingPlatformVacuumStage}
\end{figure}

\subsubsection{electrostatic wafer bonding}

\label{sec:bondingElectrostatic} To understand the electrostatic mechanism for
wafer bonding, consider a capacitor formed by a glass slab thickness~$d$
sandwiched between two metal plates \cite{wallis1969a}). At room temperature an
electric potential $V_{0}$ applied across the plates produces a constant voltage
drop and hence a electric field $E=\left.V_{0}\right/d$. At elevated temperatures
most glasses contain mobile cations in the form of Alkali metal atoms (Na+, K+);
the anions remain fixed in place. As the capacitor temperature is increased
cation mobility increases exponentially and cations following the electric field
lines accumulate in the glass at the edge of the cathode. A current~$i$ flowing
from the potential source to the cathode largely neutralizes the cations. The
glass near the anode is left depleted of cations and it is in this polarized
region where most of the capacitor voltage drop is concentrated. Imagine there
is a tiny air gap $d_{\text{gap}}$ between the anode and the glass and that the
remaining potential drop is $V_{\text{gap}}<V_{0}$. The electrostatic force
across this gap is $F=\frac{1}{2}\epsilon_{0}AE^{2}$ where here
$E=V_{\text{gap}}/d_{\text{gap}}$ and $A$ is the capacitor area. For example, for
$V_{\text{gap}}=10~V$, $d_{\text{gap}}=10$~nm,
$F=4.4\times10^6$~$N\left/m^{2}\right.=640$~psi. This force brings surfaces
into close enough contact that bonding can happen.

\begin{figure}
  \centering
  \includegraphics[width=0.9\textwidth]{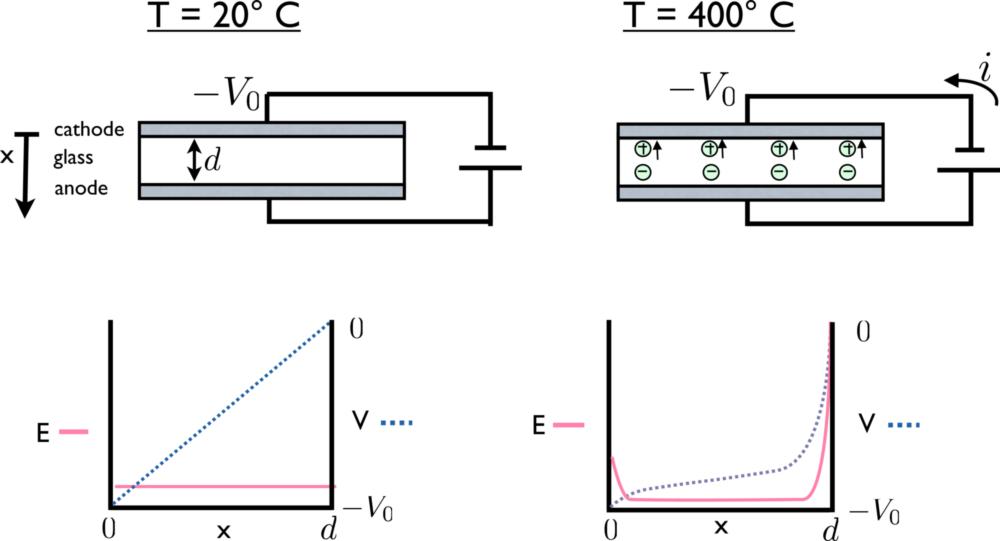}
  \caption[Figure illustrating the electrostatics responsible for anodic bonding.]
  {Figure illustrating the electrostatics responsible for anodic bonding.
  The electric potential (dashed, blue) and field (solid, pink) in the
  glass dielectric of a glass-filled capacitor are plotted at room temperature
  (20$^{\circ}$~C) and bonding temperature (400$^{\circ}$~C). Anodic
  bonding is accomplished when a silicon wafer is slipped between the
  glass and anode. Note that bonding temperatures are well below the
  softening point of both materials. Also, although ordinary silicon
  wafers have a high surface resistance (order $1\times10^5~\Omega-cm$), 
  for the purposes of bonding the wafer can be treated as an equipotential surface.}
  \label{fig:anodicBondingElectrostaticPhysics}
\end{figure}

Stress in bonded thin wafers can cause cracking if the coefficients of thermal
expansion (CTE) are too dissimilar across the full bonding and operating
temperature of a device. The CTE of Pyrex is well matched that to silicon.
Corning 7070 Lithia Potash glass is another borosilicate whose CTE is engineered
to match that of silicon but with a lower ionic content. This is disadvantageous
for wafer bonding--higher temperatures and voltages are required--but is
advantageous for RF Paul traps as its loss tangent is lower than Pyrex.

\begin{table}
  \centering
  \begin{tabular}{l|l}
    \textbf{Material} & \textbf{Loss Tangent (at 1 MHz)}\tabularnewline
    \hline
    fused silica & 0.0002\tabularnewline
    alumina (99$\%$) & 0.0002\tabularnewline
    7070 glass & 0.06\tabularnewline
    7740 glass & 0.5\tabularnewline
  \end{tabular}
  \caption[Loss tangent of Corning 7070 and 7740 borosilicate glasses.]
  {Loss tangent of Corning 7070 and 7740 borosilicate glasses \cite{comeforo1967a,corning1999a}.}
  \label{tab:lossTangentTable}
\end{table}

\begin{figure}
  \centering
  \includegraphics[width=0.80\textwidth]{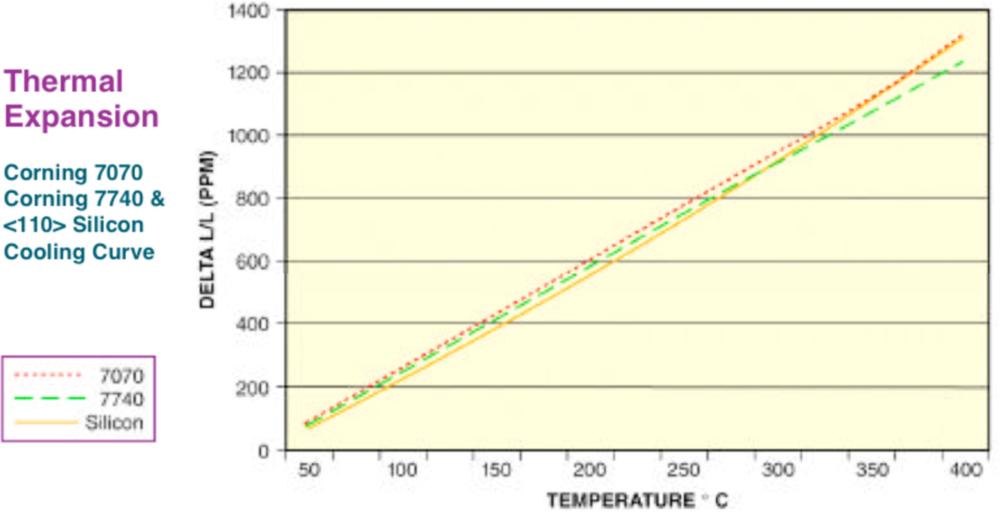}
  \caption[Thermal expansion coefficient of glasses used for anodic bonding.]
  {Thermal expansion of Corning 7070 and 7740 borosilicate glasses as
  a function of temperature \cite{corning1999a}. }
  \label{fig:expansionOf7070glass}
\end{figure}

\subsubsection{direct wafer bonding}

\label{sec:bondingDirectWafer} Room temperature wafer bonding (direct bonding)
due to the formation of hydrogen bonds at the bond interface is possible for
exceptionally flat, smooth and moreover hydrophilic surfaces. Direct bonding is
common in industrial applications but it is technically challenging in laboratory
environments as it relies on wafer geometry alone (vs electrostatic forces) to
bring wafers into close enough contact ($<1$~nm).

Silicon surfaces in air are naturally hydrophilic. This arises as follows. In air
$\sim1$~nm silicon oxide $\left(\text{SiO}_{2}\right)$ grows on silicon
surfaces. Free $\text{SiO}_{2}$ at the surface (Si-O-Si) reacts with water vapor
in air to form extended silanol (SiOH) complexes which are themselves
hydrophilic. The silanol complexes readily adsorb water vapor forming a hydrogen
bonded network of water molecules on the wafer surface
$\text{SiOH}:\left(\text{OH}_{2}\right)_{2}$. When two thusly hydrated
$\text{SiO}_{2}$ surfaces are brought into close contact, hydrogen bonds develop
between the oxygen and hydrogen atoms of the adsorbed water
$\text{SiOH}:\left(\text{OH}_{2}\right)_{2}:\left(\text{OH}_{2}\right)_{2}:\text{OHSi}$.
This mechanism alone will bond surfaces at room temperature. It is energetically
favorable for hydrogen bonds to form directly between silanols SiOH:OHSi, but
this stronger bond only forms above 300$^{\circ}$~C. At 700$^{\circ}$~C silanol
bonds give way to tightly bound (covalent) siloxane SiOSi. Details on this
chemistry and its kinetics are discussed in \cite{stengl1989a}.

\begin{SCfigure}
  \centering
  \includegraphics[width=0.60\textwidth]{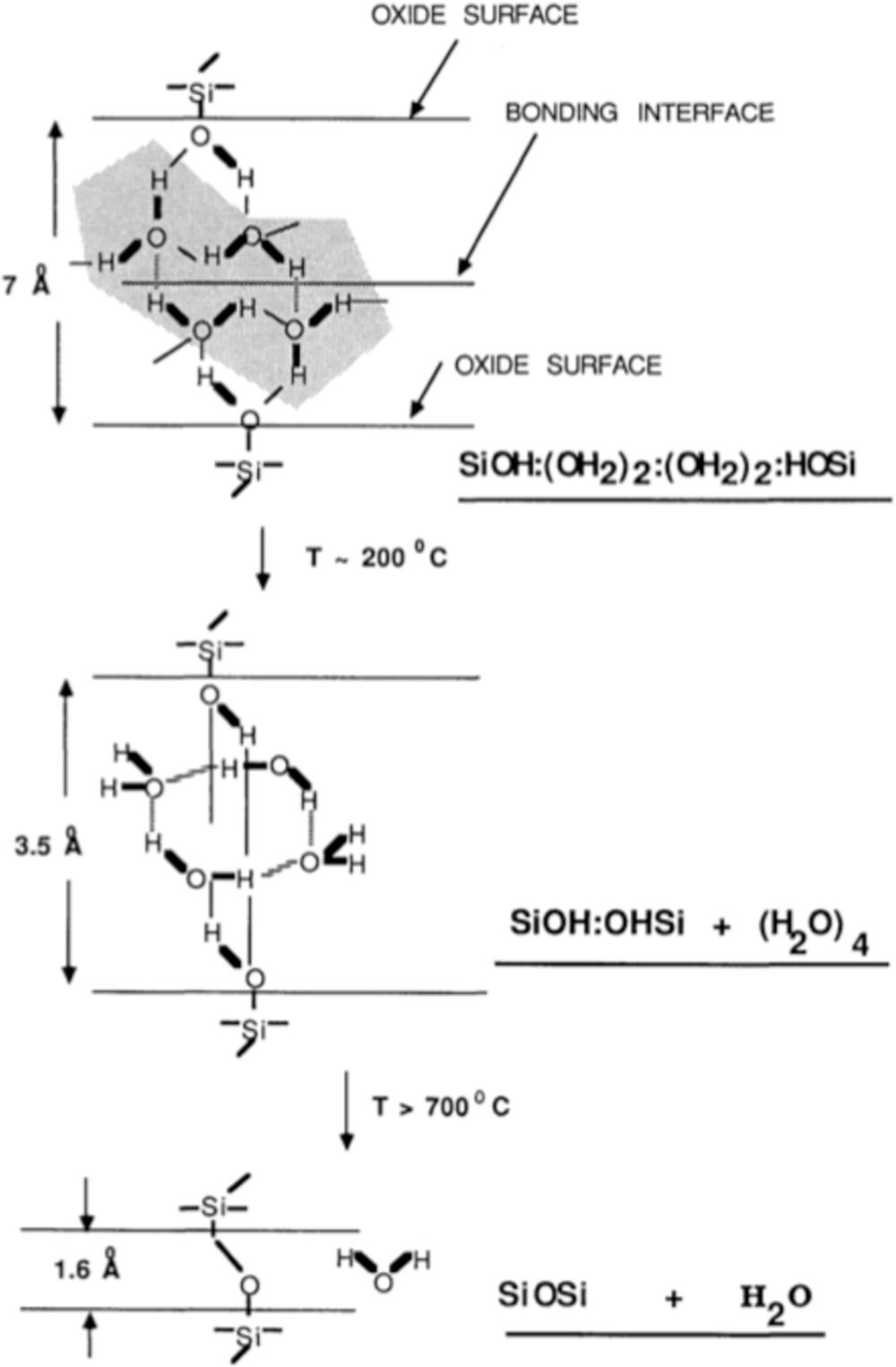}
  \caption[A proposed model for silicon wafer bonding mediated by hydrogen bonds
  between water molecules.]
  {A proposed model for silicon wafer bonding mediated by hydrogen bonds
  between water molecules. Top: Bonding across an interface mediated
  by weak hydrogen bonds between water clusters (thin dashed lines):
  $\text{SiOH}:\left(\text{OH}_{2}\right)_{2}:\left(\text{OH}_{2}\right)_{2}:\text{OHSi}$.
  Middle: Bonding via hydrogen bonds between silanol groups: SiOH:OHSi.
  Bottom: Bonding via siloxane bonds: SiOSi. Figure from \cite{stengl1989a}.}
  \label{fig:hydrogenBondMediatedBonding}
\end{SCfigure}


\subsubsection{SOI wafers}

\label{sec:SOI} 
\begin{SCfigure}
  \centering
  \includegraphics[width=0.70\textwidth]{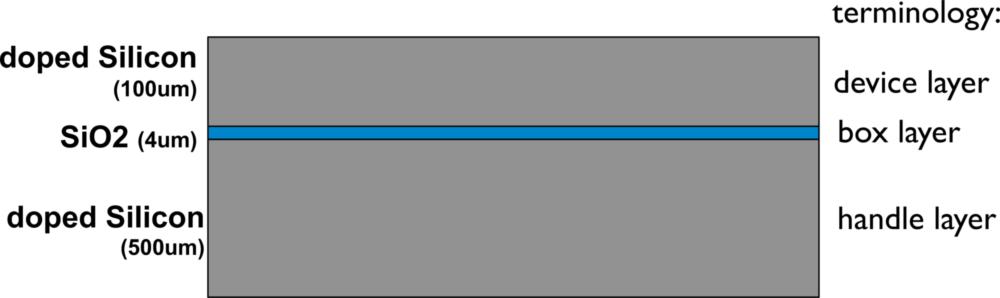}
  \caption[Typical silicon on oxide (SOI) wafer in cross section.]
  {Figure showing a typical silicon on oxide (SOI) wafer in cross section.
  Note that the illustration is not drawn to scale.}
  \label{fig:SOIcrossSection}
\end{SCfigure}

Subsequent to the successful use of in-house anodic
bonding of silicon to glass, commercial silicon on insulator (SOI)
wafers were used to make traps. The cross section of a SOI wafer is
drawn in Figure~\vref{fig:SOIcrossSection}.

Thermal oxide is the typical oxide for SOI wafers. Thermal oxide is grown by
oxidation of a silicon wafer in an oven. The silicon reactant is derived from the
wafer bulk and the oxides maximum thickness is diffusion limited to about 2~$\mu
m$. SOI is a bonded stack of two such silicon wafers, separately oxidized. Hence,
the maximum SOI box thickness is about 4~$\mu m$.

After bonding the oxide is often stripped from the exposed top and
bottom surfaces. Commonly the top of the stack, the so called device
layer, is thinned to between 100~nm and~100 $\mu m$. The remaining
backing layer is called the handle layer.

Boron doped SOI wafers for this thesis work were obtained from the
on-line inventory of \href{http://www.ultrasil.com}{Ultrasil, Inc}. 
Typical cost was $\$$100-200~per wafer in lots of~1. Example specification:

\begin{compactitem}
  \item crystal orientation: $<$100$>$
  \item overall thickness: 360 +/-5 $\mu m$
  \item TTV: $<$ 2 $\mu m$
  \item polish: double side polished (haze free)
  \item device layer: $4\pm0.5~\mu m$, boron doped (P type), 0.005-0.020 ohm-cm
  \item handle layer: $355 \pm5 ~\mu m$, boron doped (P type), $<$ 0.01 ohm-cm
  \item buried oxide layer: $4\pm 0.2~\mu m$
\end{compactitem}

Lower resistivity silicon and thicker oxide is available commercially by special
order. For example, I received a quote from \href{http://www.virginiasemi.com}{Virginia
Semiconductor, Inc.} for 30 wafers, \$375.00 each with the following
specifications.
\begin{compactitem}
  \item polish: double side polished (haze free)
  \item device layer: $10 \pm2~\mu m$, boron doped (P type), 0.0005-0.001
  ohm-cm
  \item oxide layer: $10 \pm 10\%$
  \item handle layer: $400 \pm10~\mu m$, boron doped (P type), 0.0005-0.001
  ohm-cm 
\end{compactitem}

Oxide thicker than $4~\mu m$ can be produced by stress balanced plasma enhanced
tetraethylorthosilicate (PETEOS).  PETEOS oxide is used by the semiconductor
industry as an intermetallic dielectric (IMD).


\subsubsection{wafer manufacturers}

\label{sec:waferManufacturers}
Doped silicon wafers were obtained
from \href{http://www.virginiasemi.com}{Virginia Semiconductor, Inc}. Typical wafer
cost was $\$$170 per wafer in lots of 20. Typical specifications:

\begin{compactitem}
  \item crystal orientation: $<$100$>$
  \item dopant: boron (P type) 0.0005-0.001 ohm-cm
  \item bow: $<$ 20$\mu$m
  \item center thickness: 400 $\mu$m $\pm$ 15 $\mu$m
  \item total thickness variation (5 point measurement): $<$10 $\mu$m
  \item surface: double side polished
  \item micro roughness: $\leq$5A
  \item edge rounded: Yes
\end{compactitem}

\label{sec:usonicMilling}
Corning~7740 and~7070 glass wafers were obtained from 
\href{http://www.sensorprepservices.com}{Sensor Prep 
Services Inc}. Typical cost for blank wafers
was $\$$20-50 in lots of 5-20. For ultrasonically milled wafers the
cost was $\$$200 per wafer in lots of~5. Typical specifications:

\begin{compactitem}
  \item glass: 7070
  \item surface: roughness average (Ra) $<$ 1.5 nm (aka bond quality SI superior
  finish)
  \item thickness: 150~$\mu m$ $\pm$25~$\mu m$
  \item hole diameter: $\pm$ 0.003 in
  \item hole taper: $\pm$ $<$0.0015 in
  \item hole location: $\pm$ 0.003
  \item overall pattern centering: $<$ 0.007 in
  \item missing or bad holes: 1$\%$
  \item chips around holes: $<$ 0.003 in
\end{compactitem}

Both chemical and electrostatic processes contribute to wafer bonding.
They are discussed in the next two sections.

\clearpage

\subsection{RCA wafer clean}
\label{sec:RCA} MEMS processing may contaminate the process wafer
with organic and inorganic substances. Two processes used in this
thesis work require exceptionally clean surfaces: thermal oven wafer
oxidation and anodic bonding. The industry standard cleaning process
was developed in 1960s by Werner Kern at the Radio Corporation of
America known as the RCA Standard Clean Process~\cite{reinhardt2008a}.
It is a four step chemical process. 
\begin{compactitem}[]
	\item (I) organic clean - strip photoresist residue 
 	\item (II) oxide strip - remove oxide and ionic contaminants embedded in the
 	oxide
 	\item (III) ionic clean - remove remaining ionic contaminants
 	\item (IV) solvent clean
\end{compactitem}

I skip step III because there was no indication that ionic contamination
was causing problems. I used a variant of I and II.

\paragraph{Step I}
 An acid preparation called Piranha is used for the organic clean.
The reagents are 1~part $H_{2}O_{2}$ (30-50$\%$ aqueous solution) and 3~parts
$H_{2}\text{SO}_{4}$ by volume. Add the $H_{2}O_{2}$ to the $H_{2}\text{SO}_{4}$
as this order is less likely to splatter. Piranha is highly exothermic so combine
ingredients slowly. Heat to 100$^{\circ}$~C, continuously agitating. Immerse the
wafer in piranha using a teflon basket/clamp for 30~min. The acid remains
chemically active for up to 4~hours after preparation. This step is compatible
with bare silicon, glass, sapphire and SOI wafers. Most metals (including Ti and
Au) and photoresist are rapidly degraded by piranha.

\paragraph{Step II}
A 5-10~minute BOE dip is used for the step II oxide strip. See
Section~\vref{sec:oxideEtch} for details. Be sure to rinse the wafers for at
least 5~minutes in running deionized water.

\paragraph{Step IV} 
 In this step the devices are cleaned in
 \href{http://en.wikipedia.org/wiki/VLSI}{VLSI} grade acetone then isopropanol.
 First, if working with individual chips, put them in a teflon chip carrying
 basket: one chip per carrier slot (see
Section~\vref{sec:teflonBaskets}). For full 3~inch wafers use teflon edge
grabbers. Rinse the devices with running deionized water for 30~seconds. Immerse
in acetone and agitate gently for 30~seconds. If working with individual chips
proceed one chip at a time. Squirt each chip with isopropanol and blow dry with a
dry nitrogen jet. Do not let the solvents dry on the chips in the air.

It was observed that anodic bonding of silicon to 7070 glass was easier
within several hours of cleaning. See Section~\vref{sec:waferBonding}.

\subsubsection{teflon chip holders}

\label{sec:teflonBaskets} Delicate MEMS structures can be be easily damaged
during processing in wet baths. If found that yield was higher when I used
appropriate processing equipment and processed device chips in parallel instead
of one at a time. In particular, the following Teflon structures proved helpful
during wet etching steps. They are manufactured by
\href{http://www.entegris.com}{Entegris, Inc}.

Several 3~inch wafers can be processed in the same beaker using the
following Teflon carriers.

\begin{compactitem}
  \item 3 inch wafer dipper, PFA (Teflon), p/n D14-0215, $\$$45.00/ea in
  lots of one
\end{compactitem}
Several dozen 1~$\text{cm}^{2}$ chips can be processed simultaneously
in the following Teflon basket.

\begin{compactitem}
  \item wafer holder, PFA (Teflon), p/n A14-01S-0215, $\$$71.00/ea in lots
  of one
  \item wafer holder lid, PFA (Teflon), p/n A14-02S-0215, $\$$53.50/ea in
  lots of one
  \item wafer holder handle, PFA (Teflon), p/n A029-0215, $\$$60.50/ea in
  lots of one
\end{compactitem}

\clearpage
\subsection{Electrical interconnect}

\label{sec:electricalInterconnect} Three techniques were used to
attach wires to silicon MEMS devices: resistive welding and wire bonding.
Each uses a different bonding mechanism and type of wire. Additionally,
gap welding was used for electrical connections to other ion trap
structures like Mg and Be ovens.

Resistive welding is appropriate for bonding $0.002-0.020$~inch thick metal
ribbon to flat circuit boards. It forms a bond by flowing a current between two
points on the bonders tip through the target metals. They heats locally and a
bond forms. The bonder I used is a
\href{http://www.unitekequipment.muc.miyachi.com}{Unitek, Inc.} Light Force Weld
Head. In my case the surfaces were gold ribbon and a metalized surface. Two
surfaces commonly bonded to include alumina metalized with $1-3~\mu m$ gold (by
silk screening using \href{http://www.dupont.com}{Dupont, Inc.} 5062/5063~gold
paste) and silicon metalized with $1~\mu m $ gold. Note that bonding is difficult
with materials with good thermal conductivity like thick wire and solid metal
slabs. Between bonds the tip was cleaned with a rough alumina chip (Unitek CPD
10-274-01). For the principles of bonding see~\cite{unitek2001a} and for
practical advice see~\cite{brackell2001a}.
\\ 
\begin{tabular}{l|ll}
  \centering
  \textbf{Surface} & \textbf{Tip (Unitek) } & \textbf{Force (g)}\tabularnewline
  \hline
  Au-alumina & UTM224C & 300-400\tabularnewline
  Au-silicon & UTM222L & 160 \tabularnewline
\end{tabular}
\\
The gold ribbon was supplied by \href{http://www.williams-adv.com}{Williams
Advanced Metals, Inc.}. It is $99.99\%$~Au. The sizes used are
$0.002\times0.010$~inch and $0.002\times0.020$~inch.

Current flows between two electrode points on the bonder tip which are separated
from each other by a thin dielectric layer. It was observed that with properly
tuned bonding parameters a plume of debris was emitted by the tip within several
mm of the bond point. This debris discolored during a vacuum bake
(200$^{\circ}$~C, 24~hours). This debris may be dielectric and could be
problematic if it contaminated surfaces near the ions. This was among the reasons
that wire bonding was used instead of resistive welding for MEMS ion traps after
2004.

Wire bonding is commonly used to connect metal pads on MEMS devices with pads on
a chip carrier. It forms a bond between a wire and a pad by application of
ultrasonic energy via a vibrating tip. The bonding force, power and duration are
key to good bonds as is surface and bonding tip cleanliness. The wirebonder I
used is a \href{http://www.kns.com}{Kulicke and Soffa, Inc.} KS~4523 manual
bonder. I used it with a wedge bonding tip and 0.001~inch diameter gold wire.
Good bonds to $1~\mu m$ thick gold bond pads on silicon wafers required heating
of the wafer to 80$^{\circ}$~C. Bonding parameters are device dependent. Good
starting parameters for the NIST KS~4523 with 0.001~inch gold wire bonding to
gold pads are as follows.

\begin{compactitem}
  \item static force: 18-20 grams
  \item bond parameters: Power, Force and Time set to 2.0-3.0 (arbitrary units).
  Slowly increase if necessary.
  \item use enough force to squash the wire slightly, with minimum power and
  moderate time.
\end{compactitem}

Gold wire was obtained from \href{http://www.calfinewire.com}{California Fine
Wire, Inc}. The wire specification was 99.99$\%$~Au, 0.001~inch diameter, 0.5
to 2.0$\%$ elongation and spooled 90~feet per spool.

The effect of baking (200$^{\circ}$~C in air for several hours) on wire bonds was
investigated. After a bake the bonds were still strong but the gold wires were
noticeably softer and more malleable. Air bridges 2-4~mm in length did not
collapse.

Gap welding is an alternative to resistive/ultrasonic welding when bonding to
$\sim$mm size structures whose surfaces that are not necessarily flat. It forms
bonds by running current thru two metal parts clamped between two metal
electrodes. The strongest bond is formed between dissimilar materials with
respect to alloy and conductivity. This also seems to discourage sticking to the
bonder electrodes. I used a resistive metal alloy called Advance (45$\%$ Ni,
55$\%$~Cu) as an intermediate conductor between some metals which are otherwise
difficult to bond together: Au, Cu, Al, W. Note in particular that gold doesn't
bond well to gold, nor Advance to Advance. Advance is sold under several trade
names including Constantan, Copel, Alloy~294 and Cupron. Advance is available
under then name Constantan from \href{http://www.goodfellow.com}{Goodfellow,
Inc.} (p/n 564-050-99).

\subsubsection{wafer dicing saw}

\label{sec:dicingSaw}A dicing saw was used to cut thick glass and bonded wafer
stacks. Blades appropriate for cutting silicon are immediately destroyed by
contact with glass. Instead, special purpose resin blades impregnated with tiny
diamond particles were used. These blades are
made by \href{http://dicing.com}{Thermocarbon, Inc.} under the trade name
Resinoid. Suggested part numbers are below. The cost per blade is $\$$50 in
lots of 1.

\begin{compactitem}
  \item 2 25M-5B-46RU7-3 good for glass
  \item 2 25M-5B-46R7-3 good for ceramic substrates and ferrites
  \item 2.25M: blade diameter 2.250 inch
  \item 5B: blade width 0.0050 +/- 0.0003
  \item 46: diamond particle size 46 $\mu m$
  \item RU7: resin diamond matrix is soft/medium good for glass
  \item R7: resin diamond matrix is medium/hard good for ceramic substrates
  and ferrites
\end{compactitem}

Some rules of thumb for dicing.

\begin{compactitem}
  \item Blade cut depth should be no more than 10~times the blade thickness.
  \item Dress a new blade by wearing 4-5~mil off its diameter by making
  successive cuts on a glass plate. This gives a good quality edge and prevents
  wafer chipping.
\end{compactitem}

Advice for dicing delicate structures.

\begin{compactitem}
  \item Use a silicon blade for silicon parts and a Resinoid blade for glass
  and bonded glass-silicon stacks.
  \item Use wax to attach the process wafer to a sacrificial silicon backing
  wafer. See Section~\vref{sec:DRIE:thruWaferAdvice}.
  \item Vibration during dicing can damage delicate structures. This can be
  mitigated by embedding the structures in the same wax used to attach
  the backing wafer. Cut some wax into mm size fragments. With the wafer
  stack on a hot plate at 130$^{\circ}$~C, drop the wax bits on the
  process wafer atop the delicate structures. The wax puddles should
  not protrude $>100~\mu m$ above the wafer surface else the dicing saw
  has clearance problems.
  \item Dice the wafer.
  \item Release the chips from the backing wafer in solvent as in Section~\vref{sec:DRIE:thruWaferAdvice}.
  \item Put the chips in a Teflon chip carrier.
  \item Wash the chips with acetone to remove any remaining protective wax.
  Remove one chip at a time from the acetone. Squirt each with isopropanol
  and blow dry with a dry nitrogen jet. Do not let the solvents dry
  on the chips in the air.
\end{compactitem}



\bibliographystyle{these}
\bibliography{joe_b_v2}

\end{document}